\documentclass[12pt,draftcls,onecolumn]{IEEEtran}
\IEEEoverridecommandlockouts

\usepackage[cmex10]{amsmath}
\usepackage{textcomp}
\usepackage{amssymb}

\usepackage{cite}

\usepackage{graphicx}
\usepackage{color}

\usepackage{mathtools}

\newtheorem{theorem}{Theorem}

\begin{document}
\title{Stochastic Analysis of Homogeneous Wireless Networks Assisted by Intelligent Reflecting Surfaces}



\author{Ali H. Abdollahi Bafghi, Mahtab Mirmohseni, Masoumeh Nasiri-Kenari, Behrouz Maham, and Umberto Spagnolini\\

\thanks{
Ali H. Abdollahi Bafghi and Masoumeh Nasiri-Kenari  are with the Department of Electrical Engineering, Sharif University of Technology, Tehran, Iran (email: aliabdolahi@ee.sharif.edu, mnasiri@sharif.edu).   {Mahtab Mirmohseni is with Institute for Communication Systems (ICS), University of Surrey (email: m.mirmohseni@surrey.ac.uk). Behrouz Maham is with the Department of the Electrical and Computer Engineering, Nazarbayev University (email: behrouz.maham@nu.edu.kz). Umberto Spagnolini is with the Dipartimento di Elettronica, Informazione e Bioingegneria (DEIB), Politecnico di Milan (email: umberto.spagnolini@polimi.it).}

}

}

\maketitle









\begin{abstract}
In this paper, we study the impact of the existence of multiple IRSs in a homogeneous
wireless network, in which  all BSs, users (U), and IRSs are spatially
distributed by an independent homogeneous PPP,
with density $\lambda_{{\rm BS}}\rm{[BS/m^2]}$, $\lambda_{{\rm U}}\rm{[U/m^2]}$,
and $\lambda_{{\rm IRS}}\rm{[IRS/m^2]}$, respectively.
We utilize a uniformly random serving strategy for BS and IRS to create
stochastic symmetry in the network. We analyze the performance of the network and study the effect of the existence of the IRS on the network performance.
To this end,  for a typical user
in the system, we derive analytical upper and lower bounds on the
expectation of the power (second statistical moment) of the desired
signal and the interference caused by BSs and other users. 
After that, we obtain analytical upper bounds
on the decay of the probability of the power of the desired signal
and the interference for the typical user (which results in a lower
bound for the cumulative distribution function (CDF)). Moreover, we
derive upper bounds on the decay of the probability of the capacity
of one typical user, which results in a lower bound for the outage
probability.
In the numerical results, we observe that the numerical calculation of the power of the desired signal and the interference is near the derived lower bounds and we show that the increment of the parameter ${(\lambda_{\rm IRS})}$ causes increment in   powers of both the desired and interference signals. We also observe that the increment  of the parameter ${\lambda_{\rm IRS}}$ causes the decrement of outage probability.
\end{abstract}

\section{Introduction}

\label{section 1}

Enriching the propagation through the usage of metasurfaces is an emerging
research topic in the wireless communication systems. This is referred
as the smart radio environment, where metasurfaces feature reconfigurable elements capable of manipulating incident electromagnetic waves. In particular, in conventional radio environments, physical
objects might play a disruptive role in the data transmission by causing
blockage, multipath fading, and more. In contrast, a smart radio environment can
improve the capacity of the wireless network by modifying its electromagnetic
properties \cite{Renzo}. Intelligent reflecting surfaces (IRSs) are
new facilities, which can realize smart radio environments \cite{Liaskos}.
An IRS contains an array of elements whose electromagnetic properties
can be controlled by a control unit. As a result, the IRS elements reflect
the incident wave with an arbitrary phase shift \cite{Gong}.

Besides, the extent of next-generation wireless networks causes randomness
in the spatial position of transmitters, receivers, and other elements. Hence, the
usage of point processes for modeling and analysis of next-generation
wireless networks will  be crucial \cite{azimi1}-\cite{azimi6}.
On the other hand, IRS-assisted networks have been studied from various
perspectives including channel modeling \cite{Najafi}, IRS optimization
\cite{Rui2}, and system analysis \cite{Najafi} (see \cite{Gong}
for a recent survey). Several works about the capacity of the
IRS-assisted networks are documented in \cite{Karasik}-\cite{Mu}.

In some recent works like \cite{Ayoubi} -- \cite{Kishk}, the
usage of point processes in IRS-assisted networks has been studied.
In \cite{Ayoubi}, the authors determined the statistical properties
of the power of the aggregated interference for a network with a cylindrical
array using Poisson point process (PPP). In \cite{Mizmizi}, authors obtained the phase pattern
for the approximation of the shape of the car body with cylindrical IRSs
in a system modeled by point processes for cars, using both incidence
and reflection angles and proposed a design for pre-configured IRS
to change the behavior of electromagnetic flat surfaces on car doors.
Hence, they achieved a reduction of blockage probability and an improvement
in terms of the average signal-to-noise ratio.
In  \cite{Bian} and \cite{Kishk}, authors considered randomly distributed IRSs and blockages to analyze the outage and the system performance.
 However, there has not been a comprehensive study on dense IRS-assisted networks, where all communication nodes are positioned based on PPPs, and the impact of the IRS is not clearly understood in this type of network.

\begin{figure}
\begin{centering}
\includegraphics[scale=0.6]{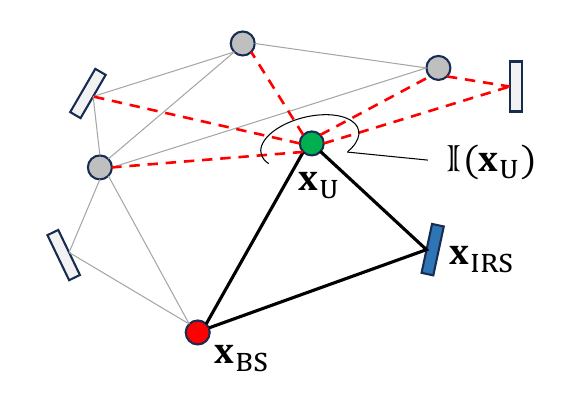}
\par\end{centering}
\caption{IRS assisted network where BS, U, and IRS are located in $\mathbf{x}_{\rm BS},\mathbf{x}_{\rm U},\mathbf{x}_{\rm IRS}$,
and $\mathbb{I}({\mathbf{x}}_{{\rm {U}}})$ is the ensemble interference
to U (dashed lines).}
\label{intro-1}
\end{figure}

Fig. \ref{intro-1} is the illustrative of the IRS-assisted network considered in this
paper where all BSs, users (U), and IRSs are randomly distributed
and interfering each other, the reference BS, U, and IRS are located
in $\mathbf{x}_{\rm BS},\mathbf{x}_{\rm U},\mathbf{x}_{\rm IRS}$, and $\mathbb{I}({\mathbf{x}}_{{\rm {U}}})$
is the ensemble interference to the U in $\mathbf{x}_{\rm U}$. 
Specifically, the goal of the paper is to investigate the impact of the IRSs on a homogeneous
wireless network where all BSs, users (U), and IRSs are spatially
distributed by an independent homogeneous PPP,
with density $\lambda_{{\rm BS}}\textrm{[BS/m]}$, $\lambda_{{\rm U}}\textrm{[U/m]}$,
and $\lambda_{{\rm IRS}}\textrm{[IRS/m]}$, respectively. For each
user U, the BS serving strategy is such that, among all the set of
BSs that can serve the user, the serving BS is randomly selected.
Similarly, for each IRS the user is selected randomly out of all the
users in the coverage of the IRS set. 

The main contributions of this paper can be summarized as follows: 
\begin{itemize}
\item We derive analytical upper and lower bounds on the expectation of the second statistical moment (i.e., power) of the signal from one Base Station (BS) to a user of interest in the downlink. This analysis includes the corresponding interference caused by all other non-serving BSs.

\item
For IRSs  with sufficiently large number of IRS elements ($Q$), we prove that the order of the upper and lower bounds on the power of the desired
signal will be $P_{\max }^{{\mathbb{S}  }} \simeq O\left( {{Q^2}{{\left( {{\lambda _{{\rm{IRS}}}}} \right)}^4}} \right)$
and $P_{\min }^{{\mathbb{S}  }} \simeq O(Q{\lambda _{{\rm{IRS}}}})$, respectively. Moreover, the order of the upper and lower bounds on the interference signal will be $P_{\max }^{{\mathbb{I}  }} \simeq O\left( {{Q^2}{\lambda _{\rm{U}}}{{\left( {{\lambda _{{\rm{IRS}}}}} \right)}^3}} \right)$ and
$P_{\min }^{{\mathbb{I}  }} \simeq O\left( {{Q^2}\frac{{\left( {{\lambda _{{\rm{IRS}}}}} \right)}^2}{{\lambda _{\rm{U}}}{\lambda _{\rm{BS}}}}} \right)$, respectively.
 In other words,   the power of the desired signal increases at least with the order of $\lambda_{\rm IRS}$ and the power of the interference signal 
increases at most with the order of ${(\lambda_{\rm IRS})}^3$, which show the slope of the increment of the power of the desired
signal and the interference.

\item We obtain analytical
upper bounds on the decay of the probability of the power of the desired
signal and the interference for the typical user located at ${\bf x}_{{\rm U}}$,
i.e., we show that the decay of the probability of the power of the
desired signal ($\Pr\left\{ {{{\mathbb{P}}\left({\mathbb{S}\left({{\mathbf{x}}_{{\rm {U}}}}\right)}\right)}>t}\right\} $
and the interference $\Pr\left\{ {{{\mathbb{P}}\left({\mathbb{I}\left({{\mathbf{x}}_{{\rm {U}}}}\right)}\right)}>t}\right\} $)
have the order of $\exp\left\{ {-\sqrt[9]{{\tau t}}}\right\} $ and
$\exp\left\{ {-\sqrt[11]{{\tau t}}}\right\} $, respectively, which will be used to obtain an upper bound on the outage probability. Moreover, the tail of the upper bound on the outage probability is $O\left(\min \left\{ {\frac{{P_{\max }^\mathbb{S}}}{{{N_0}\left( {{e^\alpha } - 1} \right)}},\frac{1}{{\left( {1 - \frac{{L{e^9}\tau }}{{{9^9}}}} \right)\mathbb{H}\left( {\tau {N_0}\left( {{e^\alpha } - 1} \right)} \right)}}} \right\}\right)$.

\end{itemize}
This paper is organized as follows: In Section \ref{section2}, the
system model and the transmission strategy are introduced. In Section
\ref{section3}, we derive the analytical upper and lower bounds on the
power of the desired and interference signals. In Section \ref{section4},
we obtain bounds on the outage probability and the cumulative distribution
function CDF of the power of the desired and interference signals.
 In
Section \ref{section5}, the proposed bounds are evaluated numerically,
and finally the paper is concluded in Section \ref{section6}.

\textbf{Notations:} $\mathbb{R}$ is the set of
real numbers. Bold lower-case and upper-case
letters denote vectors and matrices, respectively, and $x_{i,j}$
denotes the $(i,j)$ entry of matrix ${\bf X}$. Sets are indicated by
calligraphic upper case letters, $\Pr\{{\cal A}\}$ is the probability
of event ${\cal A}$,  $\mu({\cal A\mathit{)}}$ refers to the
measure of set ${\cal A}$, and $|{\cal A\mathsf{|}}$ is the cardinality.
Position of a node is ${\bf x}=(x,y),x,y\in\mathbb{R}$, ${\left\Vert {\bf {x}}\right\Vert _{2}}$
is the Euclidean norm, which is equal to $\sqrt{{x^{2}}+{y^{2}}}$.
In $\mathbb{R}^{2}$ space, the set of points in a circle at the center
of ${\bf x}_{0}$ with radius $r$ is denoted as ${\cal C}({\bf x}_{0},r)=\{(x,y):{\left\Vert {\bf {x}}-{\bf {x}}_{0}\right\Vert _{2}\le r}\}$.
The random set formalism of a homogeneous PPP of intensity $\lambda$
is $\Phi=\{{\bf x}_{1},{\bf x}_{2},...\}\subset\mathbb{R}^{2}$; the
random measure of a PPP of intensity $\lambda$ is $N({\cal A})$,
which is the number of points falling in the measurable set ${\cal A}$,
the probability measure of which obeys the following relation: 
\begin{equation}
\Pr\{N({\cal A})=k\}=\exp\{-\lambda\mu({\cal A})\}\frac{{{\left({\lambda\mu({\cal A})}\right)}^{k}}}{{k!}}
\end{equation}
Moreover, ${ X}\sim{\cal CN}({\bf 0},{\bf C})$ indicates that
${ X}$ is a zero-mean complex Gaussian vector with correlation
matrix ${\bf C}$. Moreover, we define the indicator function as follows:
\begin{equation}
{\pmb{\mathbb{I}}}(x)=\left\{ {\begin{array}{c}
{1,|x|\ne0}\\
{0,|x|=0}
\end{array}}\right..
\end{equation}
In addition, for a random variable $X$, the $L_{p}$ norm of $X$
is${\left({E\left\{ {{\left|X\right|}^{p}}\right\} }\right)^{\frac{1}{p}}}$
. Any function $f(x)$ goes to its limit $\mathop{\lim}\limits _{x\to\infty}\left|{f(x)}\right|=f^{*}$
with the order of $O(g(x))$, if 
\begin{equation}
0<\mathop{\lim}\limits _{x\to\infty}\frac{{\left|{f(x)}\right|}}{{\left|{g(x)}\right|}}<\infty.\label{order}
\end{equation}
\begin{figure}

\begin{centering}
\includegraphics[scale=0.5]{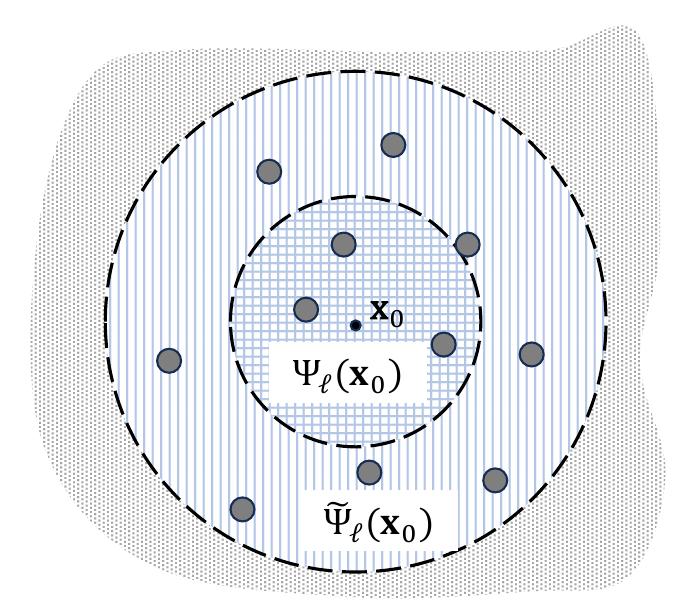}
\par\end{centering}
\caption{PPP processes in ${\cal C}({{\bf {x}}_{{0}}},{R_{{\rm {co}}}})$
and ${\cal C}({{\bf {x}}_{{0}}},{2R_{{\rm {co}}}})$ for $\ell=1$
the BSs, $\ell=2$ the Us, and $\ell=3$ the IRSs.}
\end{figure}

\section{System Model and Preliminaries}

\label{section2}

\subsection{System Model}

We consider a system, in which multi-antenna base stations (BSs),
single-antenna users (Us), and IRSs are spatially distributed with
independent PPPs $\Phi_{{\rm BS}}$, $\Phi_{{\rm U}}$, and $\Phi_{{\rm IRS}}$
with intensities $\lambda_{{\rm BS}}$, $\lambda_{{\rm U}}$, and
$\lambda_{{\rm IRS}}$, respectively, and their random measure formalism
is denoted by $N_{{\rm BS}}({\cal A})$, $N_{{\rm U}}({\cal A})$,
and $N_{{\rm IRS}}({\cal A})$, where ${\cal A}\subset\mathbb{R}^{2}$.
Moreover, each IRS is equipped with $Q$ elements. For simplicity,
we define the following processes (referring for simplicity $\ell=1$
the BSs, $\ell=2$ the Us, $\ell=3$ the IRSs): 
\begin{equation}
\Psi_{{\rm \ell}}({{\bf {x}}_{{0}}})={\Phi_{{\rm \ell}}}\bigcap{{\cal C}({{\bf {x}}_{{0}}},{R_{{\rm {co}}}})},
\end{equation}
\begin{equation}
{\tilde{\Psi}}_{{\rm \ell}}({{\bf {x}}_{{0}}})={\Phi_{{\rm {\ell}}}}\bigcap{{\cal C}({{\bf {x}}_{{0}}},2{R_{{\rm {co}}}})},
\end{equation}
\begin{equation}
n_{{\rm \ell}}({\bf x}_{0})=N_{{\rm \ell}}\left({{\cal C}({{\bf {x}}_{{0}}},{R_{{\rm {co}}}})}\right),
\end{equation}
\begin{equation}
{\tilde{n}}_{{\rm \ell}}({\bf x}_{0})=N_{{\rm \ell}}\left({{\cal C}({{\bf {x}}_{{0}}},{2R_{{\rm {co}}}})}\right),
\end{equation}
where ${R_{{\rm {co}}}}$ is the maximum distance, for which the propagated signal can be detected by receivers.

For simplicity, we consider a free space pathloss, and thus, the received
power is $P_{r}=\frac{P_{t}\lambda_{{\rm wave}}^{2}}{{(4\pi d)^{2}}}$,
where $P_{t}$ and $\lambda_{{\rm wave}}$ are the transmitted power and the wavelength, respectively.
 The link between the BS located at ${\bf x}_{\rm BS}$ and the user located at ${\bf x}_{\rm U}$ can be represented by Rician distribution as follows \cite[Eq. 2.55]{Tse}:

\begin{equation}
\mathbb{H}({{\mathbf{x}}_{{\text{BS}}}},{{\mathbf{x}}_{\text{U}}}) = \sqrt {\frac{\kappa }{{\kappa  + 1}}} \frac{{\exp \left\{ {j\theta ({{\mathbf{x}}_{{\text{BS}}}},{{\mathbf{x}}_{\text{U}}})} \right\}{\lambda _{{\text{wave}}}}\sqrt {\hat h({{\mathbf{x}}_{{\text{BS}}}},{{\mathbf{x}}_{\text{U}}})} }}{{4\pi \sqrt {{{({h_{{\text{BS}}}})}^2} + \left\| {{{\mathbf{x}}_{{\text{BS}}}} - {{\mathbf{x}}_{\text{U}}}} \right\|_2^2} }} + \sqrt {\frac{1}{{\kappa  + 1}}} H({{\mathbf{x}}_{{\text{BS}}}},{{\mathbf{x}}_{\text{U}}}),
\label{rice-m}
\end{equation}
\begin{equation}
H({{  {\bf x}}_{\rm BS}},{{  {\bf x}}_{\rm  U}})\sim {\cal CN}\left( {0,\hat h({{  {\bf x}}_{\rm BS}},{{  {\bf x}}_{\rm  U}}){{\left( {\frac{{{\lambda _{\rm wave}}}}{{4\pi \sqrt {{{({h_{\rm BS}})}^2} + \left\| {{{  {\bf x}}_{\rm BS}} - {{  {\bf x}}_{\rm U}}} \right\|_2^2} }}} \right)}^2}} \right),{{  {\bf x}}_{\rm BS}} \in {\Phi _{\rm BS}},{{  {\bf x}}_{\rm U}} \in {\Phi _{\rm U}},
\label{model3}
\end{equation}
where $h_{{\rm BS}}$ is the height of BSs, which is a constant. 
The first term of (\ref{rice-m}) indicates the specular path arriving with uniform phase and
the second term indicates the summation of the large number of scattered paths, thus, the factor $\kappa$ is the ratio of energy of the specular path and scattered paths.
We note that the height of users are assumed to be zero, and $\hat{h}({{\bf x}_{{\rm BS}}},{{\bf x}_{{\rm U}}})$
models a potential $i.i.d$ blockage in direct link between ${{\bf x}_{{\rm U}}}$
and ${{\bf x}_{{\rm BS}}}$. In particular, we assume: 
\begin{equation}
\left\{ {\begin{array}{c}
{\Pr\left\{ {\hat{h}({{\mathbf{x}}_{{\rm {BS}}}},{{\mathbf{x}}_{{\rm {U}}}})=1}\right\} =1-p_{{\rm{b}}}}\\
{\Pr\left\{ {\hat{h}({{\mathbf{x}}_{{\rm {BS}}}},{{\mathbf{x}}_{{\rm {U}}}})=\hat{h}}\right\} =p_{{\rm{b}}}}
\end{array}}\right.,\quad0<\hat{h}<<1,
\end{equation}
where $p_{\rm b}$ is the probability of the blockage and $\hat h$ is the blockage parameter. 
Moreover, for the $q$-th element of the IRS located at ${\bf x}_{\rm IRS}$, we have:
\begin{equation}
{\mathbb{H}^{[q]}}({{\mathbf{x}}_{{\text{BS}}}},{{\mathbf{x}}_{{\text{IRS}}}}) = \sqrt {\frac{\kappa }{{\kappa  + 1}}} \frac{{\exp \left\{ {j{\theta ^{[q]}}({{\mathbf{x}}_{{\text{BS}}}},{{\mathbf{x}}_{{\text{IRS}}}})} \right\}}{{\lambda _{{\text{wave}}}}}}{{4\pi \sqrt {{{({h_{{\text{BS}}}} - {h_{{\text{IRS}}}})}^2} + \left\| {{{\mathbf{x}}_{{\text{BS}}}} - {{\mathbf{x}}_{{\text{IRS}}}}} \right\|_2^2} }} + \sqrt {\frac{1}{{\kappa  + 1}}} {H^{[q]}}({{\mathbf{x}}_{{\text{BS}}}},{{\mathbf{x}}_{{\text{IRS}}}}),
\end{equation}
\begin{equation}
{H^{[q]}({{{  {\bf x}}_{\rm BS}} , {{  {\bf x}}_{\rm IRS}}})} \sim {\cal CN}\left( {0,{{\left( {\frac{{{\lambda _{\rm wave}}}}{{4\pi \sqrt {{(h_{\rm BS}-h_{\rm IRS})}^2+\left\| {{{  {\bf x}}_{\rm BS}} - {{  {\bf x}}_{\rm IRS}}} \right\|_2^2} }}} \right)}^2}} \right),{{  {\bf x}}_{\rm BS}} \in {\Phi _{\rm BS}},{{  {\bf x}}_{\rm IRS}} \in {\Phi _{\rm IRS}},
\label{model1}
\end{equation}
\begin{equation}
{\mathbb{H}^{[q]}}({{\mathbf{x}}_{{\text{IRS}}}},{{\mathbf{x}}_{\text{U}}}) = \sqrt {\frac{\kappa }{{\kappa  + 1}}} \frac{{\exp \left\{ {j{\theta ^{[q]}}({{\mathbf{x}}_{{\text{IRS}}}},{{\mathbf{x}}_{\text{U}}})} \right\}{\lambda _{{\text{wave}}}}}}{{4\pi \sqrt {{{({h_{{\text{IRS}}}})}^2} + \left\| {{{\mathbf{x}}_{{\text{IRS}}}} - {{\mathbf{x}}_{\text{U}}}} \right\|_2^2} }} + \sqrt {\frac{1}{{\kappa  + 1}}} {H^{[q]}}({{\mathbf{x}}_{{\text{IRS}}}},{{\mathbf{x}}_{\text{U}}}),
\end{equation}
\begin{equation}
H^{[q]}({{  {\bf x}}_{\rm IRS}},{{  {\bf x}}_{\rm U}})\sim {\cal CN}\left( {0,{{\left( {\frac{{{\lambda _{\rm wave}}}}{{4\pi \sqrt {{(h_{\rm IRS})}^2+\left\| {{{  {\bf x}}_{\rm IRS}} - {{  {\bf x}}_{\rm U}}} \right\|_2^2} }}} \right)}^2}} \right),{{  {\bf x}}_{\rm IRS}} \in {\Phi _{\rm IRS}},{{  {\bf x}}_{\rm U}} \in {\Phi _{\rm U}},
\label{model2}
\end{equation}
where $h_{\rm BS}$ and $h_{\rm IRS}$ are the height of BSs and IRSs, respectively. In addition, $\theta ({{\mathbf{x}}_{{\text{BS}}}},{{\mathbf{x}}_{\text{U}}})$, ${\theta ^{[q]}}({{\mathbf{x}}_{{\text{BS}}}},{{\mathbf{x}}_{\text{IRS}}})$, and ${\theta ^{[q]}}({{\mathbf{x}}_{{\text{IRS}}}},{{\mathbf{x}}_{\text{U}}})$  are  independent and uniformly distributed over $[0,2\pi)$.

\subsection{BSs and IRSs selection and serving strategy:}

\textbf{BS-U link:} For the simplicity of analysis, we assume that any
users and IRSs are affected only if propagation distance is within
$R_{{\rm co}}$, i.e., users and IRSs cannot receive a wave if the propagated
distance is larger than $R_{{\rm co}}$. Thus, each user located at ${\bf x}_{{\rm U}}$
can be served by one of BSs located at ${\bf x}_{{\rm BS}}\in\Psi_{{\rm BS}}({{\bf {x}}_{{\rm U}}})$
selected randomly from $\Psi_{{\rm BS}}({{\bf {x}}_{{\rm U}}})$
as follows: 
\begin{equation}
\Pr\left\{ {{{\mathbf{\tilde{x}}}_{{\rm {BS}}}}({{\mathbf{x}}_{{\rm {U}}}})={{\mathbf{x}}_{{\rm {BS}}}}}\right\} =\frac{1}{{{n_{{\rm {BS}}}}\left({{\mathbf{x}}_{{\rm {U}}}}\right)}},
\end{equation}
where ${{\mathbf{\tilde{x}}}_{{\rm {BS}}}}({{\mathbf{x}}_{{\rm {U}}}})$
is the selected BS for serving the specific user located at ${\bf x}_{{\rm U}}$.
Moreover, the link between the user located at ${\bf x}_{{\rm U}}$
and the BS located at ${{\mathbf{\tilde{x}}}_{{\rm {BS}}}}({{\mathbf{x}}_{{\rm {U}}}})$
can operate in the downlink mode\footnote{In a more general model, the nework is asynchronous and the interference of users in uplink mode will induce another amount of interference. However, for simplicity, we assume that the uplink mode is in a different frequency slot and does not interfere in downlink transmission.}.

As we mentioned previously, BSs are multi antenna, and thus, we can design
the antenna array such that the radiation directivity gain of the
BS located at ${\bf x}_{{\rm BS}}$ in a position ${\bf x}$ is \cite{azimi}: 
\begin{equation}
G\left({{{\mathbf{x}}_{{\rm {BS}}}},{{\mathbf{x}}_{{\rm {0}}}},{\mathbf{x}}}\right)={\begin{cases}
{1,\quad{\rm if}\quad1-\varepsilon\le\frac{{\left({{{\mathbf{x}}_{{\rm {0}}}}-{{\mathbf{x}}_{{\rm {BS}}}}}\right).\left({{\mathbf{x}}-{{\mathbf{x}}_{{\rm {BS}}}}}\right)}}{{{{\left\Vert {{{\mathbf{x}}_{{\rm {0}}}}-{{\mathbf{x}}_{{\rm {BS}}}}}\right\Vert }_{2}}{{\left\Vert {{\mathbf{x}}-{{\mathbf{x}}_{{\rm {BS}}}}}\right\Vert }_{2}}}}\le1}\\
{\delta,\quad{\rm if}\quad{\rm Otherwise}}
\end{cases}},
\label{G-func}
\end{equation}
where ${\bf x}_{{\rm 0}}-{\bf x}_{{\rm BS}}$ is the direction of
the maximum BS directivity. Because of the existence of the IRS, each
BS can divide its transmitted power into multiple directions (direct
direction to the user or indirect directions toward IRSs).

\textbf{IRS-U link:} Each IRS located at ${\bf x}_{{\rm IRS}}$ selects
randomly one of the users located at ${\bf x}_{{\rm U}}\in\Psi_{{\rm U}}({\bf x}_{{\rm IRS}})$
(if exists), i.e., if $n_{{\rm U}}\left({\bf x}_{{\rm IRS}}\right)\ne0$.
Then, we have: 
\begin{equation}
\Pr\left\{ {{{\mathbf{\tilde{x}}}_{{\rm {U}}}}({{\mathbf{x}}_{{\rm {IRS}}}})={{\mathbf{x}}_{{\rm {U}}}}}\right\} =\frac{1}{{{n_{{\rm {U}}}}\left({{\mathbf{x}}_{{\rm {IRS}}}}\right)}},
\end{equation}
where ${{\mathbf{\tilde{x}}}_{{\rm {U}}}}({{\mathbf{x}}_{{\rm {IRS}}}})$
is the user served by the IRS located at ${\bf x}_{{\rm IRS}}$. In
particular, for the $q$-th element of the IRS located at ${\bf x}_{\rm IRS}$, we assume: 
\begin{equation}
{{\mathbf{\tilde{x}}}_{{\rm {U}}}}({{\mathbf{x}}_{{\rm {IRS}}}})={{\mathbf{x}}_{{\rm {U}}}}:{\theta_{q}}({{\mathbf{x}}_{{\rm {IRS}}}})=- {\theta ^{[q]}}({{{\mathbf{\tilde x}}}_{{\text{BS}}}}({{\mathbf{x}}_{\text{U}}}),{{\mathbf{x}}_{{\text{IRS}}}}) - {\theta ^{[q]}}({{\mathbf{x}}_{{\text{IRS}}}},{{\mathbf{x}}_{\text{U}}}).
\end{equation}

\begin{figure}
\centering
\includegraphics[width=6cm]{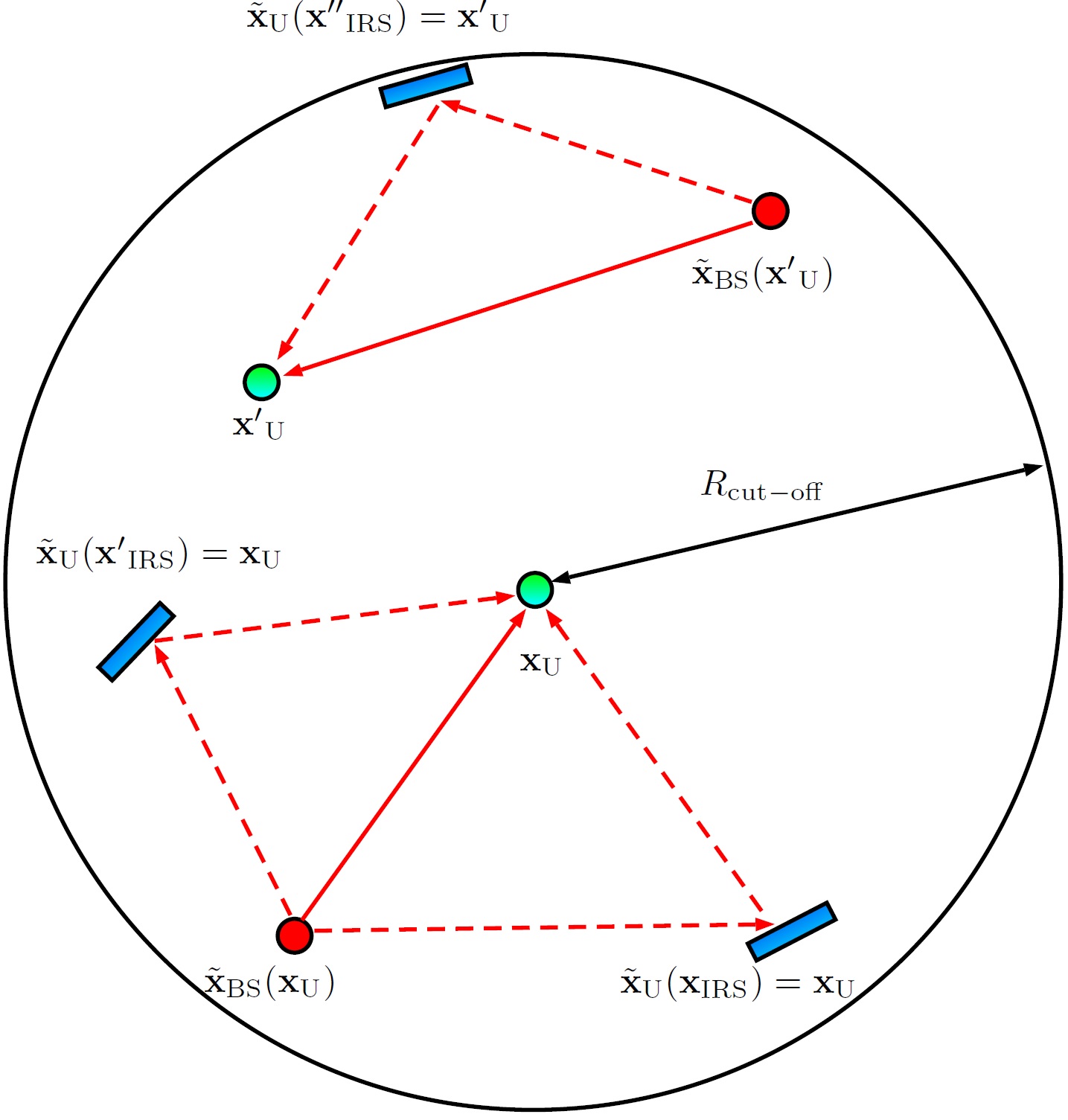}
\caption{Illustration of the system with users, BSs, and IRSs spatially distributed with PPP. In this figure, green circles,  red circles, and blue rectangles  denote users, BSs, and IRSs, respectively.}
\label{illu}
\end{figure}

Thus, the sensed signal  at position  ${\bf x}$ radiated from the BS located at ${\bf x}_{\rm BS}$ is obtained as follows:
\begin{equation*}
X({{\mathbf{x}}_{{\rm{BS}}}},{\mathbf{x}}) = \sum\limits_{{{{\mathbf{x'}}}_{\rm{U}}} \in {\Psi _{\rm{U}}}({{\mathbf{x}}_{{\rm{BS}}}}):{{{\mathbf{\tilde x}}}_{{\rm{BS}}}}({{{\mathbf{x'}}}_{\rm{U}}}) = {{\mathbf{x}}_{{\rm{BS}}}}} {} 
\end{equation*}
\begin{equation*}
 {\big( {G\left( {{{\mathbf{x}}_{{\rm{BS}}}},{{{\mathbf{x'}}}_{\rm{U}}},{\mathbf{x}}} \right) + \sum\limits_{{{\mathbf{x}}_{{\rm{IRS}}}} \in {\Psi _{{\rm{IRS}}}}({{\mathbf{x}}_{{\rm{BS}}}}):{{{\mathbf{x'}}}_{\rm{U}}} = {{{\mathbf{\tilde x}}}_{\rm{U}}}({{\mathbf{x}}_{{\rm{IRS}}}})} {G\left( {{{\mathbf{x}}_{{\rm{BS}}}},{{\mathbf{x}}_{{\rm{IRS}}}},{\mathbf{x}}} \right)} } \big)} \times{X}({{{\mathbf{x'}}}_{\rm{U}}}),
\end{equation*}
where $G(\cdot)$ is given by (\ref{G-func}) and $X({{{\mathbf{x'}}}_{\text{U}}}) \sim {\cal CN}\left( {0,\sigma _{\rm d}^2} \right)$.
The received signal at the $q$-th element of the IRS located at ${ {\bf x}}_{\rm IRS}$, denoted by $Y^{[q]}({ {\bf x}}_{\rm IRS})$, can be obtained as follows:
\begin{equation}
{Y^{[q]}}({{\mathbf{x}}_{{\text{IRS}}}}) = \underbrace {\sum\limits_{{{\mathbf{x}}_{{\text{BS}}}} \in {\Psi _{{\text{BS}}}}({{\mathbf{x}}_{{\text{IRS}}}})} {{{\mathbb{H}}^{[q]}}({{\mathbf{x}}_{{\text{BS}}}},{{\mathbf{x}}_{{\text{IRS}}}})X({{\mathbf{x}}_{{\text{BS}}}},{{\mathbf{x}}_{{\text{IRS}}}})} }_{from\ BSs} ,
\label{model44}
\end{equation}
where ${\theta _q}({ {\bf{x}} _{{\rm{IRS}}}})\in[0,2\pi)$.
An illustration of this system and transmission strategy is illustrated in Fig. \ref{illu}.
The received signal at the position of the user located at ${\bf x}_{\rm U}$ is obtained as follows:
\begin{equation}
\mathbb{Y}({{\mathbf{x}}_{\text{U}}}) = \underbrace {\sum\limits_{{{{\mathbf{x'}}}_{{\text{BS}}}} \in {\Psi _{{\text{BS}}}}({{\mathbf{x}}_{\text{U}}})} {{\mathbb{H}}({{{\mathbf{x'}}}_{{\text{BS}}}},{{\mathbf{x}}_{\text{U}}})X({{{\mathbf{x'}}}_{{\text{BS}}}},{{\mathbf{x}}_{\text{U}}})} }_{from\ BSs}  + \underbrace {\sum\limits_{{{\mathbf{x}}_{{\text{IRS}}}} \in {\Psi _{{\text{IRS}}}}({{\mathbf{x}}_{\text{U}}})} {\sum\limits_{q = 1}^Q {{{\mathbb{H}}^{[q]}}({{\mathbf{x}}_{{\text{IRS}}}},{{\mathbf{x}}_{\text{U}}}){X^{[q]}}({{\mathbf{x}}_{{\text{IRS}}}})} } }_{from\ IRSs} + Z({{\mathbf{x}}_{\text{U}}}),
\label{yy}
\end{equation}
where $Z({{\mathbf{x}}_{\rm{U}}})$ is the complex additive white Gaussian noise with variance $N_0$. And
\begin{equation}
{X^{[q]}}({{\mathbf{x}}_{{\text{IRS}}}})={Y^{[q]}}({{\mathbf{x}}_{{\text{IRS}}}}){{\theta}_{q}}({{\mathbf{x}}_{{\text{IRS}}}}).
\end{equation}

Now, from (\ref{yy}), we define $\mathbb{S}({  {\bf x}}_{\rm U})$ and ${{\mathbb{I}}}({  {\bf x}}_{\rm U})$ as the message and interference parts of the received signal of the user located at ${ {\bf x}}_{\rm U}$, respectively, where:
\begin{equation}
\mathbb{Y}({{\mathbf{x}}_{\rm{U}}}) = \mathbb{S}({{\mathbf{x}}_{\rm{U}}}) + {{\mathbb{I}}}({{\mathbf{x}}_{\rm{U}}}) + Z({{\mathbf{x}}_{\rm{U}}}),
\label{II}
\end{equation}
\begin{equation*}
\mathbb{S}({{\mathbf{x}}_{\rm{U}}}) = {\mathbb{H}}({{{\mathbf{\tilde x}}}_{{\rm{BS}}}}({{\mathbf{x}}_{\rm{U}}}),{{\mathbf{x}}_{\rm{U}}})X({{{\mathbf{\tilde x}}}_{{\rm{BS}}}}({{\mathbf{x}}_{\rm{U}}}),{{\mathbf{x}}_{\rm{U}}}) + \sum\limits_{{{\mathbf{x}}_{{\rm{IRS}}}} \in {\Psi _{{\rm{IRS}}}}({{\mathbf{x}}_{\rm{U}}})} {\sum\limits_{q = 1}^Q {{\mathbb{H}}^{[q]}({{\mathbf{x}}_{{\rm{IRS}}}},{{\mathbf{x}}_{\rm{U}}})\exp \{ j{\theta _q}({{\mathbf{x}}_{{\rm{IRS}}}})\} \times} } 
\end{equation*}
\begin{equation}
{{\mathbb{H}}^{[q]}}({{{\mathbf{\tilde x}}}_{{\rm{BS}}}}({{\mathbf{x}}_{\rm{U}}}),{{\mathbf{x}}_{{\rm{IRS}}}})X({{{\mathbf{\tilde x}}}_{{\rm{BS}}}}({{\mathbf{x}}_{\rm{U}}}),{{\mathbf{x}}_{{\rm{IRS}}}}).
\label{SS}
\end{equation}
We note that we are interested in the analysis of the statistical properties of $\mathbb{S}({{\mathbf{x}}_{\rm{U}}}) $ and ${{\mathbb{I}}}({{\mathbf{x}}_{\rm{U}}}) $ for the reference user ${\bf x}_{\rm U}$. Thus, we have assumed implicitly that ${\bf x}_{\rm U}$ is occured in point process $\Phi_{\rm U}$, i.e.,
\begin{equation}
{\bf x}_{\rm U}\in \Phi_{\rm U},\quad n_{\rm U}({\bf x}_{\rm U})>0,\quad {\tilde n}_{\rm U}({\bf x}_{\rm U})>0 .
\label{implicit1}
\end{equation}

\section{Upper and Lower Bounds on $E\left\{ {{{\left| {{\mathbb{S} }({{\mathbf{x}}_{\rm{U}}})} \right|}^2}} \right\}$ and $E\left\{ {{{\left| {\mathbb{I} ({{\mathbf{x}}_{\rm{U}}})} \right|}^2}} \right\} $}

In this section, we analyze bounds on  $E\left\{ {{{\left| {{\mathbb{S} }({{\mathbf{x}}_{\rm{U}}})} \right|}^2}} \right\}$ and  $E\left\{ {{{\left| {{\mathbb{I} }({{\mathbf{x}}_{\rm{U}}})} \right|}^2}} \right\}$, to study the relation between these variables and $\lambda_{\rm IRS}$. Moreover, these results will be used to analyze the outage probability in next sections.

\label{section3}

\subsection{Bounds on $E\left\{ {{{\left| {{\mathbb{S} }({{\mathbf{x}}_{\rm{U}}})} \right|}^2}} \right\}$}
In this subsection, we introduce lower and upper bounds for $E\left\{ {{{\left| {{\mathbb{S} }({{\mathbf{x}}_{\rm{U}}})} \right|}^2}} \right\}$.
 The upper bound is introduced in the following theorem:
\begin{theorem}
$E\left\{ {{{\left| {\mathbb{S} ({{\mathbf{x}}_{\rm{U}}})} \right|}^2}} \right\}$ is upper bounded as follows:
\begin{equation*}
E\left\{ {{{\left| {{\mathbb{S} }({{\mathbf{x}}_{\rm{U}}})} \right|}^2}} \right\} \le P^{\mathbb{S} }_{\max}=P^{\mathbb{S} }_{\max}({\rm BS})+ P^{\mathbb{S} }_{\max}({\rm IRS}),
\end{equation*}
where $P^{\mathbb{S} }_{\max}({\rm BS})$ is caused from the BS and $P^{\mathbb{S} }_{\max}({\rm IRS})$ is caused from the IRSs, which are given as follows:
\begin{equation*}
P^{\mathbb{S} }_{\max}({\rm BS})=
\end{equation*}
{{\begin{equation*}
\frac{{\pi  + \left( {3{\lambda _{{\rm{IRS}}}}\pi {{\left( {{R_{{\rm{co}}}}} \right)}^2} + {{\left( {{\lambda _{{\rm{IRS}}}}\pi {{\left( {{R_{{\rm{co}}}}} \right)}^2}} \right)}^2}} \right) \times ((1 - \delta )\arccos \left( {1 - \varepsilon } \right) + \delta \pi )}}{\pi } \times 
\end{equation*}}}
\begin{equation*}
\frac{{\left( {1 + \left( {\hat h - 1} \right){p_{{\rm{b}}}}} \right){{\left( {{\lambda _{{\rm{wave}}}}} \right)}^2}\sigma _{{\rm{d}}}^2}}{{ {{\left( {{R_{{\rm{co}}}}} \right)}^2}{{\left( {4\pi } \right)}^2}}}\times 
\end{equation*}
\begin{equation}
\ln \left( {1 + {{\left( {\frac{{{R_{{\rm{co}}}}}}{{{h_{{\rm{BS}}}}}}} \right)}^2}} \right)\times  \left( {1 - \exp \left\{- {{\lambda _{{\rm{BS}}}}\pi {{\left( {{R_{{\rm{co}}}}} \right)}^2}} \right\}} \right),
\end{equation}
\begin{equation*}
P^{\mathbb{S} }_{\max}({\rm IRS})=
\end{equation*}
\begin{equation*}
 \scalebox{1}[1]{$\left( {\frac{{3{\lambda _{{\rm{IRS}}}}\pi {{\left( {2{R_{{\rm{co}}}}} \right)}^2} + {{\left( {{\lambda _{{\rm{IRS}}}}\pi {{\left( {2{R_{{\rm{co}}}}} \right)}^2}} \right)}^2}}}{{{{\left( {4\pi } \right)}^4}}} + }  {\frac{{{{\left( {{\lambda _{{\rm{IRS}}}}\pi {{\left( {2{R_{{\rm{co}}}}} \right)}^2}} \right)}^3}\exp \left\{ {\frac{9}{{2{\lambda _{{\rm{IRS}}}}\pi {{\left( {2{R_{{\rm{co}}}}} \right)}^2}}}} \right\}}}{{{{\left( {4\pi } \right)}^4}}}} \right) \times $}
\end{equation*}
{\begin{equation*}
\frac{{3Q{{\left( {{\lambda _{{\rm{wave}}}}} \right)}^4}\left( {1 - \exp \left\{ { - {\lambda _{{\rm{BS}}}}\pi {{\left( {{R_{{\rm{co}}}}} \right)}^2}} \right\}} \right)\sigma _{{\rm{d}}}^2}}{{\sqrt {4\left( {{{({h_{{\rm{BS}}}} - {h_{{\rm{IRS}}}})}^2} + 9{{\left( {{R_{{\rm{co}}}}} \right)}^2}} \right)\left( {{{({h_{{\rm{IRS}}}})}^2} + 4{{\left( {{R_{{\rm{co}}}}} \right)}^2}} \right)} }} \times \frac{1}{{\left| {{h_{{\rm{BS}}}} - {h_{{\rm{IRS}}}}} \right|{h_{{\rm{IRS}}}}}} +
\end{equation*}}
\begin{equation*}
 \scalebox{.95}[1]{$\left( {\frac{{{{\left( {{\lambda _{{\rm{IRS}}}}\pi {{\left( {2{R_{{\rm{co}}}}} \right)}^2}} \right)}^4}\exp \left\{ {\frac{{16}}{{2\left( {{\lambda _{{\rm{IRS}}}}\pi {{\left( {2{R_{{\rm{co}}}}} \right)}^2}} \right)}}} \right\}}}{{{4^8}{\pi ^2}{{\left( {{R_{{\rm{co}}}}} \right)}^4}}}} \right. + \frac{{{{\left( {{\lambda _{{\rm{IRS}}}}\pi {{\left( {2{R_{{\rm{co}}}}} \right)}^2}} \right)}^3}\exp \left\{ {\frac{9}{{2\left( {{\lambda _{{\rm{IRS}}}}\pi {{\left( {2{R_{{\rm{co}}}}} \right)}^2}} \right)}}} \right\}}}{{{4^8}{\pi ^2}{{\left( {{R_{{\rm{co}}}}} \right)}^4}}}$}
\end{equation*}
\begin{equation*}
  \left. { - \frac{{{{\left( {{\lambda _{{\rm{IRS}}}}\pi {{\left( {2{R_{{\rm{co}}}}} \right)}^2}} \right)}^2} + 2{\lambda _{{\rm{IRS}}}}\pi {{\left( {2{R_{{\rm{co}}}}} \right)}^2}}}{{{4^8}{\pi ^2}{{\left( {{R_{{\rm{co}}}}} \right)}^4}}}} \right) \times  
\end{equation*}
\begin{equation*}
\left( {1 - \exp \left\{ { - {\lambda _{{\rm{BS}}}}\pi {{\left( {{R_{{\rm{co}}}}} \right)}^2}} \right\}} \right)Q(Q - 1){\left( {{\lambda _{{\rm{wave}}}}} \right)^4}\times 
\end{equation*}
{\begin{equation*}
\times {\frac{\kappa }{{\kappa  + 1}}} \times \ln \left( {1 + {{\left( {\frac{{3{R_{{\rm{co}}}}}}{{{h_{{\rm{BS}}}} - {h_{{\rm{IRS}}}}}}} \right)}^2}} \right) \times \ln \left( {1 + {{\left( {\frac{{2{R_{{\rm{co}}}}}}{{{h_{{\rm{IRS}}}}}}} \right)}^2}} \right)\sigma _{{\rm{d}}}^2+
\end{equation*}}
\begin{equation*}
 \frac{\left( {1 - \exp \left\{- {{\lambda _{{\rm{BS}}}}\pi{{\left( {{R_{{\rm{co}}}}} \right)}^2}} \right\}} \right){Q{{\left( {{\lambda _{{\rm{wave}}}}} \right)}^4}\sigma _{{\rm{d}}}^2}}{{{4^5}{\pi ^4}{{\left( {{R_{{\rm{co}}}}} \right)}^2}}}\times 
\end{equation*}
\begin{equation*}
 \sqrt {\frac{{{{\left( {2{R_{{\rm{co}}}}} \right)}^2}}}{{\left( {{{({h_{{\rm{IRS}}}})}^2} + {{\left( {2{R_{{\rm{co}}}}} \right)}^2}} \right){{({h_{{\rm{IRS}}}})}^2}}}}  \times \sqrt {{\frac{{{{\left( {3{R_{{\rm{co}}}}} \right)}^2}}}{{\left( {{{({h_{{\rm{BS}}}} - {h_{{\rm{IRS}}}})}^2} + {{\left( {3{R_{{\rm{co}}}}} \right)}^2}} \right){{({h_{{\rm{BS}}}} - {h_{{\rm{IRS}}}})}^2}}}} } \times
\end{equation*}
\begin{equation}
 \scalebox{.95}[1]{$\left( {3{\lambda _{{\rm{IRS}}}}{{\left( {2{R_{{\rm{co}}}}} \right)}^2} + 2{{\left( {{\lambda _{{\rm{IRS}}}}{{\left( {2{R_{{\rm{co}}}}} \right)}^2}} \right)}^2} + } \right.\left.{{{\left( {{\lambda _{{\rm{IRS}}}}{{\left( {2{R_{{\rm{co}}}}} \right)}^2}} \right)}^3}\exp \left\{ {\frac{9}{{2{\lambda _{{\rm{IRS}}}}{{\left( {2{R_{{\rm{co}}}}} \right)}^2}}}} \right\}}\right).$}
\end{equation}
\label{theorem1}

\end{theorem}

\begin{IEEEproof}
The proof is provided in Appendix A.
\end{IEEEproof}

Next, we introduce a lower bound for $E\left\{ {{{\left| {\mathbb{S} ({{\mathbf{x}}_{\rm{U}}})} \right|}^2}} \right\}$ in the following theorem:

\begin{theorem}
$E\left\{ {{{\left| {\mathbb{S} ({{\mathbf{x}}_{\rm{U}}})} \right|}^2}} \right\}$ is lower bounded as follows:
\begin{equation*}
E\left\{ {{{\left| {\mathbb{S} ({{\mathbf{x}}_{\rm{U}}})} \right|}^2}} \right\} \ge P^{\mathbb{S} }_{\min}=P^{\mathbb{S} }_{\min}({\rm BS})+ P^{\mathbb{S} }_{\min}({\rm IRS}),
\end{equation*}
where $P^{\mathbb{S} }_{\min}({\rm BS})$ is caused from the BS and $P^{\mathbb{S} }_{\min}({\rm IRS})$ is caused from the IRSs, which are given as follows:
\begin{equation*}
P^{\mathbb{S} }_{\min}({\rm BS})=
\end{equation*}
\begin{equation*}
\left( {1 - \exp \left\{ { - {\lambda _{{\rm{BS}}}}\pi {{\left( {{R_{{\rm{co}}}}} \right)}^2}} \right\}} \right){{\left( {{\lambda _{{\rm{wave}}}}} \right)}^2}\times
\end{equation*}
\begin{equation}
\frac{{\left( {1 + \left( {\hat h - 1} \right){p_{{\rm{b}}}}} \right)\sigma _{{\rm{d}}}^2}}{{{{\left( {4\pi } \right)}^2}{{\left( {{R_{{\rm{co}}}}} \right)}^2}}}\ln \left( {1 + {{\left( {\frac{{{R_{{\rm{co}}}}}}{{{h_{{\rm{BS}}}}}}} \right)}^2}} \right),
\end{equation}

\begin{equation*}
P^{\mathbb{S} }_{\min}({\rm IRS})=
\end{equation*}
{\begin{equation*}
\frac{{{\lambda _{{\rm{IRS}}}}\left( {1 - \exp \left\{ -{{\lambda _{{\rm{BS}}}}\pi {{\left( {{R_{{\rm{co}}}}} \right)}^2}} \right\}} \right){\delta ^2}Q{{\left( {{\lambda _{{\rm{wave}}}}} \right)}^4}\sigma _{{\rm{d}}}^2}}{{{{\left( {4\pi } \right)}^4}}} \times 
\end{equation*}}
\begin{equation*}
 \scalebox{.95}[1]{$\left( {\frac{{1 - {\lambda _{\rm{U}}}\pi {{\left( {{R_{{\rm{co}}}}} \right)}^2}\exp \left\{ { - {\lambda _{\rm{U}}}\pi {{\left( {{R_{{\rm{co}}}}} \right)}^2}} \right\}}}{{{\lambda _{\rm{U}}}{{\left( {{R_{{\rm{co}}}}} \right)}^2}\left( {1 - \exp \left\{ { - {\lambda _{\rm{U}}}\pi {{\left( {{R_{{\rm{co}}}}} \right)}^2}} \right\}} \right)}}} \right.\left. { - \frac{{\exp \left\{ { - {\lambda _{\rm{U}}}\pi {{\left( {{R_{{\rm{co}}}}} \right)}^2}} \right\}}}{{{\lambda _{\rm{U}}}{{\left( {{R_{{\rm{co}}}}} \right)}^2}\left( {1 - \exp \left\{ { - {\lambda _{\rm{U}}}\pi {{\left( {{R_{{\rm{co}}}}} \right)}^2}} \right\}} \right)}}} \right) \times $}
\end{equation*}
\begin{equation}
 \scalebox{.95}[1]{$\frac{{{{\left( {{R_{{\rm{co}}}}} \right)}^2}}}{{\left( {\max \left\{ {{{({h_{{\rm{BS}}}} - {h_{{\rm{IRS}}}})}^2},{{({h_{{\rm{IRS}}}})}^2}} \right\} + 3{{\left( {{R_{{\rm{co}}}}} \right)}^2}} \right)}}\times\frac{1}{{\left( {\max \left\{ {{{({h_{{\rm{BS}}}} - {h_{{\rm{IRS}}}})}^2},{{({h_{{\rm{IRS}}}})}^2}} \right\} + 4{{\left( {{R_{{\rm{co}}}}} \right)}^2}} \right)}}.$}
\end{equation}
\label{theorem2}
\end{theorem}

\begin{IEEEproof}
The proof is provided in Appendix B.
\end{IEEEproof}

\textit{Corollary 1}: Theorems \ref{theorem1} and \ref{theorem2} indicate that for a suffiently large $Q$, we obtain:
\begin{equation}
P_{\max }^{{\mathbb{S}  }} \simeq O\left( {{Q^2}{{\left( {{\lambda _{{\rm{IRS}}}}} \right)}^4}} \right),
\end{equation}
\begin{equation}
P_{\min }^{{\mathbb{S}  }} \simeq O(Q{\lambda _{{\rm{IRS}}}}).
\end{equation}

\subsection{Bounds on $E\left\{ {{{\left| {\mathbb{I} ({{\mathbf{x}}_{\rm{U}}})} \right|}^2}} \right\}$}

By (\ref{yy}), the interference denoted by ${{{\mathbb{I} }}}({\bf x}_{\rm U})$ can be obtained as follows:
\begin{equation*}
{{\mathbb{I} }}({{\mathbf{x}}_{\rm{U}}}) =
\end{equation*}
{\begin{equation*}
 \sum\limits_{{{{\mathbf{x'}}}_{{\rm{BS}}}} \in {\Psi _{{\rm{BS}}}}({{\mathbf{x}}_{\rm{U}}})} {{\mathbb{H}}({{{\mathbf{x'}}}_{{\rm{BS}}}},{{\mathbf{x}}_{\rm{U}}})\sum\limits_{{{{\mathbf{x'}}}_{\rm{U}}} \in {\Psi _{\rm{U}}}({{{\mathbf{x'}}}_{{\rm{BS}}}}):{{{\mathbf{x'}}}_{\rm{U}}} \ne {{\mathbf{x}}_{\rm{U}}},{{{\mathbf{\tilde x}}}_{{\rm{BS}}}}({{{\mathbf{x'}}}_{\rm{U}}}) = {{{\mathbf{x'}}}_{{\rm{BS}}}}}  }
\end{equation*}}
{\begin{equation*}
{\big( {G\left( {{{{\mathbf{x'}}}_{{\rm{BS}}}},{{{\mathbf{x'}}}_{\rm{U}}},{{\mathbf{x}}_{\rm{U}}}} \right) + \sum\limits_{{{\mathbf{x}}_{{\rm{IRS}}}} \in {\Psi _{{\rm{IRS}}}}({{{\mathbf{x'}}}_{{\rm{BS}}}}):{{{\mathbf{x'}}}_{\rm{U}}} = {{{\mathbf{\tilde x}}}_{\rm{U}}}({{\mathbf{x}}_{{\rm{IRS}}}})} {G\left( {{{{\mathbf{x'}}}_{{\rm{BS}}}},{{\mathbf{x}}_{{\rm{IRS}}}},{{\mathbf{x}}_{\rm{U}}}} \right)} } \big)}\times {X}({{{\mathbf{x'}}}_{\rm{U}}})
\end{equation*}}
\begin{equation*}
 + \sum\limits_{{{\mathbf{x}}_{{\rm{IRS}}}} \in {\Psi _{{\rm{IRS}}}}({{\mathbf{x}}_{\rm{U}}})} {\sum\limits_{q = 1}^Q {{{\mathbb{H}}^{[q]}}({{\mathbf{x}}_{{\rm{IRS}}}},{{\mathbf{x}}_{\rm{U}}})\exp \{ j{\theta _q}({{\mathbf{x}}_{{\rm{IRS}}}})\}  \times } } 
\end{equation*}
{\begin{equation*}
 \left( {\sum\limits_{{{{\mathbf{x'}}}_{{\rm{BS}}}} \in {\Psi _{{\rm{BS}}}}({{\mathbf{x}}_{{\rm{IRS}}}})} {\sum\limits_{{{{\mathbf{x'}}}_{\rm{U}}} \in {\Psi _{\rm{U}}}({{{\mathbf{x'}}}_{{\rm{BS}}}}):{{{\mathbf{x'}}}_{\rm{U}}} \ne {{\mathbf{x}}_{\rm{U}}},{{{\mathbf{\tilde x}}}_{{\rm{BS}}}}({{{\mathbf{x'}}}_{\rm{U}}}) = {{{\mathbf{x'}}}_{{\rm{BS}}}}} {{{\mathbb{H}}^{[q]}}({{{\mathbf{x'}}}_{{\rm{BS}}}},{{\mathbf{x}}_{{\rm{IRS}}}})\times} } } \right.
\end{equation*}}
{\begin{equation}
\left.\left( {G\left( {{{{\mathbf{x'}}}_{{\rm{BS}}}},{{{\mathbf{x'}}}_{\rm{U}}},{{\mathbf{x}}_{{\rm{IRS}}}}} \right) + \sum\limits_{{{{\mathbf{x'}}}_{{\rm{IRS}}}} \in {\Psi _{{\rm{IRS}}}}({{{\mathbf{x'}}}_{{\rm{BS}}}}):{{{\mathbf{x'}}}_{\rm{U}}} = {{{\mathbf{\tilde x}}}_{\rm{U}}}({{{\mathbf{x'}}}_{{\rm{IRS}}}})} {G\left( {{{{\mathbf{x'}}}_{{\rm{BS}}}},{{{\mathbf{x'}}}_{{\rm{IRS}}}},{{\mathbf{x}}_{{\rm{IRS}}}}} \right)} } \right)\times{X}({{{\mathbf{x'}}}_{\rm{U}}})\right)
\label{iiiiii}
\end{equation}}

The upper bound is introduced in the following theorem:

\begin{theorem}
$E\left\{ {{{\left| {{{\mathbb{I} }}({{\mathbf{x}}_{\rm{U}}})} \right|}^2}} \right\}$ is upper bounded as follows:
\begin{equation*}
E\left\{ {{{\left| {{{\mathbb{I}} }({{\mathbf{x}}_{\rm{U}}})} \right|}^2}} \right\} \le P^{\mathbb{I} }_{\max}=P^{\mathbb{I} }_{\max}({\rm BS})+ P^{\mathbb{I} }_{\max}({\rm IRS}),
\end{equation*}
where $P^{\mathbb{I} }_{\max}({\rm BS})$ is caused from the BSs  and $P^{\mathbb{I} }_{\max}({\rm IRS})$ is caused from the IRSs, which are given as follows:
\begin{equation*}
P^{\mathbb{I} }_{\max}({\rm BS})=
\end{equation*}
\begin{equation*}
 \scalebox{.95}[1]{${\lambda _{{\rm{BS}}}}\left( {1 + 3{\lambda _{{\rm{IRS}}}}\pi {{\left( {{R_{{\rm{co}}}}} \right)}^2} + {{\left( {{\lambda _{{\rm{IRS}}}}\pi {{\left( {{R_{{\rm{co}}}}} \right)}^2}} \right)}^2}} \right)\times\frac{{\left( {1 + \left( {\hat h - 1} \right){p_{{\rm{b}}}}} \right){{\left( {{\lambda _{{\rm{wave}}}}} \right)}^2}\sigma _{{\rm{d}}}^2}}{{\left( {1 - \exp \left\{ { - {\lambda _{\rm{U}}}\pi {{\left( {{R_{{\rm{co}}}}} \right)}^2}} \right\}} \right)16\pi }} \times$}
\end{equation*}
\begin{equation}
\left( {{\lambda _{\rm{U}}}\pi {{\left( {{R_{{\rm{co}}}}} \right)}^2} + \exp \left\{ { - {\lambda _{\rm{U}}}\pi {{\left( {{R_{{\rm{co}}}}} \right)}^2}} \right\} - 1} \right) \times  \ln \left( {1 + {{\left( {\frac{{{R_{{\rm{co}}}}}}{{{h_{{\rm{BS}}}}}}} \right)}^2}} \right),
\end{equation}

\begin{equation*}
P^{\mathbb{I} }_{\max}({\rm IRS})=
\end{equation*}
{\begin{equation*}
 \scalebox{.9}[1]{${{\left( {{\lambda _{{\rm{wave}}}}} \right)}^4}\left( {{\lambda _{\rm{U}}}\pi {{\left( {{R_{{\rm{co}}}}} \right)}^2} + \exp \left\{ { - {\lambda _{\rm{U}}}\pi {{\left( {{R_{{\rm{co}}}}} \right)}^2}} \right\} - 1} \right)\times\frac{{{Q^2}\left( {1 - \exp \left\{ { - {\lambda _{{\rm{BS}}}}\pi {{\left( {{R_{{\rm{co}}}}} \right)}^2}} \right\}} \right)\sigma _{{\rm{d}}}^2}}{{\left( {1 - \exp \left\{ { - {\lambda _{\rm{U}}}\pi {{\left( {{R_{{\rm{co}}}}} \right)}^2}} \right\}} \right){{\left( {4\pi } \right)}^4}}} \times $}
\end{equation*}}
\begin{equation*}
\left( {\frac{{{{\left( {{\lambda _{{\rm{IRS}}}}\pi {{\left( {2{R_{{\rm{co}}}}} \right)}^2}} \right)}^3}\exp \left\{ {\frac{9}{{2{\lambda _{{\rm{IRS}}}}\pi {{\left( {2{R_{{\rm{co}}}}} \right)}^2}}}} \right\}}}{{4{{\left( {{R_{{\rm{co}}}}} \right)}^2}}} \times } \right.\sqrt {\frac{{{{\left( {2{R_{{\rm{co}}}}} \right)}^2}}}{{\left( {{{({h_{{\rm{IRS}}}})}^2} + {{\left( {2{R_{{\rm{co}}}}} \right)}^2}} \right){{\left( {{h_{{\rm{IRS}}}}} \right)}^2}}}}  \times 
\end{equation*}
\begin{equation*}
\sqrt {\frac{{{{\left( {3{R_{{\rm{co}}}}} \right)}^2}}}{{\left( {{{({h_{{\rm{BS}}}} - {h_{{\rm{IRS}}}})}^2} + {{\left( {3{R_{{\rm{co}}}}} \right)}^2}} \right){{\left( {{h_{{\rm{BS}}}} - {h_{{\rm{IRS}}}}} \right)}^2}}}} +
\end{equation*}
\begin{equation*}
\frac{{\left( {2{{\left( {{\lambda _{{\rm{IRS}}}}\pi {{\left( {2{R_{{\rm{co}}}}} \right)}^2}} \right)}^2} + 3{\lambda _{{\rm{IRS}}}}\pi {{\left( {2{R_{{\rm{co}}}}} \right)}^2}} \right)}}{{4{{\left( {{R_{{\rm{co}}}}} \right)}^2}}}\times \sqrt {\frac{{{{\left( {2{R_{{\rm{co}}}}} \right)}^2}}}{{\left( {{{({h_{{\rm{IRS}}}})}^2} + {{\left( {2{R_{{\rm{co}}}}} \right)}^2}} \right){{\left( {{h_{{\rm{IRS}}}}} \right)}^2}}}}  \times 
\end{equation*}
\begin{equation*}
\sqrt {\frac{{{{\left( {3{R_{{\rm{co}}}}} \right)}^2}}}{{\left( {{{({h_{{\rm{BS}}}} - {h_{{\rm{IRS}}}})}^2} + {{\left( {3{R_{{\rm{co}}}}} \right)}^2}} \right){{\left( {{h_{{\rm{BS}}}} - {h_{{\rm{IRS}}}}} \right)}^2}}}} +
\end{equation*}
\begin{equation*}
\frac{{{{\left( {{\lambda _{{\rm{IRS}}}}\pi {{\left( {2{R_{{\rm{co}}}}} \right)}^2}} \right)}^4}\exp \left\{ {\frac{{16}}{{2\left( {{\lambda _{{\rm{IRS}}}}\pi {{\left( {2{R_{{\rm{co}}}}} \right)}^2}} \right)}}} \right\}}}{{4{{\left( {{R_{{\rm{co}}}}} \right)}^4}}}  \times\pi \ln \left( {1 + {{\left( {\frac{{2{R_{{\rm{co}}}}}}{{{h_{{\rm{IRS}}}}}}} \right)}^2}} \right)  \ln \left( {1 + {{\left( {\frac{{3{R_{{\rm{co}}}}}}{{{h_{{\rm{BS}}}} - {h_{{\rm{IRS}}}}}}} \right)}^2}} \right)+
\end{equation*}
\begin{equation*}
\frac{{{{\left( {{\lambda _{{\rm{IRS}}}}\pi {{\left( {2{R_{{\rm{co}}}}} \right)}^2}} \right)}^3}\exp \left\{ {\frac{9}{{2\left( {{\lambda _{{\rm{IRS}}}}\pi {{\left( {2{R_{{\rm{co}}}}} \right)}^2}} \right)}}} \right\}}}{{4{{\left( {{R_{{\rm{co}}}}} \right)}^4}}} \times\pi \ln \left( {1 + {{\left( {\frac{{2{R_{{\rm{co}}}}}}{{{h_{{\rm{IRS}}}}}}} \right)}^2}} \right)  \ln \left( {1 + {{\left( {\frac{{3{R_{{\rm{co}}}}}}{{{h_{{\rm{BS}}}} - {h_{{\rm{IRS}}}}}}} \right)}^2}} \right)-
\end{equation*}
\begin{equation*}
 {  \frac{{\left( {{{\left( {{\lambda _{{\rm{IRS}}}}\pi {{\left( {2{R_{{\rm{co}}}}} \right)}^2}} \right)}^2} + 2{\lambda _{{\rm{IRS}}}}\pi {{\left( {2{R_{{\rm{co}}}}} \right)}^2}} \right)}}{{4{{\left( {{R_{{\rm{co}}}}} \right)}^4}}}}\times\left.\pi \ln \left( {1 + {{\left( {\frac{{2{R_{{\rm{co}}}}}}{{{h_{{\rm{IRS}}}}}}} \right)}^2}} \right)  \ln \left( {1 + {{\left( {\frac{{3{R_{{\rm{co}}}}}}{{{h_{{\rm{BS}}}} - {h_{{\rm{IRS}}}}}}} \right)}^2}} \right)\right)+
\end{equation*}
\begin{equation*}
\frac{{\left( {{\lambda _{\rm{U}}}\pi {{\left( {{R_{{\rm{co}}}}} \right)}^2} + \exp \left\{ { - {\lambda _{\rm{U}}}\pi {{\left( {{R_{{\rm{co}}}}} \right)}^2}} \right\} - 1} \right){\lambda _{{\rm{BS}}}}}}{{{4^5}{\pi ^3}\left( {1 - \exp \left\{ { - {\lambda _{\rm{U}}}\pi {{\left( {{R_{{\rm{co}}}}} \right)}^2}} \right\}} \right)}} \times \frac{Q{{\left( {{\lambda _{{\rm{wave}}}}} \right)}^4}\sigma _{{\rm{d}}}^2}{{(R_{\rm co})}^2} \times
\end{equation*}
\begin{equation*}
\ln \left( {1 + {{\left( {\frac{{{R_{{\rm{co}}}}}}{{{h_{\rm BS}-h_{{\rm{IRS}}}}}}} \right)}^2}} \right)\ln \left( {1 + {{\left( {\frac{{2{R_{{\rm{co}}}}}}{{{h_{{\rm{IRS}}}}}}} \right)}^2}} \right)\times\left( {{\left( {{\lambda _{{\rm{IRS}}}}\pi {{\left( {2{R_{{\rm{co}}}}} \right)}^2}} \right)}^3}\exp \left\{ {\frac{9}{{2{\lambda _{{\rm{IRS}}}}\pi {{\left( {2{R_{{\rm{co}}}}} \right)}^2}}}} \right\}\right.  +
\end{equation*}
\begin{equation}
\left. 2{{\left( {{\lambda _{{\rm{IRS}}}}\pi {{\left( {2{R_{{\rm{co}}}}} \right)}^2}} \right)}^2} + 3{\lambda _{{\rm{IRS}}}}\pi {{\left( {2{R_{{\rm{co}}}}} \right)}^2} \right).
\label{t41}
\end{equation}
\label{theorem3}
\end{theorem}

\begin{IEEEproof}
The proof is provided in Appendix C.
\end{IEEEproof}

Next, we introduce the lower bound on $E\left\{ {{{\left| {{{\mathbb{I} }}({{\mathbf{x}}_{\rm{U}}})} \right|}^2}} \right\}$ in the following theorem:

\begin{theorem}
Consider $0<b<R_{\rm co}$ and $0<d<\frac{R_{\rm co}}{2}$. Then, 
$E\left\{ {{{\left| {{{\mathbb{I} }}({{\mathbf{x}}_{\rm{U}}})} \right|}^2}} \right\}$ can be lower bounded as follows:
\begin{equation*}
E\left\{ {{{\left| {{{\mathbb{I} }}({{\mathbf{x}}_{\rm{U}}})} \right|}^2}} \right\} \ge P^{\mathbb{I} }_{\min}=P^{\mathbb{I} }_{\min}({\rm BS})+  P^{\mathbb{I} }_{\min}({\rm IRS}),
\end{equation*}
where $P^{\mathbb{I} }_{\min}({\rm BS})$ is caused from the BSs,  $P^{\mathbb{I} }_{\min}({\rm U})$ is caused from other users, and $P^{\mathbb{I} }_{\min}({\rm IRS})$ is caused from the IRSs, which are given as follows:
\begin{equation*}
P^{\mathbb{I} }_{\min}({\rm BS})=
\end{equation*}
\begin{equation*}
\left( {1 - \exp \left( { - {\lambda _{{\rm{BS}}}}\pi \left( {4{{\left( {{R_{{\rm{co}}}}} \right)}^2} - {b^2}} \right)} \right)} \right) {\delta ^2} \sigma _{{\rm{d}}}^2 \times
\end{equation*}
\begin{equation*}
\frac{{\left( {1 - \exp \left( { - {\lambda _{{\rm{BS}}}}\pi {b^2}} \right)} \right)\left( {1 + \left( {\hat h - 1} \right){p_{{\rm{b}}}}} \right){{\left( {{\lambda _{{\rm{wave}}}}} \right)}^2}}}{{{\lambda _{{\rm{BS}}}}\pi \left( {4{{\left( {{R_{{\rm{co}}}}} \right)}^2} - {b^2}} \right){{\left( {4\pi } \right)}^2}\left( {{{({h_{{\rm{BS}}}})}^2} + {{\left( {{R_{{\rm{co}}}}} \right)}^2}} \right)}} \times 
\end{equation*}
\begin{equation}
\left( {\frac{{{\lambda _{\rm{U}}}S + \exp \left\{ { - {\lambda _{\rm{U}}}S} \right\} - 1}}{{1 - \exp \left\{ { - {\lambda _{\rm{U}}}S} \right\}}}} \right),
\end{equation}

\begin{equation*}
P^{\mathbb{I} }_{\min}({\rm IRS})=
\end{equation*}
\begin{equation*}
\frac{\kappa }{{\kappa  + 1}} \times \left( {1 - \exp \left\{ { - {\lambda _{{\rm{BS}}}}\pi {{\left( {{R_{{\rm{co}}}}} \right)}^2}} \right\}} \right)Q(Q - 1)\times
\end{equation*}
\begin{equation*}
\frac{{{{\left( {{\lambda _{{\rm{IRS}}}}\pi {{\left( {{R_{{\rm{co}}}}} \right)}^2}} \right)}^2}{{\left( {{\lambda _{{\rm{wave}}}}} \right)}^4}{\delta ^2} \sigma _{{\rm{d}}}^2}}{{{4^6}{\pi ^2}{{\left( {{R_{{\rm{co}}}}} \right)}^4}}} \times 
\end{equation*}
{\begin{equation*}
\max \left\{ {\left( {{\lambda _{\rm{U}}}\pi {{\left( {\frac{{{R_{{\rm{co}}}}}}{2} - d} \right)}^2} + \exp \left\{ { - {\lambda _{\rm{U}}}\pi {{\left( {\frac{{{R_{{\rm{co}}}}}}{2} + d} \right)}^2}} \right\} - 1} \right),0} \right\}\times
\end{equation*}}
\begin{equation*}
\frac{{2d\left( {1 - \exp \left\{ { - {\lambda _{{\rm{BS}}}}3\pi {{\left( {{R_{{\rm{co}}}}} \right)}^2}} \right\}} \right)}}{{\left( {1 - \exp \left\{ { - {\lambda _{\rm{U}}}\pi {{\left( {\frac{{{R_{{\rm{co}}}}}}{2} + d} \right)}^2}} \right\}} \right){\lambda _{{\rm{BS}}}}3\pi {{\left( {{R_{{\rm{co}}}}} \right)}^3}}} \times 
\end{equation*}
{ \begin{equation*}
\scalebox{1}[1]{$\max \left\{ {0,\frac{{1 - \exp \left\{ { - {\lambda _{\rm{U}}}\pi \left( {3{{\left( {{R_{{\rm{co}}}}} \right)}^2}} \right)} \right\} - {\lambda _{\rm{U}}}\pi \left( {4{{\left( {{R_{{\rm{co}}}}} \right)}^2}} \right)\exp \left\{ { - {\lambda _{\rm{U}}}\pi \left( {3{{\left( {{R_{{\rm{co}}}}} \right)}^2}} \right)} \right\}}}{{{{\left( {{\lambda _{\rm{U}}}\pi \left( {4{{\left( {{R_{{\rm{co}}}}} \right)}^2}} \right)} \right)}^2}}}} \right\} \times $}
\end{equation*}}
\begin{equation*}
{\left( {\ln \left( {\frac{{\max \left\{ {{{({h_{{\rm{IRS}}}})}^2},{{({h_{{\rm{BS}}}} - {h_{{\rm{IRS}}}})}^2}} \right\} + 4{{\left( {{R_{{\rm{co}}}}} \right)}^2}}}{{\max \left\{ {{{({h_{{\rm{IRS}}}})}^2},{{({h_{{\rm{BS}}}} - {h_{{\rm{IRS}}}})}^2}} \right\} + 3{{\left( {{R_{{\rm{co}}}}} \right)}^2}}}} \right)} \right)^2}+
\end{equation*}
\begin{equation*}
\left( {\frac{{{\lambda _{\rm{U}}}S + \exp \left\{ { - {\lambda _{\rm{U}}}S} \right\} - 1}}{{1 - \exp \left\{ { - {\lambda _{\rm{U}}}S} \right\}}}} \right)\ln \left( {1 + {{\left( {\frac{b}{{{h_{{\rm{IRS}}}}}}} \right)}^2}} \right) Q{{\left( {{\lambda _{{\rm{wave}}}}} \right)}^4}{\delta ^2}\times
\end{equation*}
{\begin{equation}
\frac{{{\lambda _{{\rm{IRS}}}}\left( {1 - \exp \left( { - {\lambda _{{\rm{BS}}}}\pi \left( {4{{\left( {{R_{{\rm{co}}}}} \right)}^2} - {b^2}} \right)} \right)} \right)\left( {1 - \exp \left( { - {\lambda _{{\rm{BS}}}}\pi {b^2}} \right)} \right) \sigma _{{\rm{d}}}^2}}{{{\lambda _{{\rm{BS}}}}\pi \left( {4{{\left( {{R_{{\rm{co}}}}} \right)}^2} - {b^2}} \right){4^4}{\pi ^3}\left( {{{({h_{{\rm{BS}}}} - {h_{{\rm{IRS}}}})}^2} + {{\left( {{R_{{\rm{co}}}}} \right)}^2}} \right)}} 
\end{equation}}

\label{theorem4}
\end{theorem}

\begin{IEEEproof}
The proof is provided in Appendix D.
\end{IEEEproof}

\textit{Corollary 2}: From Theorems \ref{theorem3} and \ref{theorem4}, for a suffiently large $Q$, we obtain:

\begin{equation}
P_{\max }^{{\mathbb{I}  }} \simeq O\left( {{Q^2}{\lambda _{\rm{U}}}{{\left( {{\lambda _{{\rm{IRS}}}}} \right)}^3}} \right),
\end{equation}
\begin{equation}
P_{\min }^{{\mathbb{I}  }} \simeq O\left( {{Q^2}\frac{{\left( {{\lambda _{{\rm{IRS}}}}} \right)}^2}{{\lambda _{\rm{U}}}{\lambda _{\rm{BS}}}}} \right).
\end{equation}

\section{Bounds on the CDF of the Power of Signal and Interference}

\label{section4}

In this section, we derive bounds on the CDF of the power of signal and interference. To this end, we define $\mathbb{P}\left( {\mathbb{S}\left( {{{\mathbf{x}}_{\rm{U}}}} \right)} \right)$ and $\mathbb{P}\left( {\mathbb{I}\left( {{{\mathbf{x}}_{\rm{U}}}} \right)} \right)$ as  the power of the signal and the interference, respectively:
\begin{equation*}
\mathbb{P}\left( {\mathbb{S}\left( {{{\mathbf{x}}_{\rm{U}}}} \right)} \right) = 
\end{equation*}
\begin{equation}
E\left\{ {\left. {{{\left| {\mathbb{S}\left( {{{\mathbf{x}}_{\rm{U}}}} \right)} \right|}^2}} \right|{\Phi _{\rm{U}}},{\Phi _{{\rm{BS}}}},{\Phi _{{\rm{IRS}}}},{{{\mathbf{\tilde x}}}_{{\rm{BS}}}}({{\mathbf{x}}_{\rm{U}}}),{{{\mathbf{\tilde x}}}_{\rm{U}}}({{\mathbf{x}}_{{\rm{IRS}}}}),} \right.\left. {{{\mathbb{H}}^{[q]}}({{\mathbf{x}}_{{\rm{BS}}}},{{\mathbf{x}}_{{\rm{IRS}}}}),{{\mathbb{H}}^{[q]}}({{\mathbf{x}}_{{\rm{IRS}}}},{{\mathbf{x}}_{\rm{U}}}),{\mathbb{H}}({{\mathbf{x}}_{{\rm{BS}}}},{{\mathbf{x}}_{\rm{U}}})} \right\},
\label{p1}
\end{equation}

\begin{equation*}
\mathbb{P}\left( {\mathbb{I}\left( {{{\mathbf{x}}_{\rm{U}}}} \right)} \right) = 
\end{equation*}
\begin{equation}
E\left\{ {\left. {{{\left| {\mathbb{I}\left( {{{\mathbf{x}}_{\rm{U}}}} \right)} \right|}^2}} \right|{\Phi _{\rm{U}}},{\Phi _{{\rm{BS}}}},{\Phi _{{\rm{IRS}}}},{{{\mathbf{\tilde x}}}_{{\rm{BS}}}}({{\mathbf{x}}_{\rm{U}}}),{{{\mathbf{\tilde x}}}_{\rm{U}}}({{\mathbf{x}}_{{\rm{IRS}}}}),} \right.\left. {{{\mathbb{H}}^{[q]}}({{\mathbf{x}}_{{\rm{BS}}}},{{\mathbf{x}}_{{\rm{IRS}}}}),{{\mathbb{H}}^{[q]}}({{\mathbf{x}}_{{\rm{IRS}}}},{{\mathbf{x}}_{\rm{U}}}),{\mathbb{H}}({{\mathbf{x}}_{{\rm{BS}}}},{{\mathbf{x}}_{\rm{U}}})} \right\},
\label{p2}
\end{equation}
Moreover, the capacity is defined as follows:
\begin{equation}
\mathbb{C}\left( {{{\mathbf{x}}_{\rm{U}}}} \right)=\ln \left(1+\frac{\mathbb{P}\left( {\mathbb{S}\left( {{{\mathbf{x}}_{\rm{U}}}} \right)} \right)}{\mathbb{P}\left( {\mathbb{I}\left( {{{\mathbf{x}}_{\rm{U}}}} \right)} \right)+N_0} \right).
\end{equation}

\subsection{CDF of  $\mathbb{P}\left( {\mathbb{S}\left( {{{\mathbf{x}}_{\rm{U}}}} \right)} \right) $ and $\mathbb{P}\left( {\mathbb{I}\left( {{{\mathbf{x}}_{\rm{U}}}} \right)} \right) $}

We derive the decay of the probability of $\mathbb{P}\left( {\mathbb{I}\left( {{{\mathbf{x}}_{\rm{U}}}} \right)} \right) $. To this end, fom (\ref{iiiiii}), we have:
\begin{equation*}
\left| {\mathbb{I}({{\mathbf{x}}_{\rm{U}}})} \right| \le \sum\limits_{{{{\mathbf{x'}}}_{{\rm{BS}}}} \in {\Psi _{{\rm{BS}}}}({{\mathbf{x}}_{\rm{U}}})} {\sum\limits_{{{{\mathbf{x'}}}_{\rm{U}}} \in {{\tilde \Psi }_{\rm{U}}}({{\mathbf{x}}_{\rm{U}}}):{{{\mathbf{x'}}}_{\rm{U}}} \ne {{\mathbf{x}}_{\rm{U}}}} {\left( {1 + {{\tilde n}_{{\text{IRS}}}}({{\mathbf{x}}_{\text{U}}})} \right)\left| {{\mathbb{H}}({{{\mathbf{x'}}}_{{\text{BS}}}},{{\mathbf{x}}_{\text{U}}})} \right|\left| {X({{{\mathbf{x'}}}_{\text{U}}})} \right|} } 
\end{equation*}
\begin{equation*}
+ \sum\limits_{{{\mathbf{x}}_{{\rm{IRS}}}} \in {\Psi _{{\rm{IRS}}}}({{\mathbf{x}}_{\rm{U}}})} {\sum\limits_{q = 1}^Q {\sum\limits_{{{{\mathbf{x'}}}_{{\rm{BS}}}} \in {{\tilde \Psi }_{{\rm{BS}}}}({{\mathbf{x}}_{\rm{U}}})} {\sum\limits_{{{{\mathbf{x'}}}_{\rm{U}}} \in {{\tilde \Psi }_{\rm{U}}}({{\mathbf{x}}_{\rm{U}}}):{{{\mathbf{x'}}}_{\rm{U}}} \ne {{\mathbf{x}}_{\rm{U}}}} {} } } } 
\end{equation*}
{\begin{equation*}
\left| {{{\mathbb{H}}^{[q]}}({{\mathbf{x}}_{{\text{IRS}}}},{{\mathbf{x}}_{\text{U}}}){{\mathbb{H}}^{[q]}}({{{\mathbf{x'}}}_{{\text{BS}}}},{{\mathbf{x}}_{{\text{IRS}}}})} \right|\left( {1 + {{\tilde n}_{{\text{IRS}}}}({{\mathbf{x}}_{\text{U}}})} \right)\left| {X({{{\mathbf{x'}}}_{\text{U}}})} \right|\end{equation*}}
\begin{equation*}
\le 2\sum\limits_{{{\mathbf{x}}_{{\rm{IRS}}}} \in {\Psi _{{\rm{IRS}}}}({{\mathbf{x}}_{\rm{U}}})} {\sum\limits_{q = 1}^Q {\sum\limits_{{{{\mathbf{x'}}}_{{\rm{BS}}}} \in {{\tilde \Psi }_{{\rm{BS}}}}({{\mathbf{x}}_{\rm{U}}})} {\sum\limits_{{{{\mathbf{x'}}}_{\rm{U}}} \in {{\tilde \Psi }_{\rm{U}}}({{\mathbf{x}}_{\rm{U}}}):{{{\mathbf{x'}}}_{\rm{U}}} \ne {{\mathbf{x}}_{\rm{U}}}} {} } } } 
\end{equation*}
\begin{equation*}
\left( {1 + {{\tilde n}_{{\rm{IRS}}}}({{\mathbf{x}}_{\rm{U}}})} \right)\left( {\left| {{{\mathbb{H}}^{[q]}}({{\mathbf{x}}_{{\text{IRS}}}},{{\mathbf{x}}_{\text{U}}}){{\mathbb{H}}^{[q]}}({{{\mathbf{x'}}}_{{\text{BS}}}},{{\mathbf{x}}_{{\text{IRS}}}})} \right| + \left| {{\mathbb{H}}({{{\mathbf{x'}}}_{{\text{BS}}}},{{\mathbf{x}}_{\text{U}}})} \right|} \right){\left| {X({{{\mathbf{x'}}}_{\text{U}}})} \right|}.
\end{equation*}

As we can see in \cite{vershynin}, to bound the tail of $\Pr \left\{ {\left| {\mathbb{P}\left( {\mathbb{I}\left( {{{\mathbf{x}}_{\rm{U}}}} \right)} \right)} \right| > t} \right\}$ and $\Pr \left\{ {\left| {\mathbb{P}\left( {\mathbb{S}\left( {{{\mathbf{x}}_{\rm{U}}}} \right)} \right)} \right| > t} \right\}$, we can use the similar methodology given in \cite{vershynin} for Subgaussian and Subexponential random variables by bounding  $L_p$-norm of these variables as follows:

\begin{theorem}
The $L_p$-norm ($p\in\{1,2,...\}$) of $\mathbb{P}\left( {\mathbb{I}\left( {{{\mathbf{x}}_{\rm{U}}}} \right)} \right)$ is upper bounded as follows:
\begin{equation}
{\left( {E\left\{ {{{\left( {\mathbb{P}\left( {\mathbb{I}\left( {{{\mathbf{x}}_{\rm{U}}}} \right)} \right)} \right)}^p}} \right\}} \right)^{\frac{1}{p}}} \le K{p^{11}},
\end{equation}
where
\begin{equation*}
K = Q^2{2^{6}}\left( {\frac{{ \sigma _{{\rm{d}}}^2 }}{{1 - \exp \left\{ { - {\lambda _{\rm{U}}}4\pi {{\left( {{R_{{\rm{co}}}}} \right)}^2}} \right\}}}} \right) \times 
\end{equation*}
\begin{equation*}
 \scalebox{1}[1]{${\left( {\max \left\{ {1,\frac{{{\lambda _{{\text{wave}}}}}}{{4\pi {h_{{\text{BS}}}}}}} \right\} \times \max \left\{ {1,\frac{{{\lambda _{{\text{wave}}}}}}{{4\pi \left| {{h_{{\text{BS}}}} - {h_{{\text{IRS}}}}} \right|}}} \right\} \times \max \left\{ {1,\frac{{{\lambda _{{\text{wave}}}}}}{{4\pi {h_{{\text{IRS}}}}}}} \right\}} \right)^{2p}} \times$}
\end{equation*}
\begin{equation*}
\left( {{2^{11}}{\pi ^5}{{\left( 3 \right)}^{\frac{3}{2}}}\exp \left\{ {\frac{{35}}{{12}}} \right\}} \right) \times \left( {\frac{{18}}{{\exp \{ 3\} }}} \right) \times {\left( {8\max \left\{ {1,\frac{1}{{\log \left( {1 + \frac{1}{{{\lambda _{{\rm{IRS}}}}4\pi {{\left( {{R_{{\rm{co}}}}} \right)}^2}}}} \right)}}} \right\}} \right)^4} \times 
\end{equation*}
\begin{equation}
{\left( {\frac{2}{{\log \left( {1 + \frac{1}{{{\lambda _{{\rm{BS}}}}4\pi {{\left( {{R_{{\rm{co}}}}} \right)}^2}}}} \right)}}} \right)^2} \times {\left( {\frac{2}{{\log \left( {1 + \frac{1}{{{\lambda _{\rm{U}}}4\pi {{\left( {{R_{{\rm{co}}}}} \right)}^2}}}} \right)}}} \right)^2}.
\label{K-coef}
\end{equation}
\label{II-LP}
\end{theorem}

\begin{IEEEproof}
The proof is provided in Appendix G.
\end{IEEEproof}

Now, we introduce the following theorem for the tale of the probability of ${{\mathbb{P}}\left( {\mathbb{I}\left( {{{\mathbf{x}}_{\rm{U}}}} \right)} \right)}$:
\begin{theorem}
If we define  $\mathbb{G}(x)$ as follows:
\begin{equation}
\mathbb{G}(x) = \sum\limits_{p = 0}^\infty  {\frac{{{x^{p}}}}{{\left( {11p} \right)!}}}  = \frac{1}{11}\sum\limits_{k = 0}^{10} {\exp \left\{ {\sqrt[11]{x}\exp \left\{ {\frac{{j2\pi k}}{11}} \right\}} \right\}} ,
\end{equation}
then, for every $0 < \tau  < \frac{{{{11}^{11}}}}{{\left( {{ K} } \right){e^{11}}}}$ and $t>0$, we have:
\begin{equation}
\Pr \left\{ {\left| {\mathbb{P}\left( {\mathbb{I}\left( {{{\mathbf{x}}_{\rm{U}}}} \right)} \right)} \right| > t} \right\} \le  \frac{1}{{\left( {1 - \frac{{K{e^{11}}\tau }}{{{{11}^{11}}}}} \right)\mathbb{G}\left( {\tau t} \right)}},
\end{equation}
where $ K$ is given by (\ref{K-coef}).
\label{I-tail}
\end{theorem}

\begin{IEEEproof}
The proof is provided in Appendix H.
\end{IEEEproof}

Next, we analyze the decay of the probability of $\mathbb{P}\left( {\mathbb{S}\left( {{{\mathbf{x}}_{\rm{U}}}} \right)} \right)$. Hence, we have: 
\begin{equation*}
\left| {\mathbb{S}({{\mathbf{x}}_{\rm{U}}})} \right| \le 
\end{equation*}
\begin{equation*}
\left( {1 + {{\tilde n}_{{\rm{IRS}}}}({{\mathbf{x}}_{\rm{U}}})} \right)\left| {{\mathbb{H}}({{{\mathbf{\tilde x}}}_{{\rm{BS}}}}({{\mathbf{x}}_{\rm{U}}}),{{\mathbf{x}}_{\rm{U}}})} \right|\left| {{X}({{\mathbf{x}}_{\rm{U}}})} \right| + 
\end{equation*}
\begin{equation*}
\sum\limits_{{{\mathbf{x}}_{{\rm{IRS}}}} \in {\Psi _{{\rm{IRS}}}}({{\mathbf{x}}_{\rm{U}}})} {\sum\limits_{q = 1}^Q {\left( {1 + {{\tilde n}_{{\rm{IRS}}}}({{\mathbf{x}}_{\rm{U}}})} \right)\left| {{\mathbb{H}}^{[q]}}({{\mathbf{x}}_{{\rm{IRS}}}},{{\mathbf{x}}_{\rm{U}}})\right|\times} } 
\end{equation*}
\begin{equation*}
 \left|{{\mathbb{H}}^{[q]}}({{{\mathbf{\tilde x}}}_{{\rm{BS}}}}({{\mathbf{x}}_{\rm{U}}}),{{\mathbf{x}}_{{\rm{IRS}}}}) \right|\left| {{X}({{\mathbf{x}}_{\rm{U}}})} \right| \le
\end{equation*}
\begin{equation*}
\sum\limits_{{{\mathbf{x}}_{{\rm{BS}}}} \in {\Psi _{{\rm{BS}}}}({{\mathbf{x}}_{\rm{U}}})} {\sum\limits_{{{\mathbf{x}}_{{\rm{IRS}}}} \in {\Psi _{{\rm{IRS}}}}({{\mathbf{x}}_{\rm{U}}})} {\sum\limits_{q = 1}^Q {\left( {1 + {{\tilde n}_{{\rm{IRS}}}}({{\mathbf{x}}_{\rm{U}}})} \right)\left|{X}({{\mathbf{x}}_{\rm{U}}}) \right|\times} } } 
\end{equation*}
\begin{equation*}
\left( {\left| {{{\mathbb{H}}^{[q]}}({{\mathbf{x}}_{{\rm{IRS}}}},{{\mathbf{x}}_{\rm{U}}}){{\mathbb{H}}^{[q]}}({{\mathbf{x}}_{{\rm{BS}}}},{{\mathbf{x}}_{{\rm{IRS}}}})} \right| + \left| {{\mathbb{H}}({{\mathbf{x}}_{{\rm{BS}}}},{{\mathbf{x}}_{\rm{U}}})} \right|} \right),
\end{equation*}
thus, we obtain:
\begin{equation*}
{\left| {\mathbb{S}({{\mathbf{x}}_{\rm{U}}})} \right|^2} \le 
\end{equation*}
{\small\begin{equation*}
\sum\limits_{{{\mathbf{x}}_{{\rm{BS}}}} \in {\Psi _{{\rm{BS}}}}({{\mathbf{x}}_{\rm{U}}})} {\sum\limits_{{{\mathbf{x}}_{{\rm{IRS}}}} \in {\Psi _{{\rm{IRS}}}}({{\mathbf{x}}_{\rm{U}}})} {\sum\limits_{q = 1}^Q {\sum\limits_{{{{\mathbf{x'}}}_{{\rm{BS}}}} \in {\Psi _{{\rm{BS}}}}({{\mathbf{x}}_{\rm{U}}})} {\sum\limits_{{{{\mathbf{x'}}}_{{\rm{IRS}}}} \in {\Psi _{{\rm{IRS}}}}({{\mathbf{x}}_{\rm{U}}})} {\sum\limits_{q' = 1}^Q {} } } } } } 
\end{equation*}}
\begin{equation*}
{{\left( {1 + {{\tilde n}_{{\rm{IRS}}}}({{\mathbf{x}}_{\rm{U}}})} \right)}^2}{{\left| {{X}({{\mathbf{x}}_{\rm{U}}})} \right|}^2}\times
\end{equation*}
\begin{equation*}
\left( {\left| {{{\mathbb{H}}^{[q]}}({{\mathbf{x}}_{{\rm{IRS}}}},{{\mathbf{x}}_{\rm{U}}}){{\mathbb{H}}^{[q]}}({{\mathbf{x}}_{{\rm{BS}}}},{{\mathbf{x}}_{{\rm{IRS}}}})} \right| + \left| {{\mathbb{H}}({{\mathbf{x}}_{{\rm{BS}}}},{{\mathbf{x}}_{\rm{U}}})} \right|} \right) \times 
\end{equation*}
\begin{equation*}
\left( {\left| {{{\mathbb{H}}^{[q']}}({{{\mathbf{x'}}}_{{\rm{IRS}}}},{{\mathbf{x}}_{\rm{U}}}){{\mathbb{H}}^{[q']}}({{{\mathbf{x'}}}_{{\rm{BS}}}},{{{\mathbf{x'}}}_{{\rm{IRS}}}})} \right| + \left| {{\mathbb{H}}({{{\mathbf{x'}}}_{{\rm{BS}}}},{{\mathbf{x}}_{\rm{U}}})} \right|} \right).
\end{equation*}

Then, we introduce the following theorem:

\begin{theorem}
The $L_p$-norm ($p\in\{1,2,...\}$) of $\mathbb{P}\left( {\mathbb{S}\left( {{{\mathbf{x}}_{\rm{U}}}} \right)} \right)$ is upper bounded as follows:
\begin{equation}
{\left( {E\left\{ {{{\left( {\mathbb{P}\left( {\mathbb{S}\left( {{{\mathbf{x}}_{\rm{U}}}} \right)} \right)} \right)}^p}} \right\}} \right)^{\frac{1}{p}}} \le Lp^{9},
\end{equation}
where
\begin{equation*}
L = Q^2{2^{6}}\left( {{{ \sigma _{{\rm{d}}}^2 }}} \right) \times 
\end{equation*}
\begin{equation*}
 \scalebox{1}[1]{${\left( {\max \left\{ {1,\frac{{{\lambda _{{\text{wave}}}}}}{{4\pi {h_{{\text{BS}}}}}}} \right\} \times \max \left\{ {1,\frac{{{\lambda _{{\text{wave}}}}}}{{4\pi \left| {{h_{{\text{BS}}}} - {h_{{\text{IRS}}}}} \right|}}} \right\} \times \max \left\{ {1,\frac{{{\lambda _{{\text{wave}}}}}}{{4\pi {h_{{\text{IRS}}}}}}} \right\}} \right)^{2p}} \times$}
\end{equation*}
\begin{equation*}
\left( {{2^{11}}{\pi ^5}{{\left( 3 \right)}^{\frac{3}{2}}}\exp \left\{ {\frac{{35}}{{12}}} \right\}} \right) \times \left( {\frac{{18}}{{\exp \{ 3\} }}} \right) \times {\left( {8\max \left\{ {1,\frac{1}{{\log \left( {1 + \frac{1}{{{\lambda _{{\rm{IRS}}}}4\pi {{\left( {{R_{{\rm{co}}}}} \right)}^2}}}} \right)}}} \right\}} \right)^4} \times 
\end{equation*}
\begin{equation}
{\left( {\frac{2}{{\log \left( {1 + \frac{1}{{{\lambda _{{\rm{BS}}}}4\pi {{\left( {{R_{{\rm{co}}}}} \right)}^2}}}} \right)}}} \right)^2} .
\label{lcoef}
\end{equation}

\end{theorem}

\begin{IEEEproof}
The proof is provided in Appendix I.
\end{IEEEproof}

Now, we introduce the following theorem for the tale of the probability of ${{\mathbb{P}}\left( {\mathbb{S}\left( {{{\mathbf{x}}_{\rm{U}}}} \right)} \right)}$:
\begin{theorem}
If we define $\mathbb{H}(x)$ as follows:
\begin{equation}
\mathbb{H}(x) = \sum\limits_{p = 0}^\infty  {\frac{{{x^{p}}}}{{\left( {9p} \right)!}}}  = \frac{1}{9}\sum\limits_{k = 0}^{8} {\exp \left\{ {\sqrt[9]{x}\exp \left\{ {\frac{{j2\pi k}}{9}} \right\}} \right\}} ,
\label{H-func}
\end{equation}
then, for every $0 < \tau  < \frac{{{{9}^{9}}}}{{\left( {{ L} } \right){e^{9}}}}$ and $t>0$, we have:
\begin{equation}
\Pr \left\{ {\left|{{\mathbb{P}}\left( {\mathbb{S}\left( {{{\mathbf{x}}_{\rm{U}}}} \right)} \right)}\right| > t} \right\} \le  \frac{1}{{\left( {1 - \frac{{L{e^{9}}\tau }}{{{{9}^{9}}}}} \right)\mathbb{H}\left( {\tau t} \right)}},
\end{equation}
where $ L$ is given by (\ref{lcoef}).
\label{S-tail}
\end{theorem}

\begin{IEEEproof}
The proof is provided in Appendix J.
\end{IEEEproof}

\textit{Corollary 3}: Theorems \ref{I-tail} and \ref{S-tail}, and Eq. (\ref{order}) show that the tail of $\Pr \left\{ {\left| \mathbb{P}\left( {\mathbb{I}\left( {{{\mathbf{x}}_{\rm{U}}}} \right)} \right)  \right| > t} \right\}$ goes to zero with the order of $\exp \left\{ { - \sqrt[11]{{\tau t}}} \right\}$, however, the tail of $\Pr \left\{ {\left|\mathbb{P}\left( {\mathbb{S}\left( {{{\mathbf{x}}_{\rm{U}}}} \right)} \right)  \right| > t} \right\}$ goes to zero with the order of $\exp \left\{ { - \sqrt[9]{{\tau t}}} \right\}$.
\begin{IEEEproof}
\begin{equation*}
\mathop {\lim }\limits_{t \to \infty } \frac{{\left| {\exp \left\{ { - \sqrt[11]{{\tau t}}} \right\}} \right|}}{{\frac{1}{{\left| {\mathbb{G}\left( {\tau t} \right)} \right|}}}} = \mathop {\lim }\limits_{t \to \infty } \left| {\exp \left\{ { - \sqrt[11]{{\tau t}}} \right\}} \right|\left| {\mathbb{G}\left( {\tau t} \right)} \right| 
\end{equation*}
{\small\begin{equation*}
= \mathop {\lim }\limits_{t \to \infty } \left| {\exp \left\{ { - \sqrt[11]{{\tau t}}} \right\}} \right|\left| {\frac{1}{11}\sum\limits_{k = 0}^{10 }{\exp \left\{ {\sqrt[11]{{\tau t}}\exp \left\{ {\frac{{j2\pi k}}{11}} \right\}} \right\}} } \right| = \frac{1}{11},
\end{equation*}}
\begin{equation*}
\mathop {\lim }\limits_{t \to \infty } \frac{{\left| {\exp \left\{ { - \sqrt[9]{{\tau t}}} \right\}} \right|}}{{\frac{1}{{\left| {\mathbb{H}\left( {\tau t} \right)} \right|}}}} = \mathop {\lim }\limits_{t \to \infty } \left| {\exp \left\{ { - \sqrt[9]{{\tau t}}} \right\}} \right|\left| {\mathbb{H}\left( {\tau t} \right)} \right| 
\end{equation*}
\begin{equation*}
 =\mathop {\lim }\limits_{t \to \infty } \left| {\exp \left\{ { - \sqrt[9]{{\tau t}}} \right\}} \right|\left| {\frac{1}{9}\sum\limits_{k = 0}^8{\exp \left\{ {\sqrt[9]{{\tau t}}\exp \left\{ {\frac{{j2\pi k}}{9}} \right\}} \right\}} } \right| = \frac{1}{9}.
\end{equation*}
\end{IEEEproof}

\subsection{Outage probability}

In this subsection, first, we introduce the following theorem for the tail of the probability measure $\Pr \left\{ {\mathbb{C}\left( {{{\mathbf{x}}_{\rm{U}}}} \right) > \alpha } \right\}$:

\begin{theorem} 
$\Pr \left\{ {\mathbb{C}\left( {{{\mathbf{x}}_{\rm{U}}}} \right) > \alpha } \right\}$ can be upper bounded as follows:
\begin{equation*}
\Pr \left\{ {\mathbb{C}\left( {{{\mathbf{x}}_{\rm{U}}}} \right) > \alpha } \right\} \le 
\end{equation*}
\begin{equation}
\min \left\{ {1,\frac{{P_{\max }^\mathbb{S}}}{{{N_0}\left( {{e^\alpha } - 1} \right)}},\frac{1}{{\left( {1 - \frac{{L{e^9}\tau }}{{{9^9}}}} \right)\mathbb{H}\left( {\tau {N_0}\left( {{e^\alpha } - 1} \right)} \right)}}} \right\},
\end{equation}
where
\begin{equation*}
0 < \tau  < \frac{{{{9}^{9}}}}{{\left( {{ L} } \right){e^{9}}}}.
\end{equation*}
and function $\mathbb{H}$ is given by (\ref{H-func}).
\label{theorem-C-decay}
\end{theorem}

\begin{IEEEproof}
The proof is provided in Appendix K.
\end{IEEEproof}

\section{Numerical Results}

\label{section5}

\begin{figure}
\centering
\includegraphics[width=18cm]{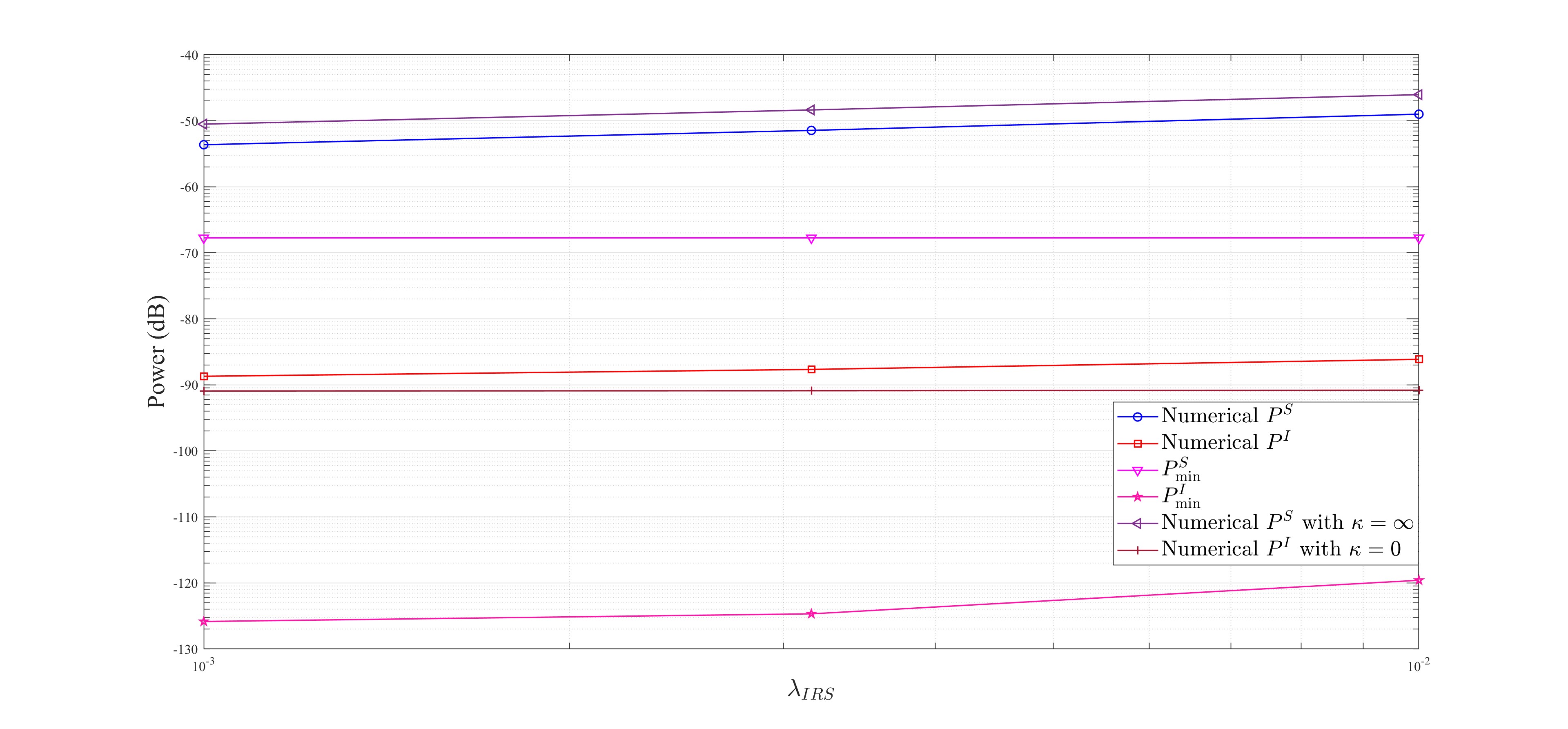}
\caption{Comparison of $P^{\mathbb{S}}_{\min},P^{\mathbb{I}}_{\min}$,  and numerical powers of the desired signal and interference for different values of $\lambda_{\rm IRS} \rm{[IRS/m^2]}$.}
\label{numeric1}
\end{figure}

\begin{figure}
\centering
\includegraphics[width=18cm]{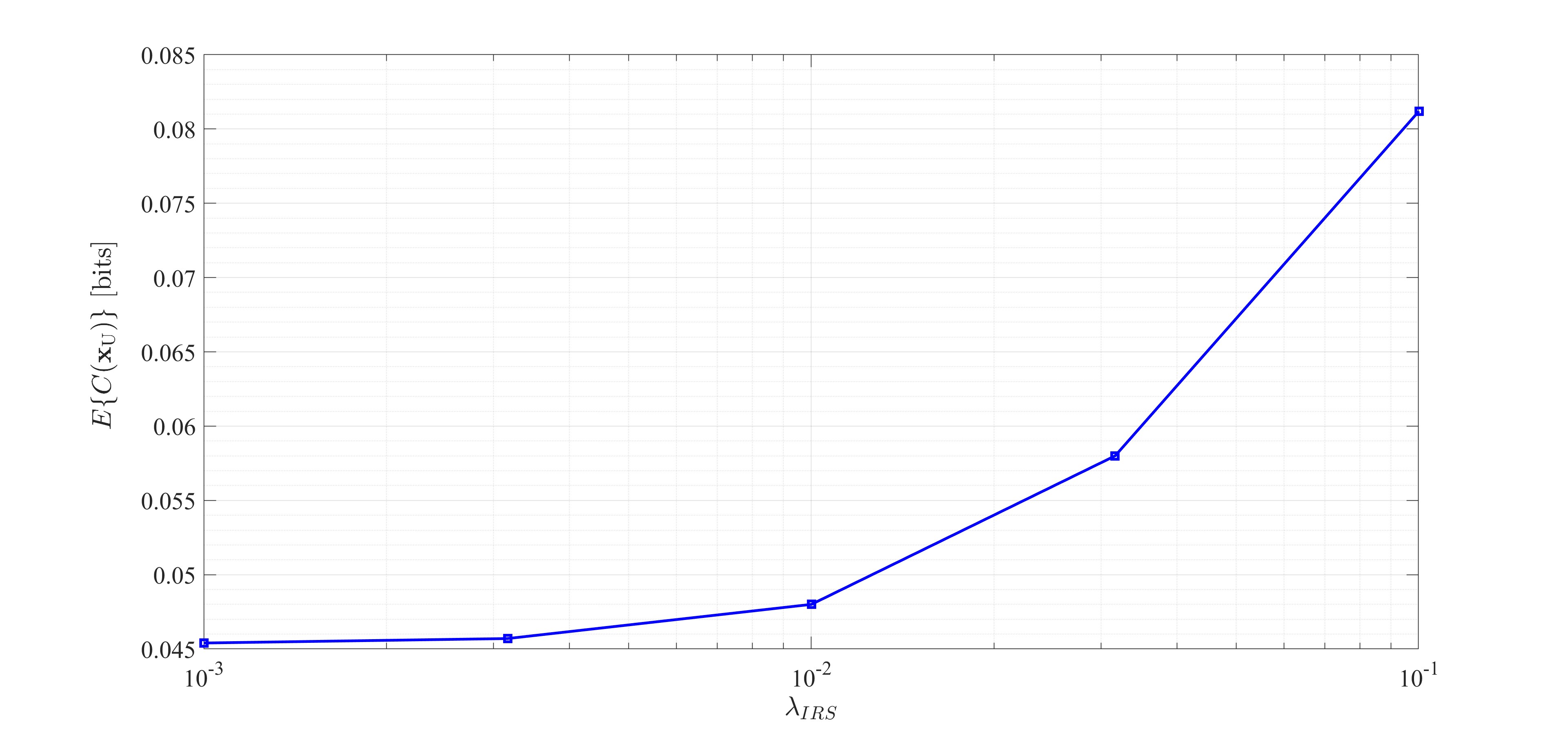}
\caption{$E\{{\mathbb{C}}({\bf x}_{\rm U})\}$ [bits] for different values of $\lambda_{\rm IRS} \rm{[IRS/m^2]}$.}
\label{numeric100}
\end{figure}


\begin{figure}
\centering
\includegraphics[width=18cm]{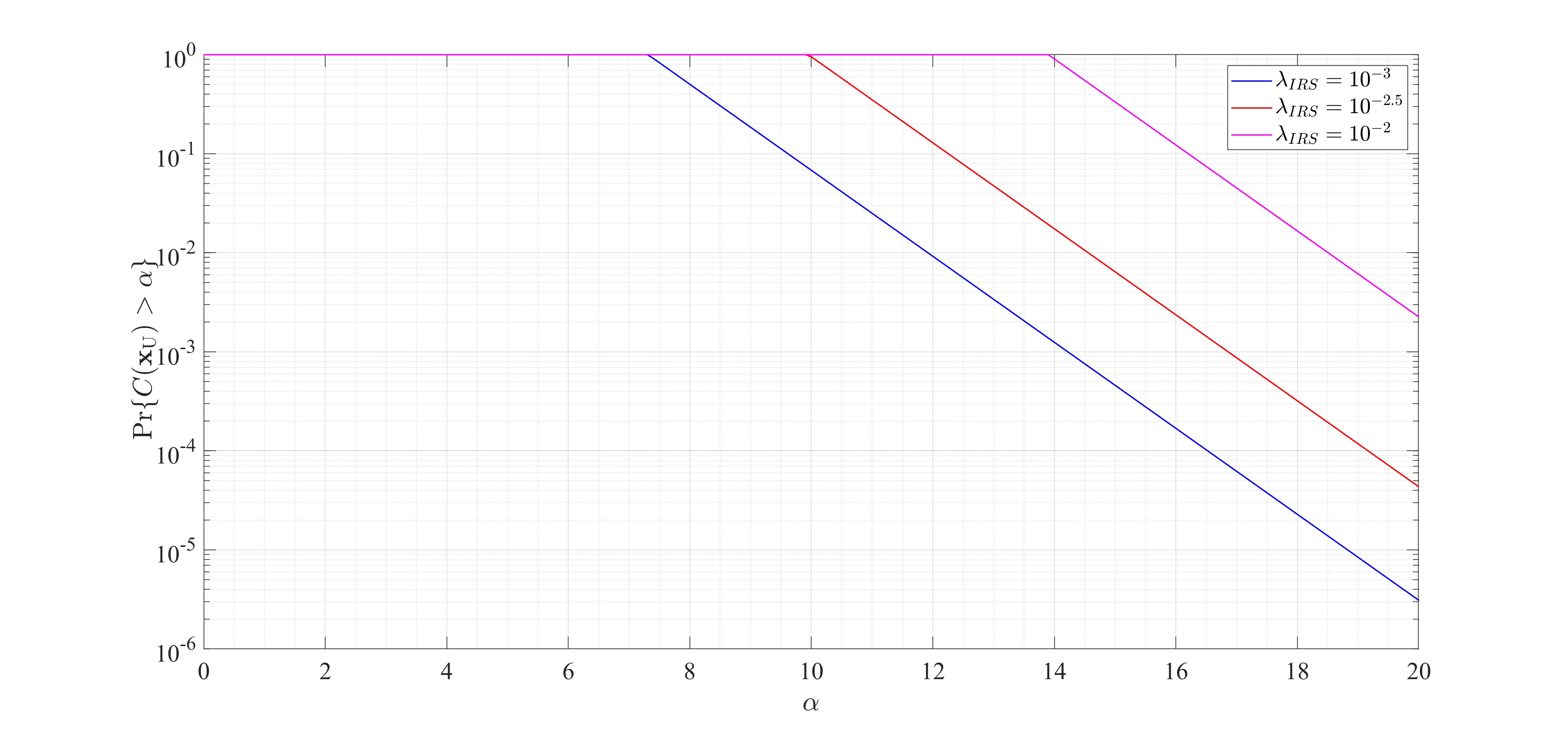}
\caption{Comparison of the upper bound on the tail of  $\Pr \left\{ {{{\mathbb{C}}\left( { {{{\mathbf{x}}_{\rm{U}}}} } \right)} > \alpha} \right\}$ for different values of $\lambda_{\rm IRS}$ with $\kappa =1$.}
\label{numeric4}
\end{figure}
\begin{figure}
\centering
\includegraphics[width=18cm]{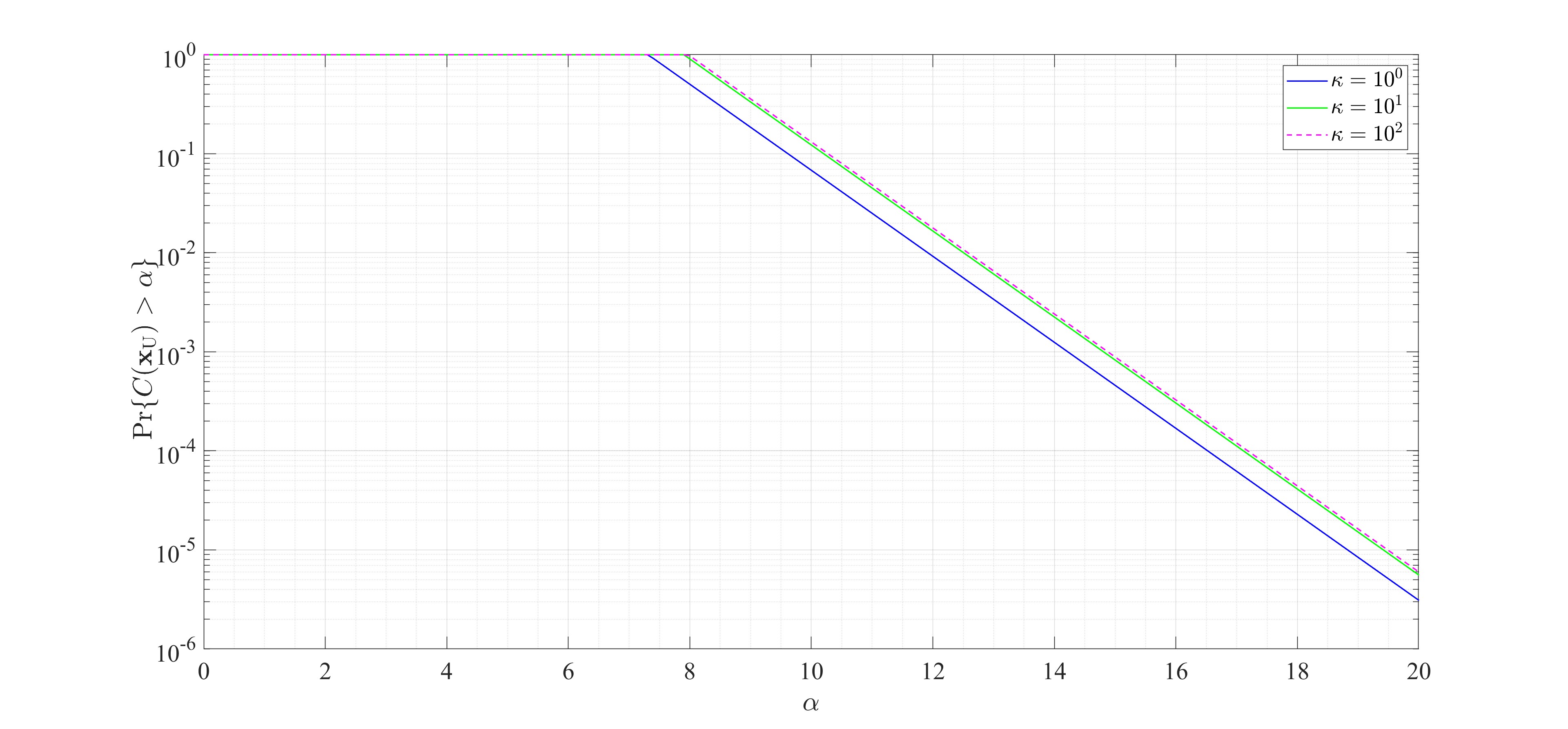}
\caption{Comparison of the upper bound on the tail of  $\Pr \left\{ {{{\mathbb{C}}\left( { {{{\mathbf{x}}_{\rm{U}}}} } \right)} > \alpha} \right\}$ for different values of $\kappa$ with $\lambda_{\rm IRS}=10^{-3}\rm{[IRS/m^2]}$.}
\label{numeric4n}
\end{figure}

In this section, we numerically evaluate the proposed analytical results. In these simulations, we consider parameters $\lambda_{\rm wave}=1{\rm cm}$, $\lambda_{\rm U}={10}^{-2}\rm{[U/m^2]}$, $\lambda_{\rm BS}={10}^{-3}\rm{[BS/m^2]}$, $R_{\rm co}=15{\rm m}$, ${\hat h}={10}^{-4}$, $p_{\rm{b}}=.5$, $\sigma_{\rm{d}}=1000$,  $h_{\rm BS}=10{\rm m}$, $h_{\rm IRS}=11{\rm m}$,  $\varepsilon=\delta=0.01$, and $Q=1000$.

In Fig. \ref{numeric1}, we compared lower and upper bounds $P^{\mathbb{S}}_{\min}$ and $P^{\mathbb{I}}_{\min}$ with the simulation results for different values of $\lambda_{\rm IRS}$. In Fig. \ref{numeric1}, we assumed that the number of IRS elements is $Q=1000$ and $\kappa=1$.  This figure shows the impact of the existence of IRS, i.e., increment of $\lambda_{\rm IRS}$ results in increment of both $E\{{|\mathbb{S}({\bf x}_{\rm U})|}\}$ and $E\{{|\mathbb{I}({\bf x}_{\rm U})|}\}$. Moreover, Fig. \ref{numeric100} shows the increment of $E\{{\mathbb{C}}({\bf x}_{\rm U})\}$ with respect to $\lambda_{\rm IRS}$.



In Figs. \ref{numeric4} and \ref{numeric4n}, we   compared the proposed upper bound on $\Pr\{\mathbb{C}({\bf x}_{\rm U})>\alpha\}$ for different values of $\lambda_{\rm IRS}$ and $\kappa$. We observed that increment of $\lambda_{\rm IRS}$ and $\kappa$ increases the upper bound on $\Pr\{\mathbb{C}({\bf x}_{\rm U})>\alpha\}$.

\section{Conclusion}

\label{section6}

In this paper, we studied a homogeneous IRS-assisted wireless network.
For a typical user in the system, we derived analytical bounds on the expectation of the power of the desired signal and the interference caused by BSs and other users.
Then, we obtained analytical upper bounds on the decay of the probability of the power of the desired signal and the interference for the typical user, which results in a lower bound for the cumulative distribution function (CDF).
At last, we derived upper bounds on the decay of the probability of the capacity of one typical user, which results in a lower bound for the outage probability. These are some of directions for future research: 1) extending the proposed results to a network, in which the BS and IRS serving strategy is not random and depends on the distance between them and users, 2) studying a network, in which users can be served by multiple BSs with coordination, and 3) modeling and analysis of an IRS-assisted network, in which users mobility is considered.

\begin{appendices}

\section{Proof of Theorem 1}

We have:

\begin{equation*}
E\left\{ {{{\left| {\mathbb{S}({{\mathbf{x}}_{\rm{U}}})} \right|}^2}\left| {{\Phi _{\rm{U}}},{\Phi _{{\rm{BS}}}},{\Phi _{{\rm{IRS}}},{{{\mathbf{\tilde x}}}_{{\rm{BS}}}}({{\mathbf{x}}_{\rm{U}}}),{{{\mathbf{\tilde x}}}_{{\rm{U}}}}({{\mathbf{x}}_{\rm{IRS}}}) }} \right.} \right\} = 
\end{equation*}
\begin{equation}
  \scalebox{.9}[1]{$\frac{{{\left( {1 + \left( {\hat h - 1} \right){p_{{\rm{b}}}}} \right)}{{\left( {{\lambda _{{\rm{wave}}}}} \right)}^2}\sigma _{{\rm{d}}}^2}}{{{{\left( {4\pi } \right)}^2}\left( {{{({h_{{\rm{BS}}}})}^2} + \left\| {{{{\mathbf{\tilde x}}}_{{\rm{BS}}}}({{\mathbf{x}}_{\rm{U}}}) - {{\mathbf{x}}_{\rm{U}}}} \right\|_2^2} \right)}} \times {\left( {G\left( {{{{\mathbf{\tilde x}}}_{{\rm{BS}}}}({{\mathbf{x}}_{\rm{U}}}),{{\mathbf{x}}_{\rm{U}}},{{\mathbf{x}}_{\rm{U}}}} \right) + \sum\limits_{{{\mathbf{x}}_{{\rm{IRS}}}} \in {\Psi _{{\rm{IRS}}}}({{{\mathbf{\tilde x}}}_{{\rm{BS}}}}({{\mathbf{x}}_{\rm{U}}})):{{\mathbf{x}}_{\rm{U}}} = {{{\mathbf{\tilde x}}}_{\rm{U}}}({{\mathbf{x}}_{{\rm{IRS}}}})} {G\left( {{{{\mathbf{\tilde x}}}_{{\rm{BS}}}}({{\mathbf{x}}_{\rm{U}}}),{{\mathbf{x}}_{{\rm{IRS}}}},{{\mathbf{x}}_{\rm{U}}}} \right)} } \right)^2} + $}
\label{t1}
\end{equation}
\begin{equation*}
{\frac{\kappa }{{\kappa  + 1}}}\times\sum\limits_{{{\mathbf{x}}_{{\rm{IRS}}}} \in {\Psi _{{\rm{IRS}}}}({{\mathbf{x}}_{\rm{U}}}):{{{\mathbf{\tilde x}}}_{\rm{U}}}({{\mathbf{x}}_{{\rm{IRS}}}}) = {{\mathbf{x}}_{\rm{U}}}} {\sum\limits_{q = 1}^Q {\sum\limits_{{{{\mathbf{x'}}}_{{\rm{IRS}}}} \in {\Psi _{{\rm{IRS}}}}({{\mathbf{x}}_{\rm{U}}}):{{{\mathbf{\tilde x}}}_{\rm{U}}}({{{\mathbf{x'}}}_{{\rm{IRS}}}}) = {{\mathbf{x}}_{\rm{U}}}} {\sum\limits_{q' = 1}^Q {} } } } 
\end{equation*}
\begin{equation*}
  \scalebox{1}[1]{$\left( {\frac{{{\delta _{q - q'}}{\delta _{{{\mathbf{x}}_{{\rm{IRS}}}} - {{\mathbf{x}}^\prime }_{{\rm{IRS}}}}}{{\left( {{\lambda _{{\rm{wave}}}}} \right)}^4}}}{{{{\left( {4\pi } \right)}^4}\left( {{{({h_{{\rm{BS}}}} - {h_{{\rm{IRS}}}})}^2} + \left\| {{{{\mathbf{\tilde x}}}_{{\rm{BS}}}}({{\mathbf{x}}_{\rm{U}}}) - {{\mathbf{x}}_{{\rm{IRS}}}}} \right\|_2^2} \right)\left( {{{({h_{{\rm{IRS}}}})}^2} + \left\| {{{\mathbf{x}}_{{\rm{IRS}}}} - {{\mathbf{x}}_{\rm{U}}}} \right\|_2^2} \right)}}} \right. + $}
\end{equation*}
\begin{equation*}
  \scalebox{.94}[1]{$\left. {\frac{{\left( {1 - {\delta _{q - q'}}} \right)\left( {1 - {\delta _{{{\mathbf{x}}_{{\rm{IRS}}}} - {{\mathbf{x}}^\prime }_{{\rm{IRS}}}}}} \right){{\left( {{\lambda _{{\rm{wave}}}}} \right)}^4} \times {{\left( {\frac{\pi }{4}} \right)}^2}}}{{{{\left( {4\pi } \right)}^4}\left( {\sqrt {{{({h_{{\rm{BS}}}} - {h_{{\rm{IRS}}}})}^2} + \left\| {{{{\mathbf{\tilde x}}}_{{\rm{BS}}}}({{\mathbf{x}}_{\rm{U}}}) - {{\mathbf{x}}_{{\rm{IRS}}}}} \right\|_2^2} } \right)\left( {\sqrt {{{({h_{{\rm{IRS}}}})}^2} + \left\| {{{\mathbf{x}}_{{\rm{IRS}}}} - {{\mathbf{x}}_{\rm{U}}}} \right\|_2^2} } \right)\left( {\sqrt {{{({h_{{\rm{BS}}}} - {h_{{\rm{IRS}}}})}^2} + \left\| {{{{\mathbf{\tilde x}}}_{{\rm{BS}}}}({{\mathbf{x}}_{\rm{U}}}) - {{{\mathbf{x'}}}_{{\rm{IRS}}}}} \right\|_2^2} } \right)\left( {\sqrt {{{({h_{{\rm{IRS}}}})}^2} + \left\| {{{{\mathbf{x'}}}_{{\rm{IRS}}}} - {{\mathbf{x}}_{\rm{U}}}} \right\|_2^2} } \right)}}} \right) \times $}
\end{equation*}
\begin{equation*}
\left( {G\left( {{{{\mathbf{\tilde x}}}_{{\rm{BS}}}}({{\mathbf{x}}_{\rm{U}}}),{{\mathbf{x}}_{\rm{U}}},{{\mathbf{x}}_{{\rm{IRS}}}}} \right) + \sum\limits_{{{{\mathbf{x''}}}_{{\rm{IRS}}}} \in {\Psi _{{\rm{IRS}}}}({{{\mathbf{\tilde x}}}_{{\rm{BS}}}}({{\mathbf{x}}_{\rm{U}}})):{{\mathbf{x}}_{\rm{U}}} = {{{\mathbf{\tilde x}}}_{\rm{U}}}({{{\mathbf{x''}}}_{{\rm{IRS}}}})} {G\left( {{{{\mathbf{\tilde x}}}_{{\rm{BS}}}}({{\mathbf{x}}_{\rm{U}}}),{{{\mathbf{x''}}}_{{\rm{IRS}}}},{{\mathbf{x}}_{{\rm{IRS}}}}} \right)} } \right) \times 
\end{equation*}
\begin{equation}
\left( {G\left( {{{{\mathbf{\tilde x}}}_{{\rm{BS}}}}({{\mathbf{x}}_{\rm{U}}}),{{\mathbf{x}}_{\rm{U}}},{{{\mathbf{x'}}}_{{\rm{IRS}}}}} \right) + \sum\limits_{{{{\mathbf{x''}}}_{{\rm{IRS}}}} \in {\Psi _{{\rm{IRS}}}}({{{\mathbf{\tilde x}}}_{{\rm{BS}}}}({{\mathbf{x}}_{\rm{U}}})):{{\mathbf{x}}_{\rm{U}}} = {{{\mathbf{\tilde x}}}_{\rm{U}}}({{{\mathbf{x''}}}_{{\rm{IRS}}}})} {G\left( {{{{\mathbf{\tilde x}}}_{{\rm{BS}}}}({{\mathbf{x}}_{\rm{U}}}),{{{\mathbf{x''}}}_{{\rm{IRS}}}},{{{\mathbf{x'}}}_{{\rm{IRS}}}}} \right)} } \right)\sigma _{{\rm{d}}}^2 + 
\label{t2}
\end{equation}

\begin{equation*}
{\frac{1 }{{\kappa  + 1}}}\times\sum\limits_{{{\mathbf{x}}_{{\rm{IRS}}}} \in {\Psi _{{\rm{IRS}}}}({{\mathbf{x}}_{\rm{U}}}):{{{\mathbf{\tilde x}}}_{\rm{U}}}({{\mathbf{x}}_{{\rm{IRS}}}}) = {{\mathbf{x}}_{\rm{U}}}} {\sum\limits_{q = 1}^Q {\sum\limits_{{{{\mathbf{x'}}}_{{\rm{IRS}}}} \in {\Psi _{{\rm{IRS}}}}({{\mathbf{x}}_{\rm{U}}}):{{{\mathbf{\tilde x}}}_{\rm{U}}}({{{\mathbf{x'}}}_{{\rm{IRS}}}}) = {{\mathbf{x}}_{\rm{U}}}} {\sum\limits_{q' = 1}^Q {} } } } 
\end{equation*}
\begin{equation*}
  \scalebox{1}[1]{$\left( {\frac{{{\delta _{q - q'}}{\delta _{{{\mathbf{x}}_{{\rm{IRS}}}} - {{\mathbf{x}}^\prime }_{{\rm{IRS}}}}}{{\left( {{\lambda _{{\rm{wave}}}}} \right)}^4}}}{{{{\left( {4\pi } \right)}^4}\left( {{{({h_{{\rm{BS}}}} - {h_{{\rm{IRS}}}})}^2} + \left\| {{{{\mathbf{\tilde x}}}_{{\rm{BS}}}}({{\mathbf{x}}_{\rm{U}}}) - {{\mathbf{x}}_{{\rm{IRS}}}}} \right\|_2^2} \right)\left( {{{({h_{{\rm{IRS}}}})}^2} + \left\| {{{\mathbf{x}}_{{\rm{IRS}}}} - {{\mathbf{x}}_{\rm{U}}}} \right\|_2^2} \right)}}} \right) \times $}
\end{equation*}
\begin{equation*}
\left( {G\left( {{{{\mathbf{\tilde x}}}_{{\rm{BS}}}}({{\mathbf{x}}_{\rm{U}}}),{{\mathbf{x}}_{\rm{U}}},{{\mathbf{x}}_{{\rm{IRS}}}}} \right) + \sum\limits_{{{{\mathbf{x''}}}_{{\rm{IRS}}}} \in {\Psi _{{\rm{IRS}}}}({{{\mathbf{\tilde x}}}_{{\rm{BS}}}}({{\mathbf{x}}_{\rm{U}}})):{{\mathbf{x}}_{\rm{U}}} = {{{\mathbf{\tilde x}}}_{\rm{U}}}({{{\mathbf{x''}}}_{{\rm{IRS}}}})} {G\left( {{{{\mathbf{\tilde x}}}_{{\rm{BS}}}}({{\mathbf{x}}_{\rm{U}}}),{{{\mathbf{x''}}}_{{\rm{IRS}}}},{{\mathbf{x}}_{{\rm{IRS}}}}} \right)} } \right) \times 
\end{equation*}

\begin{equation}
\left( {G\left( {{{{\mathbf{\tilde x}}}_{{\rm{BS}}}}({{\mathbf{x}}_{\rm{U}}}),{{\mathbf{x}}_{\rm{U}}},{{{\mathbf{x'}}}_{{\rm{IRS}}}}} \right) + \sum\limits_{{{{\mathbf{x''}}}_{{\rm{IRS}}}} \in {\Psi _{{\rm{IRS}}}}({{{\mathbf{\tilde x}}}_{{\rm{BS}}}}({{\mathbf{x}}_{\rm{U}}})):{{\mathbf{x}}_{\rm{U}}} = {{{\mathbf{\tilde x}}}_{\rm{U}}}({{{\mathbf{x''}}}_{{\rm{IRS}}}})} {G\left( {{{{\mathbf{\tilde x}}}_{{\rm{BS}}}}({{\mathbf{x}}_{\rm{U}}}),{{{\mathbf{x''}}}_{{\rm{IRS}}}},{{{\mathbf{x'}}}_{{\rm{IRS}}}}} \right)} } \right)\sigma _{{\rm{d}}}^2 + 
\label{t2new}
\end{equation}
\begin{equation*}
\sum\limits_{{{\mathbf{x}}_{{\rm{IRS}}}} \in {\Psi _{{\rm{IRS}}}}({{\mathbf{x}}_{\rm{U}}}):{{{\mathbf{\tilde x}}}_{\rm{U}}}({{\mathbf{x}}_{{\rm{IRS}}}}) \ne {{\mathbf{x}}_{\rm{U}}}} {\sum\limits_{q = 1}^Q {\frac{{{{\left( {{\lambda _{{\rm{wave}}}}} \right)}^4}}}{{{{\left( {4\pi } \right)}^4}\left( {{{({h_{{\rm{IRS}}}})}^2} + \left\| {{{\mathbf{x}}_{{\rm{IRS}}}} - {{\mathbf{x}}_{\rm{U}}}} \right\|_2^2} \right)\left( {{{({h_{{\rm{BS}}}} - {h_{{\rm{IRS}}}})}^2} + \left\| {{{{\mathbf{\tilde x}}}_{{\rm{BS}}}}({{\mathbf{x}}_{\rm{U}}}) - {{\mathbf{x}}_{{\rm{IRS}}}}} \right\|_2^2} \right)}} \times } } 
\end{equation*}
\begin{equation}
{\left( {G\left( {{{{\mathbf{\tilde x}}}_{{\rm{BS}}}}({{\mathbf{x}}_{\rm{U}}}),{{\mathbf{x}}_{\rm{U}}},{{\mathbf{x}}_{{\rm{IRS}}}}} \right) + \sum\limits_{{{{\mathbf{x'}}}_{{\rm{IRS}}}} \in {\Psi _{{\rm{IRS}}}}({{{\mathbf{\tilde x}}}_{{\rm{BS}}}}({{\mathbf{x}}_{\rm{U}}})):{{\mathbf{x}}_{\rm{U}}} = {{{\mathbf{\tilde x}}}_{\rm{U}}}({{{\mathbf{x'}}}_{{\rm{IRS}}}})} {G\left( {{{{\mathbf{\tilde x}}}_{{\rm{BS}}}}({{\mathbf{x}}_{\rm{U}}}),{{{\mathbf{x'}}}_{{\rm{IRS}}}},{{\mathbf{x}}_{{\rm{IRS}}}}} \right)} } \right)^2}\sigma _{{\rm{d}}}^2,
\label{t3}
\end{equation}
where:
\begin{equation*}
{\delta _{q - q'}} = \left\{ {\begin{array}{*{20}{c}}
  {1,q = q'} \\ 
  {0,q \ne q'} 
\end{array}} \right.,
\end{equation*}
\begin{equation*}
{\delta _{{{  {\bf{x}} }_{{\rm{IRS}}}} - {{  {\bf{x}} }^\prime }_{{\rm{IRS}}}}} = \left\{ {\begin{array}{*{20}{c}}
  {1,{{  {\bf{x}} }_{{\rm{IRS}}}} = {{  {\bf{x}} }^\prime }_{{\rm{IRS}}}} \\ 
  {0,{{  {\bf{x}} }_{{\rm{IRS}}}} \ne {{  {\bf{x}} }^\prime }_{{\rm{IRS}}}} 
\end{array}} \right..
\end{equation*}
Term (\ref{t1}) can be upper bounded as follows:
\begin{equation*}
  \scalebox{.9}[1]{$\frac{{\left( {1 + \left( {\hat h - 1} \right){p_{{\rm{b}}}}} \right){{\left( {{\lambda _{{\rm{wave}}}}} \right)}^2}\sigma _{{\rm{d}}}^2}}{{{{\left( {4\pi } \right)}^2}\left( {{{({h_{{\rm{BS}}}})}^2} + \left\| {{{{\mathbf{\tilde x}}}_{{\rm{BS}}}}({{\mathbf{x}}_{\rm{U}}}) - {{\mathbf{x}}_{\rm{U}}}} \right\|_2^2} \right)}} \times {\left( {G\left( {{{{\mathbf{\tilde x}}}_{{\rm{BS}}}}({{\mathbf{x}}_{\rm{U}}}),{{\mathbf{x}}_{\rm{U}}},{{\mathbf{x}}_{\rm{U}}}} \right) + \sum\limits_{{{\mathbf{x}}_{{\rm{IRS}}}} \in {\Psi _{{\rm{IRS}}}}({{{\mathbf{\tilde x}}}_{{\rm{BS}}}}({{\mathbf{x}}_{\rm{U}}})):{{\mathbf{x}}_{\rm{U}}} = {{{\mathbf{\tilde x}}}_{\rm{U}}}({{\mathbf{x}}_{{\rm{IRS}}}})} {G\left( {{{{\mathbf{\tilde x}}}_{{\rm{BS}}}}({{\mathbf{x}}_{\rm{U}}}),{{\mathbf{x}}_{{\rm{IRS}}}},{{\mathbf{x}}_{\rm{U}}}} \right)} } \right)^2}$}
\end{equation*}
\begin{equation*}
  \scalebox{.9}[1]{$ = \frac{{\left( {1 + \left( {\hat h - 1} \right){p_{{\rm{b}}}}} \right){{\left( {{\lambda _{{\rm{wave}}}}} \right)}^2}\sigma _{{\rm{d}}}^2}}{{{{\left( {4\pi } \right)}^2}\left( {{{({h_{{\rm{BS}}}})}^2} + \left\| {{{{\mathbf{\tilde x}}}_{{\rm{BS}}}}({{\mathbf{x}}_{\rm{U}}}) - {{\mathbf{x}}_{\rm{U}}}} \right\|_2^2} \right)}} \times {\left( {1 + \sum\limits_{{{\mathbf{x}}_{{\rm{IRS}}}} \in {\Psi _{{\rm{IRS}}}}({{{\mathbf{\tilde x}}}_{{\rm{BS}}}}({{\mathbf{x}}_{\rm{U}}}))} {\left( {1 - {\pmb{\mathbb{I}}}\left( {\left\| {{{\mathbf{x}}_{\rm{U}}} - {{{\mathbf{\tilde x}}}_{\rm{U}}}({{\mathbf{x}}_{{\rm{IRS}}}})} \right\|} \right)} \right)G\left( {{{{\mathbf{\tilde x}}}_{{\rm{BS}}}}({{\mathbf{x}}_{\rm{U}}}),{{\mathbf{x}}_{{\rm{IRS}}}},{{\mathbf{x}}_{\rm{U}}}} \right)} } \right)^2}$}
\end{equation*}
\begin{equation}
  \scalebox{1}[1]{$ \le \frac{{\left( {1 + \left( {\hat h - 1} \right){p_{{\rm{b}}}}} \right){{\left( {{\lambda _{{\rm{wave}}}}} \right)}^2}\sigma _{{\rm{d}}}^2}}{{{{\left( {4\pi } \right)}^2}\left( {{{({h_{{\rm{BS}}}})}^2} + \left\| {{{{\mathbf{\tilde x}}}_{{\rm{BS}}}}({{\mathbf{x}}_{\rm{U}}}) - {{\mathbf{x}}_{\rm{U}}}} \right\|_2^2} \right)}} \times \left( {1 + \left( {2 + {n_{{\rm{IRS}}}}({{{\mathbf{\tilde x}}}_{{\rm{BS}}}}({{\mathbf{x}}_{\rm{U}}}))} \right)\sum\limits_{{{\mathbf{x}}_{{\rm{IRS}}}} \in {\Psi _{{\rm{IRS}}}}({{{\mathbf{\tilde x}}}_{{\rm{BS}}}}({{\mathbf{x}}_{\rm{U}}}))} {G\left( {{{{\mathbf{\tilde x}}}_{{\rm{BS}}}}({{\mathbf{x}}_{\rm{U}}}),{{\mathbf{x}}_{{\rm{IRS}}}},{{\mathbf{x}}_{\rm{U}}}} \right)} } \right)$}.
\end{equation}
For each IRS located at ${\bf x}_{\rm IRS}$,  ${{\pmb{\mathbb{I}}}\left( {\left\| {{{\mathbf{x}}_{\rm{U}}} - {{{\mathbf{\tilde x}}}_{\rm{U}}}({{\mathbf{x}}_{{\rm{IRS}}}})} \right\|} \right)}$ are independent binary random variables, for which we have:
\begin{equation}
\left\{ \begin{gathered}
  \Pr \left\{ {{\pmb{\mathbb{I}}}\left( {\left\| {{{\mathbf{x}}_{\rm{U}}} - {{{\mathbf{\tilde x}}}_{\rm{U}}}({{\mathbf{x}}_{{\rm{IRS}}}})} \right\|} \right) = 0} \right\} = \frac{1}{{{n_{\rm{U}}}({{\mathbf{x}}_{{\rm{IRS}}}})}} \hfill \\
  \Pr \left\{ {{\pmb{\mathbb{I}}}\left( {\left\| {{{\mathbf{x}}_{\rm{U}}} - {{{\mathbf{\tilde x}}}_{\rm{U}}}({{\mathbf{x}}_{{\rm{IRS}}}})} \right\|} \right) = 1} \right\} = 1 - \frac{1}{{{n_{\rm{U}}}({{\mathbf{x}}_{{\rm{IRS}}}})}} \hfill \\ 
\end{gathered}  \right..
\label{IRS-select}
\end{equation}
In the following, by \cite[Box 2.4]{Haenggi}, we obtain:
\begin{equation*}
  \scalebox{.88}[1]{$E\left\{ {\left. {1 + \left( {2 + {n_{{\rm{IRS}}}}({{{\mathbf{\tilde x}}}_{{\rm{BS}}}}({{\mathbf{x}}_{\rm{U}}}))} \right)\sum\limits_{{{\mathbf{x}}_{{\rm{IRS}}}} \in {\Psi _{{\rm{IRS}}}}({{{\mathbf{\tilde x}}}_{{\rm{BS}}}}({{\mathbf{x}}_{\rm{U}}}))} {G\left( {{{{\mathbf{\tilde x}}}_{{\rm{BS}}}}({{\mathbf{x}}_{\rm{U}}}),{{\mathbf{x}}_{{\rm{IRS}}}},{{\mathbf{x}}_{\rm{U}}}} \right)} } \right|{\Phi _{\rm{U}}},{\Phi _{{\rm{BS}}}},{{{\mathbf{\tilde x}}}_{{\rm{BS}}}}({{\mathbf{x}}_{\rm{U}}}),{n_{{\rm{IRS}}}}({{{\mathbf{\tilde x}}}_{{\rm{BS}}}}({{\mathbf{x}}_{\rm{U}}}))} \right\}$}
\end{equation*}
\begin{equation*}
\scalebox{1}[1]{$ = 1 + \left( {2 + {n_{{\rm{IRS}}}}({{{\mathbf{\tilde x}}}_{{\rm{BS}}}}({{\mathbf{x}}_{\rm{U}}}))} \right) \times {n_{{\rm{IRS}}}}({{{\mathbf{\tilde x}}}_{{\rm{BS}}}}({{\mathbf{x}}_{\rm{U}}}))  \times \int\limits_{{\cal C}({{{\mathbf{\tilde x}}}_{{\rm{BS}}}}({{\mathbf{x}}_{\rm{U}}}),{R_{{\rm{co}}}})} {\frac{{G\left( {{{{\mathbf{\tilde x}}}_{{\rm{BS}}}}({{\mathbf{x}}_{\rm{U}}}),{\mathbf{x}},{{\mathbf{x}}_{\rm{U}}}} \right)}}{{\pi {{\left( {{R_{{\rm{co}}}}} \right)}^2}}}d{\mathbf{x}}} $}
\end{equation*}
\begin{equation*}
 = 1 + \frac{{\left( {\left( {2 + {n_{{\rm{IRS}}}}({{{\mathbf{\tilde x}}}_{{\rm{BS}}}}({{\mathbf{x}}_{\rm{U}}}))} \right) \times {n_{{\rm{IRS}}}}({{{\mathbf{\tilde x}}}_{{\rm{BS}}}}({{\mathbf{x}}_{\rm{U}}})) } \right) \times {(\arccos \left( {1 - \varepsilon } \right)+\delta(\pi-\arccos \left( {1 - \varepsilon } \right)))}}}{{\pi }},
\end{equation*}
thus,   we have:
\begin{equation*}
\scalebox{1}[1]{$E\left\{ {\left. {1 + \left( {2 + {n_{{\rm{IRS}}}}({{{\mathbf{\tilde x}}}_{{\rm{BS}}}}({{\mathbf{x}}_{\rm{U}}}))} \right)\sum\limits_{{{\mathbf{x}}_{{\rm{IRS}}}} \in {\Psi _{{\rm{IRS}}}}({{{\mathbf{\tilde x}}}_{{\rm{BS}}}}({{\mathbf{x}}_{\rm{U}}}))} {G\left( {{{{\mathbf{\tilde x}}}_{{\rm{BS}}}}({{\mathbf{x}}_{\rm{U}}}),{{\mathbf{x}}_{{\rm{IRS}}}},{{\mathbf{x}}_{\rm{U}}}} \right)} } \right|{\Phi _{\rm{U}}},{\Phi _{{\rm{BS}}}},{{{\mathbf{\tilde x}}}_{{\rm{BS}}}}({{\mathbf{x}}_{\rm{U}}})} \right\}$}
\end{equation*}
\begin{equation}
 \scalebox{1}[1]{$\le 1 + \frac{{\left( {\left( {3{\lambda _{{\rm{IRS}}}}\pi {{\left( {{R_{{\rm{co}}}}} \right)}^2} + {{\left( {{\lambda _{{\rm{IRS}}}}\pi {{\left( {{R_{{\rm{co}}}}} \right)}^2}} \right)}^2}} \right) } \right) \times {(\arccos \left( {1 - \varepsilon } \right)+\delta(\pi-\arccos \left( {1 - \varepsilon } \right)))}}}{{\pi }}.$}
\label{t4}
\end{equation}
Now, by (\ref{t1}) and (\ref{t4}), we obtain:
\begin{equation*}
\scalebox{.85}[1]{$E\left\{ {\left. {\frac{{\left( {1 + \left( {\hat h - 1} \right){p_{{\rm{b}}}}} \right){{\left( {{\lambda _{{\rm{wave}}}}} \right)}^2}\sigma _{{\rm{d}}}^2{{\left( {G\left( {{{{\mathbf{\tilde x}}}_{{\rm{BS}}}}({{\mathbf{x}}_{\rm{U}}}),{{\mathbf{x}}_{\rm{U}}},{{\mathbf{x}}_{\rm{U}}}} \right) + \sum\limits_{{{\mathbf{x}}_{{\rm{IRS}}}} \in {\Psi _{{\rm{IRS}}}}({{{\mathbf{\tilde x}}}_{{\rm{BS}}}}({{\mathbf{x}}_{\rm{U}}})):{{\mathbf{x}}_{\rm{U}}} = {{{\mathbf{\tilde x}}}_{\rm{U}}}({{\mathbf{x}}_{{\rm{IRS}}}})} {G\left( {{{{\mathbf{\tilde x}}}_{{\rm{BS}}}}({{\mathbf{x}}_{\rm{U}}}),{{\mathbf{x}}_{{\rm{IRS}}}},{{\mathbf{x}}_{\rm{U}}}} \right)} } \right)}^2}}}{{{{\left( {4\pi } \right)}^2}\left( {{{({h_{{\rm{BS}}}})}^2} + \left\| {{{{\mathbf{\tilde x}}}_{{\rm{BS}}}}({{\mathbf{x}}_{\rm{U}}}) - {{\mathbf{x}}_{\rm{U}}}} \right\|_2^2} \right)}}} \right|{\Phi _{\rm{U}}},{n_{{\rm{BS}}}}({{\mathbf{x}}_{\rm{U}}})} \right\}$}
\end{equation*}
\begin{equation*}
\scalebox{1}[1]{$\le\left( {1 + \frac{{ {\left( {3{\lambda _{{\rm{IRS}}}}\pi {{\left( {{R_{{\rm{co}}}}} \right)}^2} + {{\left( {{\lambda _{{\rm{IRS}}}}\pi {{\left( {{R_{{\rm{co}}}}} \right)}^2}} \right)}^2}} \right) }  \times {(\arccos \left( {1 - \varepsilon } \right)+\delta(\pi-\arccos \left( {1 - \varepsilon } \right)))}}}{{\pi }}} \right) \times {\pmb{\mathbb{I}}}\left( {{n_{{\rm{BS}}}}({{\mathbf{x}}_{\rm{U}}})} \right) \times $}
\end{equation*}
\begin{equation*}
\int\limits_{{\cal C}({{\mathbf{x}}_{\rm{U}}},{R_{{\rm{co}}}})} {\frac{{\left( {1 + \left( {\hat h - 1} \right){p_{{\rm{b}}}}} \right){{\left( {{\lambda _{{\rm{wave}}}}} \right)}^2}\sigma _{{\rm{d}}}^2d{\mathbf{x}}}}{{\pi {{\left( {{R_{{\rm{co}}}}} \right)}^2}{{\left( {4\pi } \right)}^2}\left( {{{({h_{{\rm{BS}}}})}^2} + \left\| {{\mathbf{x}} - {{\mathbf{x}}_{\rm{U}}}} \right\|_2^2} \right)}}}  
\end{equation*}
\begin{equation*}
\scalebox{1}[1]{$=\left( {1 + \frac{{ {\left( {3{\lambda _{{\rm{IRS}}}}\pi {{\left( {{R_{{\rm{co}}}}} \right)}^2} + {{\left( {{\lambda _{{\rm{IRS}}}}\pi {{\left( {{R_{{\rm{co}}}}} \right)}^2}} \right)}^2}} \right) }  \times {(\arccos \left( {1 - \varepsilon } \right)+\delta(\pi-\arccos \left( {1 - \varepsilon } \right)))}}}{{\pi }}} \right) \times {\pmb{\mathbb{I}}}\left( {{n_{{\rm{BS}}}}({{\mathbf{x}}_{\rm{U}}})} \right) \times $}
\end{equation*}
\begin{equation}
\frac{{\left( {1 + \left( {\hat h - 1} \right){p_{{\rm{b}}}}} \right){{\left( {{\lambda _{{\rm{wave}}}}} \right)}^2}\sigma _{{\rm{d}}}^2}}{{ {{\left( {{R_{{\rm{co}}}}} \right)}^2}{{\left( {4\pi } \right)}^2}}}\ln \left( {1 + {{\left( {\frac{{{R_{{\rm{co}}}}}}{{{h_{{\rm{BS}}}}}}} \right)}^2}} \right).
\end{equation}
Hence, we have:
\begin{equation*}
\scalebox{1}[1]{$E\left\{ { {\frac{{\left( {1 + \left( {\hat h - 1} \right){p_{{\rm{b}}}}} \right){{\left( {{\lambda _{{\rm{wave}}}}} \right)}^2}\sigma _{{\rm{d}}}^2{{\left( {G\left( {{{{\mathbf{\tilde x}}}_{{\rm{BS}}}}({{\mathbf{x}}_{\rm{U}}}),{{\mathbf{x}}_{\rm{U}}},{{\mathbf{x}}_{\rm{U}}}} \right) + \sum\limits_{{{\mathbf{x}}_{{\rm{IRS}}}} \in {\Psi _{{\rm{IRS}}}}({{{\mathbf{\tilde x}}}_{{\rm{BS}}}}({{\mathbf{x}}_{\rm{U}}})):{{\mathbf{x}}_{\rm{U}}} = {{{\mathbf{\tilde x}}}_{\rm{U}}}({{\mathbf{x}}_{{\rm{IRS}}}})} {G\left( {{{{\mathbf{\tilde x}}}_{{\rm{BS}}}}({{\mathbf{x}}_{\rm{U}}}),{{\mathbf{x}}_{{\rm{IRS}}}},{{\mathbf{x}}_{\rm{U}}}} \right)} } \right)}^2}}}{{{{\left( {4\pi } \right)}^2}\left( {{{({h_{{\rm{BS}}}})}^2} + \left\| {{{{\mathbf{\tilde x}}}_{{\rm{BS}}}}({{\mathbf{x}}_{\rm{U}}}) - {{\mathbf{x}}_{\rm{U}}}} \right\|_2^2} \right)}}}} \right\}$}\le
\end{equation*}
\begin{equation*}
\scalebox{1}[1]{$\left( {1 + \frac{{ {\left( {3{\lambda _{{\rm{IRS}}}}\pi {{\left( {{R_{{\rm{co}}}}} \right)}^2} + {{\left( {{\lambda _{{\rm{IRS}}}}\pi {{\left( {{R_{{\rm{co}}}}} \right)}^2}} \right)}^2}} \right) }  \times {(\arccos \left( {1 - \varepsilon } \right)+\delta(\pi-\arccos \left( {1 - \varepsilon } \right)))}}}{{\pi }}} \right) \times$}
\end{equation*}
\begin{equation}
\frac{{\left( {1 + \left( {\hat h - 1} \right){p_{{\rm{b}}}}} \right){{\left( {{\lambda _{{\rm{wave}}}}} \right)}^2}\sigma _{{\rm{d}}}^2}}{{ {{\left( {{R_{{\rm{co}}}}} \right)}^2}{{\left( {4\pi } \right)}^2}}}\ln \left( {1 + {{\left( {\frac{{{R_{{\rm{co}}}}}}{{{h_{{\rm{BS}}}}}}} \right)}^2}} \right)\times  \left( {1 - \exp \left\{ -{{\lambda _{{\rm{BS}}}}\pi {{\left( {{R_{{\rm{co}}}}} \right)}^2}} \right\}} \right).
\label{t6}
\end{equation}

Now, to bound term (\ref{t2}), first, we bound the following term:
\begin{equation*}
{\frac{\kappa }{{\kappa  + 1}}}\sum\limits_{{{\mathbf{x}}_{{\rm{IRS}}}} \in {\Psi _{{\rm{IRS}}}}({{\mathbf{x}}_{\rm{U}}}):{{{\mathbf{\tilde x}}}_{\rm{U}}}({{\mathbf{x}}_{{\rm{IRS}}}}) = {{\mathbf{x}}_{\rm{U}}}} {\sum\limits_{q = 1}^Q {\sum\limits_{{{{\mathbf{x'}}}_{{\rm{IRS}}}} \in {\Psi _{{\rm{IRS}}}}({{\mathbf{x}}_{\rm{U}}}):{{{\mathbf{\tilde x}}}_{\rm{U}}}({{{\mathbf{x'}}}_{{\rm{IRS}}}}) = {{\mathbf{x}}_{\rm{U}}}} {\sum\limits_{q' = 1}^Q {} } } } 
\end{equation*}
\begin{equation*}
\left( {\frac{{{\delta _{q - q'}}{\delta _{{{\mathbf{x}}_{{\rm{IRS}}}} - {{\mathbf{x}}^\prime }_{{\rm{IRS}}}}}{{\left( {{\lambda _{{\rm{wave}}}}} \right)}^4}}}{{{{\left( {4\pi } \right)}^4}\left( {{{({h_{{\rm{BS}}}} - {h_{{\rm{IRS}}}})}^2} + \left\| {{{{\mathbf{\tilde x}}}_{{\rm{BS}}}}({{\mathbf{x}}_{\rm{U}}}) - {{\mathbf{x}}_{{\rm{IRS}}}}} \right\|_2^2} \right)\left( {{{({h_{{\rm{IRS}}}})}^2} + \left\| {{{\mathbf{x}}_{{\rm{IRS}}}} - {{\mathbf{x}}_{\rm{U}}}} \right\|_2^2} \right)}}} \right) \times 
\end{equation*}
\begin{equation*}
\left( {G\left( {{{{\mathbf{\tilde x}}}_{{\rm{BS}}}}({{\mathbf{x}}_{\rm{U}}}),{{\mathbf{x}}_{\rm{U}}},{{\mathbf{x}}_{{\rm{IRS}}}}} \right) + \sum\limits_{{{{\mathbf{x''}}}_{{\rm{IRS}}}} \in {\Psi _{{\rm{IRS}}}}({{{\mathbf{\tilde x}}}_{{\rm{BS}}}}({{\mathbf{x}}_{\rm{U}}})):{{\mathbf{x}}_{\rm{U}}} = {{{\mathbf{\tilde x}}}_{\rm{U}}}({{{\mathbf{x''}}}_{{\rm{IRS}}}})} {G\left( {{{{\mathbf{\tilde x}}}_{{\rm{BS}}}}({{\mathbf{x}}_{\rm{U}}}),{{{\mathbf{x''}}}_{{\rm{IRS}}}},{{\mathbf{x}}_{{\rm{IRS}}}}} \right)} } \right) \times 
\end{equation*}
\begin{equation*}
\left( {G\left( {{{{\mathbf{\tilde x}}}_{{\rm{BS}}}}({{\mathbf{x}}_{\rm{U}}}),{{\mathbf{x}}_{\rm{U}}},{{{\mathbf{x'}}}_{{\rm{IRS}}}}} \right) + \sum\limits_{{{{\mathbf{x''}}}_{{\rm{IRS}}}} \in {\Psi _{{\rm{IRS}}}}({{{\mathbf{\tilde x}}}_{{\rm{BS}}}}({{\mathbf{x}}_{\rm{U}}})):{{\mathbf{x}}_{\rm{U}}} = {{{\mathbf{\tilde x}}}_{\rm{U}}}({{{\mathbf{x''}}}_{{\rm{IRS}}}})} {G\left( {{{{\mathbf{\tilde x}}}_{{\rm{BS}}}}({{\mathbf{x}}_{\rm{U}}}),{{{\mathbf{x''}}}_{{\rm{IRS}}}},{{{\mathbf{x'}}}_{{\rm{IRS}}}}} \right)} } \right)\sigma _{{\rm{d}}}^2
\end{equation*}
\begin{equation*}
 \scalebox{1}[1]{$ ={\frac{\kappa }{{\kappa  + 1}}} \sum\limits_{{{\mathbf{x}}_{{\rm{IRS}}}} \in {\Psi _{{\rm{IRS}}}}({{\mathbf{x}}_{\rm{U}}}):{{{\mathbf{\tilde x}}}_{\rm{U}}}({{\mathbf{x}}_{{\rm{IRS}}}}) = {{\mathbf{x}}_{\rm{U}}}} {\sum\limits_{q = 1}^Q {\left( {\frac{{{{\left( {{\lambda _{{\rm{wave}}}}} \right)}^4}}}{{{{\left( {4\pi } \right)}^4}\left( {{{({h_{{\rm{BS}}}} - {h_{{\rm{IRS}}}})}^2} + \left\| {{{{\mathbf{\tilde x}}}_{{\rm{BS}}}}({{\mathbf{x}}_{\rm{U}}}) - {{\mathbf{x}}_{{\rm{IRS}}}}} \right\|_2^2} \right)\left( {{{({h_{{\rm{IRS}}}})}^2} + \left\| {{{\mathbf{x}}_{{\rm{IRS}}}} - {{\mathbf{x}}_{\rm{U}}}} \right\|_2^2} \right)}}} \right) \times } } $}
\end{equation*}
\begin{equation*}
{\left( {G\left( {{{{\mathbf{\tilde x}}}_{{\rm{BS}}}}({{\mathbf{x}}_{\rm{U}}}),{{\mathbf{x}}_{\rm{U}}},{{\mathbf{x}}_{{\rm{IRS}}}}} \right) + \sum\limits_{{{{\mathbf{x''}}}_{{\rm{IRS}}}} \in {\Psi _{{\rm{IRS}}}}({{{\mathbf{\tilde x}}}_{{\rm{BS}}}}({{\mathbf{x}}_{\rm{U}}})):{{\mathbf{x}}_{\rm{U}}} = {{{\mathbf{\tilde x}}}_{\rm{U}}}({{{\mathbf{x''}}}_{{\rm{IRS}}}})} {G\left( {{{{\mathbf{\tilde x}}}_{{\rm{BS}}}}({{\mathbf{x}}_{\rm{U}}}),{{{\mathbf{x''}}}_{{\rm{IRS}}}},{{\mathbf{x}}_{{\rm{IRS}}}}} \right)} } \right)^2}\sigma _{{\rm{d}}}^2
\end{equation*}
\begin{equation*}
={\frac{\kappa }{{\kappa  + 1}}} \sum\limits_{{{\mathbf{x}}_{{\rm{IRS}}}} \in {\Psi _{{\rm{IRS}}}}({{\mathbf{x}}_{\rm{U}}})} {\sum\limits_{q = 1}^Q {\left( {\frac{{\left( {1 - {\pmb{\mathbb{I}}}\left( {\left\| {{{\mathbf{x}}_{\rm{U}}} - {{{\mathbf{\tilde x}}}_{\rm{U}}}({{\mathbf{x}}_{{\rm{IRS}}}})} \right\|} \right)} \right){{\left( {{\lambda _{{\rm{wave}}}}} \right)}^4}}}{{{{\left( {4\pi } \right)}^4}\left( {{{({h_{{\rm{BS}}}} - {h_{{\rm{IRS}}}})}^2} + \left\| {{{{\mathbf{\tilde x}}}_{{\rm{BS}}}}({{\mathbf{x}}_{\rm{U}}}) - {{\mathbf{x}}_{{\rm{IRS}}}}} \right\|_2^2} \right)\left( {{{({h_{{\rm{IRS}}}})}^2} + \left\| {{{\mathbf{x}}_{{\rm{IRS}}}} - {{\mathbf{x}}_{\rm{U}}}} \right\|_2^2} \right)}}} \right) \times } } 
\end{equation*}
\begin{equation*}
 \scalebox{.95}[1]{${\left( {G\left( {{{{\mathbf{\tilde x}}}_{{\rm{BS}}}}({{\mathbf{x}}_{\rm{U}}}),{{\mathbf{x}}_{\rm{U}}},{{\mathbf{x}}_{{\rm{IRS}}}}} \right) + \sum\limits_{{{{\mathbf{x''}}}_{{\rm{IRS}}}} \in {\Psi _{{\rm{IRS}}}}({{{\mathbf{\tilde x}}}_{{\rm{BS}}}}({{\mathbf{x}}_{\rm{U}}}))} {\left( {1 - {\pmb{\mathbb{I}}}\left( {\left\| {{{\mathbf{x}}_{\rm{U}}} - {{{\mathbf{\tilde x}}}_{\rm{U}}}({{{\mathbf{x''}}}_{{\rm{IRS}}}})} \right\|} \right)} \right)G\left( {{{{\mathbf{\tilde x}}}_{{\rm{BS}}}}({{\mathbf{x}}_{\rm{U}}}),{{{\mathbf{x''}}}_{{\rm{IRS}}}},{{\mathbf{x}}_{{\rm{IRS}}}}} \right)} } \right)^2}\sigma _{{\rm{d}}}^2$}
\end{equation*}
\begin{equation*}
 \le {\frac{\kappa }{{\kappa  + 1}}} \sum\limits_{{{\mathbf{x}}_{{\rm{IRS}}}} \in {\Psi _{{\rm{IRS}}}}({{\mathbf{x}}_{\rm{U}}})} {\sum\limits_{q = 1}^Q {\left( {\frac{{\left( {1 - {\pmb{\mathbb{I}}}\left( {\left\| {{{\mathbf{x}}_{\rm{U}}} - {{{\mathbf{\tilde x}}}_{\rm{U}}}({{\mathbf{x}}_{{\rm{IRS}}}})} \right\|} \right)} \right){{\left( {{\lambda _{{\rm{wave}}}}} \right)}^4}}}{{{{\left( {4\pi } \right)}^4}\left( {{{({h_{{\rm{BS}}}} - {h_{{\rm{IRS}}}})}^2} + \left\| {{{{\mathbf{\tilde x}}}_{{\rm{BS}}}}({{\mathbf{x}}_{\rm{U}}}) - {{\mathbf{x}}_{{\rm{IRS}}}}} \right\|_2^2} \right)\left( {{{({h_{{\rm{IRS}}}})}^2} + \left\| {{{\mathbf{x}}_{{\rm{IRS}}}} - {{\mathbf{x}}_{\rm{U}}}} \right\|_2^2} \right)}}} \right) \times } } 
\end{equation*}
\begin{equation*}
\scalebox{.83}[1]{$\left( {G\left( {{{{\mathbf{\tilde x}}}_{{\rm{BS}}}}({{\mathbf{x}}_{\rm{U}}}),{{\mathbf{x}}_{\rm{U}}},{{\mathbf{x}}_{{\rm{IRS}}}}} \right) + \left( {2 + {n_{{\rm{IRS}}}}({{{\mathbf{\tilde x}}}_{{\rm{BS}}}}({{\mathbf{x}}_{\rm{U}}}))} \right)\sum\limits_{{{{\mathbf{x''}}}_{{\rm{IRS}}}} \in {\Psi _{{\rm{IRS}}}}({{{\mathbf{\tilde x}}}_{{\rm{BS}}}}({{\mathbf{x}}_{\rm{U}}}))} {\left( {1 - {\pmb{\mathbb{I}}}\left( {\left\| {{{\mathbf{x}}_{\rm{U}}} - {{{\mathbf{\tilde x}}}_{\rm{U}}}({{{\mathbf{x''}}}_{{\rm{IRS}}}})} \right\|} \right)} \right)G\left( {{{{\mathbf{\tilde x}}}_{{\rm{BS}}}}({{\mathbf{x}}_{\rm{U}}}),{{{\mathbf{x''}}}_{{\rm{IRS}}}},{{\mathbf{x}}_{{\rm{IRS}}}}} \right)} } \right)\sigma _{{\rm{d}}}^2$}
\end{equation*}
\begin{equation*}
 \le  {\frac{\kappa }{{\kappa  + 1}}}\sum\limits_{{{\mathbf{x}}_{{\rm{IRS}}}} \in {\Psi _{{\rm{IRS}}}}({{\mathbf{x}}_{\rm{U}}})} {\sum\limits_{q = 1}^Q {\left( {\frac{{\left( {1 - {\pmb{\mathbb{I}}}\left( {\left\| {{{\mathbf{x}}_{\rm{U}}} - {{{\mathbf{\tilde x}}}_{\rm{U}}}({{\mathbf{x}}_{{\rm{IRS}}}})} \right\|} \right)} \right){{\left( {{\lambda _{{\rm{wave}}}}} \right)}^4}}}{{{{\left( {4\pi } \right)}^4}\left( {{{({h_{{\rm{BS}}}} - {h_{{\rm{IRS}}}})}^2} + \left\| {{{{\mathbf{\tilde x}}}_{{\rm{BS}}}}({{\mathbf{x}}_{\rm{U}}}) - {{\mathbf{x}}_{{\rm{IRS}}}}} \right\|_2^2} \right)\left( {{{({h_{{\rm{IRS}}}})}^2} + \left\| {{{\mathbf{x}}_{{\rm{IRS}}}} - {{\mathbf{x}}_{\rm{U}}}} \right\|_2^2} \right)}}} \right) \times } } 
\end{equation*}
\begin{equation*}
\left( {G\left( {{{{\mathbf{\tilde x}}}_{{\rm{BS}}}}({{\mathbf{x}}_{\rm{U}}}),{{\mathbf{x}}_{\rm{U}}},{{\mathbf{x}}_{{\rm{IRS}}}}} \right) + \left( {2 + {n_{{\rm{IRS}}}}({{{\mathbf{\tilde x}}}_{{\rm{BS}}}}({{\mathbf{x}}_{\rm{U}}}))} \right)\sum\limits_{{{{\mathbf{x''}}}_{{\rm{IRS}}}} \in {\Psi _{{\rm{IRS}}}}({{{\mathbf{\tilde x}}}_{{\rm{BS}}}}({{\mathbf{x}}_{\rm{U}}}))} {G\left( {{{{\mathbf{\tilde x}}}_{{\rm{BS}}}}({{\mathbf{x}}_{\rm{U}}}),{{{\mathbf{x''}}}_{{\rm{IRS}}}},{{\mathbf{x}}_{{\rm{IRS}}}}} \right)} } \right)\sigma _{{\rm{d}}}^2
\end{equation*}

\begin{equation*}
\le {\frac{\kappa }{{\kappa  + 1}}} \sum\limits_{{{\mathbf{x}}_{{\rm{IRS}}}} \in {\Psi _{{\rm{IRS}}}}({{\mathbf{x}}_{\rm{U}}})} {\left( {\frac{{Q\left( {1 - {\pmb{\mathbb{I}}}\left( {\left\| {{{\mathbf{x}}_{\rm{U}}} - {{{\mathbf{\tilde x}}}_{\rm{U}}}({{\mathbf{x}}_{{\rm{IRS}}}})} \right\|} \right)} \right){{\left( {{\lambda _{{\rm{wave}}}}} \right)}^4}}}{{{{\left( {4\pi } \right)}^4}\left( {{{({h_{{\rm{BS}}}} - {h_{{\rm{IRS}}}})}^2} + \left\| {{{{\mathbf{\tilde x}}}_{{\rm{BS}}}}({{\mathbf{x}}_{\rm{U}}}) - {{\mathbf{x}}_{{\rm{IRS}}}}} \right\|_2^2} \right)\left( {{{({h_{{\rm{IRS}}}})}^2} + \left\| {{{\mathbf{x}}_{{\rm{IRS}}}} - {{\mathbf{x}}_{\rm{U}}}} \right\|_2^2} \right)}}} \right) \times } 
\end{equation*}
\begin{equation*}
\left( {G\left( {{{{\mathbf{\tilde x}}}_{{\rm{BS}}}}({{\mathbf{x}}_{\rm{U}}}),{{\mathbf{x}}_{\rm{U}}},{{\mathbf{x}}_{{\rm{IRS}}}}} \right) + \left( {2 + {n_{{\rm{IRS}}}}({{{\mathbf{\tilde x}}}_{{\rm{BS}}}}({{\mathbf{x}}_{\rm{U}}}))} \right)\sum\limits_{{{{\mathbf{x''}}}_{{\rm{IRS}}}} \in {\Psi _{{\rm{IRS}}}}({{{\mathbf{\tilde x}}}_{{\rm{BS}}}}({{\mathbf{x}}_{\rm{U}}}))} {G\left( {{{{\mathbf{\tilde x}}}_{{\rm{BS}}}}({{\mathbf{x}}_{\rm{U}}}),{{{\mathbf{x''}}}_{{\rm{IRS}}}},{{\mathbf{x}}_{{\rm{IRS}}}}} \right)} } \right)\sigma _{{\rm{d}}}^2
\end{equation*}
\begin{equation*}
 \le {\frac{\kappa }{{\kappa  + 1}}} \sum\limits_{{{\mathbf{x}}_{{\rm{IRS}}}} \in {{\tilde \Psi }_{{\rm{IRS}}}}({{\mathbf{x}}_{\rm{U}}})} {\frac{{Q{{\left( {{\lambda _{{\rm{wave}}}}} \right)}^4}{\pmb{\mathbb{I}}}\left( {{n_{{\rm{BS}}}}({{\mathbf{x}}_{\rm{U}}})} \right)\left( {1 + 2{{\tilde n}_{{\rm{IRS}}}}({{\mathbf{x}}_{\rm{U}}}) + {{\left( {{{\tilde n}_{{\rm{IRS}}}}({{\mathbf{x}}_{\rm{U}}})} \right)}^2}} \right)\sigma _{{\rm{d}}}^2}}{{{{\left( {4\pi } \right)}^4}\left( {{{({h_{{\rm{BS}}}} - {h_{{\rm{IRS}}}})}^2} + \left\| {{{{\mathbf{\tilde x}}}_{{\rm{BS}}}}({{\mathbf{x}}_{\rm{U}}}) - {{\mathbf{x}}_{{\rm{IRS}}}}} \right\|_2^2} \right)\left( {{{({h_{{\rm{IRS}}}})}^2} + \left\| {{{\mathbf{x}}_{{\rm{IRS}}}} - {{\mathbf{x}}_{\rm{U}}}} \right\|_2^2} \right)}}} .
\end{equation*}

Thus, we obtain:
\begin{equation*}
\scalebox{.95}[1]{$E\left\{{\frac{\kappa }{{\kappa  + 1}}}{\left. {\sum\limits_{{{\mathbf{x}}_{{\rm{IRS}}}} \in {{\tilde \Psi }_{{\rm{IRS}}}}({{\mathbf{x}}_{\rm{U}}})} {\frac{{Q{{\left( {{\lambda _{{\rm{wave}}}}} \right)}^4}{\pmb{\mathbb{I}}}\left( {{n_{{\rm{BS}}}}({{\mathbf{x}}_{\rm{U}}})} \right)\left( {1 + 2{{\tilde n}_{{\rm{IRS}}}}({{\mathbf{x}}_{\rm{U}}}) + {{\left( {{{\tilde n}_{{\rm{IRS}}}}({{\mathbf{x}}_{\rm{U}}})} \right)}^2}} \right)\sigma _{{\rm{d}}}^2}}{{{{\left( {4\pi } \right)}^4}\left( {{{({h_{{\rm{BS}}}} - {h_{{\rm{IRS}}}})}^2} + \left\| {{{{\mathbf{\tilde x}}}_{{\rm{BS}}}}({{\mathbf{x}}_{\rm{U}}}) - {{\mathbf{x}}_{{\rm{IRS}}}}} \right\|_2^2} \right)\left( {{{({h_{{\rm{IRS}}}})}^2} + \left\| {{{\mathbf{x}}_{{\rm{IRS}}}} - {{\mathbf{x}}_{\rm{U}}}} \right\|_2^2} \right)}}} } \right|{\Phi _{\rm{U}}},{\Phi _{{\rm{BS}}}},{{\tilde n}_{{\rm{IRS}}}}({{\mathbf{x}}_{\rm{U}}}),{{{\mathbf{\tilde x}}}_{{\rm{BS}}}}({{\mathbf{x}}_{\rm{U}}})} \right\}$}
\end{equation*}
\begin{equation*}
 ={\frac{\kappa }{{\kappa  + 1}}}\times \frac{{Q{{\left( {{\lambda _{{\rm{wave}}}}} \right)}^4}{\pmb{\mathbb{I}}}\left( {{n_{{\rm{BS}}}}({{\mathbf{x}}_{\rm{U}}})} \right)\left( {{{\tilde n}_{{\rm{IRS}}}}({{\mathbf{x}}_{\rm{U}}}) + 2{{\left( {{{\tilde n}_{{\rm{IRS}}}}({{\mathbf{x}}_{\rm{U}}})} \right)}^2} + {{\left( {{{\tilde n}_{{\rm{IRS}}}}({{\mathbf{x}}_{\rm{U}}})} \right)}^3}} \right)\sigma _{{\rm{d}}}^2}}{{{{\left( {4\pi } \right)}^5}{{\left( {{R_{{\rm{co}}}}} \right)}^2}}} \times 
\end{equation*}
\begin{equation*}
\int\limits_{{\cal C}({{\mathbf{x}}_{\rm{U}}},2{R_{{\rm{co}}}})} {\frac{{d{\mathbf{x}}}}{{\left( {{{({h_{{\rm{BS}}}} - {h_{{\rm{IRS}}}})}^2} + \left\| {{\mathbf{x}} - {{{\mathbf{\tilde x}}}_{{\rm{BS}}}}({{\mathbf{x}}_{\rm{U}}})} \right\|_2^2} \right)\left( {{{({h_{{\rm{IRS}}}})}^2} + \left\| {{\mathbf{x}} - {{\mathbf{x}}_{\rm{U}}}} \right\|_2^2} \right)}}} 
\end{equation*}
\begin{equation*}
\le {\frac{\kappa }{{\kappa  + 1}}}\times \frac{{Q{{\left( {{\lambda _{{\rm{wave}}}}} \right)}^4}{\pmb{\mathbb{I}}}\left( {{n_{{\rm{BS}}}}({{\mathbf{x}}_{\rm{U}}})} \right)\left( {{{\tilde n}_{{\rm{IRS}}}}({{\mathbf{x}}_{\rm{U}}}) + 2{{\left( {{{\tilde n}_{{\rm{IRS}}}}({{\mathbf{x}}_{\rm{U}}})} \right)}^2} + {{\left( {{{\tilde n}_{{\rm{IRS}}}}({{\mathbf{x}}_{\rm{U}}})} \right)}^3}} \right)\sigma _{{\rm{d}}}^2}}{{{{\left( {4\pi } \right)}^5}{{\left( {{R_{{\rm{co}}}}} \right)}^2}}} \times 
\end{equation*}
\begin{equation*}
\sqrt {\int\limits_{{\cal C}({{\mathbf{x}}_{\rm{U}}},2{R_{{\rm{co}}}})} {\frac{{d{\mathbf{x}}}}{{{{\left( {{{({h_{{\rm{BS}}}} - {h_{{\rm{IRS}}}})}^2} + \left\| {{\mathbf{x}} - {{{\mathbf{\tilde x}}}_{{\rm{BS}}}}({{\mathbf{x}}_{\rm{U}}})} \right\|_2^2} \right)}^2}}}} }  \times \sqrt {\int\limits_{{\cal C}({{\mathbf{x}}_{\rm{U}}},2{R_{{\rm{co}}}})} {\frac{{d{\mathbf{x}}}}{{{{\left( {{{({h_{{\rm{IRS}}}})}^2} + \left\| {{\mathbf{x}} - {{\mathbf{x}}_{\rm{U}}}} \right\|_2^2} \right)}^2}}}} } 
\end{equation*}
\begin{equation*}
\le  {\frac{\kappa }{{\kappa  + 1}}}\times \frac{{Q{{\left( {{\lambda _{{\rm{wave}}}}} \right)}^4}{\pmb{\mathbb{I}}}\left( {{n_{{\rm{BS}}}}({{\mathbf{x}}_{\rm{U}}})} \right)\left( {{{\tilde n}_{{\rm{IRS}}}}({{\mathbf{x}}_{\rm{U}}}) + 2{{\left( {{{\tilde n}_{{\rm{IRS}}}}({{\mathbf{x}}_{\rm{U}}})} \right)}^2} + {{\left( {{{\tilde n}_{{\rm{IRS}}}}({{\mathbf{x}}_{\rm{U}}})} \right)}^3}} \right)\sigma _{{\rm{d}}}^2}}{{{{\left( {4\pi } \right)}^5}{{\left( {{R_{{\rm{co}}}}} \right)}^2}}} \times 
\end{equation*}
\begin{equation*}
\sqrt {\int\limits_{{\cal C}({{{\mathbf{\tilde x}}}_{{\rm{BS}}}}({{\mathbf{x}}_{\rm{U}}}),3{R_{{\rm{co}}}})} {\frac{{d{\mathbf{x}}}}{{{{\left( {{{({h_{{\rm{BS}}}} - {h_{{\rm{IRS}}}})}^2} + \left\| {{\mathbf{x}} - {{{\mathbf{\tilde x}}}_{{\rm{BS}}}}({{\mathbf{x}}_{\rm{U}}})} \right\|_2^2} \right)}^2}}}} }  \times \sqrt {\int\limits_{{\cal C}({{\mathbf{x}}_{\rm{U}}},2{R_{{\rm{co}}}})} {\frac{{d{\mathbf{x}}}}{{{{\left( {{{({h_{{\rm{IRS}}}})}^2} + \left\| {{\mathbf{x}} - {{\mathbf{x}}_{\rm{U}}}} \right\|_2^2} \right)}^2}}}} } 
\end{equation*}
\begin{equation*}
 =  {\frac{\kappa }{{\kappa  + 1}}}\times \frac{{Q{{\left( {{\lambda _{{\rm{wave}}}}} \right)}^4}{\pmb{\mathbb{I}}}\left( {{n_{{\rm{BS}}}}({{\mathbf{x}}_{\rm{U}}})} \right)\left( {{{\tilde n}_{{\rm{IRS}}}}({{\mathbf{x}}_{\rm{U}}}) + 2{{\left( {{{\tilde n}_{{\rm{IRS}}}}({{\mathbf{x}}_{\rm{U}}})} \right)}^2} + {{\left( {{{\tilde n}_{{\rm{IRS}}}}({{\mathbf{x}}_{\rm{U}}})} \right)}^3}} \right)\sigma _{{\rm{d}}}^2}}{{{{\left( {4\pi } \right)}^5}{{\left( {{R_{{\rm{co}}}}} \right)}^2}}} \times 
\end{equation*}
\begin{equation*}
\sqrt {\frac{{36{\pi ^2}{{\left( {{R_{{\rm{co}}}}} \right)}^4}}}{{\left( {{{({h_{{\rm{BS}}}} - {h_{{\rm{IRS}}}})}^2} + 9{{\left( {{R_{{\rm{co}}}}} \right)}^2}} \right)\left( {{{({h_{{\rm{IRS}}}})}^2} + 4{{\left( {{R_{{\rm{co}}}}} \right)}^2}} \right){{({h_{{\rm{BS}}}} - {h_{{\rm{IRS}}}})}^2}{{({h_{{\rm{IRS}}}})}^2}}}} 
\end{equation*}
\begin{equation*}
 = {\frac{\kappa }{{\kappa  + 1}}}\times \frac{{Q{{\left( {{\lambda _{{\rm{wave}}}}} \right)}^4}{\pmb{\mathbb{I}}}\left( {{n_{{\rm{BS}}}}({{\mathbf{x}}_{\rm{U}}})} \right)\left( {{{\tilde n}_{{\rm{IRS}}}}({{\mathbf{x}}_{\rm{U}}}) + 2{{\left( {{{\tilde n}_{{\rm{IRS}}}}({{\mathbf{x}}_{\rm{U}}})} \right)}^2} + {{\left( {{{\tilde n}_{{\rm{IRS}}}}({{\mathbf{x}}_{\rm{U}}})} \right)}^3}} \right)\sigma _{{\rm{d}}}^2}}{{{{\left( {4\pi } \right)}^4}}} \times 
\end{equation*}
\begin{equation*}
\sqrt {\frac{9}{{4\left( {{{({h_{{\rm{BS}}}} - {h_{{\rm{IRS}}}})}^2} + 9{{\left( {{R_{{\rm{co}}}}} \right)}^2}} \right)\left( {{{({h_{{\rm{IRS}}}})}^2} + 4{{\left( {{R_{{\rm{co}}}}} \right)}^2}} \right){{({h_{{\rm{BS}}}} - {h_{{\rm{IRS}}}})}^2}{{({h_{{\rm{IRS}}}})}^2}}}} .
\end{equation*}

By \cite{Ahle},
for a poisson random variable  $W$ with parameter $\lambda$ and distribution  $P_{\lambda}(w)=\frac{{{\lambda ^w}{e^{ - \lambda }}}}{{w!}}$, we have:
\begin{equation}
E\left\{ {{W^k}} \right\} \le {\left( {\frac{k}{{\log \left( {1 + \frac{k}{\lambda }} \right)}}} \right)^k} \le {\lambda ^k}\exp \left\{ {\frac{{{k^2}}}{{2\lambda }}} \right\},
\label{lambda-k}
\end{equation}
hence, we obtain:

\begin{equation*}
E\left\{ {\frac{\kappa }{{\kappa  + 1}}}\times {\frac{{Q{{\left( {{\lambda _{{\rm{wave}}}}} \right)}^4}{\pmb{\mathbb{I}}}\left( {{n_{{\rm{BS}}}}({{\mathbf{x}}_{\rm{U}}})} \right)\left( {{{\tilde n}_{{\rm{IRS}}}}({{\mathbf{x}}_{\rm{U}}}) + 2{{\left( {{{\tilde n}_{{\rm{IRS}}}}({{\mathbf{x}}_{\rm{U}}})} \right)}^2} + {{\left( {{{\tilde n}_{{\rm{IRS}}}}({{\mathbf{x}}_{\rm{U}}})} \right)}^3}} \right)\sigma _{{\rm{d}}}^2}}{{{{\left( {4\pi } \right)}^4}}} \times } \right.
\end{equation*}
\begin{equation*}
 \scalebox{.95}[1]{$\left. {\left. {\sqrt {\frac{9}{{4\left( {{{({h_{{\rm{BS}}}} - {h_{{\rm{IRS}}}})}^2} + 9{{\left( {{R_{{\rm{co}}}}} \right)}^2}} \right)\left( {{{({h_{{\rm{IRS}}}})}^2} + 4{{\left( {{R_{{\rm{co}}}}} \right)}^2}} \right){{({h_{{\rm{BS}}}} - {h_{{\rm{IRS}}}})}^2}{{({h_{{\rm{IRS}}}})}^2}}}} } \right|{\Phi _{\rm{U}}},{\Phi _{{\rm{BS}}}},{{{\mathbf{\tilde x}}}_{{\rm{BS}}}}({{\mathbf{x}}_{\rm{U}}})} \right\}$}
\end{equation*}
\begin{equation*}
\le {\frac{\kappa }{{\kappa  + 1}}}\times\frac{{3{\lambda _{{\rm{IRS}}}}\pi {{\left( {2{R_{{\rm{co}}}}} \right)}^2} + {{\left( {{\lambda _{{\rm{IRS}}}}\pi {{\left( {2{R_{{\rm{co}}}}} \right)}^2}} \right)}^2} + {{\left( {{\lambda _{{\rm{IRS}}}}\pi {{\left( {2{R_{{\rm{co}}}}} \right)}^2}} \right)}^3}\exp \left\{ {\frac{9}{{2{\lambda _{{\rm{IRS}}}}\pi {{\left( {2{R_{{\rm{co}}}}} \right)}^2}}}} \right\}}}{{{{\left( {4\pi } \right)}^4}}} \times 
\end{equation*}
\begin{equation*}
Q{\left( {{\lambda _{{\rm{wave}}}}} \right)^4}{\pmb{\mathbb{I}}}\left( {{n_{{\rm{BS}}}}({{\mathbf{x}}_{\rm{U}}})} \right)\sigma _{{\rm{d}}}^2\times 
\end{equation*}
\begin{equation*}
\sqrt {\frac{9}{{4\left( {{{({h_{{\rm{BS}}}} - {h_{{\rm{IRS}}}})}^2} + 9{{\left( {{R_{{\rm{co}}}}} \right)}^2}} \right)\left( {{{({h_{{\rm{IRS}}}})}^2} + 4{{\left( {{R_{{\rm{co}}}}} \right)}^2}} \right){{({h_{{\rm{BS}}}} - {h_{{\rm{IRS}}}})}^2}{{({h_{{\rm{IRS}}}})}^2}}}},
\end{equation*}

thus, we have:
\begin{equation*}
E\left\{ {\frac{\kappa }{{\kappa  + 1}}}\times{\frac{{3{\lambda _{{\rm{IRS}}}}\pi {{\left( {2{R_{{\rm{co}}}}} \right)}^2} + {{\left( {{\lambda _{{\rm{IRS}}}}\pi {{\left( {2{R_{{\rm{co}}}}} \right)}^2}} \right)}^2} + {{\left( {{\lambda _{{\rm{IRS}}}}\pi {{\left( {2{R_{{\rm{co}}}}} \right)}^2}} \right)}^3}\exp \left\{ {\frac{9}{{2{\lambda _{{\rm{IRS}}}}\pi {{\left( {2{R_{{\rm{co}}}}} \right)}^2}}}} \right\}}}{{{{\left( {4\pi } \right)}^4}}} \times } \right.
\end{equation*}
\begin{equation*}
\left. {\frac{{3Q{{\left( {{\lambda _{{\rm{wave}}}}} \right)}^4}{\pmb{\mathbb{I}}}\left( {{n_{{\rm{BS}}}}({{\mathbf{x}}_{\rm{U}}})} \right)\sigma _{{\rm{d}}}^2}}{{\sqrt {4\left( {{{({h_{{\rm{BS}}}} - {h_{{\rm{IRS}}}})}^2} + 9{{\left( {{R_{{\rm{co}}}}} \right)}^2}} \right)\left( {{{({h_{{\rm{IRS}}}})}^2} + 4{{\left( {{R_{{\rm{co}}}}} \right)}^2}} \right){{({h_{{\rm{BS}}}} - {h_{{\rm{IRS}}}})}^2}{{({h_{{\rm{IRS}}}})}^2}} }}} \right\} \le
\end{equation*}
\begin{equation*}
{\frac{\kappa }{{\kappa  + 1}}}\times\frac{{3{\lambda _{{\rm{IRS}}}}\pi {{\left( {2{R_{{\rm{co}}}}} \right)}^2} + {{\left( {{\lambda _{{\rm{IRS}}}}\pi {{\left( {2{R_{{\rm{co}}}}} \right)}^2}} \right)}^2} + {{\left( {{\lambda _{{\rm{IRS}}}}\pi {{\left( {2{R_{{\rm{co}}}}} \right)}^2}} \right)}^3}\exp \left\{ {\frac{9}{{2{\lambda _{{\rm{IRS}}}}\pi {{\left( {2{R_{{\rm{co}}}}} \right)}^2}}}} \right\}}}{{{{\left( {4\pi } \right)}^4}}} \times 
\end{equation*}
\begin{equation}
\frac{{3Q{{\left( {{\lambda _{{\rm{wave}}}}} \right)}^4}\left( {1 - \exp \left\{ { - {\lambda _{{\rm{BS}}}}\pi {{\left( {{R_{{\rm{co}}}}} \right)}^2}} \right\}} \right)\sigma _{{\rm{d}}}^2}}{{\sqrt {4\left( {{{({h_{{\rm{BS}}}} - {h_{{\rm{IRS}}}})}^2} + 9{{\left( {{R_{{\rm{co}}}}} \right)}^2}} \right)\left( {{{({h_{{\rm{IRS}}}})}^2} + 4{{\left( {{R_{{\rm{co}}}}} \right)}^2}} \right){{({h_{{\rm{BS}}}} - {h_{{\rm{IRS}}}})}^2}{{({h_{{\rm{IRS}}}})}^2}} }}.
\label{t8}
\end{equation}

Next, in (\ref{t2}), we bound the following term:
\begin{equation*}
{\frac{\kappa }{{\kappa  + 1}}}\times\sum\limits_{{{\mathbf{x}}_{{\rm{IRS}}}} \in {\Psi _{{\rm{IRS}}}}({{\mathbf{x}}_{\rm{U}}}):{{{\mathbf{\tilde x}}}_{\rm{U}}}({{\mathbf{x}}_{{\rm{IRS}}}}) = {{\mathbf{x}}_{\rm{U}}}} {\sum\limits_{q = 1}^Q {\sum\limits_{{{{\mathbf{x'}}}_{{\rm{IRS}}}} \in {\Psi _{{\rm{IRS}}}}({{\mathbf{x}}_{\rm{U}}}):{{{\mathbf{\tilde x}}}_{\rm{U}}}({{{\mathbf{x'}}}_{{\rm{IRS}}}}) = {{\mathbf{x}}_{\rm{U}}}} {\sum\limits_{q' = 1}^Q {} } } } 
\end{equation*}
\begin{equation*}
\scalebox{.89}[1]{$\left( {\frac{{\left( {1 - {\delta _{q - q'}}} \right)\left( {1 - {\delta _{{{\mathbf{x}}_{{\rm{IRS}}}} - {{\mathbf{x}}^\prime }_{{\rm{IRS}}}}}} \right){{\left( {{\lambda _{{\rm{wave}}}}} \right)}^4} \times {{\left( {\frac{\pi }{4}} \right)}^2}}}{{{{\left( {4\pi } \right)}^4}\left( {\sqrt {{{({h_{{\rm{BS}}}} - {h_{{\rm{IRS}}}})}^2} + \left\| {{{{\mathbf{\tilde x}}}_{{\rm{BS}}}}({{\mathbf{x}}_{\rm{U}}}) - {{\mathbf{x}}_{{\rm{IRS}}}}} \right\|_2^2} } \right)\left( {\sqrt {{{({h_{{\rm{IRS}}}})}^2} + \left\| {{{\mathbf{x}}_{{\rm{IRS}}}} - {{\mathbf{x}}_{\rm{U}}}} \right\|_2^2} } \right)\left( {\sqrt {{{({h_{{\rm{BS}}}} - {h_{{\rm{IRS}}}})}^2} + \left\| {{{{\mathbf{\tilde x}}}_{{\rm{BS}}}}({{\mathbf{x}}_{\rm{U}}}) - {{{\mathbf{x'}}}_{{\rm{IRS}}}}} \right\|_2^2} } \right)\left( {\sqrt {{{({h_{{\rm{IRS}}}})}^2} + \left\| {{{{\mathbf{x'}}}_{{\rm{IRS}}}} - {{\mathbf{x}}_{\rm{U}}}} \right\|_2^2} } \right)}}} \right) \times $}
\end{equation*}
\begin{equation*}
\left( {G\left( {{{{\mathbf{\tilde x}}}_{{\rm{BS}}}}({{\mathbf{x}}_{\rm{U}}}),{{\mathbf{x}}_{\rm{U}}},{{\mathbf{x}}_{{\rm{IRS}}}}} \right) + \sum\limits_{{{{\mathbf{x''}}}_{{\rm{IRS}}}} \in {\Psi _{{\rm{IRS}}}}({{{\mathbf{\tilde x}}}_{{\rm{BS}}}}({{\mathbf{x}}_{\rm{U}}})):{{\mathbf{x}}_{\rm{U}}} = {{{\mathbf{\tilde x}}}_{\rm{U}}}({{{\mathbf{x''}}}_{{\rm{IRS}}}})} {G\left( {{{{\mathbf{\tilde x}}}_{{\rm{BS}}}}({{\mathbf{x}}_{\rm{U}}}),{{{\mathbf{x''}}}_{{\rm{IRS}}}},{{\mathbf{x}}_{{\rm{IRS}}}}} \right)} } \right) \times 
\end{equation*}
\begin{equation*}
\left( {G\left( {{{{\mathbf{\tilde x}}}_{{\rm{BS}}}}({{\mathbf{x}}_{\rm{U}}}),{{\mathbf{x}}_{\rm{U}}},{{{\mathbf{x'}}}_{{\rm{IRS}}}}} \right) + \sum\limits_{{{{\mathbf{x''}}}_{{\rm{IRS}}}} \in {\Psi _{{\rm{IRS}}}}({{{\mathbf{\tilde x}}}_{{\rm{BS}}}}({{\mathbf{x}}_{\rm{U}}})):{{\mathbf{x}}_{\rm{U}}} = {{{\mathbf{\tilde x}}}_{\rm{U}}}({{{\mathbf{x''}}}_{{\rm{IRS}}}})} {G\left( {{{{\mathbf{\tilde x}}}_{{\rm{BS}}}}({{\mathbf{x}}_{\rm{U}}}),{{{\mathbf{x''}}}_{{\rm{IRS}}}},{{{\mathbf{x'}}}_{{\rm{IRS}}}}} \right)} } \right)\sigma _{{\rm{d}}}^2
\end{equation*}
\begin{equation*}
\le{\frac{\kappa }{{\kappa  + 1}}}\times \sum\limits_{{{\mathbf{x}}_{{\rm{IRS}}}} \in {{\tilde \Psi }_{{\rm{IRS}}}}({{\mathbf{x}}_{\rm{U}}})} {\sum\limits_{{{{\mathbf{x'}}}_{{\rm{IRS}}}} \in {{\tilde \Psi }_{{\rm{IRS}}}}({{\mathbf{x}}_{\rm{U}}})} {\frac{{{{\left( {1 + {{\tilde n}_{{\rm{IRS}}}}({{\mathbf{x}}_{\rm{U}}})} \right)}^2}\sigma _{{\rm{d}}}^2}}{{{4^6}{\pi ^2}\left( {\sqrt {{{({h_{{\rm{BS}}}} - {h_{{\rm{IRS}}}})}^2} + \left\| {{{{\mathbf{\tilde x}}}_{{\rm{BS}}}}({{\mathbf{x}}_{\rm{U}}}) - {{\mathbf{x}}_{{\rm{IRS}}}}} \right\|_2^2} } \right)}} \times } } 
\end{equation*}
\begin{equation*}
 \scalebox{.95}[1]{$\frac{{{\pmb{\mathbb{I}}}\left( {{n_{{\rm{BS}}}}({{\mathbf{x}}_{\rm{U}}})} \right)Q(Q - 1)\left( {1 - {\delta _{{{\mathbf{x}}_{{\rm{IRS}}}} - {{\mathbf{x}}^\prime }_{{\rm{IRS}}}}}} \right){{\left( {{\lambda _{{\rm{wave}}}}} \right)}^4}}}{{\left( {\sqrt {{{({h_{{\rm{BS}}}} - {h_{{\rm{IRS}}}})}^2} + \left\| {{{{\mathbf{\tilde x}}}_{{\rm{BS}}}}({{\mathbf{x}}_{\rm{U}}}) - {{{\mathbf{x'}}}_{{\rm{IRS}}}}} \right\|_2^2} } \right)\left( {\sqrt {{{({h_{{\rm{IRS}}}})}^2} + \left\| {{{\mathbf{x}}_{{\rm{IRS}}}} - {{\mathbf{x}}_{\rm{U}}}} \right\|_2^2} } \right)\left( {\sqrt {{{({h_{{\rm{IRS}}}})}^2} + \left\| {{{{\mathbf{x'}}}_{{\rm{IRS}}}} - {{\mathbf{x}}_{\rm{U}}}} \right\|_2^2} } \right)}}$}
\end{equation*}

Then, we obtain:
\begin{equation*}
E\left\{ {\frac{\kappa }{{\kappa  + 1}}}\times{ \sum\limits_{{{\mathbf{x}}_{{\rm{IRS}}}} \in {{\tilde \Psi }_{{\rm{IRS}}}}({{\mathbf{x}}_{\rm{U}}})} {\sum\limits_{{{{\mathbf{x'}}}_{{\rm{IRS}}}} \in {{\tilde \Psi }_{{\rm{IRS}}}}({{\mathbf{x}}_{\rm{U}}})} {\frac{{{{\left( {1 + {{\tilde n}_{{\rm{IRS}}}}({{\mathbf{x}}_{\rm{U}}})} \right)}^2}\sigma _{{\rm{d}}}^2}}{{{4^6}{\pi ^2}\left( {\sqrt {{{({h_{{\rm{BS}}}} - {h_{{\rm{IRS}}}})}^2} + \left\| {{{{\mathbf{\tilde x}}}_{{\rm{BS}}}}({{\mathbf{x}}_{\rm{U}}}) - {{\mathbf{x}}_{{\rm{IRS}}}}} \right\|_2^2} } \right)}} \times } } } \right.
\end{equation*}
\begin{equation*}
\frac{1}{\left( {\sqrt {{{({h_{{\rm{BS}}}} - {h_{{\rm{IRS}}}})}^2} + \left\| {{{{\mathbf{\tilde x}}}_{{\rm{BS}}}}({{\mathbf{x}}_{\rm{U}}}) - {{{\mathbf{x'}}}_{{\rm{IRS}}}}} \right\|_2^2} } \right)}\times
\end{equation*}
\begin{equation*}
\left. {\left. {\frac{{{\pmb{\mathbb{I}}}\left( {{n_{{\rm{BS}}}}({{\mathbf{x}}_{\rm{U}}})} \right)Q(Q - 1)\left( {1 - {\delta _{{{\mathbf{x}}_{{\rm{IRS}}}} - {{\mathbf{x}}^\prime }_{{\rm{IRS}}}}}} \right){{\left( {{\lambda _{{\rm{wave}}}}} \right)}^4}}}{{\left( {\sqrt {{{({h_{{\rm{IRS}}}})}^2} + \left\| {{{\mathbf{x}}_{{\rm{IRS}}}} - {{\mathbf{x}}_{\rm{U}}}} \right\|_2^2} } \right)\left( {\sqrt {{{({h_{{\rm{IRS}}}})}^2} + \left\| {{{{\mathbf{x'}}}_{{\rm{IRS}}}} - {{\mathbf{x}}_{\rm{U}}}} \right\|_2^2} } \right)}}} \right|{\Phi _{\rm{U}}},{\Phi _{{\rm{BS}}}},{{\tilde n}_{{\rm{IRS}}}}({{\mathbf{x}}_{\rm{U}}}),{{{\mathbf{\tilde x}}}_{{\rm{BS}}}}({{\mathbf{x}}_{\rm{U}}})} \right\} = 
\end{equation*}
\begin{equation*}
{\frac{\kappa }{{\kappa  + 1}}}\times\frac{{{\pmb{\mathbb{I}}}\left( {{n_{{\rm{BS}}}}({{\mathbf{x}}_{\rm{U}}})} \right)Q(Q - 1){{\left( {{\lambda _{{\rm{wave}}}}} \right)}^4}{{\left( {1 + {{\tilde n}_{{\rm{IRS}}}}({{\mathbf{x}}_{\rm{U}}})} \right)}^2}\left( {{{\tilde n}_{{\rm{IRS}}}}({{\mathbf{x}}_{\rm{U}}}) - 1} \right){{\tilde n}_{{\rm{IRS}}}}({{\mathbf{x}}_{\rm{U}}})\sigma _{{\rm{d}}}^2}}{{{4^8}{\pi ^4}{{\left( {{R_{{\rm{co}}}}} \right)}^4}}} \times 
\end{equation*}
\begin{equation*}
\int\limits_{{\mathbf{x}},{\mathbf{x'}} \in{\cal C}({{\mathbf{x}}_{\rm{U}}},2{R_{{\rm{co}}}})} {\frac{1}{{\left( {\sqrt {{{({h_{{\rm{BS}}}} - {h_{{\rm{IRS}}}})}^2} + \left\| {{{{\mathbf{\tilde x}}}_{{\rm{BS}}}}({{\mathbf{x}}_{\rm{U}}}) - {\mathbf{x}}} \right\|_2^2} } \right)\left( {\sqrt {{{({h_{{\rm{BS}}}} - {h_{{\rm{IRS}}}})}^2} + \left\| {{{{\mathbf{\tilde x}}}_{{\rm{BS}}}}({{\mathbf{x}}_{\rm{U}}}) - {\mathbf{x'}}} \right\|_2^2} } \right)}}}  \times 
\end{equation*}
\begin{equation*}
\frac{{d{\mathbf{x}}d{\mathbf{x'}}}}{{\left( {\sqrt {{{({h_{{\rm{IRS}}}})}^2} + \left\| {{\mathbf{x}} - {{\mathbf{x}}_{\rm{U}}}} \right\|_2^2} } \right)\left( {\sqrt {{{({h_{{\rm{IRS}}}})}^2} + \left\| {{\mathbf{x'}} - {{\mathbf{x}}_{\rm{U}}}} \right\|_2^2} } \right)}} \le
\end{equation*}
\begin{equation*}
{\frac{\kappa }{{\kappa  + 1}}}\times\frac{{{\pmb{\mathbb{I}}}\left( {{n_{{\rm{BS}}}}({{\mathbf{x}}_{\rm{U}}})} \right)Q(Q - 1){{\left( {{\lambda _{{\rm{wave}}}}} \right)}^4}{{\left( {1 + {{\tilde n}_{{\rm{IRS}}}}({{\mathbf{x}}_{\rm{U}}})} \right)}^2}\left( {{{\tilde n}_{{\rm{IRS}}}}({{\mathbf{x}}_{\rm{U}}}) - 1} \right){{\tilde n}_{{\rm{IRS}}}}({{\mathbf{x}}_{\rm{U}}})\sigma _{{\rm{d}}}^2}}{{{4^8}{\pi ^4}{{\left( {{R_{{\rm{co}}}}} \right)}^4}}} \times 
\end{equation*}
\begin{equation*}
\sqrt {\int\limits_{{\mathbf{x}},{\mathbf{x'}} \in {\cal C}({{\mathbf{x}}_{\rm{U}}},2{R_{{\rm{co}}}})} {\frac{{d{\mathbf{x}}d{\mathbf{x'}}}}{{{{\left( {\sqrt {{{({h_{{\rm{BS}}}} - {h_{{\rm{IRS}}}})}^2} + \left\| {{{{\mathbf{\tilde x}}}_{{\rm{BS}}}}({{\mathbf{x}}_{\rm{U}}}) - {\mathbf{x}}} \right\|_2^2} } \right)}^2}{{\left( {\sqrt {{{({h_{{\rm{BS}}}} - {h_{{\rm{IRS}}}})}^2} + \left\| {{{{\mathbf{\tilde x}}}_{{\rm{BS}}}}({{\mathbf{x}}_{\rm{U}}}) - {\mathbf{x'}}} \right\|_2^2} } \right)}^2}}}} }  \times 
\end{equation*}
\begin{equation*}
\sqrt {\int\limits_{{\mathbf{x}},{\mathbf{x'}} \in {\cal C}({{\mathbf{x}}_{\rm{U}}},2{R_{{\rm{co}}}})} {\frac{{d{\mathbf{x}}d{\mathbf{x'}}}}{{{{\left( {\sqrt {{{({h_{{\rm{IRS}}}})}^2} + \left\| {{\mathbf{x}} - {{\mathbf{x}}_{\rm{U}}}} \right\|_2^2} } \right)}^2}{{\left( {\sqrt {{{({h_{{\rm{IRS}}}})}^2} + \left\| {{\mathbf{x'}} - {{\mathbf{x}}_{\rm{U}}}} \right\|_2^2} } \right)}^2}}}} }  \le
\end{equation*}
\begin{equation*}
{\frac{\kappa }{{\kappa  + 1}}}\times\frac{{{\pmb{\mathbb{I}}}\left( {{n_{{\rm{BS}}}}({{\mathbf{x}}_{\rm{U}}})} \right)Q(Q - 1){{\left( {{\lambda _{{\rm{wave}}}}} \right)}^4}{{\left( {1 + {{\tilde n}_{{\rm{IRS}}}}({{\mathbf{x}}_{\rm{U}}})} \right)}^2}\left( {{{\tilde n}_{{\rm{IRS}}}}({{\mathbf{x}}_{\rm{U}}}) - 1} \right){{\tilde n}_{{\rm{IRS}}}}({{\mathbf{x}}_{\rm{U}}})\sigma _{{\rm{d}}}^2}}{{{4^8}{\pi ^4}{{\left( {{R_{{\rm{co}}}}} \right)}^4}}} \times 
\end{equation*}
\begin{equation*}
\sqrt {\int\limits_{{\mathbf{x}} \in {\cal C}({{{\mathbf{\tilde x}}}_{{\rm{BS}}}}({{\mathbf{x}}_{\rm{U}}}),3{R_{{\rm{co}}}})} {\frac{{d{\mathbf{x}}}}{{{{\left( {\sqrt {{{({h_{{\rm{BS}}}} - {h_{{\rm{IRS}}}})}^2} + \left\| {{{{\mathbf{\tilde x}}}_{{\rm{BS}}}}({{\mathbf{x}}_{\rm{U}}}) - {\mathbf{x}}} \right\|_2^2} } \right)}^2}}}} }  \times  
\end{equation*}
\begin{equation*}
\sqrt {\int\limits_{{\mathbf{x'}} \in {\cal C}({{{\mathbf{\tilde x}}}_{{\rm{BS}}}}({{\mathbf{x}}_{\rm{U}}}),3{R_{{\rm{co}}}})} {\frac{{d{\mathbf{x'}}}}{{{{\left( {\sqrt {{{({h_{{\rm{BS}}}} - {h_{{\rm{IRS}}}})}^2} + \left\| {{{{\mathbf{\tilde x}}}_{{\rm{BS}}}}({{\mathbf{x}}_{\rm{U}}}) - {\mathbf{x'}}} \right\|_2^2} } \right)}^2}}}} }\times
\end{equation*}
\begin{equation*}
 \times \sqrt {\int\limits_{{\mathbf{x}} \in {\cal C}({{\mathbf{x}}_{\rm{U}}},2{R_{{\rm{co}}}})} {\frac{{d{\mathbf{x}}}}{{{{\left( {\sqrt {{{({h_{{\rm{IRS}}}})}^2} + \left\| {{\mathbf{x}} - {{\mathbf{x}}_{\rm{U}}}} \right\|_2^2} } \right)}^2}}}} }  \times \sqrt {\int\limits_{{\mathbf{x'}} \in {\cal C}({{\mathbf{x}}_{\rm{U}}},2{R_{{\rm{co}}}})} {\frac{{d{\mathbf{x'}}}}{{{{\left( {\sqrt {{{({h_{{\rm{IRS}}}})}^2} + \left\| {{\mathbf{x'}} - {{\mathbf{x}}_{\rm{U}}}} \right\|_2^2} } \right)}^2}}}} }  = 
\end{equation*}
\begin{equation*}
{\frac{\kappa }{{\kappa  + 1}}}\times\frac{{{\pmb{\mathbb{I}}}\left( {{n_{{\rm{BS}}}}({{\mathbf{x}}_{\rm{U}}})} \right)Q(Q - 1){{\left( {{\lambda _{{\rm{wave}}}}} \right)}^4}{{\left( {1 + {{\tilde n}_{{\rm{IRS}}}}({{\mathbf{x}}_{\rm{U}}})} \right)}^2}\left( {{{\tilde n}_{{\rm{IRS}}}}({{\mathbf{x}}_{\rm{U}}}) - 1} \right){{\tilde n}_{{\rm{IRS}}}}({{\mathbf{x}}_{\rm{U}}})\sigma _{{\rm{d}}}^2}}{{{4^8}{\pi ^4}{{\left( {{R_{{\rm{co}}}}} \right)}^4}}} \times 
\end{equation*}
\begin{equation*}
\left( {\int\limits_{{\mathbf{x}} \in {\cal C}({{{\mathbf{\tilde x}}}_{{\rm{BS}}}}({{\mathbf{x}}_{\rm{U}}}),3{R_{{\rm{co}}}})} {\frac{{d{\mathbf{x}}}}{{{{({h_{{\rm{BS}}}} - {h_{{\rm{IRS}}}})}^2} + \left\| {{{{\mathbf{\tilde x}}}_{{\rm{BS}}}}({{\mathbf{x}}_{\rm{U}}}) - {\mathbf{x}}} \right\|_2^2}}} } \right) \times \left( {\int\limits_{{\mathbf{x}} \in {\cal C}({{\mathbf{x}}_{\rm{U}}},2{R_{{\rm{co}}}})} {\frac{{d{\mathbf{x}}}}{{{{({h_{{\rm{IRS}}}})}^2} + \left\| {{\mathbf{x}} - {{\mathbf{x}}_{\rm{U}}}} \right\|_2^2}}} } \right) = 
\end{equation*}
\begin{equation*}
{\frac{\kappa }{{\kappa  + 1}}}\times\frac{{{\pmb{\mathbb{I}}}\left( {{n_{{\rm{BS}}}}({{\mathbf{x}}_{\rm{U}}})} \right)Q(Q - 1){{\left( {{\lambda _{{\rm{wave}}}}} \right)}^4}{{\left( {1 + {{\tilde n}_{{\rm{IRS}}}}({{\mathbf{x}}_{\rm{U}}})} \right)}^2}\left( {{{\tilde n}_{{\rm{IRS}}}}({{\mathbf{x}}_{\rm{U}}}) - 1} \right){{\tilde n}_{{\rm{IRS}}}}({{\mathbf{x}}_{\rm{U}}})\sigma _{{\rm{d}}}^2}}{{{4^8}{\pi ^2}{{\left( {{R_{{\rm{co}}}}} \right)}^4}}} \times 
\end{equation*}
\begin{equation*}
\ln \left( {1 + {{\left( {\frac{{3{R_{{\rm{co}}}}}}{{{h_{{\rm{BS}}}} - {h_{{\rm{IRS}}}}}}} \right)}^2}} \right) \times \ln \left( {1 + {{\left( {\frac{{2{R_{{\rm{co}}}}}}{{{h_{{\rm{IRS}}}}}}} \right)}^2}} \right).
\end{equation*}

Now, by (\ref{lambda-k}), we obtain:
\begin{equation*}
E\left\{{\frac{\kappa }{{\kappa  + 1}}}\times {\frac{{\mathbb{I}\left( {{n_{{\rm{BS}}}}({{\mathbf{x}}_{\rm{U}}})} \right)Q(Q - 1){{\left( {{\lambda _{{\rm{wave}}}}} \right)}^4}{{\left( {1 + {{\tilde n}_{{\rm{IRS}}}}({{\mathbf{x}}_{\rm{U}}})} \right)}^2}\left( {{{\tilde n}_{{\rm{IRS}}}}({{\mathbf{x}}_{\rm{U}}}) - 1} \right){{\tilde n}_{{\rm{IRS}}}}({{\mathbf{x}}_{\rm{U}}})\sigma _{{\rm{d}}}^2}}{{{4^8}{\pi ^2}{{\left( {{R_{{\rm{co}}}}} \right)}^4}}} \times } \right.
\end{equation*}
\begin{equation*}
\left. {\ln \left( {1 + {{\left( {\frac{{3{R_{{\rm{co}}}}}}{{{h_{{\rm{BS}}}} - {h_{{\rm{IRS}}}}}}} \right)}^2}} \right) \times \ln \left( {1 + {{\left( {\frac{{2{R_{{\rm{co}}}}}}{{{h_{{\rm{IRS}}}}}}} \right)}^2}} \right)} \right\} = 
\end{equation*}
\begin{equation*}
\le {\frac{\kappa }{{\kappa  + 1}}}\times \scalebox{.9}[1]{$ \frac{{{{\left( {{\lambda _{{\rm{IRS}}}}\pi {{\left( {2{R_{{\rm{co}}}}} \right)}^2}} \right)}^4}\exp \left\{ {\frac{{16}}{{2\left( {{\lambda _{{\rm{IRS}}}}\pi {{\left( {2{R_{{\rm{co}}}}} \right)}^2}} \right)}}} \right\} + {{\left( {{\lambda _{{\rm{IRS}}}}\pi {{\left( {2{R_{{\rm{co}}}}} \right)}^2}} \right)}^3}\exp \left\{ {\frac{9}{{2\left( {{\lambda _{{\rm{IRS}}}}\pi {{\left( {2{R_{{\rm{co}}}}} \right)}^2}} \right)}}} \right\} - {{\left( {{\lambda _{{\rm{IRS}}}}\pi {{\left( {2{R_{{\rm{co}}}}} \right)}^2}} \right)}^2} - 2{\lambda _{{\rm{IRS}}}}\pi {{\left( {2{R_{{\rm{co}}}}} \right)}^2}}}{{{4^8}{\pi ^2}{{\left( {{R_{{\rm{co}}}}} \right)}^4}}} \times $}
\end{equation*}
\begin{equation}
  \scalebox{.95}[1]{$Q(Q - 1){\left( {{\lambda _{{\rm{wave}}}}} \right)^4}\sigma _{{\rm{d}}}^2\left( {1 - \exp \left\{ { - {\lambda _{{\rm{BS}}}}\pi {{\left( {{R_{{\rm{co}}}}} \right)}^2}} \right\}} \right) \times \ln \left( {1 + {{\left( {\frac{{3{R_{{\rm{co}}}}}}{{{h_{{\rm{BS}}}} - {h_{{\rm{IRS}}}}}}} \right)}^2}} \right) \times \ln \left( {1 + {{\left( {\frac{{2{R_{{\rm{co}}}}}}{{{h_{{\rm{IRS}}}}}}} \right)}^2}} \right).$}
\label{t15}
\end{equation}

To bound term (\ref{t3}), we have:
\begin{equation*}
\sum\limits_{{{\mathbf{x}}_{{\rm{IRS}}}} \in {\Psi _{{\rm{IRS}}}}({{\mathbf{x}}_{\rm{U}}}):{{{\mathbf{\tilde x}}}_{\rm{U}}}({{\mathbf{x}}_{{\rm{IRS}}}}) \ne {{\mathbf{x}}_{\rm{U}}}} {\sum\limits_{q = 1}^Q {\frac{{{{\left( {{\lambda _{{\rm{wave}}}}} \right)}^4}}}{{{{\left( {4\pi } \right)}^4}\left( {{{({h_{{\rm{IRS}}}})}^2} + \left\| {{{\mathbf{x}}_{{\rm{IRS}}}} - {{\mathbf{x}}_{\rm{U}}}} \right\|_2^2} \right)\left( {{{({h_{{\rm{BS}}}} - {h_{{\rm{IRS}}}})}^2} + \left\| {{{{\mathbf{\tilde x}}}_{{\rm{BS}}}}({{\mathbf{x}}_{\rm{U}}}) - {{\mathbf{x}}_{{\rm{IRS}}}}} \right\|_2^2} \right)}} \times } } 
\end{equation*}
\begin{equation*}
{\left( {G\left( {{{{\mathbf{\tilde x}}}_{{\rm{BS}}}}({{\mathbf{x}}_{\rm{U}}}),{{\mathbf{x}}_{\rm{U}}},{{\mathbf{x}}_{{\rm{IRS}}}}} \right) + \sum\limits_{{{{\mathbf{x'}}}_{{\rm{IRS}}}} \in {\Psi _{{\rm{IRS}}}}({{{\mathbf{\tilde x}}}_{{\rm{BS}}}}({{\mathbf{x}}_{\rm{U}}})):{{\mathbf{x}}_{\rm{U}}} = {{{\mathbf{\tilde x}}}_{\rm{U}}}({{{\mathbf{x'}}}_{{\rm{IRS}}}})} {G\left( {{{{\mathbf{\tilde x}}}_{{\rm{BS}}}}({{\mathbf{x}}_{\rm{U}}}),{{{\mathbf{x'}}}_{{\rm{IRS}}}},{{\mathbf{x}}_{{\rm{IRS}}}}} \right)} } \right)^2}\sigma _{{\rm{d}}}^2
\end{equation*}
\begin{equation*}
 = \sum\limits_{{{\mathbf{x}}_{{\rm{IRS}}}} \in {\Psi _{{\rm{IRS}}}}({{\mathbf{x}}_{\rm{U}}})} {\frac{{{\pmb{\mathbb{I}}}\left( {\left\| {{{\mathbf{x}}_{\rm{U}}} - {{{\mathbf{\tilde x}}}_{\rm{U}}}({{\mathbf{x}}_{{\rm{IRS}}}})} \right\|} \right)Q{{\left( {{\lambda _{{\rm{wave}}}}} \right)}^4}}}{{{{\left( {4\pi } \right)}^4}\left( {{{({h_{{\rm{IRS}}}})}^2} + \left\| {{{\mathbf{x}}_{{\rm{IRS}}}} - {{\mathbf{x}}_{\rm{U}}}} \right\|_2^2} \right)\left( {{{({h_{{\rm{BS}}}} - {h_{{\rm{IRS}}}})}^2} + \left\| {{{{\mathbf{\tilde x}}}_{{\rm{BS}}}}({{\mathbf{x}}_{\rm{U}}}) - {{\mathbf{x}}_{{\rm{IRS}}}}} \right\|_2^2} \right)}} \times } 
\end{equation*}
\begin{equation*}
{\left( {G\left( {{{{\mathbf{\tilde x}}}_{{\rm{BS}}}}({{\mathbf{x}}_{\rm{U}}}),{{\mathbf{x}}_{\rm{U}}},{{\mathbf{x}}_{{\rm{IRS}}}}} \right) + \sum\limits_{{{{\mathbf{x'}}}_{{\rm{IRS}}}} \in {\Psi _{{\rm{IRS}}}}({{{\mathbf{\tilde x}}}_{{\rm{BS}}}}({{\mathbf{x}}_{\rm{U}}})):{{\mathbf{x}}_{\rm{U}}} = {{{\mathbf{\tilde x}}}_{\rm{U}}}({{{\mathbf{x'}}}_{{\rm{IRS}}}})} {G\left( {{{{\mathbf{\tilde x}}}_{{\rm{BS}}}}({{\mathbf{x}}_{\rm{U}}}),{{{\mathbf{x'}}}_{{\rm{IRS}}}},{{\mathbf{x}}_{{\rm{IRS}}}}} \right)} } \right)^2}\sigma _{{\rm{d}}}^2
\end{equation*}
\begin{equation*}
 \le \sum\limits_{{{\mathbf{x}}_{{\rm{IRS}}}} \in {{\tilde \Psi }_{{\rm{IRS}}}}({{\mathbf{x}}_{\rm{U}}})} {\frac{{{\pmb{\mathbb{I}}}\left( {{n_{{\rm{BS}}}}({{\mathbf{x}}_{\rm{U}}})} \right)Q{{\left( {{\lambda _{{\rm{wave}}}}} \right)}^4}{{\left( {1 + {{\tilde n}_{{\rm{IRS}}}}({{\mathbf{x}}_{\rm{U}}})} \right)}^2}\sigma _{{\rm{d}}}^2}}{{{{\left( {4\pi } \right)}^4}\left( {{{({h_{{\rm{IRS}}}})}^2} + \left\| {{{\mathbf{x}}_{{\rm{IRS}}}} - {{\mathbf{x}}_{\rm{U}}}} \right\|_2^2} \right)\left( {{{({h_{{\rm{BS}}}} - {h_{{\rm{IRS}}}})}^2} + \left\| {{{{\mathbf{\tilde x}}}_{{\rm{BS}}}}({{\mathbf{x}}_{\rm{U}}}) - {{\mathbf{x}}_{{\rm{IRS}}}}} \right\|_2^2} \right)}}} .
\end{equation*}

Thus, we obtain:
\begin{equation*}
\scalebox{1}[1]{$E\left\{ {\left. {\sum\limits_{{{\mathbf{x}}_{{\rm{IRS}}}} \in {{\tilde \Psi }_{{\rm{IRS}}}}({{\mathbf{x}}_{\rm{U}}})} {\frac{{{\pmb{\mathbb{I}}}\left( {{n_{{\rm{BS}}}}({{\mathbf{x}}_{\rm{U}}})} \right)Q{{\left( {{\lambda _{{\rm{wave}}}}} \right)}^4}{{\left( {1 + {{\tilde n}_{{\rm{IRS}}}}({{\mathbf{x}}_{\rm{U}}})} \right)}^2}\sigma _{{\rm{d}}}^2}}{{{{\left( {4\pi } \right)}^4}\left( {{{({h_{{\rm{IRS}}}})}^2} + \left\| {{{\mathbf{x}}_{{\rm{IRS}}}} - {{\mathbf{x}}_{\rm{U}}}} \right\|_2^2} \right)\left( {{{({h_{{\rm{BS}}}} - {h_{{\rm{IRS}}}})}^2} + \left\| {{{{\mathbf{\tilde x}}}_{{\rm{BS}}}}({{\mathbf{x}}_{\rm{U}}}) - {{\mathbf{x}}_{{\rm{IRS}}}}} \right\|_2^2} \right)}}} } \right|{\Phi _{\rm{U}}},{\Phi _{{\rm{BS}}}},{{\tilde n}_{{\rm{IRS}}}}({{\mathbf{x}}_{\rm{U}}}),{{{\mathbf{\tilde x}}}_{{\rm{BS}}}}({{\mathbf{x}}_{\rm{U}}})} \right\}$}
\end{equation*}
\begin{equation*}
 = \frac{{{\pmb{\mathbb{I}}}\left( {{n_{{\rm{BS}}}}({{\mathbf{x}}_{\rm{U}}})} \right)Q{{\left( {{\lambda _{{\rm{wave}}}}} \right)}^4}{{\left( {1 + {{\tilde n}_{{\rm{IRS}}}}({{\mathbf{x}}_{\rm{U}}})} \right)}^2}{{\tilde n}_{{\rm{IRS}}}}({{\mathbf{x}}_{\rm{U}}})\sigma _{{\rm{d}}}^2}}{{{{\left( {4\pi } \right)}^5}{{\left( {{R_{{\rm{co}}}}} \right)}^2}}} \times 
\end{equation*}
\begin{equation*}
\int\limits_{{\cal C}({{\mathbf{x}}_{\rm{U}}},2{R_{{\rm{co}}}})} {\frac{{d{\mathbf{x}}}}{{\left( {{{({h_{{\rm{IRS}}}})}^2} + \left\| {{\mathbf{x}} - {{\mathbf{x}}_{\rm{U}}}} \right\|_2^2} \right)\left( {{{({h_{{\rm{BS}}}} - {h_{{\rm{IRS}}}})}^2} + \left\| {{{{\mathbf{\tilde x}}}_{{\rm{BS}}}}({{\mathbf{x}}_{\rm{U}}}) - {\mathbf{x}}} \right\|_2^2} \right)}}} 
\end{equation*}
\begin{equation*}
 \le \frac{{{\pmb{\mathbb{I}}}\left( {{n_{{\rm{BS}}}}({{\mathbf{x}}_{\rm{U}}})} \right)Q{{\left( {{\lambda _{{\rm{wave}}}}} \right)}^4}{{\left( {1 + {{\tilde n}_{{\rm{IRS}}}}({{\mathbf{x}}_{\rm{U}}})} \right)}^2}{{\tilde n}_{{\rm{IRS}}}}({{\mathbf{x}}_{\rm{U}}})\sigma _{{\rm{d}}}^2}}{{{{\left( {4\pi } \right)}^5}{{\left( {{R_{{\rm{co}}}}} \right)}^2}}} \times 
\end{equation*}
\begin{equation*}
\sqrt {\int\limits_{{\cal C}({{\mathbf{x}}_{\rm{U}}},2{R_{{\rm{co}}}})} {\frac{{d{\mathbf{x}}}}{{{{\left( {{{({h_{{\rm{IRS}}}})}^2} + \left\| {{\mathbf{x}} - {{\mathbf{x}}_{\rm{U}}}} \right\|_2^2} \right)}^2}}}} }  \times \sqrt {\int\limits_{{\cal C}({{\mathbf{x}}_{\rm{U}}},2{R_{{\rm{co}}}})} {\frac{{d{\mathbf{x}}}}{{{{\left( {{{({h_{{\rm{BS}}}} - {h_{{\rm{IRS}}}})}^2} + \left\| {{{{\mathbf{\tilde x}}}_{{\rm{BS}}}}({{\mathbf{x}}_{\rm{U}}}) - {\mathbf{x}}} \right\|_2^2} \right)}^2}}}} } 
\end{equation*}
\begin{equation*}
\le \frac{{{\pmb{\mathbb{I}}}\left( {{n_{{\rm{BS}}}}({{\mathbf{x}}_{\rm{U}}})} \right)Q{{\left( {{\lambda _{{\rm{wave}}}}} \right)}^4}{{\left( {1 + {{\tilde n}_{{\rm{IRS}}}}({{\mathbf{x}}_{\rm{U}}})} \right)}^2}{{\tilde n}_{{\rm{IRS}}}}({{\mathbf{x}}_{\rm{U}}})\sigma _{{\rm{d}}}^2}}{{{{\left( {4\pi } \right)}^5}{{\left( {{R_{{\rm{co}}}}} \right)}^2}}} \times 
\end{equation*}
\begin{equation*}
\sqrt {\int\limits_{{\cal C}({{\mathbf{x}}_{\rm{U}}},2{R_{{\rm{co}}}})} {\frac{{d{\mathbf{x}}}}{{{{\left( {{{({h_{{\rm{IRS}}}})}^2} + \left\| {{\mathbf{x}} - {{\mathbf{x}}_{\rm{U}}}} \right\|_2^2} \right)}^2}}}} }  \times \sqrt {\int\limits_{{\cal C}({{{\mathbf{\tilde x}}}_{{\rm{BS}}}}({{\mathbf{x}}_{\rm{U}}}),3{R_{{\rm{co}}}})} {\frac{{d{\mathbf{x}}}}{{{{\left( {{{({h_{{\rm{BS}}}} - {h_{{\rm{IRS}}}})}^2} + \left\| {{{{\mathbf{\tilde x}}}_{{\rm{BS}}}}({{\mathbf{x}}_{\rm{U}}}) - {\mathbf{x}}} \right\|_2^2} \right)}^2}}}} } 
\end{equation*}
\begin{equation*}
 = \frac{{{\pmb{\mathbb{I}}}\left( {{n_{{\rm{BS}}}}({{\mathbf{x}}_{\rm{U}}})} \right)Q{{\left( {{\lambda _{{\rm{wave}}}}} \right)}^4}{{\left( {1 + {{\tilde n}_{{\rm{IRS}}}}({{\mathbf{x}}_{\rm{U}}})} \right)}^2}{{\tilde n}_{{\rm{IRS}}}}({{\mathbf{x}}_{\rm{U}}})\sigma _{{\rm{d}}}^2}}{{{{\left( 4 \right)}^5}{\pi ^4}{{\left( {{R_{{\rm{co}}}}} \right)}^2}}} \times 
\end{equation*}
\begin{equation*}
\sqrt {\frac{{{{\left( {2{R_{{\rm{co}}}}} \right)}^2}}}{{\left( {{{({h_{{\rm{IRS}}}})}^2} + {{\left( {2{R_{{\rm{co}}}}} \right)}^2}} \right){{({h_{{\rm{IRS}}}})}^2}}}}  \times \sqrt {\frac{{{{\left( {3{R_{{\rm{co}}}}} \right)}^2}}}{{\left( {{{({h_{{\rm{BS}}}} - {h_{{\rm{IRS}}}})}^2} + {{\left( {3{R_{{\rm{co}}}}} \right)}^2}} \right){{({h_{{\rm{BS}}}} - {h_{{\rm{IRS}}}})}^2}}}} .
\end{equation*}

Hence, we obtain:
\begin{equation*}
E\left\{ {\frac{{{\pmb{\mathbb{I}}}\left( {{n_{{\rm{BS}}}}({{\mathbf{x}}_{\rm{U}}})} \right)Q{{\left( {{\lambda _{{\rm{wave}}}}} \right)}^4}{{\left( {1 + {{\tilde n}_{{\rm{IRS}}}}({{\mathbf{x}}_{\rm{U}}})} \right)}^2}{{\tilde n}_{{\rm{IRS}}}}({{\mathbf{x}}_{\rm{U}}})\sigma _{{\rm{d}}}^2}}{{{{\left( 4 \right)}^5}{\pi ^4}{{\left( {{R_{{\rm{co}}}}} \right)}^2}}} \times } \right.
\end{equation*}
\begin{equation*}
\left. {\sqrt {\frac{{{{\left( {2{R_{{\rm{co}}}}} \right)}^2}}}{{\left( {{{({h_{{\rm{IRS}}}})}^2} + {{\left( {2{R_{{\rm{co}}}}} \right)}^2}} \right){{({h_{{\rm{IRS}}}})}^2}}}}  \times \sqrt {\frac{{{{\left( {3{R_{{\rm{co}}}}} \right)}^2}}}{{\left( {{{({h_{{\rm{BS}}}} - {h_{{\rm{IRS}}}})}^2} + {{\left( {3{R_{{\rm{co}}}}} \right)}^2}} \right){{({h_{{\rm{BS}}}} - {h_{{\rm{IRS}}}})}^2}}}} } \right\}
\end{equation*}
\begin{equation*}
 = \left( {3{\lambda _{{\rm{IRS}}}}{{\left( {2{R_{{\rm{co}}}}} \right)}^2} + 2{{\left( {{\lambda _{{\rm{IRS}}}}{{\left( {2{R_{{\rm{co}}}}} \right)}^2}} \right)}^2} + {{\left( {{\lambda _{{\rm{IRS}}}}{{\left( {2{R_{{\rm{co}}}}} \right)}^2}} \right)}^3}\exp \left\{ {\frac{9}{{2{\lambda _{{\rm{IRS}}}}{{\left( {2{R_{{\rm{co}}}}} \right)}^2}}}} \right\}} \right) \times 
\end{equation*}
\begin{equation*}
\frac{{\left( {1 - \exp \left\{ { - {\lambda _{{\rm{BS}}}}\pi {{\left( {{R_{{\rm{co}}}}} \right)}^2}} \right\}} \right)Q{{\left( {{\lambda _{{\rm{wave}}}}} \right)}^4}\sigma _{{\rm{d}}}^2}}{{{{\left( 4 \right)}^5}{\pi ^4}{{\left( {{R_{{\rm{co}}}}} \right)}^2}}} \times 
\end{equation*}
\begin{equation}
 {\sqrt {\frac{{{{\left( {2{R_{{\rm{co}}}}} \right)}^2}}}{{\left( {{{({h_{{\rm{IRS}}}})}^2} + {{\left( {2{R_{{\rm{co}}}}} \right)}^2}} \right){{({h_{{\rm{IRS}}}})}^2}}}}  \times \sqrt {\frac{{{{\left( {3{R_{{\rm{co}}}}} \right)}^2}}}{{\left( {{{({h_{{\rm{BS}}}} - {h_{{\rm{IRS}}}})}^2} + {{\left( {3{R_{{\rm{co}}}}} \right)}^2}} \right){{({h_{{\rm{BS}}}} - {h_{{\rm{IRS}}}})}^2}}}} } .
\label{t12}
\end{equation}
Therefore, (\ref{t6}), (\ref{t8}), (\ref{t15}), and (\ref{t12}) completes the proof.

\section{Proof of Theorem 2}

For (\ref{t1}), we obtain:
\begin{equation*}
  \scalebox{.9}[1]{$\frac{{{\left( {1 + \left( {\hat h - 1} \right){p_{{\rm{b}}}}} \right)}{{\left( {{\lambda _{{\rm{wave}}}}} \right)}^2}\sigma _{{\rm{d}}}^2}}{{{{\left( {4\pi } \right)}^2}\left( {{{({h_{{\rm{BS}}}})}^2} + \left\| {{{{\mathbf{\tilde x}}}_{{\rm{BS}}}}({{\mathbf{x}}_{\rm{U}}}) - {{\mathbf{x}}_{\rm{U}}}} \right\|_2^2} \right)}} \times {\left( {G\left( {{{{\mathbf{\tilde x}}}_{{\rm{BS}}}}({{\mathbf{x}}_{\rm{U}}}),{{\mathbf{x}}_{\rm{U}}},{{\mathbf{x}}_{\rm{U}}}} \right) + \sum\limits_{{{\mathbf{x}}_{{\rm{IRS}}}} \in {\Psi _{{\rm{IRS}}}}({{{\mathbf{\tilde x}}}_{{\rm{BS}}}}({{\mathbf{x}}_{\rm{U}}})):{{\mathbf{x}}_{\rm{U}}} = {{{\mathbf{\tilde x}}}_{\rm{U}}}({{\mathbf{x}}_{{\rm{IRS}}}})} {G\left( {{{{\mathbf{\tilde x}}}_{{\rm{BS}}}}({{\mathbf{x}}_{\rm{U}}}),{{\mathbf{x}}_{{\rm{IRS}}}},{{\mathbf{x}}_{\rm{U}}}} \right)} } \right)^2}  $}
\end{equation*}
\begin{equation*}
\ge \frac{{\left( {1 + \left( {\hat h - 1} \right){p_{{\rm{b}}}}} \right){{\left( {{\lambda _{{\rm{wave}}}}} \right)}^2}\sigma _{{\rm{d}}}^2}}{{{{\left( {4\pi } \right)}^2}\left( {{{({h_{{\rm{BS}}}})}^2} + \left\| {{{{\mathbf{\tilde x}}}_{{\rm{BS}}}}({{\mathbf{x}}_{\rm{U}}}) - {{\mathbf{x}}_{\rm{U}}}} \right\|_2^2} \right)}}
\end{equation*}
Then, we obtain:
\begin{equation*}
E\left\{ {\left. {\frac{{\left( {1 + \left( {\hat h - 1} \right){p_{{\rm{b}}}}} \right){{\left( {{\lambda _{{\rm{wave}}}}} \right)}^2}\sigma _{{\rm{d}}}^2}}{{{{\left( {4\pi } \right)}^2}\left( {{{({h_{{\rm{BS}}}})}^2} + \left\| {{{{\mathbf{\tilde x}}}_{{\rm{BS}}}}({{\mathbf{x}}_{\rm{U}}}) - {{\mathbf{x}}_{\rm{U}}}} \right\|_2^2} \right)}}} \right|{n_{{\rm{BS}}}}({{\mathbf{x}}_{\rm{U}}})} \right\}
\end{equation*}
\begin{equation*}
 = \frac{{{\pmb{\mathbb{I}}}\left( {{n_{{\rm{BS}}}}({{\mathbf{x}}_{\rm{U}}})} \right)\left( {1 + \left( {\hat h - 1} \right){p_{{\rm{b}}}}} \right){{\left( {{\lambda _{{\rm{wave}}}}} \right)}^2}\sigma _{{\rm{d}}}^2}}{{{{\left( {4\pi } \right)}^2}\pi {{\left( {{R_{{\rm{co}}}}} \right)}^2}}}\int\limits_{{\cal C}({{\mathbf{x}}_{\rm{U}}},{R_{{\rm{co}}}})} {\frac{{d{\mathbf{x}}}}{{\left( {{{({h_{{\rm{BS}}}})}^2} + \left\| {{\mathbf{x}} - {{\mathbf{x}}_{\rm{U}}}} \right\|_2^2} \right)}}} 
\end{equation*}
\begin{equation*}
 = \frac{{{\pmb{\mathbb{I}}}\left( {{n_{{\rm{BS}}}}({{\mathbf{x}}_{\rm{U}}})} \right)\left( {1 + \left( {\hat h - 1} \right){p_{{\rm{b}}}}} \right){{\left( {{\lambda _{{\rm{wave}}}}} \right)}^2}\sigma _{{\rm{d}}}^2}}{{{{\left( {4\pi } \right)}^2}{{\left( {{R_{{\rm{co}}}}} \right)}^2}}}\ln \left( {1 + {{\left( {\frac{{{R_{{\rm{co}}}}}}{{{h_{{\rm{BS}}}}}}} \right)}^2}} \right),
\end{equation*}
hence, we have:
\begin{equation*}
E\left\{ { {\frac{{\left( {1 + \left( {\hat h - 1} \right){p_{{\rm{b}}}}} \right){{\left( {{\lambda _{{\rm{wave}}}}} \right)}^2}\sigma _{{\rm{d}}}^2}}{{{{\left( {4\pi } \right)}^2}\left( {{{({h_{{\rm{BS}}}})}^2} + \left\| {{{{\mathbf{\tilde x}}}_{{\rm{BS}}}}({{\mathbf{x}}_{\rm{U}}}) - {{\mathbf{x}}_{\rm{U}}}} \right\|_2^2} \right)}}} } \right\}
\end{equation*}
\begin{equation}
= \frac{{\left( {1 - \exp \left\{ { - {\lambda _{{\rm{BS}}}}\pi {{\left( {{R_{{\rm{co}}}}} \right)}^2}} \right\}} \right)\left( {1 + \left( {\hat h - 1} \right){p_{{\rm{b}}}}} \right){{\left( {{\lambda _{{\rm{wave}}}}} \right)}^2}\sigma _{{\rm{d}}}^2}}{{{{\left( {4\pi } \right)}^2}{{\left( {{R_{{\rm{co}}}}} \right)}^2}}}\ln \left( {1 + {{\left( {\frac{{{R_{{\rm{co}}}}}}{{{h_{{\rm{BS}}}}}}} \right)}^2}} \right).
\label{t21}
\end{equation}
For the first term of (\ref{t2}), we obtain:
\begin{equation*}
\sum\limits_{{{\mathbf{x}}_{{\rm{IRS}}}} \in {\Psi _{{\rm{IRS}}}}({{\mathbf{x}}_{\rm{U}}}):{{{\mathbf{\tilde x}}}_{\rm{U}}}({{\mathbf{x}}_{{\rm{IRS}}}}) = {{\mathbf{x}}_{\rm{U}}}} {\sum\limits_{q = 1}^Q {\sum\limits_{{{{\mathbf{x'}}}_{{\rm{IRS}}}} \in {\Psi _{{\rm{IRS}}}}({{\mathbf{x}}_{\rm{U}}}):{{{\mathbf{\tilde x}}}_{\rm{U}}}({{{\mathbf{x'}}}_{{\rm{IRS}}}}) = {{\mathbf{x}}_{\rm{U}}}} {\sum\limits_{q' = 1}^Q {} } } } 
\end{equation*}
\begin{equation*}
\frac{{{\delta _{q - q'}}{\delta _{{{\mathbf{x}}_{{\rm{IRS}}}} - {{\mathbf{x}}^\prime }_{{\rm{IRS}}}}}{{\left( {{\lambda _{{\rm{wave}}}}} \right)}^4}}}{{{{\left( {4\pi } \right)}^4}\left( {{{({h_{{\rm{BS}}}} - {h_{{\rm{IRS}}}})}^2} + \left\| {{{{\mathbf{\tilde x}}}_{{\rm{BS}}}}({{\mathbf{x}}_{\rm{U}}}) - {{\mathbf{x}}_{{\rm{IRS}}}}} \right\|_2^2} \right)\left( {{{({h_{{\rm{IRS}}}})}^2} + \left\| {{{\mathbf{x}}_{{\rm{IRS}}}} - {{\mathbf{x}}_{\rm{U}}}} \right\|_2^2} \right)}} \times 
\end{equation*}
\begin{equation*}
{\left( {G\left( {{{{\mathbf{\tilde x}}}_{{\rm{BS}}}}({{\mathbf{x}}_{\rm{U}}}),{{\mathbf{x}}_{\rm{U}}},{{\mathbf{x}}_{{\rm{IRS}}}}} \right) + \sum\limits_{{{{\mathbf{x''}}}_{{\rm{IRS}}}} \in {\Psi _{{\rm{IRS}}}}({{{\mathbf{\tilde x}}}_{{\rm{BS}}}}({{\mathbf{x}}_{\rm{U}}})):{{\mathbf{x}}_{\rm{U}}} = {{{\mathbf{\tilde x}}}_{\rm{U}}}({{{\mathbf{x''}}}_{{\rm{IRS}}}})} {G\left( {{{{\mathbf{\tilde x}}}_{{\rm{BS}}}}({{\mathbf{x}}_{\rm{U}}}),{{{\mathbf{x''}}}_{{\rm{IRS}}}},{{\mathbf{x}}_{{\rm{IRS}}}}} \right)} } \right)^2}\sigma _{{\rm{d}}}^2 \ge
\end{equation*}
\begin{equation*}
\sum\limits_{{{\mathbf{x}}_{{\rm{IRS}}}} \in {\Psi _{{\rm{IRS}}}}({{\mathbf{x}}_{\rm{U}}})} {\frac{{{\delta ^2}\left( {1 - {\pmb{\mathbb{I}}}\left( {\left\| {{{{\mathbf{\tilde x}}}_{\rm{U}}}({{\mathbf{x}}_{{\rm{IRS}}}}) - {{\mathbf{x}}_{\rm{U}}}} \right\|} \right)} \right)Q{{\left( {{\lambda _{{\rm{wave}}}}} \right)}^4}\sigma _{{\rm{d}}}^2}}{{{{\left( {4\pi } \right)}^4}\left( {{{({h_{{\rm{BS}}}} - {h_{{\rm{IRS}}}})}^2} + \left\| {{{{\mathbf{\tilde x}}}_{{\rm{BS}}}}({{\mathbf{x}}_{\rm{U}}}) - {{\mathbf{x}}_{{\rm{IRS}}}}} \right\|_2^2} \right)\left( {{{({h_{{\rm{IRS}}}})}^2} + \left\| {{{\mathbf{x}}_{{\rm{IRS}}}} - {{\mathbf{x}}_{\rm{U}}}} \right\|_2^2} \right)}}}  \ge
\end{equation*}
\begin{equation*}
\sum\limits_{{{\mathbf{x}}_{{\rm{IRS}}}} \in {\Psi _{{\rm{IRS}}}}({{\mathbf{x}}_{\rm{U}}})} {\frac{{{\pmb{\mathbb{I}}}\left( {{n_{{\rm{BS}}}}({{\mathbf{x}}_{\rm{U}}})} \right){\delta ^2}\left( {1 - {\pmb{\mathbb{I}}}\left( {\left\| {{{{\mathbf{\tilde x}}}_{\rm{U}}}({{\mathbf{x}}_{{\rm{IRS}}}}) - {{\mathbf{x}}_{\rm{U}}}} \right\|} \right)} \right)Q{{\left( {{\lambda _{{\rm{wave}}}}} \right)}^4}\sigma _{{\rm{d}}}^2}}{{{{\left( {4\pi } \right)}^4}{{\left( {\max \left\{ {{{({h_{{\rm{BS}}}} - {h_{{\rm{IRS}}}})}^2},{{({h_{{\rm{IRS}}}})}^2}} \right\} + 3{{\left( {{R_{{\rm{co}}}}} \right)}^2} + \left\| {{{\mathbf{x}}_{{\rm{IRS}}}} - {{\mathbf{x}}_{\rm{U}}}} \right\|_2^2} \right)}^2}}}}  .
\end{equation*}
Then, we have:
\begin{equation*}
 \scalebox{.95}[1]{$E\left\{ {\left. {\sum\limits_{{{\mathbf{x}}_{{\rm{IRS}}}} \in {\Psi _{{\rm{IRS}}}}({{\mathbf{x}}_{\rm{U}}})} {\frac{{{\pmb{\mathbb{I}}}\left( {{n_{{\rm{BS}}}}({{\mathbf{x}}_{\rm{U}}})} \right){\delta ^2}\left( {1 - {\pmb{\mathbb{I}}}\left( {\left\| {{{{\mathbf{\tilde x}}}_{\rm{U}}}({{\mathbf{x}}_{{\rm{IRS}}}}) - {{\mathbf{x}}_{\rm{U}}}} \right\|} \right)} \right)Q{{\left( {{\lambda _{{\rm{wave}}}}} \right)}^4}\sigma _{{\rm{d}}}^2}}{{{{\left( {4\pi } \right)}^4}{{\left( {\max \left\{ {{{({h_{{\rm{BS}}}} - {h_{{\rm{IRS}}}})}^2},{{({h_{{\rm{IRS}}}})}^2}} \right\} + 3{{\left( {{R_{{\rm{co}}}}} \right)}^2} + \left\| {{{\mathbf{x}}_{{\rm{IRS}}}} - {{\mathbf{x}}_{\rm{U}}}} \right\|_2^2} \right)}^2}}}} } \right|{\Phi _{\rm{U}}},{\Phi _{{\rm{IRS}}}}} \right\} = $}
\end{equation*}
\begin{equation*}
\sum\limits_{{{\mathbf{x}}_{{\rm{IRS}}}} \in {\Psi _{{\rm{IRS}}}}({{\mathbf{x}}_{\rm{U}}})} {\frac{{\left( {1 - \exp \left\{- {{\lambda _{{\rm{BS}}}}\pi {{\left( {{R_{{\rm{co}}}}} \right)}^2}} \right\}} \right){\delta ^2}Q{{\left( {{\lambda _{{\rm{wave}}}}} \right)}^4}\sigma _{{\rm{d}}}^2}}{{{n_{\rm{U}}}({{\mathbf{x}}_{{\rm{IRS}}}}){{\left( {4\pi } \right)}^4}{{\left( {\max \left\{ {{{({h_{{\rm{BS}}}} - {h_{{\rm{IRS}}}})}^2},{{({h_{{\rm{IRS}}}})}^2}} \right\} + 3{{\left( {{R_{{\rm{co}}}}} \right)}^2} + \left\| {{{\mathbf{x}}_{{\rm{IRS}}}} - {{\mathbf{x}}_{\rm{U}}}} \right\|_2^2} \right)}^2}}}}  .
\end{equation*}

Now, one of remaining randoms variables is ${ n}_{\rm U}({\bf x}_{\rm U})$.
To bound the expectation of this term,
 consider $W$ as a poisson random variable with parameter $\lambda$, whose distribution is $P_{\lambda}(w)=\frac{{{\lambda ^w}{e^{ - \lambda }}}}{{w!}}$. Then, we have to calculate the distribution of random variable $\tilde W$  ($P({\tilde w}=w)=P(w|w\ge 1)$) as follows:
\begin{equation}
P({\tilde w}=w)=P(w|w \ge 1) = \frac{{\Pr \left\{ {W = w,W \ge 1} \right\}}}{{\Pr \left\{ {W \ge 1} \right\}}} = \left\{ {\begin{array}{*{20}{c}}
  {\frac{{{P_\lambda }(w)}}{{1 - {e^{ - \lambda }}}},w \ge 1} \\ 
  {0,w < 1} 
\end{array}} \right..
\label{condition-on-x}
\end{equation}
Then, we obtain:
\begin{equation*}
E\left\{ {\frac{1}{{\tilde W}}} \right\} = \sum\limits_{\tilde w = 1}^\infty  {\frac{{{\lambda ^{\tilde w}}{e^{ - \lambda }}}}{{\tilde w\left( {\tilde w!} \right)\left( {1 - {e^{ - \lambda }}} \right)}}}  \ge \sum\limits_{\tilde w = 1}^\infty  {\frac{{{\lambda ^{\tilde w}}{e^{ - \lambda }}}}{{\left( {\tilde w + 1} \right)!\left( {1 - {e^{ - \lambda }}} \right)}}}  = \sum\limits_{\tilde w = 1}^\infty  {\frac{{{\lambda ^{\tilde w + 1}}{e^{ - \lambda }}}}{{\lambda \left( {\tilde w + 1} \right)!\left( {1 - {e^{ - \lambda }}} \right)}}} 
\end{equation*}
\begin{equation}
 = \sum\limits_{u = 2}^\infty  {\frac{{{\lambda ^u}{e^{ - \lambda }}}}{{\lambda \left( u \right)!\left( {1 - {e^{ - \lambda }}} \right)}}}  = \frac{{{e^{ - \lambda }}\left( {{e^\lambda } - \lambda  - 1} \right)}}{{\lambda \left( {1 - {e^{ - \lambda }}} \right)}} = \frac{{1 - \lambda {e^{ - \lambda }} - {e^{ - \lambda }}}}{{\lambda \left( {1 - {e^{ - \lambda }}} \right)}},
\label{1/w}
\end{equation}
hence, we have:
\begin{equation}
E\left\{ \frac{1}{{{ n}_{\rm{U}}}({{\mathbf{x}}_{\rm{U}}})} \right\} \ge \frac{{1 - {\lambda _{\rm{U}}}\pi {{\left( {{R_{{\rm{co}}}}} \right)}^2}\exp \left\{ { - {\lambda _{\rm{U}}}\pi {{\left( {{R_{{\rm{co}}}}} \right)}^2}} \right\} - \exp \left\{ { - {\lambda _{\rm{U}}}\pi {{\left( {{R_{{\rm{co}}}}} \right)}^2}} \right\}}}{{{\lambda _{\rm{U}}}\pi {{\left( {{R_{{\rm{co}}}}} \right)}^2}\left( {1 - \exp \left\{ { - {\lambda _{\rm{U}}}\pi {{\left( {{R_{{\rm{co}}}}} \right)}^2}} \right\}} \right)}}.
\label{t7}
\end{equation}
Thus, we have:
\begin{equation*}
 \scalebox{.95}[1]{$E\left\{ {\left. {\sum\limits_{{{\mathbf{x}}_{{\rm{IRS}}}} \in {\Psi _{{\rm{IRS}}}}({{\mathbf{x}}_{\rm{U}}})} {\frac{{\left( {1 - \exp \left\{- {{\lambda _{{\rm{BS}}}}\pi {{\left( {{R_{{\rm{co}}}}} \right)}^2}} \right\}} \right){\delta ^2}Q{{\left( {{\lambda _{{\rm{wave}}}}} \right)}^4}\sigma _{{\rm{d}}}^2}}{{{n_{\rm{U}}}({{\mathbf{x}}_{{\rm{IRS}}}}){{\left( {4\pi } \right)}^4}{{\left( {\max \left\{ {{{({h_{{\rm{BS}}}} - {h_{{\rm{IRS}}}})}^2},{{({h_{{\rm{IRS}}}})}^2}} \right\} + 3{{\left( {{R_{{\rm{co}}}}} \right)}^2} + \left\| {{{\mathbf{x}}_{{\rm{IRS}}}} - {{\mathbf{x}}_{\rm{U}}}} \right\|_2^2} \right)}^2}}}} } \right|{\Phi _{{\rm{IRS}}}}} \right\}$}
\end{equation*}
\begin{equation*}
 \ge \sum\limits_{{{\mathbf{x}}_{{\rm{IRS}}}} \in {\Psi _{{\rm{IRS}}}}({{\mathbf{x}}_{\rm{U}}})} {\frac{{\left( {1 - \exp \left\{- {{\lambda _{{\rm{BS}}}}\pi {{\left( {{R_{{\rm{co}}}}} \right)}^2}} \right\}} \right){\delta ^2}Q{{\left( {{\lambda _{{\rm{wave}}}}} \right)}^4}\sigma _{{\rm{d}}}^2}}{{{{\left( {4\pi } \right)}^4}{{\left( {\max \left\{ {{{({h_{{\rm{BS}}}} - {h_{{\rm{IRS}}}})}^2},{{({h_{{\rm{IRS}}}})}^2}} \right\} + 3{{\left( {{R_{{\rm{co}}}}} \right)}^2} + \left\| {{{\mathbf{x}}_{{\rm{IRS}}}} - {{\mathbf{x}}_{\rm{U}}}} \right\|_2^2} \right)}^2}}}}  \times 
\end{equation*}
\begin{equation*}
\frac{{1 - {\lambda _{\rm{U}}}\pi {{\left( {{R_{{\rm{co}}}}} \right)}^2}\exp \left\{ { - {\lambda _{\rm{U}}}\pi {{\left( {{R_{{\rm{co}}}}} \right)}^2}} \right\} - \exp \left\{ { - {\lambda _{\rm{U}}}\pi {{\left( {{R_{{\rm{co}}}}} \right)}^2}} \right\}}}{{{\lambda _{\rm{U}}}\pi {{\left( {{R_{{\rm{co}}}}} \right)}^2}\left( {1 - \exp \left\{ { - {\lambda _{\rm{U}}}\pi {{\left( {{R_{{\rm{co}}}}} \right)}^2}} \right\}} \right)}}.
\end{equation*}
We note that we had explicitly assumed that ${{\mathbf{x}}_{\rm{U}}}$ has been occurred in the process ${{\Phi}_{\rm{U}}}$, thus, ${{\mathbf{x}}_{\rm{U}}}$, is deterministic and did not disappear in the above term. And by   Campbell’s theorem \cite[Theorem 4.1]{Haenggi}, we have:
\begin{equation*}
E\left\{ {\sum\limits_{{{\mathbf{x}}_{{\rm{IRS}}}} \in {\Psi _{{\rm{IRS}}}}({{\mathbf{x}}_{\rm{U}}})} {\frac{{\left( {1 - \exp \left\{- {{\lambda _{{\rm{BS}}}}\pi {{\left( {{R_{{\rm{co}}}}} \right)}^2}} \right\}} \right){\delta ^2}Q{{\left( {{\lambda _{{\rm{wave}}}}} \right)}^4}\sigma _{{\rm{d}}}^2}}{{{{\left( {4\pi } \right)}^4}{{\left( {\max \left\{ {{{({h_{{\rm{BS}}}} - {h_{{\rm{IRS}}}})}^2},{{({h_{{\rm{IRS}}}})}^2}} \right\} + 3{{\left( {{R_{{\rm{co}}}}} \right)}^2} + \left\| {{{\mathbf{x}}_{{\rm{IRS}}}} - {{\mathbf{x}}_{\rm{U}}}} \right\|_2^2} \right)}^2}}}}  \times } \right.
\end{equation*}
\begin{equation*}
\left. {\frac{{1 - {\lambda _{\rm{U}}}\pi {{\left( {{R_{{\rm{co}}}}} \right)}^2}\exp \left\{ { - {\lambda _{\rm{U}}}\pi {{\left( {{R_{{\rm{co}}}}} \right)}^2}} \right\} - \exp \left\{ { - {\lambda _{\rm{U}}}\pi {{\left( {{R_{{\rm{co}}}}} \right)}^2}} \right\}}}{{{\lambda _{\rm{U}}}\pi {{\left( {{R_{{\rm{co}}}}} \right)}^2}\left( {1 - \exp \left\{ { - {\lambda _{\rm{U}}}\pi {{\left( {{R_{{\rm{co}}}}} \right)}^2}} \right\}} \right)}}} \right\} = 
\end{equation*}
\begin{equation*}
\frac{{{\lambda _{{\rm{IRS}}}}\left( {1 - \exp \left\{- {{\lambda _{{\rm{BS}}}}\pi {{\left( {{R_{{\rm{co}}}}} \right)}^2}} \right\}} \right){\delta ^2}Q{{\left( {{\lambda _{{\rm{wave}}}}} \right)}^4}\sigma _{{\rm{d}}}^2}}{{{{\left( {4\pi } \right)}^4}}} \times 
\end{equation*}
\begin{equation*}
\frac{{1 - {\lambda _{\rm{U}}}\pi {{\left( {{R_{{\rm{co}}}}} \right)}^2}\exp \left\{ { - {\lambda _{\rm{U}}}\pi {{\left( {{R_{{\rm{co}}}}} \right)}^2}} \right\} - \exp \left\{ { - {\lambda _{\rm{U}}}\pi {{\left( {{R_{{\rm{co}}}}} \right)}^2}} \right\}}}{{{\lambda _{\rm{U}}}\pi {{\left( {{R_{{\rm{co}}}}} \right)}^2}\left( {1 - \exp \left\{ { - {\lambda _{\rm{U}}}\pi {{\left( {{R_{{\rm{co}}}}} \right)}^2}} \right\}} \right)}} \times 
\end{equation*}
\begin{equation*}
\int\limits_{{\cal C}\left( {{{\mathbf{x}}_{\rm{U}}},{R_{{\rm{co}}}}} \right)} {\frac{{d{\mathbf{x}}}}{{{{\left( {\max \left\{ {{{({h_{{\rm{BS}}}} - {h_{{\rm{IRS}}}})}^2},{{({h_{{\rm{IRS}}}})}^2}} \right\} + 3{{\left( {{R_{{\rm{co}}}}} \right)}^2} + \left\| {{\mathbf{x}} - {{\mathbf{x}}_{\rm{U}}}} \right\|_2^2} \right)}^2}}}}  = 
\end{equation*}
\begin{equation*}
\frac{{{\lambda _{{\rm{IRS}}}}\left( {1 - \exp \left\{- {{\lambda _{{\rm{BS}}}}\pi {{\left( {{R_{{\rm{co}}}}} \right)}^2}} \right\}} \right){\delta ^2}Q{{\left( {{\lambda _{{\rm{wave}}}}} \right)}^4}\sigma _{{\rm{d}}}^2}}{{{{\left( {4\pi } \right)}^4}}} \times 
\end{equation*}
\begin{equation*}
\frac{{1 - {\lambda _{\rm{U}}}\pi {{\left( {{R_{{\rm{co}}}}} \right)}^2}\exp \left\{ { - {\lambda _{\rm{U}}}\pi {{\left( {{R_{{\rm{co}}}}} \right)}^2}} \right\} - \exp \left\{ { - {\lambda _{\rm{U}}}\pi {{\left( {{R_{{\rm{co}}}}} \right)}^2}} \right\}}}{{{\lambda _{\rm{U}}}{{\left( {{R_{{\rm{co}}}}} \right)}^2}\left( {1 - \exp \left\{ { - {\lambda _{\rm{U}}}\pi {{\left( {{R_{{\rm{co}}}}} \right)}^2}} \right\}} \right)}} \times 
\end{equation*}
\begin{equation*}
\frac{{{{\left( {{R_{{\rm{co}}}}} \right)}^2}}}{{\left( {\max \left\{ {{{({h_{{\rm{BS}}}} - {h_{{\rm{IRS}}}})}^2},{{({h_{{\rm{IRS}}}})}^2}} \right\} + 3{{\left( {{R_{{\rm{co}}}}} \right)}^2}} \right)\left( {\max \left\{ {{{({h_{{\rm{BS}}}} - {h_{{\rm{IRS}}}})}^2},{{({h_{{\rm{IRS}}}})}^2}} \right\} + 4{{\left( {{R_{{\rm{co}}}}} \right)}^2}} \right)}}.
\end{equation*}
This completes the proof.

\section{Proof of Theorem 3}

We have:

\begin{equation*}
{{\mathbb{I}}}({{\mathbf{x}}_{\rm{U}}}) =
\end{equation*}
\begin{equation*}
  \scalebox{.92}[1]{$ \sum\limits_{{{{\mathbf{x'}}}_{{\rm{BS}}}} \in {\Psi _{{\rm{BS}}}}({{\mathbf{x}}_{\rm{U}}})} {{\mathbb{H}}({{{\mathbf{x'}}}_{{\rm{BS}}}},{{\mathbf{x}}_{\rm{U}}})\sum\limits_{{{{\mathbf{x'}}}_{\rm{U}}} \in {\Psi _{\rm{U}}}({{{\mathbf{x'}}}_{{\rm{BS}}}}):{{{\mathbf{x'}}}_{\rm{U}}} \ne {{\mathbf{x}}_{\rm{U}}},{{{\mathbf{\tilde x}}}_{{\rm{BS}}}}({{{\mathbf{x'}}}_{\rm{U}}}) = {{{\mathbf{x'}}}_{{\rm{BS}}}}}  } $}
\end{equation*}
\begin{equation*}
  \scalebox{1}[1]{${\left( {G\left( {{{{\mathbf{x'}}}_{{\rm{BS}}}},{{{\mathbf{x'}}}_{\rm{U}}},{{\mathbf{x}}_{\rm{U}}}} \right) + \sum\limits_{{{\mathbf{x}}_{{\rm{IRS}}}} \in {\Psi _{{\rm{IRS}}}}({{{\mathbf{x'}}}_{{\rm{BS}}}}):{{{\mathbf{x'}}}_{\rm{U}}} = {{{\mathbf{\tilde x}}}_{\rm{U}}}({{\mathbf{x}}_{{\rm{IRS}}}})} {G\left( {{{{\mathbf{x'}}}_{{\rm{BS}}}},{{\mathbf{x}}_{{\rm{IRS}}}},{{\mathbf{x}}_{\rm{U}}}} \right)} } \right)}\times {X}({{{\mathbf{x'}}}_{\rm{U}}})$}
\end{equation*}
\begin{equation*}
 + \sum\limits_{{{\mathbf{x}}_{{\rm{IRS}}}} \in {\Psi _{{\rm{IRS}}}}({{\mathbf{x}}_{\rm{U}}})} {\sum\limits_{q = 1}^Q {{{\mathbb{H}}^{[q]}}({{\mathbf{x}}_{{\rm{IRS}}}},{{\mathbf{x}}_{\rm{U}}})\exp \{ j{\theta _q}({{\mathbf{x}}_{{\rm{IRS}}}})\}  \times } } 
\end{equation*}
\begin{equation*}
  \scalebox{1}[1]{$\left( {\sum\limits_{{{{\mathbf{x'}}}_{{\rm{BS}}}} \in {\Psi _{{\rm{BS}}}}({{\mathbf{x}}_{{\rm{IRS}}}})} {\sum\limits_{{{{\mathbf{x'}}}_{\rm{U}}} \in {\Psi _{\rm{U}}}({{{\mathbf{x'}}}_{{\rm{BS}}}}):{{{\mathbf{x'}}}_{\rm{U}}} \ne {{\mathbf{x}}_{\rm{U}}},{{{\mathbf{\tilde x}}}_{{\rm{BS}}}}({{{\mathbf{x'}}}_{\rm{U}}}) = {{{\mathbf{x'}}}_{{\rm{BS}}}}} {{{\mathbb{H}}^{[q]}}({{{\mathbf{x'}}}_{{\rm{BS}}}},{{\mathbf{x}}_{{\rm{IRS}}}})\times} } } \right.$}
\end{equation*}
\begin{equation*}
  \scalebox{1}[1]{$\left.\left( {G\left( {{{{\mathbf{x'}}}_{{\rm{BS}}}},{{{\mathbf{x'}}}_{\rm{U}}},{{\mathbf{x}}_{{\rm{IRS}}}}} \right) + \sum\limits_{{{{\mathbf{x'}}}_{{\rm{IRS}}}} \in {\Psi _{{\rm{IRS}}}}({{{\mathbf{x'}}}_{{\rm{BS}}}}):{{{\mathbf{x'}}}_{\rm{U}}} = {{{\mathbf{\tilde x}}}_{\rm{U}}}({{{\mathbf{x'}}}_{{\rm{IRS}}}})} {G\left( {{{{\mathbf{x'}}}_{{\rm{BS}}}},{{{\mathbf{x'}}}_{{\rm{IRS}}}},{{\mathbf{x}}_{{\rm{IRS}}}}} \right)} } \right){X}({{{\mathbf{x'}}}_{\rm{U}}})\right) $}
\end{equation*}

Hence:
\begin{equation*}
E\left\{ {{{\left| {{{\mathbb{I}}}({{\mathbf{x}}_{\rm{U}}})} \right|}^2}\left| {{\Phi _{\rm{U}}},{\Phi _{{\rm{BS}}}},{\Phi _{{\rm{IRS}}}},{{{\mathbf{\tilde x}}}_{{\rm{BS}}}}({{\mathbf{x}}_{\rm{U}}}),{{{\mathbf{\tilde x}}}_{\rm{U}}}({{\mathbf{x}}_{{\rm{IRS}}}})} \right.} \right\} \le
\end{equation*}
\begin{equation*}
\sum\limits_{{{{\mathbf{x'}}}_{{\rm{BS}}}} \in {\Psi _{{\rm{BS}}}}({{\mathbf{x}}_{\rm{U}}})} {\begin{array}{*{20}{c}}
  {}&{} 
\end{array}\sum\limits_{{{{\mathbf{x'}}}_{\rm{U}}} \in {\Psi _{\rm{U}}}({{{\mathbf{x'}}}_{{\rm{BS}}}}):{{{\mathbf{x'}}}_{\rm{U}}} \ne {{\mathbf{x}}_{\rm{U}}},{{{\mathbf{\tilde x}}}_{{\rm{BS}}}}({{{\mathbf{x'}}}_{\rm{U}}}) = {{{\mathbf{x'}}}_{{\rm{BS}}}}} {\frac{{\left( {1 + \left( {\hat h - 1} \right){p_{{\rm{b}}}}} \right){{\left( {{\lambda _{{\rm{wave}}}}} \right)}^2}\sigma _{{\rm{d}}}^2}}{{{{\left( {4\pi } \right)}^2}\left( {{{({h_{{\rm{BS}}}})}^2} + \left\| {{{{\mathbf{x'}}}_{{\rm{BS}}}} - {{\mathbf{x}}_{\rm{U}}}} \right\|_2^2} \right)}} \times } } 
\end{equation*}
\begin{equation}
{\left( {G\left( {{{{\mathbf{x'}}}_{{\rm{BS}}}},{{{\mathbf{x'}}}_{\rm{U}}},{{\mathbf{x}}_{\rm{U}}}} \right) + \sum\limits_{{{\mathbf{x}}_{{\rm{IRS}}}} \in {\Psi _{{\rm{IRS}}}}({{{\mathbf{x'}}}_{{\rm{BS}}}}):{{{\mathbf{x'}}}_{\rm{U}}} = {{{\mathbf{\tilde x}}}_{\rm{U}}}({{\mathbf{x}}_{{\rm{IRS}}}})} {G\left( {{{{\mathbf{x'}}}_{{\rm{BS}}}},{{\mathbf{x}}_{{\rm{IRS}}}},{{\mathbf{x}}_{\rm{U}}}} \right)} } \right)^2} + 
\label{t25}
\end{equation}
\begin{equation*}
\sum\limits_{{{\mathbf{x}}_{{\rm{IRS}}}} \in {\Psi _{{\rm{IRS}}}}({{\mathbf{x}}_{\rm{U}}}):{{{\mathbf{\tilde x}}}_{\rm{U}}}({{\mathbf{x}}_{{\rm{IRS}}}}) = {{\mathbf{x}}_{\rm{U}}}} {\begin{array}{*{20}{c}}
  {}&{} 
\end{array}\sum\limits_{{{{\mathbf{x'}}}_{{\rm{IRS}}}} \in {\Psi _{{\rm{IRS}}}}({{\mathbf{x}}_{\rm{U}}}):{{{\mathbf{\tilde x}}}_{\rm{U}}}({{{\mathbf{x'}}}_{{\rm{IRS}}}}) = {{\mathbf{x}}_{\rm{U}}}} {\begin{array}{*{20}{c}}
  {}&{} 
\end{array}\sum\limits_{{{{\mathbf{x'}}}_{\rm{U}}} \in {\Psi _{\rm{U}}}({{{\mathbf{\tilde x}}}_{{\rm{BS}}}}({{\mathbf{x}}_{\rm{U}}})):{{{\mathbf{x'}}}_{\rm{U}}} \ne {{\mathbf{x}}_{\rm{U}}},{{{\mathbf{\tilde x}}}_{{\rm{BS}}}}({{{\mathbf{x'}}}_{\rm{U}}}) = {{{\mathbf{\tilde x}}}_{{\rm{BS}}}}({{\mathbf{x}}_{\rm{U}}})} {} } } 
\end{equation*}
\begin{equation*}
   \scalebox{.9}[1]{$\frac{{{Q^2{\left( {{\lambda _{{\rm{wave}}}}} \right)}^4}}}{{{{\left( {4\pi } \right)}^4}\sqrt {{{({h_{{\rm{IRS}}}})}^2} + \left\| {{{\mathbf{x}}_{{\rm{IRS}}}} - {{\mathbf{x}}_{\rm{U}}}} \right\|_2^2}  \times \sqrt {{{({h_{{\rm{IRS}}}})}^2} + \left\| {{{{\mathbf{x'}}}_{{\rm{IRS}}}} - {{\mathbf{x}}_{\rm{U}}}} \right\|_2^2}  \times \sqrt {{{({h_{{\rm{BS}}}} - {h_{{\rm{IRS}}}})}^2} + \left\| {{{{\mathbf{\tilde x}}}_{{\rm{BS}}}}({{\mathbf{x}}_{\rm{U}}}) - {{\mathbf{x}}_{{\rm{IRS}}}}} \right\|_2^2}  \times \sqrt {{{({h_{{\rm{BS}}}} - {h_{{\rm{IRS}}}})}^2} + \left\| {{{{\mathbf{\tilde x}}}_{{\rm{BS}}}}({{\mathbf{x}}_{\rm{U}}}) - {{{\mathbf{x'}}}_{{\rm{IRS}}}}} \right\|_2^2} }} \times $}
\end{equation*}
\begin{equation*}
\left( {G\left( {{{{\mathbf{\tilde x}}}_{{\rm{BS}}}}({{\mathbf{x}}_{\rm{U}}}),{{{\mathbf{x'}}}_{\rm{U}}},{{\mathbf{x}}_{{\rm{IRS}}}}} \right) + \sum\limits_{{{{\mathbf{x''}}}_{{\rm{IRS}}}} \in {\Psi _{{\rm{IRS}}}}({{{\mathbf{x'}}}_{{\rm{BS}}}}):{{{\mathbf{x'}}}_{\rm{U}}} = {{{\mathbf{\tilde x}}}_{\rm{U}}}({{{\mathbf{x''}}}_{{\rm{IRS}}}})} {G\left( {{{{\mathbf{\tilde x}}}_{{\rm{BS}}}}({{\mathbf{x}}_{\rm{U}}}),{{{\mathbf{x''}}}_{{\rm{IRS}}}},{{\mathbf{x}}_{{\rm{IRS}}}}} \right)} } \right) \times 
\end{equation*}
\begin{equation}
\left( {G\left( {{{{\mathbf{\tilde x}}}_{{\rm{BS}}}}({{\mathbf{x}}_{\rm{U}}}),{{{\mathbf{x'}}}_{\rm{U}}},{{{\mathbf{x'}}}_{{\rm{IRS}}}}} \right) + \sum\limits_{{{{\mathbf{x''}}}_{{\rm{IRS}}}} \in {\Psi _{{\rm{IRS}}}}({{{\mathbf{x'}}}_{{\rm{BS}}}}):{{{\mathbf{x'}}}_{\rm{U}}} = {{{\mathbf{\tilde x}}}_{\rm{U}}}({{{\mathbf{x''}}}_{{\rm{IRS}}}})} {G\left( {{{{\mathbf{\tilde x}}}_{{\rm{BS}}}}({{\mathbf{x}}_{\rm{U}}}),{{{\mathbf{x''}}}_{{\rm{IRS}}}},{{{\mathbf{x'}}}_{{\rm{IRS}}}}} \right)} } \right) \sigma _{{\rm{d}}}^2 + 
\label{t26}
\end{equation}
\begin{equation*}
\sum\limits_{{{\mathbf{x}}_{{\rm{IRS}}}} \in {\Psi _{{\rm{IRS}}}}({{\mathbf{x}}_{\rm{U}}}):{{{\mathbf{\tilde x}}}_{\rm{U}}}({{\mathbf{x}}_{{\rm{IRS}}}}) = {{\mathbf{x}}_{\rm{U}}}} {\begin{array}{*{20}{c}}
  {}&{} 
\end{array}\sum\limits_{{{{\mathbf{x'}}}_{{\rm{BS}}}} \in {\Psi _{{\rm{BS}}}}({{\mathbf{x}}_{{\rm{IRS}}}}):{{{\mathbf{x'}}}_{{\rm{BS}}}} \ne {{{\mathbf{\tilde x}}}_{{\rm{BS}}}}({{\mathbf{x}}_{\rm{U}}})} {\begin{array}{*{20}{c}}
  {}&{} 
\end{array}\sum\limits_{{{{\mathbf{x'}}}_{\rm{U}}} \in {\Psi _{\rm{U}}}({{{\mathbf{x'}}}_{{\rm{BS}}}}):{{{\mathbf{x'}}}_{\rm{U}}} \ne {{\mathbf{x}}_{\rm{U}}},{{{\mathbf{\tilde x}}}_{{\rm{BS}}}}({{{\mathbf{x'}}}_{\rm{U}}}) = {{{\mathbf{x'}}}_{{\rm{BS}}}}} {} } } 
\end{equation*}
\begin{equation*}
\frac{{Q{{\left( {{\lambda _{{\rm{wave}}}}} \right)}^4}}}{{{{\left( {4\pi } \right)}^4}\left( {{{({h_{{\rm{IRS}}}})}^2} + \left\| {{{\mathbf{x}}_{{\rm{IRS}}}} - {{\mathbf{x}}_{\rm{U}}}} \right\|_2^2} \right)\left( {{{({h_{{\rm{BS}}}} - {h_{{\rm{IRS}}}})}^2} + \left\| {{{{\mathbf{x'}}}_{{\rm{BS}}}} - {{\mathbf{x}}_{{\rm{IRS}}}}} \right\|_2^2} \right)}} \times 
\end{equation*}
\begin{equation}
{\left( {G\left( {{{{\mathbf{x'}}}_{{\rm{BS}}}},{{{\mathbf{x'}}}_{\rm{U}}},{{\mathbf{x}}_{{\rm{IRS}}}}} \right) + \sum\limits_{{{{\mathbf{x'}}}_{{\rm{IRS}}}} \in {\Psi _{{\rm{IRS}}}}({{{\mathbf{x'}}}_{{\rm{BS}}}}):{{{\mathbf{x'}}}_{\rm{U}}} = {{{\mathbf{\tilde x}}}_{\rm{U}}}({{{\mathbf{x'}}}_{{\rm{IRS}}}})} {G\left( {{{{\mathbf{x'}}}_{{\rm{BS}}}},{{{\mathbf{x'}}}_{{\rm{IRS}}}},{{\mathbf{x}}_{{\rm{IRS}}}}} \right)} } \right)^2} \sigma _{{\rm{d}}}^2 + 
\label{t27}
\end{equation}
\begin{equation*}
\sum\limits_{{{\mathbf{x}}_{{\rm{IRS}}}} \in {\Psi _{{\rm{IRS}}}}({{\mathbf{x}}_{\rm{U}}}):{{{\mathbf{\tilde x}}}_{\rm{U}}}({{\mathbf{x}}_{{\rm{IRS}}}}) \ne {{\mathbf{x}}_{\rm{U}}}} {\begin{array}{*{20}{c}}
  {}&{} 
\end{array}\sum\limits_{{{{\mathbf{x'}}}_{{\rm{BS}}}} \in {\Psi _{{\rm{BS}}}}({{\mathbf{x}}_{{\rm{IRS}}}})} {\begin{array}{*{20}{c}}
  {}&{} 
\end{array}\sum\limits_{{{{\mathbf{x'}}}_{\rm{U}}} \in {\Psi _{\rm{U}}}({{{\mathbf{x'}}}_{{\rm{BS}}}}):{{{\mathbf{x'}}}_{\rm{U}}} \ne {{\mathbf{x}}_{\rm{U}}},{{{\mathbf{\tilde x}}}_{{\rm{BS}}}}({{{\mathbf{x'}}}_{\rm{U}}}) = {{{\mathbf{x'}}}_{{\rm{BS}}}}} {} } } 
\end{equation*}
\begin{equation*}
\frac{{Q{{\left( {{\lambda _{{\rm{wave}}}}} \right)}^4}}}{{{{\left( {4\pi } \right)}^4}\left( {{{({h_{{\rm{IRS}}}})}^2} + \left\| {{{\mathbf{x}}_{{\rm{IRS}}}} - {{\mathbf{x}}_{\rm{U}}}} \right\|_2^2} \right)\left( {{{({h_{{\rm{BS}}}} - {h_{{\rm{IRS}}}})}^2} + \left\| {{{{\mathbf{x'}}}_{{\rm{BS}}}} - {{\mathbf{x}}_{{\rm{IRS}}}}} \right\|_2^2} \right)}} \times 
\end{equation*}
\begin{equation}
{\left( {G\left( {{{{\mathbf{x'}}}_{{\rm{BS}}}},{{{\mathbf{x'}}}_{\rm{U}}},{{\mathbf{x}}_{{\rm{IRS}}}}} \right) + \sum\limits_{{{{\mathbf{x'}}}_{{\rm{IRS}}}} \in {\Psi _{{\rm{IRS}}}}({{{\mathbf{x'}}}_{{\rm{BS}}}}):{{{\mathbf{x'}}}_{\rm{U}}} = {{{\mathbf{\tilde x}}}_{\rm{U}}}({{{\mathbf{x'}}}_{{\rm{IRS}}}})} {G\left( {{{{\mathbf{x'}}}_{{\rm{BS}}}},{{{\mathbf{x'}}}_{{\rm{IRS}}}},{{\mathbf{x}}_{{\rm{IRS}}}}} \right)} } \right)^2} \sigma _{{\rm{d}}}^2 . 
\label{t28}
\end{equation}

Now, to bound term (\ref{t25}), we have:
\begin{equation*}
\sum\limits_{{{{\mathbf{x'}}}_{{\rm{BS}}}} \in {\Psi _{{\rm{BS}}}}({{\mathbf{x}}_{\rm{U}}})} {\begin{array}{*{20}{c}}
  {}&{} 
\end{array}\sum\limits_{{{{\mathbf{x'}}}_{\rm{U}}} \in {\Psi _{\rm{U}}}({{{\mathbf{x'}}}_{{\rm{BS}}}}):{{{\mathbf{x'}}}_{\rm{U}}} \ne {{\mathbf{x}}_{\rm{U}}},{{{\mathbf{\tilde x}}}_{{\rm{BS}}}}({{{\mathbf{x'}}}_{\rm{U}}}) = {{{\mathbf{x'}}}_{{\rm{BS}}}}} {\frac{{\left( {1 + \left( {\hat h - 1} \right){p_{{\rm{b}}}}} \right){{\left( {{\lambda _{{\rm{wave}}}}} \right)}^2} \sigma _{{\rm{d}}}^2}}{{{{\left( {4\pi } \right)}^2}\left( {{{({h_{{\rm{BS}}}})}^2} + \left\| {{{{\mathbf{x'}}}_{{\rm{BS}}}} - {{\mathbf{x}}_{\rm{U}}}} \right\|_2^2} \right)}} \times } } 
\end{equation*}
\begin{equation*}
{\left( {G\left( {{{{\mathbf{x'}}}_{{\rm{BS}}}},{{{\mathbf{x'}}}_{\rm{U}}},{{\mathbf{x}}_{\rm{U}}}} \right) + \sum\limits_{{{\mathbf{x}}_{{\rm{IRS}}}} \in {\Psi _{{\rm{IRS}}}}({{{\mathbf{x'}}}_{{\rm{BS}}}}):{{{\mathbf{x'}}}_{\rm{U}}} = {{{\mathbf{\tilde x}}}_{\rm{U}}}({{\mathbf{x}}_{{\rm{IRS}}}})} {G\left( {{{{\mathbf{x'}}}_{{\rm{BS}}}},{{\mathbf{x}}_{{\rm{IRS}}}},{{\mathbf{x}}_{\rm{U}}}} \right)} } \right)^2} = 
\end{equation*}
\begin{equation*}
\sum\limits_{{{{\mathbf{x'}}}_{{\rm{BS}}}} \in {\Psi _{{\rm{BS}}}}({{\mathbf{x}}_{\rm{U}}})} {\begin{array}{*{20}{c}}
  {}&{} 
\end{array}\sum\limits_{{{{\mathbf{x'}}}_{\rm{U}}} \in {\Psi _{\rm{U}}}({{{\mathbf{x'}}}_{{\rm{BS}}}}):{{{\mathbf{x'}}}_{\rm{U}}} \ne {{\mathbf{x}}_{\rm{U}}}} {\frac{{\left( {1 + \left( {\hat h - 1} \right){p_{{\rm{b}}}}} \right){{\left( {{\lambda _{{\rm{wave}}}}} \right)}^2} \sigma _{{\rm{d}}}^2}}{{{{\left( {4\pi } \right)}^2}\left( {{{({h_{{\rm{BS}}}})}^2} + \left\| {{{{\mathbf{x'}}}_{{\rm{BS}}}} - {{\mathbf{x}}_{\rm{U}}}} \right\|_2^2} \right)}} \times } } 
\end{equation*}
\begin{equation*}
{\left( {G\left( {{{{\mathbf{x'}}}_{{\rm{BS}}}},{{{\mathbf{x'}}}_{\rm{U}}},{{\mathbf{x}}_{\rm{U}}}} \right) + \sum\limits_{{{\mathbf{x}}_{{\rm{IRS}}}} \in {\Psi _{{\rm{IRS}}}}({{{\mathbf{x'}}}_{{\rm{BS}}}})} {G\left( {{{{\mathbf{x'}}}_{{\rm{BS}}}},{{\mathbf{x}}_{{\rm{IRS}}}},{{\mathbf{x}}_{\rm{U}}}} \right)} } \right)^2} \le 
\end{equation*}
\begin{equation*}
\sum\limits_{{{{\mathbf{x'}}}_{{\rm{BS}}}} \in {\Psi _{{\rm{BS}}}}({{\mathbf{x}}_{\rm{U}}})} {\frac{{{{\left( {1 + {n_{{\rm{IRS}}}}({{{\mathbf{x'}}}_{{\rm{BS}}}})} \right)}^2}\left( {{n_{\rm{U}}}({{{\mathbf{x'}}}_{{\rm{BS}}}}) - 1} \right)\left( {1 + \left( {\hat h - 1} \right){p_{{\rm{b}}}}} \right){{\left( {{\lambda _{{\rm{wave}}}}} \right)}^2} \sigma _{{\rm{d}}}^2}}{{{{\left( {4\pi } \right)}^2}\left( {{{({h_{{\rm{BS}}}})}^2} + \left\| {{{{\mathbf{x'}}}_{{\rm{BS}}}} - {{\mathbf{x}}_{\rm{U}}}} \right\|_2^2} \right)}}} ,
\end{equation*}
then, we obtain:
\begin{equation*}
 \scalebox{.95}[1]{$E\left\{ {\left. {\sum\limits_{{{{\mathbf{x'}}}_{{\rm{BS}}}} \in {\Psi _{{\rm{BS}}}}({{\mathbf{x}}_{\rm{U}}})} {\frac{{{{\left( {1 + {n_{{\rm{IRS}}}}({{{\mathbf{x'}}}_{{\rm{BS}}}})} \right)}^2}\left( {{n_{\rm{U}}}({{{\mathbf{x'}}}_{{\rm{BS}}}}) - 1} \right)\left( {1 + \left( {\hat h - 1} \right){p_{{\rm{b}}}}} \right){{\left( {{\lambda _{{\rm{wave}}}}} \right)}^2} \sigma _{{\rm{d}}}^2}}{{{{\left( {4\pi } \right)}^2}\left( {{{({h_{{\rm{BS}}}})}^2} + \left\| {{{{\mathbf{x'}}}_{{\rm{BS}}}} - {{\mathbf{x}}_{\rm{U}}}} \right\|_2^2} \right)}}} } \right|{\Phi _{{\rm{BS}}}}} \right\} = $}
\end{equation*}
\begin{equation*}
 \scalebox{.95}[1]{$\sum\limits_{{{{\mathbf{x'}}}_{{\rm{BS}}}} \in {\Psi _{{\rm{BS}}}}({{\mathbf{x}}_{\rm{U}}})} {\frac{{\left( {1 + 3{\lambda _{{\rm{IRS}}}}\pi {{\left( {{R_{{\rm{co}}}}} \right)}^2} + {{\left( {{\lambda _{{\rm{IRS}}}}\pi {{\left( {{R_{{\rm{co}}}}} \right)}^2}} \right)}^2}} \right)\left( {1 + \left( {\hat h - 1} \right){p_{{\rm{b}}}}} \right){{\left( {{\lambda _{{\rm{wave}}}}} \right)}^2} \sigma _{{\rm{d}}}^2}}{{{{\left( {4\pi } \right)}^2}\left( {{{({h_{{\rm{BS}}}})}^2} + \left\| {{{{\mathbf{x'}}}_{{\rm{BS}}}} - {{\mathbf{x}}_{\rm{U}}}} \right\|_2^2} \right)}} \times } $}
\end{equation*}
\begin{equation*}
\frac{{{\lambda _{\rm{U}}}\pi {{\left( {{R_{{\rm{co}}}}} \right)}^2} + \exp \left\{ { - {\lambda _{\rm{U}}}\pi {{\left( {{R_{{\rm{co}}}}} \right)}^2}} \right\} - 1}}{{1 - \exp \left\{ { - {\lambda _{\rm{U}}}\pi {{\left( {{R_{{\rm{co}}}}} \right)}^2}} \right\}}}.
\end{equation*}
Hence, we have:
\begin{equation*}
 \scalebox{.95}[1]{$E\left\{ {\sum\limits_{{{{\mathbf{x'}}}_{{\rm{BS}}}} \in {\Psi _{{\rm{BS}}}}({{\mathbf{x}}_{\rm{U}}})} {\frac{{\left( {1 + 3{\lambda _{{\rm{IRS}}}}\pi {{\left( {{R_{{\rm{co}}}}} \right)}^2} + {{\left( {{\lambda _{{\rm{IRS}}}}\pi {{\left( {{R_{{\rm{co}}}}} \right)}^2}} \right)}^2}} \right)\left( {1 + \left( {\hat h - 1} \right){p_{{\rm{b}}}}} \right){{\left( {{\lambda _{{\rm{wave}}}}} \right)}^2} \sigma _{{\rm{d}}}^2}}{{{{\left( {4\pi } \right)}^2}\left( {{{({h_{{\rm{BS}}}})}^2} + \left\| {{{{\mathbf{x'}}}_{{\rm{BS}}}} - {{\mathbf{x}}_{\rm{U}}}} \right\|_2^2} \right)}} \times } } \right.$}
\end{equation*}
\begin{equation*}
\left. {\frac{{{\lambda _{\rm{U}}}\pi {{\left( {{R_{{\rm{co}}}}} \right)}^2} + \exp \left\{ { - {\lambda _{\rm{U}}}\pi {{\left( {{R_{{\rm{co}}}}} \right)}^2}} \right\} - 1}}{{1 - \exp \left\{ { - {\lambda _{\rm{U}}}\pi {{\left( {{R_{{\rm{co}}}}} \right)}^2}} \right\}}}} \right\} =
\end{equation*}
\begin{equation*}
\frac{{{\lambda _{{\rm{BS}}}}\left( {1 + 3{\lambda _{{\rm{IRS}}}}\pi {{\left( {{R_{{\rm{co}}}}} \right)}^2} + {{\left( {{\lambda _{{\rm{IRS}}}}\pi {{\left( {{R_{{\rm{co}}}}} \right)}^2}} \right)}^2}} \right)\left( {1 + \left( {\hat h - 1} \right){p_{{\rm{b}}}}} \right){{\left( {{\lambda _{{\rm{wave}}}}} \right)}^2} \sigma _{{\rm{d}}}^2}}{{\left( {1 - \exp \left\{ { - {\lambda _{\rm{U}}}\pi {{\left( {{R_{{\rm{co}}}}} \right)}^2}} \right\}} \right){{\left( {4\pi } \right)}^2}}} \times 
\end{equation*}
\begin{equation*}
\left( {{\lambda _{\rm{U}}}\pi {{\left( {{R_{{\rm{co}}}}} \right)}^2} + \exp \left\{ { - {\lambda _{\rm{U}}}\pi {{\left( {{R_{{\rm{co}}}}} \right)}^2}} \right\} - 1} \right) \times \int\limits_{{\cal C}\left( {{{\mathbf{x}}_{\rm{U}}},{R_{{\rm{co}}}}} \right)} {\frac{{d{\mathbf{x}}}}{{{{({h_{{\rm{BS}}}})}^2} + \left\| {{\mathbf{x}} - {{\mathbf{x}}_{\rm{U}}}} \right\|_2^2}}}  = 
\end{equation*}
\begin{equation*}
\frac{{{\lambda _{{\rm{BS}}}}\left( {1 + 3{\lambda _{{\rm{IRS}}}}\pi {{\left( {{R_{{\rm{co}}}}} \right)}^2} + {{\left( {{\lambda _{{\rm{IRS}}}}\pi {{\left( {{R_{{\rm{co}}}}} \right)}^2}} \right)}^2}} \right)\left( {1 + \left( {\hat h - 1} \right){p_{{\rm{b}}}}} \right){{\left( {{\lambda _{{\rm{wave}}}}} \right)}^2} \sigma _{{\rm{d}}}^2}}{{\left( {1 - \exp \left\{ { - {\lambda _{\rm{U}}}\pi {{\left( {{R_{{\rm{co}}}}} \right)}^2}} \right\}} \right)16\pi }} \times 
\end{equation*}
\begin{equation}
\left( {{\lambda _{\rm{U}}}\pi {{\left( {{R_{{\rm{co}}}}} \right)}^2} + \exp \left\{ { - {\lambda _{\rm{U}}}\pi {{\left( {{R_{{\rm{co}}}}} \right)}^2}} \right\} - 1} \right) \times \ln \left( {1 + {{\left( {\frac{{{R_{{\rm{co}}}}}}{{{h_{{\rm{BS}}}}}}} \right)}^2}} \right).
\label{t34}
\end{equation}

To bound term (\ref{t26}), we have:
\begin{equation*}
\sum\limits_{{{\mathbf{x}}_{{\rm{IRS}}}} \in {\Psi _{{\rm{IRS}}}}({{\mathbf{x}}_{\rm{U}}}):{{{\mathbf{\tilde x}}}_{\rm{U}}}({{\mathbf{x}}_{{\rm{IRS}}}}) = {{\mathbf{x}}_{\rm{U}}}} {\begin{array}{*{20}{c}}
  {}&{} 
\end{array}\sum\limits_{{{{\mathbf{x'}}}_{{\rm{IRS}}}} \in {\Psi _{{\rm{IRS}}}}({{\mathbf{x}}_{\rm{U}}}):{{{\mathbf{\tilde x}}}_{\rm{U}}}({{{\mathbf{x'}}}_{{\rm{IRS}}}}) = {{\mathbf{x}}_{\rm{U}}}} {\begin{array}{*{20}{c}}
  {}&{} 
\end{array}\sum\limits_{{{{\mathbf{x'}}}_{\rm{U}}} \in {\Psi _{\rm{U}}}({{{\mathbf{\tilde x}}}_{{\rm{BS}}}}({{\mathbf{x}}_{\rm{U}}})):{{{\mathbf{x'}}}_{\rm{U}}} \ne {{\mathbf{x}}_{\rm{U}}},{{{\mathbf{\tilde x}}}_{{\rm{BS}}}}({{{\mathbf{x'}}}_{\rm{U}}}) = {{{\mathbf{\tilde x}}}_{{\rm{BS}}}}({{\mathbf{x}}_{\rm{U}}})} {} } } 
\end{equation*}
\begin{equation*}
   \scalebox{.9}[1]{$\frac{{{Q^2}{{\left( {{\lambda _{{\rm{wave}}}}} \right)}^4}}}{{{{\left( {4\pi } \right)}^4}\sqrt {{{({h_{{\rm{IRS}}}})}^2} + \left\| {{{\mathbf{x}}_{{\rm{IRS}}}} - {{\mathbf{x}}_{\rm{U}}}} \right\|_2^2}  \times \sqrt {{{({h_{{\rm{IRS}}}})}^2} + \left\| {{{{\mathbf{x'}}}_{{\rm{IRS}}}} - {{\mathbf{x}}_{\rm{U}}}} \right\|_2^2}  \times \sqrt {{{({h_{{\rm{BS}}}} - {h_{{\rm{IRS}}}})}^2} + \left\| {{{{\mathbf{\tilde x}}}_{{\rm{BS}}}}({{\mathbf{x}}_{\rm{U}}}) - {{\mathbf{x}}_{{\rm{IRS}}}}} \right\|_2^2}  \times \sqrt {{{({h_{{\rm{BS}}}} - {h_{{\rm{IRS}}}})}^2} + \left\| {{{{\mathbf{\tilde x}}}_{{\rm{BS}}}}({{\mathbf{x}}_{\rm{U}}}) - {{{\mathbf{x'}}}_{{\rm{IRS}}}}} \right\|_2^2} }} \times $}
\end{equation*}
\begin{equation*}
\left( {G\left( {{{{\mathbf{\tilde x}}}_{{\rm{BS}}}}({{\mathbf{x}}_{\rm{U}}}),{{{\mathbf{x'}}}_{\rm{U}}},{{\mathbf{x}}_{{\rm{IRS}}}}} \right) + \sum\limits_{{{{\mathbf{x''}}}_{{\rm{IRS}}}} \in {\Psi _{{\rm{IRS}}}}({{{\mathbf{x'}}}_{{\rm{BS}}}}):{{{\mathbf{x'}}}_{\rm{U}}} = {{{\mathbf{\tilde x}}}_{\rm{U}}}({{{\mathbf{x''}}}_{{\rm{IRS}}}})} {G\left( {{{{\mathbf{\tilde x}}}_{{\rm{BS}}}}({{\mathbf{x}}_{\rm{U}}}),{{{\mathbf{x''}}}_{{\rm{IRS}}}},{{\mathbf{x}}_{{\rm{IRS}}}}} \right)} } \right) \times 
\end{equation*}
\begin{equation*}
\left( {G\left( {{{{\mathbf{\tilde x}}}_{{\rm{BS}}}}({{\mathbf{x}}_{\rm{U}}}),{{{\mathbf{x'}}}_{\rm{U}}},{{{\mathbf{x'}}}_{{\rm{IRS}}}}} \right) + \sum\limits_{{{{\mathbf{x''}}}_{{\rm{IRS}}}} \in {\Psi _{{\rm{IRS}}}}({{{\mathbf{x'}}}_{{\rm{BS}}}}):{{{\mathbf{x'}}}_{\rm{U}}} = {{{\mathbf{\tilde x}}}_{\rm{U}}}({{{\mathbf{x''}}}_{{\rm{IRS}}}})} {G\left( {{{{\mathbf{\tilde x}}}_{{\rm{BS}}}}({{\mathbf{x}}_{\rm{U}}}),{{{\mathbf{x''}}}_{{\rm{IRS}}}},{{{\mathbf{x'}}}_{{\rm{IRS}}}}} \right)} } \right) \sigma _{{\rm{d}}}^2
\end{equation*}
\begin{equation*}
 \le \sum\limits_{{{\mathbf{x}}_{{\rm{IRS}}}} \in {{\tilde \Psi }_{{\rm{IRS}}}}({{\mathbf{x}}_{\rm{U}}})} {\sum\limits_{{{{\mathbf{x'}}}_{{\rm{IRS}}}} \in {{\tilde \Psi }_{{\rm{IRS}}}}({{\mathbf{x}}_{\rm{U}}})} {\frac{{{\pmb{\mathbb{I}}}\left( {{n_{{\rm{BS}}}}({{\mathbf{x}}_{\rm{U}}})} \right){Q^2}{{\left( {{\lambda _{{\rm{wave}}}}} \right)}^4}{{\left( {1 + {{\tilde n}_{{\rm{IRS}}}}({{\mathbf{x}}_{\rm{U}}})} \right)}^2}\left( {{{\tilde n}_{\rm{U}}}({{\mathbf{x}}_{\rm{U}}}) - 1} \right) \sigma _{{\rm{d}}}^2}}{{{{\left( {4\pi } \right)}^4}\sqrt {{{({h_{{\rm{IRS}}}})}^2} + \left\| {{{\mathbf{x}}_{{\rm{IRS}}}} - {{\mathbf{x}}_{\rm{U}}}} \right\|_2^2}  \times \sqrt {{{({h_{{\rm{IRS}}}})}^2} + \left\| {{{{\mathbf{x'}}}_{{\rm{IRS}}}} - {{\mathbf{x}}_{\rm{U}}}} \right\|_2^2} }}} }  \times 
\end{equation*}
\begin{equation*}
\frac{1}{{\sqrt {{{({h_{{\rm{BS}}}} - {h_{{\rm{IRS}}}})}^2} + \left\| {{{{\mathbf{\tilde x}}}_{{\rm{BS}}}}({{\mathbf{x}}_{\rm{U}}}) - {{\mathbf{x}}_{{\rm{IRS}}}}} \right\|_2^2}  \times \sqrt {{{({h_{{\rm{BS}}}} - {h_{{\rm{IRS}}}})}^2} + \left\| {{{{\mathbf{\tilde x}}}_{{\rm{BS}}}}({{\mathbf{x}}_{\rm{U}}}) - {{{\mathbf{x'}}}_{{\rm{IRS}}}}} \right\|_2^2} }}
\end{equation*}
\begin{equation*}
 = \sum\limits_{{{\mathbf{x}}_{{\rm{IRS}}}} \in {{\tilde \Psi }_{{\rm{IRS}}}}({{\mathbf{x}}_{\rm{U}}})} {\frac{{{\pmb{\mathbb{I}}}\left( {{n_{{\rm{BS}}}}({{\mathbf{x}}_{\rm{U}}})} \right){Q^2}{{\left( {{\lambda _{{\rm{wave}}}}} \right)}^4}{{\left( {1 + {{\tilde n}_{{\rm{IRS}}}}({{\mathbf{x}}_{\rm{U}}})} \right)}^2}\left( {{{\tilde n}_{\rm{U}}}({{\mathbf{x}}_{\rm{U}}}) - 1} \right) \sigma _{{\rm{d}}}^2}}{{{{\left( {4\pi } \right)}^4}\left( {{{({h_{{\rm{IRS}}}})}^2} + \left\| {{{\mathbf{x}}_{{\rm{IRS}}}} - {{\mathbf{x}}_{\rm{U}}}} \right\|_2^2} \right)\left( {{{({h_{{\rm{BS}}}} - {h_{{\rm{IRS}}}})}^2} + \left\| {{{{\mathbf{\tilde x}}}_{{\rm{BS}}}}({{\mathbf{x}}_{\rm{U}}}) - {{\mathbf{x}}_{{\rm{IRS}}}}} \right\|_2^2} \right)}}} 
\end{equation*}
\begin{equation*}
 + \sum\limits_{{{\mathbf{x}}_{{\rm{IRS}}}} \in {{\tilde \Psi }_{{\rm{IRS}}}}({{\mathbf{x}}_{\rm{U}}})} {\sum\limits_{{{{\mathbf{x'}}}_{{\rm{IRS}}}} \in {{\tilde \Psi }_{{\rm{IRS}}}}({{\mathbf{x}}_{\rm{U}}})} {\frac{{{\pmb{\mathbb{I}}}\left( {{n_{{\rm{BS}}}}({{\mathbf{x}}_{\rm{U}}})} \right){Q^2}{{\left( {{\lambda _{{\rm{wave}}}}} \right)}^4}{{\left( {1 + {{\tilde n}_{{\rm{IRS}}}}({{\mathbf{x}}_{\rm{U}}})} \right)}^2}\left( {{{\tilde n}_{\rm{U}}}({{\mathbf{x}}_{\rm{U}}}) - 1} \right) \sigma _{{\rm{d}}}^2}}{{{{\left( {4\pi } \right)}^4}\sqrt {{{({h_{{\rm{IRS}}}})}^2} + \left\| {{{\mathbf{x}}_{{\rm{IRS}}}} - {{\mathbf{x}}_{\rm{U}}}} \right\|_2^2}  \times \sqrt {{{({h_{{\rm{IRS}}}})}^2} + \left\| {{{{\mathbf{x'}}}_{{\rm{IRS}}}} - {{\mathbf{x}}_{\rm{U}}}} \right\|_2^2} }}} }  \times 
\end{equation*}
\begin{equation*}
\frac{{\left( {1 - {\delta _{{{\mathbf{x}}_{{\rm{IRS}}}} - {{{\mathbf{x'}}}_{{\rm{IRS}}}}}}} \right)}}{{\sqrt {{{({h_{{\rm{BS}}}} - {h_{{\rm{IRS}}}})}^2} + \left\| {{{{\mathbf{\tilde x}}}_{{\rm{BS}}}}({{\mathbf{x}}_{\rm{U}}}) - {{\mathbf{x}}_{{\rm{IRS}}}}} \right\|_2^2}  \times \sqrt {{{({h_{{\rm{BS}}}} - {h_{{\rm{IRS}}}})}^2} + \left\| {{{{\mathbf{\tilde x}}}_{{\rm{BS}}}}({{\mathbf{x}}_{\rm{U}}}) - {{{\mathbf{x'}}}_{{\rm{IRS}}}}} \right\|_2^2} }}.
\end{equation*}
thus, for the first term, we have:
\begin{equation*}
   \scalebox{1}[1]{$E\left\{ {\left. {\sum\limits_{{{\mathbf{x}}_{{\rm{IRS}}}} \in {{\tilde \Psi }_{{\rm{IRS}}}}({{\mathbf{x}}_{\rm{U}}})} {\frac{{{\pmb{\mathbb{I}}}\left( {{n_{{\rm{BS}}}}({{\mathbf{x}}_{\rm{U}}})} \right){Q^2}{{\left( {{\lambda _{{\rm{wave}}}}} \right)}^4}{{\left( {1 + {{\tilde n}_{{\rm{IRS}}}}({{\mathbf{x}}_{\rm{U}}})} \right)}^2}\left( {{{\tilde n}_{\rm{U}}}({{\mathbf{x}}_{\rm{U}}}) - 1} \right) \sigma _{{\rm{d}}}^2}}{{{{\left( {4\pi } \right)}^4}\left( {{{({h_{{\rm{IRS}}}})}^2} + \left\| {{{\mathbf{x}}_{{\rm{IRS}}}} - {{\mathbf{x}}_{\rm{U}}}} \right\|_2^2} \right)\left( {{{({h_{{\rm{BS}}}} - {h_{{\rm{IRS}}}})}^2} + \left\| {{{{\mathbf{\tilde x}}}_{{\rm{BS}}}}({{\mathbf{x}}_{\rm{U}}}) - {{\mathbf{x}}_{{\rm{IRS}}}}} \right\|_2^2} \right)}}} } \right|{\Phi _{\rm{U}}},{\Phi _{{\rm{BS}}}},{{\tilde n}_{{\rm{IRS}}}}({{\mathbf{x}}_{\rm{U}}}),{{{\mathbf{\tilde x}}}_{{\rm{BS}}}}({{\mathbf{x}}_{\rm{U}}})} \right\}$}
\end{equation*}
\begin{equation*}
 = \frac{{{\pmb{\mathbb{I}}}\left( {{n_{{\rm{BS}}}}({{\mathbf{x}}_{\rm{U}}})} \right){Q^2}{{\left( {{\lambda _{{\rm{wave}}}}} \right)}^4}{{\left( {1 + {{\tilde n}_{{\rm{IRS}}}}({{\mathbf{x}}_{\rm{U}}})} \right)}^2}{{\tilde n}_{{\rm{IRS}}}}({{\mathbf{x}}_{\rm{U}}})\left( {{{\tilde n}_{\rm{U}}}({{\mathbf{x}}_{\rm{U}}}) - 1} \right) \sigma _{{\rm{d}}}^2}}{{{{\left( {4\pi } \right)}^5}{{\left( {{R_{{\rm{co}}}}} \right)}^2}}} \times 
\end{equation*}
\begin{equation*}
\int\limits_{{\cal C}({{\mathbf{x}}_{\rm{U}}},2{R_{{\rm{co}}}})} {\frac{{d{\mathbf{x}}}}{{\left( {{{({h_{{\rm{IRS}}}})}^2} + \left\| {{\mathbf{x}} - {{\mathbf{x}}_{\rm{U}}}} \right\|_2^2} \right)\left( {{{({h_{{\rm{BS}}}} - {h_{{\rm{IRS}}}})}^2} + \left\| {{{{\mathbf{\tilde x}}}_{{\rm{BS}}}}({{\mathbf{x}}_{\rm{U}}}) - {\mathbf{x}}} \right\|_2^2} \right)}}} 
\end{equation*}
\begin{equation*}
\le \frac{{{\pmb{\mathbb{I}}}\left( {{n_{{\rm{BS}}}}({{\mathbf{x}}_{\rm{U}}})} \right){Q^2}{{\left( {{\lambda _{{\rm{wave}}}}} \right)}^4}{{\left( {1 + {{\tilde n}_{{\rm{IRS}}}}({{\mathbf{x}}_{\rm{U}}})} \right)}^2}{{\tilde n}_{{\rm{IRS}}}}({{\mathbf{x}}_{\rm{U}}})\left( {{{\tilde n}_{\rm{U}}}({{\mathbf{x}}_{\rm{U}}}) - 1} \right) \sigma _{{\rm{d}}}^2}}{{{{\left( {4\pi } \right)}^5}{{\left( {{R_{{\rm{co}}}}} \right)}^2}}} \times 
\end{equation*}
\begin{equation*}
\sqrt {\int\limits_{{\cal C}({{\mathbf{x}}_{\rm{U}}},2{R_{{\rm{co}}}})} {\frac{{d{\mathbf{x}}}}{{{{\left( {{{({h_{{\rm{IRS}}}})}^2} + \left\| {{\mathbf{x}} - {{\mathbf{x}}_{\rm{U}}}} \right\|_2^2} \right)}^2}}}} }  \times \sqrt {\int\limits_{{\cal C}({{\mathbf{x}}_{\rm{U}}},2{R_{{\rm{co}}}})} {\frac{{d{\mathbf{x}}}}{{{{\left( {{{({h_{{\rm{BS}}}} - {h_{{\rm{IRS}}}})}^2} + \left\| {{{{\mathbf{\tilde x}}}_{{\rm{BS}}}}({{\mathbf{x}}_{\rm{U}}}) - {\mathbf{x}}} \right\|_2^2} \right)}^2}}}} } 
\end{equation*}
\begin{equation*}
\le \frac{{{\pmb{\mathbb{I}}}\left( {{n_{{\rm{BS}}}}({{\mathbf{x}}_{\rm{U}}})} \right){Q^2}{{\left( {{\lambda _{{\rm{wave}}}}} \right)}^4}{{\left( {1 + {{\tilde n}_{{\rm{IRS}}}}({{\mathbf{x}}_{\rm{U}}})} \right)}^2}{{\tilde n}_{{\rm{IRS}}}}({{\mathbf{x}}_{\rm{U}}})\left( {{{\tilde n}_{\rm{U}}}({{\mathbf{x}}_{\rm{U}}}) - 1} \right) \sigma _{{\rm{d}}}^2}}{{{{\left( {4\pi } \right)}^5}{{\left( {{R_{{\rm{co}}}}} \right)}^2}}} \times 
\end{equation*}
\begin{equation*}
\sqrt {\int\limits_{{\cal C}({{\mathbf{x}}_{\rm{U}}},2{R_{{\rm{co}}}})} {\frac{{d{\mathbf{x}}}}{{{{\left( {{{({h_{{\rm{IRS}}}})}^2} + \left\| {{\mathbf{x}} - {{\mathbf{x}}_{\rm{U}}}} \right\|_2^2} \right)}^2}}}} }  \times \sqrt {\int\limits_{{\cal C}({{{\mathbf{\tilde x}}}_{{\rm{BS}}}}({{\mathbf{x}}_{\rm{U}}}),3{R_{{\rm{co}}}})} {\frac{{d{\mathbf{x}}}}{{{{\left( {{{({h_{{\rm{BS}}}} - {h_{{\rm{IRS}}}})}^2} + \left\| {{{{\mathbf{\tilde x}}}_{{\rm{BS}}}}({{\mathbf{x}}_{\rm{U}}}) - {\mathbf{x}}} \right\|_2^2} \right)}^2}}}} } 
\end{equation*}
\begin{equation*}
 = \frac{{{\pmb{\mathbb{I}}}\left( {{n_{{\rm{BS}}}}({{\mathbf{x}}_{\rm{U}}})} \right){Q^2}{{\left( {{\lambda _{{\rm{wave}}}}} \right)}^4}{{\left( {1 + {{\tilde n}_{{\rm{IRS}}}}({{\mathbf{x}}_{\rm{U}}})} \right)}^2}{{\tilde n}_{{\rm{IRS}}}}({{\mathbf{x}}_{\rm{U}}})\left( {{{\tilde n}_{\rm{U}}}({{\mathbf{x}}_{\rm{U}}}) - 1} \right) \sigma _{{\rm{d}}}^2}}{{{{\left( 4 \right)}^5}{\pi ^4}{{\left( {{R_{{\rm{co}}}}} \right)}^2}}} \times 
\end{equation*}
\begin{equation*}
\sqrt {\frac{{{{\left( {2{R_{{\rm{co}}}}} \right)}^2}}}{{\left( {{{({h_{{\rm{IRS}}}})}^2} + {{\left( {2{R_{{\rm{co}}}}} \right)}^2}} \right){{\left( {{h_{{\rm{IRS}}}}} \right)}^2}}}}  \times \sqrt {\frac{{{{\left( {3{R_{{\rm{co}}}}} \right)}^2}}}{{\left( {{{({h_{{\rm{BS}}}} - {h_{{\rm{IRS}}}})}^2} + {{\left( {3{R_{{\rm{co}}}}} \right)}^2}} \right){{\left( {{h_{{\rm{BS}}}} - {h_{{\rm{IRS}}}}} \right)}^2}}}} .
\end{equation*}

hence, we obtain:
\begin{equation*}
E\left\{ {\sum\limits_{{{\mathbf{x}}_{{\rm{IRS}}}} \in {{\tilde \Psi }_{{\rm{IRS}}}}({{\mathbf{x}}_{\rm{U}}})} {\frac{{{\pmb{\mathbb{I}}}\left( {{n_{{\rm{BS}}}}({{\mathbf{x}}_{\rm{U}}})} \right){Q^2}{{\left( {{\lambda _{{\rm{wave}}}}} \right)}^4}{{\left( {1 + {{\tilde n}_{{\rm{IRS}}}}({{\mathbf{x}}_{\rm{U}}})} \right)}^2}\left( {{{\tilde n}_{\rm{U}}}({{\mathbf{x}}_{\rm{U}}}) - 1} \right) \sigma _{{\rm{d}}}^2}}{{{{\left( {4\pi } \right)}^4}\left( {{{({h_{{\rm{IRS}}}})}^2} + \left\| {{{\mathbf{x}}_{{\rm{IRS}}}} - {{\mathbf{x}}_{\rm{U}}}} \right\|_2^2} \right)\left( {{{({h_{{\rm{BS}}}} - {h_{{\rm{IRS}}}})}^2} + \left\| {{{{\mathbf{\tilde x}}}_{{\rm{BS}}}}({{\mathbf{x}}_{\rm{U}}}) - {{\mathbf{x}}_{{\rm{IRS}}}}} \right\|_2^2} \right)}}} } \right\} = 
\end{equation*}
\begin{equation*}
 \scalebox{.95}[1]{$\left( {{{\left( {{\lambda _{{\rm{IRS}}}}\pi {{\left( {2{R_{{\rm{co}}}}} \right)}^2}} \right)}^3}\exp \left\{ {\frac{9}{{2{\lambda _{{\rm{IRS}}}}\pi {{\left( {2{R_{{\rm{co}}}}} \right)}^2}}}} \right\} + 2{{\left( {{\lambda _{{\rm{IRS}}}}\pi {{\left( {2{R_{{\rm{co}}}}} \right)}^2}} \right)}^2} + 3{\lambda _{{\rm{IRS}}}}\pi {{\left( {2{R_{{\rm{co}}}}} \right)}^2}} \right) \times $}
\end{equation*}
\begin{equation*}
 \scalebox{.95}[1]{$\frac{{{Q^2}{{\left( {{\lambda _{{\rm{wave}}}}} \right)}^4}\left( {{\lambda _{\rm{U}}}\pi {{\left( {{R_{{\rm{co}}}}} \right)}^2} + \exp \left\{ { - {\lambda _{\rm{U}}}\pi {{\left( {{R_{{\rm{co}}}}} \right)}^2}} \right\} - 1} \right)\left( {1 - \exp \left\{ { - {\lambda _{{\rm{BS}}}}\pi {{\left( {{R_{{\rm{co}}}}} \right)}^2}} \right\}} \right) \sigma _{{\rm{d}}}^2}}{{\left( {1 - \exp \left\{ { - {\lambda _{\rm{U}}}\pi {{\left( {{R_{{\rm{co}}}}} \right)}^2}} \right\}} \right){{\left( 4 \right)}^5}{\pi ^4}{{\left( {{R_{{\rm{co}}}}} \right)}^2}}} \times $}
\end{equation*}
\begin{equation}
\sqrt {\frac{{{{\left( {2{R_{{\rm{co}}}}} \right)}^2}}}{{\left( {{{({h_{{\rm{IRS}}}})}^2} + {{\left( {2{R_{{\rm{co}}}}} \right)}^2}} \right){{\left( {{h_{{\rm{IRS}}}}} \right)}^2}}}}  \times \sqrt {\frac{{{{\left( {3{R_{{\rm{co}}}}} \right)}^2}}}{{\left( {{{({h_{{\rm{BS}}}} - {h_{{\rm{IRS}}}})}^2} + {{\left( {3{R_{{\rm{co}}}}} \right)}^2}} \right){{\left( {{h_{{\rm{BS}}}} - {h_{{\rm{IRS}}}}} \right)}^2}}}} .
\end{equation}
For the second term, we have:
\begin{equation*}
E\left\{ {\sum\limits_{{{\mathbf{x}}_{{\rm{IRS}}}} \in {{\tilde \Psi }_{{\rm{IRS}}}}({{\mathbf{x}}_{\rm{U}}})} {\sum\limits_{{{{\mathbf{x'}}}_{{\rm{IRS}}}} \in {{\tilde \Psi }_{{\rm{IRS}}}}({{\mathbf{x}}_{\rm{U}}})} {\frac{{{\pmb{\mathbb{I}}}\left( {{n_{{\rm{BS}}}}({{\mathbf{x}}_{\rm{U}}})} \right){Q^2}{{\left( {{\lambda _{{\rm{wave}}}}} \right)}^4}{{\left( {1 + {{\tilde n}_{{\rm{IRS}}}}({{\mathbf{x}}_{\rm{U}}})} \right)}^2}\left( {{{\tilde n}_{\rm{U}}}({{\mathbf{x}}_{\rm{U}}}) - 1} \right) \sigma _{{\rm{d}}}^2}}{{{{\left( {4\pi } \right)}^4}\sqrt {{{({h_{{\rm{IRS}}}})}^2} + \left\| {{{\mathbf{x}}_{{\rm{IRS}}}} - {{\mathbf{x}}_{\rm{U}}}} \right\|_2^2}  \times \sqrt {{{({h_{{\rm{IRS}}}})}^2} + \left\| {{{{\mathbf{x'}}}_{{\rm{IRS}}}} - {{\mathbf{x}}_{\rm{U}}}} \right\|_2^2} }}} }  \times } \right. 
\end{equation*}
\begin{equation*}
 \scalebox{.95}[1]{$\left. {\left. {\frac{{\left( {1 - {\delta _{{{\mathbf{x}}_{{\rm{IRS}}}} - {{{\mathbf{x'}}}_{{\rm{IRS}}}}}}} \right)}}{{\sqrt {{{({h_{{\rm{BS}}}} - {h_{{\rm{IRS}}}})}^2} + \left\| {{{{\mathbf{\tilde x}}}_{{\rm{BS}}}}({{\mathbf{x}}_{\rm{U}}}) - {{\mathbf{x}}_{{\rm{IRS}}}}} \right\|_2^2}  \times \sqrt {{{({h_{{\rm{BS}}}} - {h_{{\rm{IRS}}}})}^2} + \left\| {{{{\mathbf{\tilde x}}}_{{\rm{BS}}}}({{\mathbf{x}}_{\rm{U}}}) - {{{\mathbf{x'}}}_{{\rm{IRS}}}}} \right\|_2^2} }}} \right|{\Phi _{\rm{U}}},{\Phi _{{\rm{BS}}}},{{\tilde n}_{{\rm{IRS}}}}({{\mathbf{x}}_{\rm{U}}}),{{{\mathbf{\tilde x}}}_{{\rm{BS}}}}({{\mathbf{x}}_{\rm{U}}})} \right\}$}
\end{equation*}
\begin{equation*}
 = \frac{{{\pmb{\mathbb{I}}}\left( {{n_{{\rm{BS}}}}({{\mathbf{x}}_{\rm{U}}})} \right){Q^2}{{\left( {{\lambda _{{\rm{wave}}}}} \right)}^4}{{\left( {1 + {{\tilde n}_{{\rm{IRS}}}}({{\mathbf{x}}_{\rm{U}}})} \right)}^2}{{\tilde n}_{{\rm{IRS}}}}({{\mathbf{x}}_{\rm{U}}})\left( {{{\tilde n}_{{\rm{IRS}}}}({{\mathbf{x}}_{\rm{U}}}) - 1} \right)\left( {{{\tilde n}_{\rm{U}}}({{\mathbf{x}}_{\rm{U}}}) - 1} \right) \sigma _{{\rm{d}}}^2}}{{{{\left( {4\pi } \right)}^5}{{\left( {{R_{{\rm{co}}}}} \right)}^2}}} \times 
\end{equation*}
\begin{equation*}
\int\limits_{{\cal C}({{\mathbf{x}}_{\rm{U}}},2{R_{{\rm{co}}}})} {\frac{1}{{\sqrt {{{({h_{{\rm{IRS}}}})}^2} + \left\| {{\mathbf{x}} - {{\mathbf{x}}_{\rm{U}}}} \right\|_2^2}  \times \sqrt {{{({h_{{\rm{IRS}}}})}^2} + \left\| {{\mathbf{x'}} - {{\mathbf{x}}_{\rm{U}}}} \right\|_2^2} }} \times } 
\end{equation*}
\begin{equation*}
\frac{{d{\mathbf{x}}d{\mathbf{x'}}}}{{\sqrt {{{({h_{{\rm{BS}}}} - {h_{{\rm{IRS}}}})}^2} + \left\| {{{{\mathbf{\tilde x}}}_{{\rm{BS}}}}({{\mathbf{x}}_{\rm{U}}}) - {\mathbf{x}}} \right\|_2^2}  \times \sqrt {{{({h_{{\rm{BS}}}} - {h_{{\rm{IRS}}}})}^2} + \left\| {{{{\mathbf{\tilde x}}}_{{\rm{BS}}}}({{\mathbf{x}}_{\rm{U}}}) - {\mathbf{x'}}} \right\|_2^2} }}
\end{equation*}
\begin{equation*}
 \le \frac{{{\pmb{\mathbb{I}}}\left( {{n_{{\rm{BS}}}}({{\mathbf{x}}_{\rm{U}}})} \right){Q^2}{{\left( {{\lambda _{{\rm{wave}}}}} \right)}^4}{{\left( {1 + {{\tilde n}_{{\rm{IRS}}}}({{\mathbf{x}}_{\rm{U}}})} \right)}^2}{{\tilde n}_{{\rm{IRS}}}}({{\mathbf{x}}_{\rm{U}}})\left( {{{\tilde n}_{{\rm{IRS}}}}({{\mathbf{x}}_{\rm{U}}}) - 1} \right)\left( {{{\tilde n}_{\rm{U}}}({{\mathbf{x}}_{\rm{U}}}) - 1} \right) \sigma _{{\rm{d}}}^2}}{{{{\left( {4\pi } \right)}^5}{{\left( {{R_{{\rm{co}}}}} \right)}^2}}} \times 
\end{equation*}
\begin{equation*}
\sqrt {\int\limits_{{\mathbf{x}},{\mathbf{x'}} \in {\cal C}({{\mathbf{x}}_{\rm{U}}},2{R_{{\rm{co}}}})} {\frac{{d{\mathbf{x}}d{\mathbf{x'}}}}{{\left( {{{({h_{{\rm{IRS}}}})}^2} + \left\| {{\mathbf{x}} - {{\mathbf{x}}_{\rm{U}}}} \right\|_2^2} \right) \times \left( {{{({h_{{\rm{IRS}}}})}^2} + \left\| {{\mathbf{x'}} - {{\mathbf{x}}_{\rm{U}}}} \right\|_2^2} \right)}}} }  \times 
\end{equation*}
\begin{equation*}
\sqrt {\int\limits_{{\mathbf{x}},{\mathbf{x'}} \in {\cal C}({{\mathbf{x}}_{\rm{U}}},2{R_{{\rm{co}}}})} {\frac{{d{\mathbf{x}}d{\mathbf{x'}}}}{{\left( {{{({h_{{\rm{BS}}}} - {h_{{\rm{IRS}}}})}^2} + \left\| {{{{\mathbf{\tilde x}}}_{{\rm{BS}}}}({{\mathbf{x}}_{\rm{U}}}) - {\mathbf{x}}} \right\|_2^2} \right) \times \left( {{{({h_{{\rm{BS}}}} - {h_{{\rm{IRS}}}})}^2} + \left\| {{{{\mathbf{\tilde x}}}_{{\rm{BS}}}}({{\mathbf{x}}_{\rm{U}}}) - {\mathbf{x'}}} \right\|_2^2} \right)}}} }  = 
\end{equation*}
\begin{equation*}
 \frac{{{\pmb{\mathbb{I}}}\left( {{n_{{\rm{BS}}}}({{\mathbf{x}}_{\rm{U}}})} \right){Q^2}{{\left( {{\lambda _{{\rm{wave}}}}} \right)}^4}{{\left( {1 + {{\tilde n}_{{\rm{IRS}}}}({{\mathbf{x}}_{\rm{U}}})} \right)}^2}{{\tilde n}_{{\rm{IRS}}}}({{\mathbf{x}}_{\rm{U}}})\left( {{{\tilde n}_{{\rm{IRS}}}}({{\mathbf{x}}_{\rm{U}}}) - 1} \right)\left( {{{\tilde n}_{\rm{U}}}({{\mathbf{x}}_{\rm{U}}}) - 1} \right) \sigma _{{\rm{d}}}^2}}{{{{\left( {4\pi } \right)}^5}{{\left( {{R_{{\rm{co}}}}} \right)}^2}}} \times 
\end{equation*}
\begin{equation*}
\int\limits_{{\cal C}({{\mathbf{x}}_{\rm{U}}},2{R_{{\rm{co}}}})} {\frac{{d{\mathbf{x}}}}{{\left( {{{({h_{{\rm{IRS}}}})}^2} + \left\| {{\mathbf{x}} - {{\mathbf{x}}_{\rm{U}}}} \right\|_2^2} \right)}}}  \times \int\limits_{{\cal C}({{\mathbf{x}}_{\rm{U}}},2{R_{{\rm{co}}}})} {\frac{{d{\mathbf{x}}}}{{\left( {{{({h_{{\rm{BS}}}} - {h_{{\rm{IRS}}}})}^2} + \left\| {{{{\mathbf{\tilde x}}}_{{\rm{BS}}}}({{\mathbf{x}}_{\rm{U}}}) - {\mathbf{x}}} \right\|_2^2} \right)}}}  \le
\end{equation*}
\begin{equation*}
\frac{{{\pmb{\mathbb{I}}}\left( {{n_{{\rm{BS}}}}({{\mathbf{x}}_{\rm{U}}})} \right){Q^2}{{\left( {{\lambda _{{\rm{wave}}}}} \right)}^4}{{\left( {1 + {{\tilde n}_{{\rm{IRS}}}}({{\mathbf{x}}_{\rm{U}}})} \right)}^2}{{\tilde n}_{{\rm{IRS}}}}({{\mathbf{x}}_{\rm{U}}})\left( {{{\tilde n}_{{\rm{IRS}}}}({{\mathbf{x}}_{\rm{U}}}) - 1} \right)\left( {{{\tilde n}_{\rm{U}}}({{\mathbf{x}}_{\rm{U}}}) - 1} \right) \sigma _{{\rm{d}}}^2}}{{{{\left( {4\pi } \right)}^5}{{\left( {{R_{{\rm{co}}}}} \right)}^2}}} \times 
\end{equation*}
\begin{equation*}
\int\limits_{{\cal C}({{\mathbf{x}}_{\rm{U}}},2{R_{{\rm{co}}}})} {\frac{{d{\mathbf{x}}}}{{\left( {{{({h_{{\rm{IRS}}}})}^2} + \left\| {{\mathbf{x}} - {{\mathbf{x}}_{\rm{U}}}} \right\|_2^2} \right)}}}  \times \int\limits_{{\cal C}({{{\mathbf{\tilde x}}}_{{\rm{BS}}}}({{\mathbf{x}}_{\rm{U}}}),3{R_{{\rm{co}}}})} {\frac{{d{\mathbf{x}}}}{{\left( {{{({h_{{\rm{BS}}}} - {h_{{\rm{IRS}}}})}^2} + \left\| {{{{\mathbf{\tilde x}}}_{{\rm{BS}}}}({{\mathbf{x}}_{\rm{U}}}) - {\mathbf{x}}} \right\|_2^2} \right)}}} 
\end{equation*}
\begin{equation*}
 = \frac{{{\pmb{\mathbb{I}}}\left( {{n_{{\rm{BS}}}}({{\mathbf{x}}_{\rm{U}}})} \right){Q^2}{{\left( {{\lambda _{{\rm{wave}}}}} \right)}^4}{{\left( {1 + {{\tilde n}_{{\rm{IRS}}}}({{\mathbf{x}}_{\rm{U}}})} \right)}^2}{{\tilde n}_{{\rm{IRS}}}}({{\mathbf{x}}_{\rm{U}}})\left( {{{\tilde n}_{{\rm{IRS}}}}({{\mathbf{x}}_{\rm{U}}}) - 1} \right)\left( {{{\tilde n}_{\rm{U}}}({{\mathbf{x}}_{\rm{U}}}) - 1} \right) \sigma _{{\rm{d}}}^2}}{{{{\left( 4 \right)}^5}{\pi ^3}{{\left( {{R_{{\rm{co}}}}} \right)}^2}}} \times 
\end{equation*}
\begin{equation*}
\ln \left( {1 + {{\left( {\frac{{2{R_{{\rm{co}}}}}}{{{h_{{\rm{IRS}}}}}}} \right)}^2}} \right) \times \ln \left( {1 + {{\left( {\frac{{3{R_{{\rm{co}}}}}}{{{h_{{\rm{BS}}}} - {h_{{\rm{IRS}}}}}}} \right)}^2}} \right).
\end{equation*}
therefore, we obtain:

\begin{equation*}
E\left\{ {\frac{{{\pmb{\mathbb{I}}}\left( {{n_{{\rm{BS}}}}({{\mathbf{x}}_{\rm{U}}})} \right){Q^2}{{\left( {{\lambda _{{\rm{wave}}}}} \right)}^4}{{\left( {1 + {{\tilde n}_{{\rm{IRS}}}}({{\mathbf{x}}_{\rm{U}}})} \right)}^2}{{\tilde n}_{{\rm{IRS}}}}({{\mathbf{x}}_{\rm{U}}})\left( {{{\tilde n}_{{\rm{IRS}}}}({{\mathbf{x}}_{\rm{U}}}) - 1} \right)\left( {{{\tilde n}_{\rm{U}}}({{\mathbf{x}}_{\rm{U}}}) - 1} \right) \sigma _{{\rm{d}}}^2}}{{{{\left( 4 \right)}^5}{\pi ^3}{{\left( {{R_{{\rm{co}}}}} \right)}^2}}} \times } \right.
\end{equation*}
\begin{equation*}
\left. {\ln \left( {1 + {{\left( {\frac{{2{R_{{\rm{co}}}}}}{{{h_{{\rm{IRS}}}}}}} \right)}^2}} \right) \times \ln \left( {1 + {{\left( {\frac{{3{R_{{\rm{co}}}}}}{{{h_{{\rm{BS}}}} - {h_{{\rm{IRS}}}}}}} \right)}^2}} \right)} \right\} \le
\end{equation*}
\begin{equation*}
 \scalebox{.99}[1]{$\frac{{{Q^2}{{\left( {{\lambda _{{\rm{wave}}}}} \right)}^4}\left( {{\lambda _{\rm{U}}}\pi {{\left( {{R_{{\rm{co}}}}} \right)}^2} + \exp \left\{ { - {\lambda _{\rm{U}}}\pi {{\left( {{R_{{\rm{co}}}}} \right)}^2}} \right\} - 1} \right)\left( {1 - \exp \left\{ { - {\lambda _{{\rm{BS}}}}\pi {{\left( {{R_{{\rm{co}}}}} \right)}^2}} \right\}} \right) \sigma _{{\rm{d}}}^2}}{{\left( {1 - \exp \left\{ { - {\lambda _{\rm{U}}}\pi {{\left( {{R_{{\rm{co}}}}} \right)}^2}} \right\}} \right){{\left( 4 \right)}^5}{\pi ^3}{{\left( {{R_{{\rm{co}}}}} \right)}^2}}} \times $}
\end{equation*}
{\tiny\begin{equation*}
\left( {{{\left( {{\lambda _{{\rm{IRS}}}}\pi {{\left( {2{R_{{\rm{co}}}}} \right)}^2}} \right)}^4}\exp \left\{ {\frac{{16}}{{2\left( {{\lambda _{{\rm{IRS}}}}\pi {{\left( {2{R_{{\rm{co}}}}} \right)}^2}} \right)}}} \right\} + {{\left( {{\lambda _{{\rm{IRS}}}}\pi {{\left( {2{R_{{\rm{co}}}}} \right)}^2}} \right)}^3}\exp \left\{ {\frac{9}{{2\left( {{\lambda _{{\rm{IRS}}}}\pi {{\left( {2{R_{{\rm{co}}}}} \right)}^2}} \right)}}} \right\} - {{\left( {{\lambda _{{\rm{IRS}}}}\pi {{\left( {2{R_{{\rm{co}}}}} \right)}^2}} \right)}^2} - 2{\lambda _{{\rm{IRS}}}}\pi {{\left( {2{R_{{\rm{co}}}}} \right)}^2}} \right) \times 
\end{equation*}}
\begin{equation}
\ln \left( {1 + {{\left( {\frac{{2{R_{{\rm{co}}}}}}{{{h_{{\rm{IRS}}}}}}} \right)}^2}} \right) \times \ln \left( {1 + {{\left( {\frac{{3{R_{{\rm{co}}}}}}{{{h_{{\rm{BS}}}} - {h_{{\rm{IRS}}}}}}} \right)}^2}} \right).
\label{t36}
\end{equation}

For summation of terms (\ref{t27}) and (\ref{t28}), we have:
\begin{equation*}
\sum\limits_{{{\mathbf{x}}_{{\rm{IRS}}}} \in {\Psi _{{\rm{IRS}}}}({{\mathbf{x}}_{\rm{U}}}):{{{\mathbf{\tilde x}}}_{\rm{U}}}({{\mathbf{x}}_{{\rm{IRS}}}}) = {{\mathbf{x}}_{\rm{U}}}} {\begin{array}{*{20}{c}}
  {}&{} 
\end{array}\sum\limits_{{{{\mathbf{x'}}}_{{\rm{BS}}}} \in {\Psi _{{\rm{BS}}}}({{\mathbf{x}}_{{\rm{IRS}}}}):{{{\mathbf{x'}}}_{{\rm{BS}}}} \ne {{{\mathbf{\tilde x}}}_{{\rm{BS}}}}({{\mathbf{x}}_{\rm{U}}})} {\begin{array}{*{20}{c}}
  {}&{} 
\end{array}\sum\limits_{{{{\mathbf{x'}}}_{\rm{U}}} \in {\Psi _{\rm{U}}}({{{\mathbf{x'}}}_{{\rm{BS}}}}):{{{\mathbf{x'}}}_{\rm{U}}} \ne {{\mathbf{x}}_{\rm{U}}},{{{\mathbf{\tilde x}}}_{{\rm{BS}}}}({{{\mathbf{x'}}}_{\rm{U}}}) = {{{\mathbf{x'}}}_{{\rm{BS}}}}} {} } } 
\end{equation*}
\begin{equation*}
\frac{{Q{{\left( {{\lambda _{{\rm{wave}}}}} \right)}^4}}}{{{{\left( {4\pi } \right)}^4}\left( {{{({h_{{\rm{IRS}}}})}^2} + \left\| {{{\mathbf{x}}_{{\rm{IRS}}}} - {{\mathbf{x}}_{\rm{U}}}} \right\|_2^2} \right)\left( {{{({h_{{\rm{BS}}}} - {h_{{\rm{IRS}}}})}^2} + \left\| {{{{\mathbf{x'}}}_{{\rm{BS}}}} - {{\mathbf{x}}_{{\rm{IRS}}}}} \right\|_2^2} \right)}} \times 
\end{equation*}
\begin{equation*}
{\left( {G\left( {{{{\mathbf{x'}}}_{{\rm{BS}}}},{{{\mathbf{x'}}}_{\rm{U}}},{{\mathbf{x}}_{{\rm{IRS}}}}} \right) + \sum\limits_{{{{\mathbf{x'}}}_{{\rm{IRS}}}} \in {\Psi _{{\rm{IRS}}}}({{{\mathbf{x'}}}_{{\rm{BS}}}}):{{{\mathbf{x'}}}_{\rm{U}}} = {{{\mathbf{\tilde x}}}_{\rm{U}}}({{{\mathbf{x'}}}_{{\rm{IRS}}}})} {G\left( {{{{\mathbf{x'}}}_{{\rm{BS}}}},{{{\mathbf{x'}}}_{{\rm{IRS}}}},{{\mathbf{x}}_{{\rm{IRS}}}}} \right)} } \right)^2} \sigma _{{\rm{d}}}^2 + 
\end{equation*}
\begin{equation*}
\sum\limits_{{{\mathbf{x}}_{{\rm{IRS}}}} \in {\Psi _{{\rm{IRS}}}}({{\mathbf{x}}_{\rm{U}}}):{{{\mathbf{\tilde x}}}_{\rm{U}}}({{\mathbf{x}}_{{\rm{IRS}}}}) \ne {{\mathbf{x}}_{\rm{U}}}} {\begin{array}{*{20}{c}}
  {}&{} 
\end{array}\sum\limits_{{{{\mathbf{x'}}}_{{\rm{BS}}}} \in {\Psi _{{\rm{BS}}}}({{\mathbf{x}}_{{\rm{IRS}}}})} {\begin{array}{*{20}{c}}
  {}&{} 
\end{array}\sum\limits_{{{{\mathbf{x'}}}_{\rm{U}}} \in {\Psi _{\rm{U}}}({{{\mathbf{x'}}}_{{\rm{BS}}}}):{{{\mathbf{x'}}}_{\rm{U}}} \ne {{\mathbf{x}}_{\rm{U}}},{{{\mathbf{\tilde x}}}_{{\rm{BS}}}}({{{\mathbf{x'}}}_{\rm{U}}}) = {{{\mathbf{x'}}}_{{\rm{BS}}}}} {} } } 
\end{equation*}
\begin{equation*}
\frac{{Q{{\left( {{\lambda _{{\rm{wave}}}}} \right)}^4}}}{{{{\left( {4\pi } \right)}^4}\left( {{{({h_{{\rm{IRS}}}})}^2} + \left\| {{{\mathbf{x}}_{{\rm{IRS}}}} - {{\mathbf{x}}_{\rm{U}}}} \right\|_2^2} \right)\left( {{{({h_{{\rm{BS}}}} - {h_{{\rm{IRS}}}})}^2} + \left\| {{{{\mathbf{x'}}}_{{\rm{BS}}}} - {{\mathbf{x}}_{{\rm{IRS}}}}} \right\|_2^2} \right)}} \times 
\end{equation*}
\begin{equation*}
{\left( {G\left( {{{{\mathbf{x'}}}_{{\rm{BS}}}},{{{\mathbf{x'}}}_{\rm{U}}},{{\mathbf{x}}_{{\rm{IRS}}}}} \right) + \sum\limits_{{{{\mathbf{x'}}}_{{\rm{IRS}}}} \in {\Psi _{{\rm{IRS}}}}({{{\mathbf{x'}}}_{{\rm{BS}}}}):{{{\mathbf{x'}}}_{\rm{U}}} = {{{\mathbf{\tilde x}}}_{\rm{U}}}({{{\mathbf{x'}}}_{{\rm{IRS}}}})} {G\left( {{{{\mathbf{x'}}}_{{\rm{BS}}}},{{{\mathbf{x'}}}_{{\rm{IRS}}}},{{\mathbf{x}}_{{\rm{IRS}}}}} \right)} } \right)^2} \sigma _{{\rm{d}}}^2 \le
\end{equation*}
\begin{equation*}
\sum\limits_{{{\mathbf{x}}_{{\rm{IRS}}}} \in {\Psi _{{\rm{IRS}}}}({{\mathbf{x}}_{\rm{U}}})} {\begin{array}{*{20}{c}}
  {}&{} 
\end{array}\sum\limits_{{{{\mathbf{x'}}}_{{\rm{BS}}}} \in {\Psi _{{\rm{BS}}}}({{\mathbf{x}}_{{\rm{IRS}}}})} {\begin{array}{*{20}{c}}
  {}&{} 
\end{array}\sum\limits_{{{{\mathbf{x'}}}_{\rm{U}}} \in {\Psi _{\rm{U}}}({{{\mathbf{x'}}}_{{\rm{BS}}}}):{{{\mathbf{x'}}}_{\rm{U}}} \ne {{\mathbf{x}}_{\rm{U}}}} {} } } 
\end{equation*}
\begin{equation*}
\frac{{Q{{\left( {{\lambda _{{\rm{wave}}}}} \right)}^4}}}{{{{\left( {4\pi } \right)}^4}\left( {{{({h_{{\rm{IRS}}}})}^2} + \left\| {{{\mathbf{x}}_{{\rm{IRS}}}} - {{\mathbf{x}}_{\rm{U}}}} \right\|_2^2} \right)\left( {{{({h_{{\rm{BS}}}} - {h_{{\rm{IRS}}}})}^2} + \left\| {{{{\mathbf{x'}}}_{{\rm{BS}}}} - {{\mathbf{x}}_{{\rm{IRS}}}}} \right\|_2^2} \right)}} \times 
\end{equation*}
\begin{equation*}
{\left( {G\left( {{{{\mathbf{x'}}}_{{\rm{BS}}}},{{{\mathbf{x'}}}_{\rm{U}}},{{\mathbf{x}}_{{\rm{IRS}}}}} \right) + \sum\limits_{{{{\mathbf{x'}}}_{{\rm{IRS}}}} \in {\Psi _{{\rm{IRS}}}}({{{\mathbf{x'}}}_{{\rm{BS}}}}):{{{\mathbf{x'}}}_{\rm{U}}} = {{{\mathbf{\tilde x}}}_{\rm{U}}}({{{\mathbf{x'}}}_{{\rm{IRS}}}})} {G\left( {{{{\mathbf{x'}}}_{{\rm{BS}}}},{{{\mathbf{x'}}}_{{\rm{IRS}}}},{{\mathbf{x}}_{{\rm{IRS}}}}} \right)} } \right)^2} \sigma _{{\rm{d}}}^2 \le
\end{equation*}
\begin{equation*}
 \le \sum\limits_{{{\mathbf{x}}_{{\rm{IRS}}}} \in {{\tilde \Psi }_{{\rm{IRS}}}}({{\mathbf{x}}_{\rm{U}}})} {\sum\limits_{{{{\mathbf{x'}}}_{{\rm{BS}}}} \in {\Psi _{{\rm{BS}}}}({{\mathbf{x}}_{{\rm{IRS}}}})} {\frac{{Q{{\left( {{\lambda _{{\rm{wave}}}}} \right)}^4} \sigma _{{\rm{d}}}^2{{\left( {1 + {{\tilde n}_{{\rm{IRS}}}}({{\mathbf{x}}_{\rm{U}}})} \right)}^2}\left( {{n_{\rm{U}}}({{{\mathbf{x'}}}_{{\rm{BS}}}}) - 1} \right)}}{{{{\left( {4\pi } \right)}^4}\left( {{{({h_{{\rm{IRS}}}})}^2} + \left\| {{{\mathbf{x}}_{{\rm{IRS}}}} - {{\mathbf{x}}_{\rm{U}}}} \right\|_2^2} \right)\left( {{{({h_{{\rm{BS}}}} - {h_{{\rm{IRS}}}})}^2} + \left\| {{{{\mathbf{x'}}}_{{\rm{BS}}}} - {{\mathbf{x}}_{{\rm{IRS}}}}} \right\|_2^2} \right)}}} }  .
\end{equation*}

Then, we obtain:
\begin{equation*}
 \scalebox{.95}[1]{$E\left\{ {\left. {\sum\limits_{{{\mathbf{x}}_{{\rm{IRS}}}} \in {{\tilde \Psi }_{{\rm{IRS}}}}({{\mathbf{x}}_{\rm{U}}})} {\sum\limits_{{{{\mathbf{x'}}}_{{\rm{BS}}}} \in {\Psi _{{\rm{BS}}}}({{\mathbf{x}}_{{\rm{IRS}}}})} {\frac{{Q{{\left( {{\lambda _{{\rm{wave}}}}} \right)}^4} \sigma _{{\rm{d}}}^2{{\left( {1 + {{\tilde n}_{{\rm{IRS}}}}({{\mathbf{x}}_{\rm{U}}})} \right)}^2}\left( {{n_{\rm{U}}}({{{\mathbf{x'}}}_{{\rm{BS}}}}) - 1} \right)}}{{{{\left( {4\pi } \right)}^4}\left( {{{({h_{{\rm{IRS}}}})}^2} + \left\| {{{\mathbf{x}}_{{\rm{IRS}}}} - {{\mathbf{x}}_{\rm{U}}}} \right\|_2^2} \right)\left( {{{({h_{{\rm{BS}}}} - {h_{{\rm{IRS}}}})}^2} + \left\| {{{{\mathbf{x'}}}_{{\rm{BS}}}} - {{\mathbf{x}}_{{\rm{IRS}}}}} \right\|_2^2} \right)}}} } } \right|{\Phi _{{\rm{IRS}}}}} \right\}$}
\end{equation*}
\begin{equation*}
 \scalebox{.95}[1]{$ = \sum\limits_{{{\mathbf{x}}_{{\rm{IRS}}}} \in {{\tilde \Psi }_{{\rm{IRS}}}}({{\mathbf{x}}_{\rm{U}}})} {\frac{{Q{{\left( {{\lambda _{{\rm{wave}}}}} \right)}^4} \sigma _{{\rm{d}}}^2{{\left( {1 + {{\tilde n}_{{\rm{IRS}}}}({{\mathbf{x}}_{\rm{U}}})} \right)}^2}\left( {{\lambda _{\rm{U}}}\pi {{\left( {{R_{{\rm{co}}}}} \right)}^2} + \exp \left\{ { - {\lambda _{\rm{U}}}\pi {{\left( {{R_{{\rm{co}}}}} \right)}^2}} \right\} - 1} \right)}}{{{{\left( 4 \right)}^4}{\pi ^4}\left( {{{({h_{{\rm{IRS}}}})}^2} + \left\| {{{\mathbf{x}}_{{\rm{IRS}}}} - {{\mathbf{x}}_{\rm{U}}}} \right\|_2^2} \right)\left( {1 - \exp \left\{ { - {\lambda _{\rm{U}}}\pi {{\left( {{R_{{\rm{co}}}}} \right)}^2}} \right\}} \right)}}}  \times $}
\end{equation*}
\begin{equation*}
{\lambda _{{\rm{BS}}}}\times\int\limits_{{\cal C}({{\mathbf{x}}_{{\rm{IRS}}}},{R_{{\rm{co}}}})} {\frac{{d{\mathbf{x}}}}{{{{({h_{{\rm{BS}}}} - {h_{{\rm{IRS}}}})}^2} + \left\| {{\mathbf{x}} - {{\mathbf{x}}_{{\rm{IRS}}}}} \right\|_2^2}}} 
\end{equation*}
\begin{equation*}
 \scalebox{.95}[1]{$ = \sum\limits_{{{\mathbf{x}}_{{\rm{IRS}}}} \in {{\tilde \Psi }_{{\rm{IRS}}}}({{\mathbf{x}}_{\rm{U}}})} {\frac{{Q{{\left( {{\lambda _{{\rm{wave}}}}} \right)}^4} \sigma _{{\rm{d}}}^2{{\left( {1 + {{\tilde n}_{{\rm{IRS}}}}({{\mathbf{x}}_{\rm{U}}})} \right)}^2}\left( {{\lambda _{\rm{U}}}\pi {{\left( {{R_{{\rm{co}}}}} \right)}^2} + \exp \left\{ { - {\lambda _{\rm{U}}}\pi {{\left( {{R_{{\rm{co}}}}} \right)}^2}} \right\} - 1} \right)}}{{{{\left( 4 \right)}^4}{\pi ^3}\left( {{{({h_{{\rm{IRS}}}})}^2} + \left\| {{{\mathbf{x}}_{{\rm{IRS}}}} - {{\mathbf{x}}_{\rm{U}}}} \right\|_2^2} \right)\left( {1 - \exp \left\{ { - {\lambda _{\rm{U}}}\pi {{\left( {{R_{{\rm{co}}}}} \right)}^2}} \right\}} \right)}}}  \times$} 
\end{equation*}
\begin{equation*}
{\lambda _{{\rm{BS}}}}\ln \left( {1 + {{\left( {\frac{{{R_{{\rm{co}}}}}}{{{h_{{\rm{BS}}}} - {h_{{\rm{IRS}}}}}}} \right)}^2}} \right).
\end{equation*}

Thus, we have:
\begin{equation*}
 \scalebox{.95}[1]{$E\left\{ {\sum\limits_{{{\mathbf{x}}_{{\rm{IRS}}}} \in {{\tilde \Psi }_{{\rm{IRS}}}}({{\mathbf{x}}_{\rm{U}}})} {\frac{{Q{{\left( {{\lambda _{{\rm{wave}}}}} \right)}^4} \sigma _{{\rm{d}}}^2{{\left( {1 + {{\tilde n}_{{\rm{IRS}}}}({{\mathbf{x}}_{\rm{U}}})} \right)}^2}\left( {{\lambda _{\rm{U}}}\pi {{\left( {{R_{{\rm{co}}}}} \right)}^2} + \exp \left\{ { - {\lambda _{\rm{U}}}\pi {{\left( {{R_{{\rm{co}}}}} \right)}^2}} \right\} - 1} \right)}}{{{{\left( 4 \right)}^4}{\pi ^3}\left( {{{({h_{{\rm{IRS}}}})}^2} + \left\| {{{\mathbf{x}}_{{\rm{IRS}}}} - {{\mathbf{x}}_{\rm{U}}}} \right\|_2^2} \right)\left( {1 - \exp \left\{ { - {\lambda _{\rm{U}}}\pi {{\left( {{R_{{\rm{co}}}}} \right)}^2}} \right\}} \right)}}}  \times } \right.$}
\end{equation*}
\begin{equation*}
\left. {\left. {{\lambda _{{\rm{BS}}}}\ln \left( {1 + {{\left( {\frac{{{R_{{\rm{co}}}}}}{{{h_{{\rm{BS}}}} - {h_{{\rm{IRS}}}}}}} \right)}^2}} \right)} \right|{{\tilde n}_{{\rm{IRS}}}}({{\mathbf{x}}_{\rm{U}}})} \right\} = 
\end{equation*}
\begin{equation*}
 \scalebox{.95}[1]{$\frac{{{\lambda _{{\rm{BS}}}}Q{{\left( {{\lambda _{{\rm{wave}}}}} \right)}^4} \sigma _{{\rm{d}}}^2{{\left( {1 + {{\tilde n}_{{\rm{IRS}}}}({{\mathbf{x}}_{\rm{U}}})} \right)}^2}{{\tilde n}_{{\rm{IRS}}}}({{\mathbf{x}}_{\rm{U}}})\left( {{\lambda _{\rm{U}}}\pi {{\left( {{R_{{\rm{co}}}}} \right)}^2} + \exp \left\{ { - {\lambda _{\rm{U}}}\pi {{\left( {{R_{{\rm{co}}}}} \right)}^2}} \right\} - 1} \right)}}{{{{\left( 4 \right)}^5}{\pi ^4}\left( {{{({h_{{\rm{IRS}}}})}^2} + \left\| {{{\mathbf{x}}_{{\rm{IRS}}}} - {{\mathbf{x}}_{\rm{U}}}} \right\|_2^2} \right)\left( {1 - \exp \left\{ { - {\lambda _{\rm{U}}}\pi {{\left( {{R_{{\rm{co}}}}} \right)}^2}} \right\}} \right){{\left( {{R_{{\rm{co}}}}} \right)}^2}}} \times $}
\end{equation*}
\begin{equation*}
\ln \left( {1 + {{\left( {\frac{{{R_{{\rm{co}}}}}}{{{h_{{\rm{BS}}}} - {h_{{\rm{IRS}}}}}}} \right)}^2}} \right) \times \int\limits_{{\cal C}({{\mathbf{x}}_{{\rm{IRS}}}},{2R_{{\rm{co}}}})} {\frac{{d{\mathbf{x}}}}{{{{({h_{{\rm{IRS}}}})}^2} + \left\| {{\mathbf{x}} - {{\mathbf{x}}_{\rm{U}}}} \right\|_2^2}}}  = 
\end{equation*}
\begin{equation*}
 \scalebox{.95}[1]{$\frac{{{\lambda _{{\rm{BS}}}}Q{{\left( {{\lambda _{{\rm{wave}}}}} \right)}^4} \sigma _{{\rm{d}}}^2{{\left( {1 + {{\tilde n}_{{\rm{IRS}}}}({{\mathbf{x}}_{\rm{U}}})} \right)}^2}{{\tilde n}_{{\rm{IRS}}}}({{\mathbf{x}}_{\rm{U}}})\left( {{\lambda _{\rm{U}}}\pi {{\left( {{R_{{\rm{co}}}}} \right)}^2} + \exp \left\{ { - {\lambda _{\rm{U}}}\pi {{\left( {{R_{{\rm{co}}}}} \right)}^2}} \right\} - 1} \right)}}{{{{\left( 4 \right)}^5}{\pi ^3}\left( {1 - \exp \left\{ { - {\lambda _{\rm{U}}}\pi {{\left( {{R_{{\rm{co}}}}} \right)}^2}} \right\}} \right){{\left( {{R_{{\rm{co}}}}} \right)}^2}}} \times $}
\end{equation*}
\begin{equation*}
\ln \left( {1 + {{\left( {\frac{{{R_{{\rm{co}}}}}}{{{h_{{\rm{BS}}}} - {h_{{\rm{IRS}}}}}}} \right)}^2}} \right) \times \ln \left( {1 + {{\left( {\frac{{2{R_{{\rm{co}}}}}}{{{h_{{\rm{IRS}}}}}}} \right)}^2}} \right),
\end{equation*}

which yields:
\begin{equation*}
 \scalebox{.95}[1]{$E\left\{ {\sum\limits_{{{\mathbf{x}}_{{\rm{IRS}}}} \in {{\tilde \Psi }_{{\rm{IRS}}}}({{\mathbf{x}}_{\rm{U}}})} {\frac{{Q{{\left( {{\lambda _{{\rm{wave}}}}} \right)}^4} \sigma _{{\rm{d}}}^2{{\left( {1 + {{\tilde n}_{{\rm{IRS}}}}({{\mathbf{x}}_{\rm{U}}})} \right)}^2}\left( {{\lambda _{\rm{U}}}\pi {{\left( {{R_{{\rm{co}}}}} \right)}^2} + \exp \left\{ { - {\lambda _{\rm{U}}}\pi {{\left( {{R_{{\rm{co}}}}} \right)}^2}} \right\} - 1} \right)}}{{{{\left( 4 \right)}^4}{\pi ^3}\left( {{{({h_{{\rm{IRS}}}})}^2} + \left\| {{{\mathbf{x}}_{{\rm{IRS}}}} - {{\mathbf{x}}_{\rm{U}}}} \right\|_2^2} \right)\left( {1 - \exp \left\{ { - {\lambda _{\rm{U}}}\pi {{\left( {{R_{{\rm{co}}}}} \right)}^2}} \right\}} \right)}}}  \times } \right.$}
\end{equation*}
\begin{equation*}
\left. {{\lambda _{{\rm{BS}}}}\ln \left( {1 + {{\left( {\frac{{{R_{{\rm{co}}}}}}{{{h_{{\rm{BS}}}} - {h_{{\rm{IRS}}}}}}} \right)}^2}} \right)} \right\} = 
\end{equation*}
\begin{equation*}
 \scalebox{1}[1]{$\frac{{{\lambda _{{\rm{BS}}}}Q{{\left( {{\lambda _{{\rm{wave}}}}} \right)}^4} \sigma _{{\rm{d}}}^2\left( {{\lambda _{\rm{U}}}\pi {{\left( {{R_{{\rm{co}}}}} \right)}^2} + \exp \left\{ { - {\lambda _{\rm{U}}}\pi {{\left( {{R_{{\rm{co}}}}} \right)}^2}} \right\} - 1} \right)}}{{{{\left( 4 \right)}^5}{\pi ^3}\left( {1 - \exp \left\{ { - {\lambda _{\rm{U}}}\pi {{\left( {{R_{{\rm{co}}}}} \right)}^2}} \right\}} \right){{\left( {{R_{{\rm{co}}}}} \right)}^2}}} \times $}
\end{equation*}
\begin{equation*}
\ln \left( {1 + {{\left( {\frac{{{R_{{\rm{co}}}}}}{{{h_{{\rm{BS}}}} - {h_{{\rm{IRS}}}}}}} \right)}^2}} \right) \times \ln \left( {1 + {{\left( {\frac{{2{R_{{\rm{co}}}}}}{{{h_{{\rm{IRS}}}}}}} \right)}^2}} \right) \times 
\end{equation*}
\begin{equation}
 \scalebox{.95}[1]{$\left( {{{\left( {{\lambda _{{\rm{IRS}}}}\pi {{\left( {2{R_{{\rm{co}}}}} \right)}^2}} \right)}^3}\exp \left\{ {\frac{9}{{2{\lambda _{{\rm{IRS}}}}\pi {{\left( {2{R_{{\rm{co}}}}} \right)}^2}}}} \right\} + 2{{\left( {{\lambda _{{\rm{IRS}}}}\pi {{\left( {2{R_{{\rm{co}}}}} \right)}^2}} \right)}^2} + 3{\lambda _{{\rm{IRS}}}}\pi {{\left( {2{R_{{\rm{co}}}}} \right)}^2}} \right).$}
\label{t37}
\end{equation}
which completes the proof.

\section{Proof of Theorem 4}

We have:
\begin{equation*}
E\left\{ {{{\left| {\mathbb{I}({{\mathbf{x}}_{\rm{U}}})} \right|}^2}\left| {{\Phi _{\rm{U}}},{\Phi _{{\rm{BS}}}},{\Phi _{{\rm{IRS}}}},{{{\mathbf{\tilde x}}}_{{\rm{BS}}}}({{\mathbf{x}}_{\rm{U}}}),{{{\mathbf{\tilde x}}}_{\rm{U}}}({{\mathbf{x}}_{{\rm{IRS}}}})} \right.} \right\} = 
\end{equation*}
\begin{equation*}
\sum\limits_{{{{\mathbf{x'}}}_{{\rm{BS}}}} \in {\Psi _{{\rm{BS}}}}({{\mathbf{x}}_{\rm{U}}})} {\begin{array}{*{20}{c}}
  {}&{} 
\end{array}\sum\limits_{{{{\mathbf{x'}}}_{\rm{U}}} \in {\Psi _{\rm{U}}}({{{\mathbf{x'}}}_{{\rm{BS}}}}):{{{\mathbf{x'}}}_{\rm{U}}} \ne {{\mathbf{x}}_{\rm{U}}},{{{\mathbf{\tilde x}}}_{{\rm{BS}}}}({{{\mathbf{x'}}}_{\rm{U}}}) = {{{\mathbf{x'}}}_{{\rm{BS}}}}} {\frac{{\left( {1 + \left( {\hat h - 1} \right){p_{{\rm{b}}}}} \right){{\left( {{\lambda _{{\rm{wave}}}}} \right)}^2} \sigma _{{\rm{d}}}^2}}{{{{\left( {4\pi } \right)}^2}\left( {{{({h_{{\rm{BS}}}})}^2} + \left\| {{{{\mathbf{x'}}}_{{\rm{BS}}}} - {{\mathbf{x}}_{\rm{U}}}} \right\|_2^2} \right)}} \times } } 
\end{equation*}
\begin{equation*}
{\left( {G\left( {{{{\mathbf{x'}}}_{{\rm{BS}}}},{{{\mathbf{x'}}}_{\rm{U}}},{{\mathbf{x}}_{\rm{U}}}} \right) + \sum\limits_{{{\mathbf{x}}_{{\rm{IRS}}}} \in {\Psi _{{\rm{IRS}}}}({{{\mathbf{x'}}}_{{\rm{BS}}}}):{{{\mathbf{x'}}}_{\rm{U}}} = {{{\mathbf{\tilde x}}}_{\rm{U}}}({{\mathbf{x}}_{{\rm{IRS}}}})} {G\left( {{{{\mathbf{x'}}}_{{\rm{BS}}}},{{\mathbf{x}}_{{\rm{IRS}}}},{{\mathbf{x}}_{\rm{U}}}} \right)} } \right)^2} + 
\end{equation*}
\begin{equation*}
\sum\limits_{{{\mathbf{x}}_{{\rm{IRS}}}} \in {\Psi _{{\rm{IRS}}}}({{\mathbf{x}}_{\rm{U}}}):{{{\mathbf{\tilde x}}}_{\rm{U}}}({{\mathbf{x}}_{{\rm{IRS}}}}) = {{\mathbf{x}}_{\rm{U}}}} {\sum\limits_{{{{\mathbf{x'}}}_{{\rm{IRS}}}} \in {\Psi _{{\rm{IRS}}}}({{\mathbf{x}}_{\rm{U}}}):{{{\mathbf{\tilde x}}}_{\rm{U}}}({{{\mathbf{x'}}}_{{\rm{IRS}}}}) = {{\mathbf{x}}_{\rm{U}}}} {\sum\limits_q {\sum\limits_{q'} {\sum\limits_{{{{\mathbf{x'}}}_{\rm{U}}} \in {\Psi _{\rm{U}}}({{{\mathbf{\tilde x}}}_{{\rm{BS}}}}({{\mathbf{x}}_{\rm{U}}})):{{{\mathbf{x'}}}_{\rm{U}}} \ne {{\mathbf{x}}_{\rm{U}}},{{{\mathbf{\tilde x}}}_{{\rm{BS}}}}({{{\mathbf{x'}}}_{\rm{U}}}) = {{{\mathbf{\tilde x}}}_{{\rm{BS}}}}({{\mathbf{x}}_{\rm{U}}})} {} } } } } 
\end{equation*}
\begin{equation*}
\left( {\frac{{{\delta _{q - {q^\prime }}}{\delta _{{{\mathbf{x}}_{{\rm{IRS}}}} - {{\mathbf{x}}^\prime }_{{\rm{IRS}}}}}{{\left( {{\lambda _{{\rm{wave}}}}} \right)}^4}}}{{{{\left( {4\pi } \right)}^4}\left( {{{({h_{{\rm{IRS}}}})}^2} + \left\| {{{\mathbf{x}}_{{\rm{IRS}}}} - {{\mathbf{x}}_{\rm{U}}}} \right\|_2^2} \right) \times \left( {{{({h_{{\rm{BS}}}} - {h_{{\rm{IRS}}}})}^2} + \left\| {{{{\mathbf{\tilde x}}}_{{\rm{BS}}}}({{\mathbf{x}}_{\rm{U}}}) - {{\mathbf{x}}_{{\rm{IRS}}}}} \right\|_2^2} \right)}} + } \right.
\end{equation*}
\begin{equation*}
   \scalebox{.9}[1]{$\left. {\frac{\frac{\kappa }{{\kappa  + 1}} \times {\left( {1 - {\delta _{q - {q^\prime }}}} \right)\left( {1 - {\delta _{{{\mathbf{x}}_{{\rm{IRS}}}} - {{\mathbf{x}}^\prime }_{{\rm{IRS}}}}}} \right){{\left( {{\lambda _{{\rm{wave}}}}} \right)}^4}{{\left( {\frac{\pi }{4}} \right)}^2}}}{{{{\left( {4\pi } \right)}^4}\sqrt {{{({h_{{\rm{IRS}}}})}^2} + \left\| {{{\mathbf{x}}_{{\rm{IRS}}}} - {{\mathbf{x}}_{\rm{U}}}} \right\|_2^2}  \times \sqrt {{{({h_{{\rm{IRS}}}})}^2} + \left\| {{{{\mathbf{x'}}}_{{\rm{IRS}}}} - {{\mathbf{x}}_{\rm{U}}}} \right\|_2^2}  \times \sqrt {{{({h_{{\rm{BS}}}} - {h_{{\rm{IRS}}}})}^2} + \left\| {{{{\mathbf{\tilde x}}}_{{\rm{BS}}}}({{\mathbf{x}}_{\rm{U}}}) - {{\mathbf{x}}_{{\rm{IRS}}}}} \right\|_2^2}  \times \sqrt {{{({h_{{\rm{BS}}}} - {h_{{\rm{IRS}}}})}^2} + \left\| {{{{\mathbf{\tilde x}}}_{{\rm{BS}}}}({{\mathbf{x}}_{\rm{U}}}) - {{{\mathbf{x'}}}_{{\rm{IRS}}}}} \right\|_2^2} }}} \right) \times $}
\end{equation*}
\begin{equation*}
\left( {G\left( {{{{\mathbf{\tilde x}}}_{{\rm{BS}}}}({{\mathbf{x}}_{\rm{U}}}),{{{\mathbf{x'}}}_{\rm{U}}},{{\mathbf{x}}_{{\rm{IRS}}}}} \right) + \sum\limits_{{{{\mathbf{x''}}}_{{\rm{IRS}}}} \in {\Psi _{{\rm{IRS}}}}({{{\mathbf{x'}}}_{{\rm{BS}}}}):{{{\mathbf{x'}}}_{\rm{U}}} = {{{\mathbf{\tilde x}}}_{\rm{U}}}({{{\mathbf{x''}}}_{{\rm{IRS}}}})} {G\left( {{{{\mathbf{\tilde x}}}_{{\rm{BS}}}}({{\mathbf{x}}_{\rm{U}}}),{{{\mathbf{x''}}}_{{\rm{IRS}}}},{{\mathbf{x}}_{{\rm{IRS}}}}} \right)} } \right) \times 
\end{equation*}
\begin{equation*}
\left( {G\left( {{{{\mathbf{\tilde x}}}_{{\rm{BS}}}}({{\mathbf{x}}_{\rm{U}}}),{{{\mathbf{x'}}}_{\rm{U}}},{{{\mathbf{x'}}}_{{\rm{IRS}}}}} \right) + \sum\limits_{{{{\mathbf{x''}}}_{{\rm{IRS}}}} \in {\Psi _{{\rm{IRS}}}}({{{\mathbf{x'}}}_{{\rm{BS}}}}):{{{\mathbf{x'}}}_{\rm{U}}} = {{{\mathbf{\tilde x}}}_{\rm{U}}}({{{\mathbf{x''}}}_{{\rm{IRS}}}})} {G\left( {{{{\mathbf{\tilde x}}}_{{\rm{BS}}}}({{\mathbf{x}}_{\rm{U}}}),{{{\mathbf{x''}}}_{{\rm{IRS}}}},{{{\mathbf{x'}}}_{{\rm{IRS}}}}} \right)} } \right) \sigma _{{\rm{d}}}^2 + 
\end{equation*}
\begin{equation*}
\sum\limits_{{{\mathbf{x}}_{{\rm{IRS}}}} \in {\Psi _{{\rm{IRS}}}}({{\mathbf{x}}_{\rm{U}}}):{{{\mathbf{\tilde x}}}_{\rm{U}}}({{\mathbf{x}}_{{\rm{IRS}}}}) = {{\mathbf{x}}_{\rm{U}}}} {\begin{array}{*{20}{c}}
  {}&{} 
\end{array}\sum\limits_{{{{\mathbf{x'}}}_{{\rm{BS}}}} \in {\Psi _{{\rm{BS}}}}({{\mathbf{x}}_{{\rm{IRS}}}}):{{{\mathbf{x'}}}_{{\rm{BS}}}} \ne {{{\mathbf{\tilde x}}}_{{\rm{BS}}}}({{\mathbf{x}}_{\rm{U}}})} {\begin{array}{*{20}{c}}
  {}&{} 
\end{array}\sum\limits_{{{{\mathbf{x'}}}_{\rm{U}}} \in {\Psi _{\rm{U}}}({{{\mathbf{x'}}}_{{\rm{BS}}}}):{{{\mathbf{x'}}}_{\rm{U}}} \ne {{\mathbf{x}}_{\rm{U}}},{{{\mathbf{\tilde x}}}_{{\rm{BS}}}}({{{\mathbf{x'}}}_{\rm{U}}}) = {{{\mathbf{x'}}}_{{\rm{BS}}}}} {} } } 
\end{equation*}
\begin{equation*}
\frac{{Q{{\left( {{\lambda _{{\rm{wave}}}}} \right)}^4}}}{{{{\left( {4\pi } \right)}^4}\left( {{{({h_{{\rm{IRS}}}})}^2} + \left\| {{{\mathbf{x}}_{{\rm{IRS}}}} - {{\mathbf{x}}_{\rm{U}}}} \right\|_2^2} \right)\left( {{{({h_{{\rm{BS}}}} - {h_{{\rm{IRS}}}})}^2} + \left\| {{{{\mathbf{x'}}}_{{\rm{BS}}}} - {{\mathbf{x}}_{{\rm{IRS}}}}} \right\|_2^2} \right)}} \times 
\end{equation*}
\begin{equation*}
{\left( {G\left( {{{{\mathbf{x'}}}_{{\rm{BS}}}},{{{\mathbf{x'}}}_{\rm{U}}},{{\mathbf{x}}_{{\rm{IRS}}}}} \right) + \sum\limits_{{{{\mathbf{x'}}}_{{\rm{IRS}}}} \in {\Psi _{{\rm{IRS}}}}({{{\mathbf{x'}}}_{{\rm{BS}}}}):{{{\mathbf{x'}}}_{\rm{U}}} = {{{\mathbf{\tilde x}}}_{\rm{U}}}({{{\mathbf{x'}}}_{{\rm{IRS}}}})} {G\left( {{{{\mathbf{x'}}}_{{\rm{BS}}}},{{{\mathbf{x'}}}_{{\rm{IRS}}}},{{\mathbf{x}}_{{\rm{IRS}}}}} \right)} } \right)^2} \sigma _{{\rm{d}}}^2 + 
\end{equation*}
\begin{equation*}
\sum\limits_{{{\mathbf{x}}_{{\rm{IRS}}}} \in {\Psi _{{\rm{IRS}}}}({{\mathbf{x}}_{\rm{U}}}):{{{\mathbf{\tilde x}}}_{\rm{U}}}({{\mathbf{x}}_{{\rm{IRS}}}}) \ne {{\mathbf{x}}_{\rm{U}}}} {\begin{array}{*{20}{c}}
  {}&{} 
\end{array}\sum\limits_{{{{\mathbf{x'}}}_{{\rm{BS}}}} \in {\Psi _{{\rm{BS}}}}({{\mathbf{x}}_{{\rm{IRS}}}})} {\begin{array}{*{20}{c}}
  {}&{} 
\end{array}\sum\limits_{{{{\mathbf{x'}}}_{\rm{U}}} \in {\Psi _{\rm{U}}}({{{\mathbf{x'}}}_{{\rm{BS}}}}):{{{\mathbf{x'}}}_{\rm{U}}} \ne {{\mathbf{x}}_{\rm{U}}},{{{\mathbf{\tilde x}}}_{{\rm{BS}}}}({{{\mathbf{x'}}}_{\rm{U}}}) = {{{\mathbf{x'}}}_{{\rm{BS}}}}} {} } } 
\end{equation*}
\begin{equation*}
\frac{{Q{{\left( {{\lambda _{{\rm{wave}}}}} \right)}^4}}}{{{{\left( {4\pi } \right)}^4}\left( {{{({h_{{\rm{IRS}}}})}^2} + \left\| {{{\mathbf{x}}_{{\rm{IRS}}}} - {{\mathbf{x}}_{\rm{U}}}} \right\|_2^2} \right)\left( {{{({h_{{\rm{BS}}}} - {h_{{\rm{IRS}}}})}^2} + \left\| {{{{\mathbf{x'}}}_{{\rm{BS}}}} - {{\mathbf{x}}_{{\rm{IRS}}}}} \right\|_2^2} \right)}} \times 
\end{equation*}
\begin{equation*}
{\left( {G\left( {{{{\mathbf{x'}}}_{{\rm{BS}}}},{{{\mathbf{x'}}}_{\rm{U}}},{{\mathbf{x}}_{{\rm{IRS}}}}} \right) + \sum\limits_{{{{\mathbf{x'}}}_{{\rm{IRS}}}} \in {\Psi _{{\rm{IRS}}}}({{{\mathbf{x'}}}_{{\rm{BS}}}}):{{{\mathbf{x'}}}_{\rm{U}}} = {{{\mathbf{\tilde x}}}_{\rm{U}}}({{{\mathbf{x'}}}_{{\rm{IRS}}}})} {G\left( {{{{\mathbf{x'}}}_{{\rm{BS}}}},{{{\mathbf{x'}}}_{{\rm{IRS}}}},{{\mathbf{x}}_{{\rm{IRS}}}}} \right)} } \right)^2} \sigma _{{\rm{d}}}^2 \ge
\end{equation*}
\begin{equation*}
 \scalebox{1}[1]{$\sum\limits_{{{{\mathbf{x'}}}_{{\rm{BS}}}} \in {\Psi _{{\rm{BS}}}}({{\mathbf{x}}_{\rm{U}}})} {\sum\limits_{{{{\mathbf{x'}}}_{\rm{U}}} \in {\Psi _{\rm{U}}}({{{\mathbf{x'}}}_{{\rm{BS}}}}):{{{\mathbf{x'}}}_{\rm{U}}} \ne {{\mathbf{x}}_{\rm{U}}}} {\frac{{\left( {1 - {\pmb{\mathbb{I}}}\left( {\left\| {{{{\mathbf{\tilde x}}}_{{\rm{BS}}}}({{{\mathbf{x'}}}_{\rm{U}}}) - {{{\mathbf{x'}}}_{{\rm{BS}}}}} \right\|} \right)} \right)\left( {1 + \left( {\hat h - 1} \right){p_{{\rm{b}}}}} \right){{\left( {{\lambda _{{\rm{wave}}}}} \right)}^2}{\delta ^2} \sigma _{{\rm{d}}}^2}}{{{{\left( {4\pi } \right)}^2}\left( {{{({h_{{\rm{BS}}}})}^2} + \left\| {{{{\mathbf{x'}}}_{{\rm{BS}}}} - {{\mathbf{x}}_{\rm{U}}}} \right\|_2^2} \right)}}} }  + $}
\end{equation*}
\begin{equation*}
\sum\limits_{{{\mathbf{x}}_{{\rm{IRS}}}} \in {\Psi _{{\rm{IRS}}}}({{\mathbf{x}}_{\rm{U}}}):{{{\mathbf{\tilde x}}}_{\rm{U}}}({{\mathbf{x}}_{{\rm{IRS}}}}) = {{\mathbf{x}}_{\rm{U}}}} {\sum\limits_{{{{\mathbf{x'}}}_{{\rm{IRS}}}} \in {\Psi _{{\rm{IRS}}}}({{\mathbf{x}}_{\rm{U}}}):{{{\mathbf{\tilde x}}}_{\rm{U}}}({{{\mathbf{x'}}}_{{\rm{IRS}}}}) = {{\mathbf{x}}_{\rm{U}}}} {\sum\limits_q {\sum\limits_{q'} {\sum\limits_{{{{\mathbf{x'}}}_{\rm{U}}} \in {\Psi _{\rm{U}}}({{{\mathbf{\tilde x}}}_{{\rm{BS}}}}({{\mathbf{x}}_{\rm{U}}})):{{{\mathbf{x'}}}_{\rm{U}}} \ne {{\mathbf{x}}_{\rm{U}}},{{{\mathbf{\tilde x}}}_{{\rm{BS}}}}({{{\mathbf{x'}}}_{\rm{U}}}) = {{{\mathbf{\tilde x}}}_{{\rm{BS}}}}({{\mathbf{x}}_{\rm{U}}})} {} } } } } 
\end{equation*}
\begin{equation*}
\left( {\frac{{{\delta _{q - {q^\prime }}}{\delta _{{{\mathbf{x}}_{{\rm{IRS}}}} - {{\mathbf{x}}^\prime }_{{\rm{IRS}}}}}{{\left( {{\lambda _{{\rm{wave}}}}} \right)}^4}{\delta ^2} \sigma _{{\rm{d}}}^2}}{{{{\left( {4\pi } \right)}^4}\left( {{{({h_{{\rm{IRS}}}})}^2} + \left\| {{{\mathbf{x}}_{{\rm{IRS}}}} - {{\mathbf{x}}_{\rm{U}}}} \right\|_2^2} \right) \times \left( {{{({h_{{\rm{BS}}}} - {h_{{\rm{IRS}}}})}^2} + \left\| {{{{\mathbf{\tilde x}}}_{{\rm{BS}}}}({{\mathbf{x}}_{\rm{U}}}) - {{\mathbf{x}}_{{\rm{IRS}}}}} \right\|_2^2} \right)}} + } \right.
\end{equation*}
\begin{equation*}
   \scalebox{.9}[1]{$\left. {\frac{\frac{\kappa }{{\kappa  + 1}} \times {\left( {1 - {\delta _{q - {q^\prime }}}} \right)\left( {1 - {\delta _{{{\mathbf{x}}_{{\rm{IRS}}}} - {{\mathbf{x}}^\prime }_{{\rm{IRS}}}}}} \right){{\left( {{\lambda _{{\rm{wave}}}}} \right)}^4}{{\left( {\frac{\pi }{4}} \right)}^2}{\delta ^2} \sigma _{{\rm{d}}}^2}}{{{{\left( {4\pi } \right)}^4}\sqrt {{{({h_{{\rm{IRS}}}})}^2} + \left\| {{{\mathbf{x}}_{{\rm{IRS}}}} - {{\mathbf{x}}_{\rm{U}}}} \right\|_2^2}  \times \sqrt {{{({h_{{\rm{IRS}}}})}^2} + \left\| {{{{\mathbf{x'}}}_{{\rm{IRS}}}} - {{\mathbf{x}}_{\rm{U}}}} \right\|_2^2}  \times \sqrt {{{({h_{{\rm{BS}}}} - {h_{{\rm{IRS}}}})}^2} + \left\| {{{{\mathbf{\tilde x}}}_{{\rm{BS}}}}({{\mathbf{x}}_{\rm{U}}}) - {{\mathbf{x}}_{{\rm{IRS}}}}} \right\|_2^2}  \times \sqrt {{{({h_{{\rm{BS}}}} - {h_{{\rm{IRS}}}})}^2} + \left\| {{{{\mathbf{\tilde x}}}_{{\rm{BS}}}}({{\mathbf{x}}_{\rm{U}}}) - {{{\mathbf{x'}}}_{{\rm{IRS}}}}} \right\|_2^2} }}} \right) + $}
\end{equation*}
\begin{equation*}
\sum\limits_{{{\mathbf{x}}_{{\rm{IRS}}}} \in {\Psi _{{\rm{IRS}}}}({{\mathbf{x}}_{\rm{U}}}):{{{\mathbf{\tilde x}}}_{\rm{U}}}({{\mathbf{x}}_{{\rm{IRS}}}}) = {{\mathbf{x}}_{\rm{U}}}} {\begin{array}{*{20}{c}}
  {}&{} 
\end{array}\sum\limits_{{{{\mathbf{x'}}}_{{\rm{BS}}}} \in {\Psi _{{\rm{BS}}}}({{\mathbf{x}}_{{\rm{IRS}}}}):{{{\mathbf{x'}}}_{{\rm{BS}}}} \ne {{{\mathbf{\tilde x}}}_{{\rm{BS}}}}({{\mathbf{x}}_{\rm{U}}})} {\begin{array}{*{20}{c}}
  {}&{} 
\end{array}\sum\limits_{{{{\mathbf{x'}}}_{\rm{U}}} \in {\Psi _{\rm{U}}}({{{\mathbf{x'}}}_{{\rm{BS}}}}):{{{\mathbf{x'}}}_{\rm{U}}} \ne {{\mathbf{x}}_{\rm{U}}},{{{\mathbf{\tilde x}}}_{{\rm{BS}}}}({{{\mathbf{x'}}}_{\rm{U}}}) = {{{\mathbf{x'}}}_{{\rm{BS}}}}} {} } } 
\end{equation*}
\begin{equation*}
\frac{{Q{{\left( {{\lambda _{{\rm{wave}}}}} \right)}^4}{\delta ^2} \sigma _{{\rm{d}}}^2}}{{{{\left( {4\pi } \right)}^4}\left( {{{({h_{{\rm{IRS}}}})}^2} + \left\| {{{\mathbf{x}}_{{\rm{IRS}}}} - {{\mathbf{x}}_{\rm{U}}}} \right\|_2^2} \right)\left( {{{({h_{{\rm{BS}}}} - {h_{{\rm{IRS}}}})}^2} + \left\| {{{{\mathbf{x'}}}_{{\rm{BS}}}} - {{\mathbf{x}}_{{\rm{IRS}}}}} \right\|_2^2} \right)}} + 
\end{equation*}
\begin{equation*}
\sum\limits_{{{\mathbf{x}}_{{\rm{IRS}}}} \in {\Psi _{{\rm{IRS}}}}({{\mathbf{x}}_{\rm{U}}}):{{{\mathbf{\tilde x}}}_{\rm{U}}}({{\mathbf{x}}_{{\rm{IRS}}}}) \ne {{\mathbf{x}}_{\rm{U}}}} {\begin{array}{*{20}{c}}
  {}&{} 
\end{array}\sum\limits_{{{{\mathbf{x'}}}_{{\rm{BS}}}} \in {\Psi _{{\rm{BS}}}}({{\mathbf{x}}_{{\rm{IRS}}}})} {\begin{array}{*{20}{c}}
  {}&{} 
\end{array}\sum\limits_{{{{\mathbf{x'}}}_{\rm{U}}} \in {\Psi _{\rm{U}}}({{{\mathbf{x'}}}_{{\rm{BS}}}}):{{{\mathbf{x'}}}_{\rm{U}}} \ne {{\mathbf{x}}_{\rm{U}}},{{{\mathbf{\tilde x}}}_{{\rm{BS}}}}({{{\mathbf{x'}}}_{\rm{U}}}) = {{{\mathbf{x'}}}_{{\rm{BS}}}}} {} } } 
\end{equation*}
\begin{equation*}
\frac{{Q{{\left( {{\lambda _{{\rm{wave}}}}} \right)}^4}{\delta ^2} \sigma _{{\rm{d}}}^2}}{{{{\left( {4\pi } \right)}^4}\left( {{{({h_{{\rm{IRS}}}})}^2} + \left\| {{{\mathbf{x}}_{{\rm{IRS}}}} - {{\mathbf{x}}_{\rm{U}}}} \right\|_2^2} \right)\left( {{{({h_{{\rm{BS}}}} - {h_{{\rm{IRS}}}})}^2} + \left\| {{{{\mathbf{x'}}}_{{\rm{BS}}}} - {{\mathbf{x}}_{{\rm{IRS}}}}} \right\|_2^2} \right)}} =
\end{equation*}
\begin{equation}
 \scalebox{.95}[1]{$\sum\limits_{{{{\mathbf{x'}}}_{{\rm{BS}}}} \in {\Psi _{{\rm{BS}}}}({{\mathbf{x}}_{\rm{U}}})} {\sum\limits_{{{{\mathbf{x'}}}_{\rm{U}}} \in {\Psi _{\rm{U}}}({{{\mathbf{x'}}}_{{\rm{BS}}}}):{{{\mathbf{x'}}}_{\rm{U}}} \ne {{\mathbf{x}}_{\rm{U}}}} {\frac{{\left( {1 - {\pmb{\mathbb{I}}}\left( {\left\| {{{{\mathbf{\tilde x}}}_{{\rm{BS}}}}({{{\mathbf{x'}}}_{\rm{U}}}) - {{{\mathbf{x'}}}_{{\rm{BS}}}}} \right\|} \right)} \right)\left( {1 + \left( {\hat h - 1} \right){p_{{\rm{b}}}}} \right){{\left( {{\lambda _{{\rm{wave}}}}} \right)}^2}{\delta ^2} \sigma _{{\rm{d}}}^2}}{{{{\left( {4\pi } \right)}^2}\left( {{{({h_{{\rm{BS}}}})}^2} + \left\| {{{{\mathbf{x'}}}_{{\rm{BS}}}} - {{\mathbf{x}}_{\rm{U}}}} \right\|_2^2} \right)}}} }  + $}
\label{t48}
\end{equation}
\begin{equation*}
\sum\limits_{{{\mathbf{x}}_{{\rm{IRS}}}} \in {\Psi _{{\rm{IRS}}}}({{\mathbf{x}}_{\rm{U}}})} {\begin{array}{*{20}{c}}
  {}&{} 
\end{array}\sum\limits_{{{{\mathbf{x'}}}_{{\rm{IRS}}}} \in {\Psi _{{\rm{IRS}}}}({{\mathbf{x}}_{\rm{U}}})} {\begin{array}{*{20}{c}}
  {}&{} 
\end{array}\sum\limits_{{{{\mathbf{x'}}}_{\rm{U}}} \in {\Psi _{\rm{U}}}({{{\mathbf{\tilde x}}}_{{\rm{BS}}}}({{\mathbf{x}}_{\rm{U}}})):{{{\mathbf{x'}}}_{\rm{U}}} \ne {{\mathbf{x}}_{\rm{U}}},{{{\mathbf{\tilde x}}}_{{\rm{BS}}}}({{{\mathbf{x'}}}_{\rm{U}}}) = {{{\mathbf{\tilde x}}}_{{\rm{BS}}}}({{\mathbf{x}}_{\rm{U}}})} {} } } 
\end{equation*}
\begin{equation}
   \scalebox{.9}[1]{$\frac{\frac{\kappa }{{\kappa  + 1}} \times {\left( {1 -{\pmb{\mathbb{I}}}\left( {\left\| {{{{\mathbf{\tilde x}}}_{\rm{U}}}({{\mathbf{x}}_{{\rm{IRS}}}}) - {{\mathbf{x}}_{\rm{U}}}} \right\|} \right)} \right)\left( {1 - {\pmb{\mathbb{I}}}\left( {\left\| {{{{\mathbf{\tilde x}}}_{\rm{U}}}({{{\mathbf{x'}}}_{{\rm{IRS}}}}) - {{\mathbf{x}}_{\rm{U}}}} \right\|} \right)} \right)Q(Q - 1)\left( {1 - {\delta _{{{\mathbf{x}}_{{\rm{IRS}}}} - {{\mathbf{x}}^\prime }_{{\rm{IRS}}}}}} \right){{\left( {{\lambda _{{\rm{wave}}}}} \right)}^4}{{\left( {\frac{\pi }{4}} \right)}^2}{\delta ^2} \sigma _{{\rm{d}}}^2}}{{{{\left( {4\pi } \right)}^4}\sqrt {{{({h_{{\rm{IRS}}}})}^2} + \left\| {{{\mathbf{x}}_{{\rm{IRS}}}} - {{\mathbf{x}}_{\rm{U}}}} \right\|_2^2}  \times \sqrt {{{({h_{{\rm{IRS}}}})}^2} + \left\| {{{{\mathbf{x'}}}_{{\rm{IRS}}}} - {{\mathbf{x}}_{\rm{U}}}} \right\|_2^2}  \times \sqrt {{{({h_{{\rm{BS}}}} - {h_{{\rm{IRS}}}})}^2} + \left\| {{{{\mathbf{\tilde x}}}_{{\rm{BS}}}}({{\mathbf{x}}_{\rm{U}}}) - {{\mathbf{x}}_{{\rm{IRS}}}}} \right\|_2^2}  \times \sqrt {{{({h_{{\rm{BS}}}} - {h_{{\rm{IRS}}}})}^2} + \left\| {{{{\mathbf{\tilde x}}}_{{\rm{BS}}}}({{\mathbf{x}}_{\rm{U}}}) - {{{\mathbf{x'}}}_{{\rm{IRS}}}}} \right\|_2^2} }} + $}
\label{t42}
\end{equation}
\begin{equation*}
\sum\limits_{{{\mathbf{x}}_{{\rm{IRS}}}} \in {\Psi _{{\rm{IRS}}}}({{\mathbf{x}}_{\rm{U}}})} {\begin{array}{*{20}{c}}
  {}&{} 
\end{array}\sum\limits_{{{{\mathbf{x'}}}_{{\rm{BS}}}} \in {\Psi _{{\rm{BS}}}}({{\mathbf{x}}_{{\rm{IRS}}}})} {\begin{array}{*{20}{c}}
  {}&{} 
\end{array}\sum\limits_{{{{\mathbf{x'}}}_{\rm{U}}} \in {\Psi _{\rm{U}}}({{{\mathbf{x'}}}_{{\rm{BS}}}}):{{{\mathbf{x'}}}_{\rm{U}}} \ne {{\mathbf{x}}_{\rm{U}}},{{{\mathbf{\tilde x}}}_{{\rm{BS}}}}({{{\mathbf{x'}}}_{\rm{U}}}) = {{{\mathbf{x'}}}_{{\rm{BS}}}}} {} } } 
\end{equation*}
\begin{equation}
\frac{{Q{{\left( {{\lambda _{{\rm{wave}}}}} \right)}^4}{\delta ^2} \sigma _{{\rm{d}}}^2}}{{{{\left( {4\pi } \right)}^4}\left( {{{({h_{{\rm{IRS}}}})}^2} + \left\| {{{\mathbf{x}}_{{\rm{IRS}}}} - {{\mathbf{x}}_{\rm{U}}}} \right\|_2^2} \right)\left( {{{({h_{{\rm{BS}}}} - {h_{{\rm{IRS}}}})}^2} + \left\| {{{{\mathbf{x'}}}_{{\rm{BS}}}} - {{\mathbf{x}}_{{\rm{IRS}}}}} \right\|_2^2} \right)}} + 
\label{t44}
\end{equation}
 For term (\ref{t48}), first, for an arbitrary $b<R_{\rm co}$, we define the following process:
\begin{equation}
{{\tilde {\tilde{ \Psi}} }_{\rm{U}}}({{\mathbf{x}}_{\rm{U}}}) = {\Phi _{\rm{U}}}\bigcap {{\cal C}\left( {{{\mathbf{x}}_{\rm{U}}},{R_{{\rm{co}}}} - b} \right)} .
\end{equation}
Then, we have:
\begin{equation*}
 \scalebox{.95}[1]{$E\left\{ {\left. {\sum\limits_{{{{\mathbf{x'}}}_{{\rm{BS}}}} \in {\Psi _{{\rm{BS}}}}({{\mathbf{x}}_{\rm{U}}})} {\sum\limits_{{{{\mathbf{x'}}}_{\rm{U}}} \in {\Psi _{\rm{U}}}({{{\mathbf{x'}}}_{{\rm{BS}}}}):{{{\mathbf{x'}}}_{\rm{U}}} \ne {{\mathbf{x}}_{\rm{U}}}} {\frac{{\left( {1 - {\pmb{\mathbb{I}}}\left( {\left\| {{{{\mathbf{\tilde x}}}_{{\rm{BS}}}}({{{\mathbf{x'}}}_{\rm{U}}}) - {{{\mathbf{x'}}}_{{\rm{BS}}}}} \right\|} \right)} \right)\left( {1 + \left( {\hat h - 1} \right){p_{{\rm{b}}}}} \right){{\left( {{\lambda _{{\rm{wave}}}}} \right)}^2}{\delta ^2} \sigma _{{\rm{d}}}^2}}{{{{\left( {4\pi } \right)}^2}\left( {{{({h_{{\rm{BS}}}})}^2} + \left\| {{{{\mathbf{x'}}}_{{\rm{BS}}}} - {{\mathbf{x}}_{\rm{U}}}} \right\|_2^2} \right)}}} } } \right|} \right.$}
\end{equation*}
\begin{equation*}
\left. {\begin{array}{*{20}{c}}
  {} \\ 
  {} 
\end{array}{\Phi _{\rm{U}}},{\Phi _{{\rm{BS}}}},{\Phi _{{\rm{IRS}}}},{{{\mathbf{\tilde x}}}_{\rm{U}}}({{\mathbf{x}}_{{\rm{IRS}}}})} \right\} = 
\end{equation*}
\begin{equation*}
\sum\limits_{{{{\mathbf{x'}}}_{{\rm{BS}}}} \in {\Psi _{{\rm{BS}}}}({{\mathbf{x}}_{\rm{U}}})} {\sum\limits_{{{{\mathbf{x'}}}_{\rm{U}}} \in {\Psi _{\rm{U}}}({{{\mathbf{x'}}}_{{\rm{BS}}}}):{{{\mathbf{x'}}}_{\rm{U}}} \ne {{\mathbf{x}}_{\rm{U}}}} {\frac{{{\pmb{\mathbb{I}}}\left( {{n_{{\rm{BS}}}}({{{\mathbf{x'}}}_{\rm{U}}})} \right)\left( {1 + \left( {\hat h - 1} \right){p_{{\rm{b}}}}} \right){{\left( {{\lambda _{{\rm{wave}}}}} \right)}^2}{\delta ^2} \sigma _{{\rm{d}}}^2}}{{{n_{{\rm{BS}}}}({{{\mathbf{x'}}}_{\rm{U}}}){{\left( {4\pi } \right)}^2}\left( {{{({h_{{\rm{BS}}}})}^2} + \left\| {{{{\mathbf{x'}}}_{{\rm{BS}}}} - {{\mathbf{x}}_{\rm{U}}}} \right\|_2^2} \right)}}} }  \ge
\end{equation*}
\begin{equation*}
\sum\limits_{{{{\mathbf{x'}}}_{{\rm{BS}}}} \in {\Psi _{{\rm{BS}}}}({{\mathbf{x}}_{\rm{U}}})} {\sum\limits_{{{{\mathbf{x'}}}_{\rm{U}}} \in {\Psi _{\rm{U}}}({{{\mathbf{x'}}}_{{\rm{BS}}}})\bigcap {{{\tilde {\tilde{ \Psi }}}_{\rm{U}}}({{\mathbf{x}}_{\rm{U}}})} :{{{\mathbf{x'}}}_{\rm{U}}} \ne {{\mathbf{x}}_{\rm{U}}}} {\frac{{{\pmb{\mathbb{I}}}\left( {{n_{{\rm{BS}}}}({{{\mathbf{x'}}}_{\rm{U}}})} \right)\left( {1 + \left( {\hat h - 1} \right){p_{{\rm{b}}}}} \right){{\left( {{\lambda _{{\rm{wave}}}}} \right)}^2}{\delta ^2} \sigma _{{\rm{d}}}^2}}{{{n_{{\rm{BS}}}}({{{\mathbf{x'}}}_{\rm{U}}}){{\left( {4\pi } \right)}^2}\left( {{{({h_{{\rm{BS}}}})}^2} + {{\left( {{R_{{\rm{co}}}}} \right)}^2}} \right)}}} }  \ge
\end{equation*}
\begin{equation*}
   \scalebox{.9}[1]{$\sum\limits_{{{{\mathbf{x'}}}_{{\rm{BS}}}} \in {\Psi _{{\rm{BS}}}}({{\mathbf{x}}_{\rm{U}}})} {\sum\limits_{{{{\mathbf{x'}}}_{\rm{U}}} \in {\Psi _{\rm{U}}}({{{\mathbf{x'}}}_{{\rm{BS}}}})\bigcap {{{\tilde {\tilde {\Psi}} }_{\rm{U}}}({{\mathbf{x}}_{\rm{U}}})} :{{{\mathbf{x'}}}_{\rm{U}}} \ne {{\mathbf{x}}_{\rm{U}}}} {\frac{{{\pmb{\mathbb{I}}}\left( {\prod\limits_{{{{\mathbf{x'}}}_{\rm{U}}} \in {\Psi _{\rm{U}}}({{{\mathbf{x'}}}_{{\rm{BS}}}})\bigcap {{{\tilde {\tilde {\Psi}} }_{\rm{U}}}({{\mathbf{x}}_{\rm{U}}})} :{{{\mathbf{x'}}}_{\rm{U}}} \ne {{\mathbf{x}}_{\rm{U}}}} {{n_{{\rm{BS}}}}({{{\mathbf{x'}}}_{\rm{U}}})} } \right)\left( {1 + \left( {\hat h - 1} \right){p_{{\rm{b}}}}} \right){{\left( {{\lambda _{{\rm{wave}}}}} \right)}^2}{\delta ^2} \sigma _{{\rm{d}}}^2}}{{{{\tilde n}_{{\rm{BS}}}}({{\mathbf{x}}_{\rm{U}}}){{\left( {4\pi } \right)}^2}\left( {{{({h_{{\rm{BS}}}})}^2} + {{\left( {{R_{{\rm{co}}}}} \right)}^2}} \right)}}} }  \ge$}
\end{equation*}
\begin{equation*}
 \scalebox{.95}[1]{$\sum\limits_{{{{\mathbf{x'}}}_{{\rm{BS}}}} \in {\Psi _{{\rm{BS}}}}({{\mathbf{x}}_{\rm{U}}})} {\sum\limits_{{{{\mathbf{x'}}}_{\rm{U}}} \in {\Psi _{\rm{U}}}({{{\mathbf{x'}}}_{{\rm{BS}}}})\bigcap {{{\tilde {\tilde {\Psi}} }_{\rm{U}}}({{\mathbf{x}}_{\rm{U}}})} :{{{\mathbf{x'}}}_{\rm{U}}} \ne {{\mathbf{x}}_{\rm{U}}}} {\frac{{{\pmb{\mathbb{I}}}\left( {{N_{{\rm{BS}}}}\left( {{\cal C}\left( {{{\mathbf{x}}_{\rm{U}}},b} \right)} \right)} \right)\left( {1 + \left( {\hat h - 1} \right){p_{{\rm{b}}}}} \right){{\left( {{\lambda _{{\rm{wave}}}}} \right)}^2}{\delta ^2} \sigma _{{\rm{d}}}^2}}{{{{\tilde n}_{{\rm{BS}}}}({{\mathbf{x}}_{\rm{U}}}){{\left( {4\pi } \right)}^2}\left( {{{({h_{{\rm{BS}}}})}^2} + {{\left( {{R_{{\rm{co}}}}} \right)}^2}} \right)}}} }  \ge$}
\end{equation*}
\begin{equation*}
   \scalebox{1}[1]{$\sum\limits_{{{{\mathbf{x'}}}_{{\rm{BS}}}} \in {\Psi _{{\rm{BS}}}}({{\mathbf{x}}_{\rm{U}}})} {\frac{{{\pmb{\mathbb{I}}}\left( {{N_{{\rm{BS}}}}\left( {{\cal C}\left( {{{\mathbf{x}}_{\rm{U}}},b} \right)} \right)} \right)\left( {{N_{\rm{U}}}\left( {{\cal C}\left( {{{\mathbf{x}}_{\rm{U}}},b} \right)\bigcap {{\cal C}\left( {{{{\mathbf{x'}}}_{{\rm{BS}}}},{R_{{\rm{co}}}}} \right)} } \right) - 1} \right)\left( {1 + \left( {\hat h - 1} \right){p_{{\rm{b}}}}} \right){{\left( {{\lambda _{{\rm{wave}}}}} \right)}^2}{\delta ^2} \sigma _{{\rm{d}}}^2}}{{{{\tilde n}_{{\rm{BS}}}}({{\mathbf{x}}_{\rm{U}}}){{\left( {4\pi } \right)}^2}\left( {{{({h_{{\rm{BS}}}})}^2} + {{\left( {{R_{{\rm{co}}}}} \right)}^2}} \right)}}}  \ge$}
\end{equation*}
\begin{equation*}
\sum\limits_{{{{\mathbf{x'}}}_{{\rm{BS}}}} \in {\Psi _{{\rm{BS}}}}({{\mathbf{x}}_{\rm{U}}})} {\frac{{{\pmb{\mathbb{I}}}\left( {{N_{{\rm{BS}}}}\left( {{\cal C}\left( {{{\mathbf{x}}_{\rm{U}}},b} \right)} \right)} \right) \times \left( {{N_{\rm{U}}}\left( {{\cal C}\left( {{{\mathbf{x}}_{\rm{U}}},b} \right)\bigcap {{\cal C}\left( {{{\mathbf{x}}_{\rm{U}}} + \frac{{{R_{{\rm{co}}}}}}{{\left\| {{{{\mathbf{x'}}}_{{\rm{BS}}}} - {{\mathbf{x}}_{\rm{U}}}} \right\|}}\left( {{{{\mathbf{x'}}}_{{\rm{BS}}}} - {{\mathbf{x}}_{\rm{U}}}} \right),{R_{{\rm{co}}}}} \right)} } \right) - 1} \right)}}{{{{\tilde n}_{{\rm{BS}}}}({{\mathbf{x}}_{\rm{U}}}){{\left( {4\pi } \right)}^2}\left( {{{({h_{{\rm{BS}}}})}^2} + {{\left( {{R_{{\rm{co}}}}} \right)}^2}} \right)}} \times } 
\end{equation*}
\begin{equation*}
\left( {1 + \left( {\hat h - 1} \right){p_{{\rm{b}}}}} \right){\left( {{\lambda _{{\rm{wave}}}}} \right)^2}{\delta ^2} \sigma _{{\rm{d}}}^2,
\end{equation*}
hence, we obtain:
\begin{equation*}
 \scalebox{.95}[1]{$E\left\{ {\sum\limits_{{{{\mathbf{x'}}}_{{\rm{BS}}}} \in {\Psi _{{\rm{BS}}}}({{\mathbf{x}}_{\rm{U}}})} {\frac{{{\pmb{\mathbb{I}}}\left( {{N_{{\rm{BS}}}}\left( {{\cal C}\left( {{{\mathbf{x}}_{\rm{U}}},b} \right)} \right)} \right) \times \left( {{N_{\rm{U}}}\left( {{\cal C}\left( {{{\mathbf{x}}_{\rm{U}}},b} \right)\bigcap {{\cal C}\left( {{{\mathbf{x}}_{\rm{U}}} + \frac{{{R_{{\rm{co}}}}}}{{\left\| {{{{\mathbf{x'}}}_{{\rm{BS}}}} - {{\mathbf{x}}_{\rm{U}}}} \right\|}}\left( {{{{\mathbf{x'}}}_{{\rm{BS}}}} - {{\mathbf{x}}_{\rm{U}}}} \right),{R_{{\rm{co}}}}} \right)} } \right) - 1} \right)}}{{{{\tilde n}_{{\rm{BS}}}}({{\mathbf{x}}_{\rm{U}}}){{\left( {4\pi } \right)}^2}\left( {{{({h_{{\rm{BS}}}})}^2} + {{\left( {{R_{{\rm{co}}}}} \right)}^2}} \right)}} \times } } \right.$}
\end{equation*}
\begin{equation*}
\left. {\left. {\begin{array}{*{20}{c}}
  {} \\ 
  {} 
\end{array}\left( {1 + \left( {\hat h - 1} \right){p_{{\rm{b}}}}} \right){{\left( {{\lambda _{{\rm{wave}}}}} \right)}^2}{\delta ^2} \sigma _{{\rm{d}}}^2} \right|{\Phi _{{\rm{BS}}}}} \right\} = 
\end{equation*}
\begin{equation*}
\sum\limits_{{{{\mathbf{x'}}}_{{\rm{BS}}}} \in {\Psi _{{\rm{BS}}}}({{\mathbf{x}}_{\rm{U}}})} {\frac{{{\pmb{\mathbb{I}}}\left( {{N_{{\rm{BS}}}}\left( {{\cal C}\left( {{{\mathbf{x}}_{\rm{U}}},b} \right)} \right)} \right)\left( {1 + \left( {\hat h - 1} \right){p_{{\rm{b}}}}} \right){{\left( {{\lambda _{{\rm{wave}}}}} \right)}^2}{\delta ^2} \sigma _{{\rm{d}}}^2}}{{{{\tilde n}_{{\rm{BS}}}}({{\mathbf{x}}_{\rm{U}}}){{\left( {4\pi } \right)}^2}\left( {{{({h_{{\rm{BS}}}})}^2} + {{\left( {{R_{{\rm{co}}}}} \right)}^2}} \right)}} \times } 
\end{equation*}
\begin{equation*}
   \scalebox{.9}[1]{$\frac{{{\lambda _{\rm{U}}}\mu \left( {{\cal C}\left( {{{\mathbf{x}}_{\rm{U}}},b} \right)\bigcap {{\cal C}\left( {{{\mathbf{x}}_{\rm{U}}} + \frac{{{R_{{\rm{co}}}}}}{{\left\| {{{{\mathbf{x'}}}_{{\rm{BS}}}} - {{\mathbf{x}}_{\rm{U}}}} \right\|}}\left( {{{{\mathbf{x'}}}_{{\rm{BS}}}} - {{\mathbf{x}}_{\rm{U}}}} \right),{R_{{\rm{co}}}}} \right)} } \right) + \exp \left\{ { - {\lambda _{\rm{U}}}\mu \left( {{\cal C}\left( {{{\mathbf{x}}_{\rm{U}}},b} \right)\bigcap {{\cal C}\left( {{{\mathbf{x}}_{\rm{U}}} + \frac{{{R_{{\rm{co}}}}}}{{\left\| {{{{\mathbf{x'}}}_{{\rm{BS}}}} - {{\mathbf{x}}_{\rm{U}}}} \right\|}}\left( {{{{\mathbf{x'}}}_{{\rm{BS}}}} - {{\mathbf{x}}_{\rm{U}}}} \right),{R_{{\rm{co}}}}} \right)} } \right)} \right\} - 1}}{{1 - \exp \left\{ { - {\lambda _{\rm{U}}}\mu \left( {{\cal C}\left( {{{\mathbf{x}}_{\rm{U}}},b} \right)\bigcap {{\cal C}\left( {{{\mathbf{x}}_{\rm{U}}} + \frac{{{R_{{\rm{co}}}}}}{{\left\| {{{{\mathbf{x'}}}_{{\rm{BS}}}} - {{\mathbf{x}}_{\rm{U}}}} \right\|}}\left( {{{{\mathbf{x'}}}_{{\rm{BS}}}} - {{\mathbf{x}}_{\rm{U}}}} \right),{R_{{\rm{co}}}}} \right)} } \right)} \right\}}} = $}
\end{equation*}
\begin{equation*}
\sum\limits_{{{{\mathbf{x'}}}_{{\rm{BS}}}} \in {\Psi _{{\rm{BS}}}}({{\mathbf{x}}_{\rm{U}}})} {\frac{{{\pmb{\mathbb{I}}}\left( {{N_{{\rm{BS}}}}\left( {{\cal C}\left( {{{\mathbf{x}}_{\rm{U}}},b} \right)} \right)} \right)\left( {1 + \left( {\hat h - 1} \right){p_{{\rm{b}}}}} \right){{\left( {{\lambda _{{\rm{wave}}}}} \right)}^2}{\delta ^2} \sigma _{{\rm{d}}}^2}}{{{{\tilde n}_{{\rm{BS}}}}({{\mathbf{x}}_{\rm{U}}}){{\left( {4\pi } \right)}^2}\left( {{{({h_{{\rm{BS}}}})}^2} + {{\left( {{R_{{\rm{co}}}}} \right)}^2}} \right)}} \times } 
\end{equation*}
\begin{equation*}
   \scalebox{1}[1]{$\frac{{{\lambda _{\rm{U}}}\mu \left( {{\cal C}\left( {{{\mathbf{x}}_{\rm{U}}},b} \right)\bigcap {{\cal C}\left( {{{\mathbf{x}}_{\rm{U}}} - \frac{{{R_{{\rm{co}}}}}}{{\left\| {{{\mathbf{x}}_{\rm{U}}}} \right\|}}{{\mathbf{x}}_{\rm{U}}},{R_{{\rm{co}}}}} \right)} } \right) + \exp \left\{ { - {\lambda _{\rm{U}}}\mu \left( {{\cal C}\left( {{{\mathbf{x}}_{\rm{U}}},b} \right)\bigcap {{\cal C}\left( {{{\mathbf{x}}_{\rm{U}}} - \frac{{{R_{{\rm{co}}}}}}{{\left\| {{{\mathbf{x}}_{\rm{U}}}} \right\|}}{{\mathbf{x}}_{\rm{U}}},{R_{{\rm{co}}}}} \right)} } \right)} \right\} - 1}}{{1 - \exp \left\{ { - {\lambda _{\rm{U}}}\mu \left( {{\cal C}\left( {{{\mathbf{x}}_{\rm{U}}},b} \right)\bigcap {{\cal C}\left( {{{\mathbf{x}}_{\rm{U}}} - \frac{{{R_{{\rm{co}}}}}}{{\left\| {{{\mathbf{x}}_{\rm{U}}}} \right\|}}{{\mathbf{x}}_{\rm{U}}},{R_{{\rm{co}}}}} \right)} } \right)} \right\}}} = $}
\end{equation*}
\begin{equation*}
\frac{{{\pmb{\mathbb{I}}}\left( {{N_{{\rm{BS}}}}\left( {{\cal C}\left( {{{\mathbf{x}}_{\rm{U}}},b} \right)} \right)} \right){n_{{\rm{BS}}}}({{\mathbf{x}}_{\rm{U}}})\left( {1 + \left( {\hat h - 1} \right){p_{{\rm{b}}}}} \right){{\left( {{\lambda _{{\rm{wave}}}}} \right)}^2}{\delta ^2} \sigma _{{\rm{d}}}^2}}{{{{\tilde n}_{{\rm{BS}}}}({{\mathbf{x}}_{\rm{U}}}){{\left( {4\pi } \right)}^2}\left( {{{({h_{{\rm{BS}}}})}^2} + {{\left( {{R_{{\rm{co}}}}} \right)}^2}} \right)}} \times 
\end{equation*}
\begin{equation*}
   \scalebox{.9}[1]{$\left( {\frac{{{\lambda _{\rm{U}}}\left( {\left( {\frac{{2{{\left( {{R_{{\rm{co}}}}} \right)}^2}}}{\pi } - \frac{{{b^2}}}{\pi }} \right)\arcsin \left( {\frac{b}{{2{R_{{\rm{co}}}}}}} \right) - \frac{{{{\left( {{R_{{\rm{co}}}}} \right)}^2}}}{2}\sin \left( {4\arcsin \left( {\frac{b}{{2{R_{{\rm{co}}}}}}} \right)} \right) + \frac{{{b^2}}}{2}\left( {1 - \sin \left( {\pi  - 2\arcsin \left( {\frac{b}{{2{R_{{\rm{co}}}}}}} \right)} \right)} \right)} \right)}}{{1 - \exp \left\{ { - {\lambda _{\rm{U}}}\left( {\left( {\frac{{2{{\left( {{R_{{\rm{co}}}}} \right)}^2}}}{\pi } - \frac{{{b^2}}}{\pi }} \right)\arcsin \left( {\frac{b}{{2{R_{{\rm{co}}}}}}} \right) - \frac{{{{\left( {{R_{{\rm{co}}}}} \right)}^2}}}{2}\sin \left( {4\arcsin \left( {\frac{b}{{2{R_{{\rm{co}}}}}}} \right)} \right) + \frac{{{b^2}}}{2}\left( {1 - \sin \left( {\pi  - 2\arcsin \left( {\frac{b}{{2{R_{{\rm{co}}}}}}} \right)} \right)} \right)} \right)} \right\}}} + } \right.$}
\end{equation*}
\begin{equation*}
   \scalebox{.9}[1]{$\left. {\frac{{\exp \left\{ { - {\lambda _{\rm{U}}}\left( {\left( {\frac{{2{{\left( {{R_{{\rm{co}}}}} \right)}^2}}}{\pi } - \frac{{{b^2}}}{\pi }} \right)\arcsin \left( {\frac{b}{{2{R_{{\rm{co}}}}}}} \right) - \frac{{{{\left( {{R_{{\rm{co}}}}} \right)}^2}}}{2}\sin \left( {4\arcsin \left( {\frac{b}{{2{R_{{\rm{co}}}}}}} \right)} \right) + \frac{{{b^2}}}{2}\left( {1 - \sin \left( {\pi  - 2\arcsin \left( {\frac{b}{{2{R_{{\rm{co}}}}}}} \right)} \right)} \right)} \right)} \right\} - 1}}{{1 - \exp \left\{ { - {\lambda _{\rm{U}}}\left( {\left( {\frac{{2{{\left( {{R_{{\rm{co}}}}} \right)}^2}}}{\pi } - \frac{{{b^2}}}{\pi }} \right)\arcsin \left( {\frac{b}{{2{R_{{\rm{co}}}}}}} \right) - \frac{{{{\left( {{R_{{\rm{co}}}}} \right)}^2}}}{2}\sin \left( {4\arcsin \left( {\frac{b}{{2{R_{{\rm{co}}}}}}} \right)} \right) + \frac{{{b^2}}}{2}\left( {1 - \sin \left( {\pi  - 2\arcsin \left( {\frac{b}{{2{R_{{\rm{co}}}}}}} \right)} \right)} \right)} \right)} \right\}}}} \right) \ge$}
\end{equation*}
\begin{equation*}
\frac{{{\pmb{\mathbb{I}}}\left( {{N_{{\rm{BS}}}}\left( {{\cal C}\left( {{{\mathbf{x}}_{\rm{U}}},b} \right)} \right)} \right){{N_{{\rm{BS}}}}\left( {{\cal C}\left( {{{\mathbf{x}}_{\rm{U}}},b} \right)} \right)}\left( {1 + \left( {\hat h - 1} \right){p_{{\rm{b}}}}} \right){{\left( {{\lambda _{{\rm{wave}}}}} \right)}^2}{\delta ^2} \sigma _{{\rm{d}}}^2}}{{\left( {{N_{{\rm{BS}}}}\left( {{\cal C}\left( {{{\mathbf{x}}_{\rm{U}}},b} \right)} \right) + {N_{{\rm{BS}}}}\left( {{\cal C}\left( {{{\mathbf{x}}_{\rm{U}}},2{R_{{\rm{co}}}}} \right) - {\cal C}\left( {{{\mathbf{x}}_{\rm{U}}},b} \right)} \right)} \right){{\left( {4\pi } \right)}^2}\left( {{{({h_{{\rm{BS}}}})}^2} + {{\left( {{R_{{\rm{co}}}}} \right)}^2}} \right)}} \times 
\end{equation*}
\begin{equation*}
   \scalebox{.9}[1]{$\left( {\frac{{{\lambda _{\rm{U}}}\left( {\left( {\frac{{2{{\left( {{R_{{\rm{co}}}}} \right)}^2}}}{\pi } - \frac{{{b^2}}}{\pi }} \right)\arcsin \left( {\frac{b}{{2{R_{{\rm{co}}}}}}} \right) - \frac{{{{\left( {{R_{{\rm{co}}}}} \right)}^2}}}{2}\sin \left( {4\arcsin \left( {\frac{b}{{2{R_{{\rm{co}}}}}}} \right)} \right) + \frac{{{b^2}}}{2}\left( {1 - \sin \left( {\pi  - 2\arcsin \left( {\frac{b}{{2{R_{{\rm{co}}}}}}} \right)} \right)} \right)} \right)}}{{1 - \exp \left\{ { - {\lambda _{\rm{U}}}\left( {\left( {\frac{{2{{\left( {{R_{{\rm{co}}}}} \right)}^2}}}{\pi } - \frac{{{b^2}}}{\pi }} \right)\arcsin \left( {\frac{b}{{2{R_{{\rm{co}}}}}}} \right) - \frac{{{{\left( {{R_{{\rm{co}}}}} \right)}^2}}}{2}\sin \left( {4\arcsin \left( {\frac{b}{{2{R_{{\rm{co}}}}}}} \right)} \right) + \frac{{{b^2}}}{2}\left( {1 - \sin \left( {\pi  - 2\arcsin \left( {\frac{b}{{2{R_{{\rm{co}}}}}}} \right)} \right)} \right)} \right)} \right\}}} + } \right.$}
\end{equation*}
\begin{equation*}
   \scalebox{.9}[1]{$\left. {\frac{{\exp \left\{ { - {\lambda _{\rm{U}}}\left( {\left( {\frac{{2{{\left( {{R_{{\rm{co}}}}} \right)}^2}}}{\pi } - \frac{{{b^2}}}{\pi }} \right)\arcsin \left( {\frac{b}{{2{R_{{\rm{co}}}}}}} \right) - \frac{{{{\left( {{R_{{\rm{co}}}}} \right)}^2}}}{2}\sin \left( {4\arcsin \left( {\frac{b}{{2{R_{{\rm{co}}}}}}} \right)} \right) + \frac{{{b^2}}}{2}\left( {1 - \sin \left( {\pi  - 2\arcsin \left( {\frac{b}{{2{R_{{\rm{co}}}}}}} \right)} \right)} \right)} \right)} \right\} - 1}}{{1 - \exp \left\{ { - {\lambda _{\rm{U}}}\left( {\left( {\frac{{2{{\left( {{R_{{\rm{co}}}}} \right)}^2}}}{\pi } - \frac{{{b^2}}}{\pi }} \right)\arcsin \left( {\frac{b}{{2{R_{{\rm{co}}}}}}} \right) - \frac{{{{\left( {{R_{{\rm{co}}}}} \right)}^2}}}{2}\sin \left( {4\arcsin \left( {\frac{b}{{2{R_{{\rm{co}}}}}}} \right)} \right) + \frac{{{b^2}}}{2}\left( {1 - \sin \left( {\pi  - 2\arcsin \left( {\frac{b}{{2{R_{{\rm{co}}}}}}} \right)} \right)} \right)} \right)} \right\}}}} \right) \ge$}
\end{equation*}
\begin{equation*}
\frac{{{\pmb{\mathbb{I}}}\left( {{N_{{\rm{BS}}}}\left( {{\cal C}\left( {{{\mathbf{x}}_{\rm{U}}},b} \right)} \right)} \right)\left( {1 + \left( {\hat h - 1} \right){p_{{\rm{b}}}}} \right){{\left( {{\lambda _{{\rm{wave}}}}} \right)}^2}{\delta ^2} \sigma _{{\rm{d}}}^2}}{{\left( {1 + {N_{{\rm{BS}}}}\left( {{\cal C}\left( {{{\mathbf{x}}_{\rm{U}}},2{R_{{\rm{co}}}}} \right) - {\cal C}\left( {{{\mathbf{x}}_{\rm{U}}},b} \right)} \right)} \right){{\left( {4\pi } \right)}^2}\left( {{{({h_{{\rm{BS}}}})}^2} + {{\left( {{R_{{\rm{co}}}}} \right)}^2}} \right)}} \times 
\end{equation*}
\begin{equation*}
   \scalebox{.9}[1]{$\left( {\frac{{{\lambda _{\rm{U}}}\left( {\left( {\frac{{2{{\left( {{R_{{\rm{co}}}}} \right)}^2}}}{\pi } - \frac{{{b^2}}}{\pi }} \right)\arcsin \left( {\frac{b}{{2{R_{{\rm{co}}}}}}} \right) - \frac{{{{\left( {{R_{{\rm{co}}}}} \right)}^2}}}{2}\sin \left( {4\arcsin \left( {\frac{b}{{2{R_{{\rm{co}}}}}}} \right)} \right) + \frac{{{b^2}}}{2}\left( {1 - \sin \left( {\pi  - 2\arcsin \left( {\frac{b}{{2{R_{{\rm{co}}}}}}} \right)} \right)} \right)} \right)}}{{1 - \exp \left\{ { - {\lambda _{\rm{U}}}\left( {\left( {\frac{{2{{\left( {{R_{{\rm{co}}}}} \right)}^2}}}{\pi } - \frac{{{b^2}}}{\pi }} \right)\arcsin \left( {\frac{b}{{2{R_{{\rm{co}}}}}}} \right) - \frac{{{{\left( {{R_{{\rm{co}}}}} \right)}^2}}}{2}\sin \left( {4\arcsin \left( {\frac{b}{{2{R_{{\rm{co}}}}}}} \right)} \right) + \frac{{{b^2}}}{2}\left( {1 - \sin \left( {\pi  - 2\arcsin \left( {\frac{b}{{2{R_{{\rm{co}}}}}}} \right)} \right)} \right)} \right)} \right\}}} + } \right.$}
\end{equation*}
\begin{equation*}
   \scalebox{.9}[1]{$\left. {\frac{{\exp \left\{ { - {\lambda _{\rm{U}}}\left( {\left( {\frac{{2{{\left( {{R_{{\rm{co}}}}} \right)}^2}}}{\pi } - \frac{{{b^2}}}{\pi }} \right)\arcsin \left( {\frac{b}{{2{R_{{\rm{co}}}}}}} \right) - \frac{{{{\left( {{R_{{\rm{co}}}}} \right)}^2}}}{2}\sin \left( {4\arcsin \left( {\frac{b}{{2{R_{{\rm{co}}}}}}} \right)} \right) + \frac{{{b^2}}}{2}\left( {1 - \sin \left( {\pi  - 2\arcsin \left( {\frac{b}{{2{R_{{\rm{co}}}}}}} \right)} \right)} \right)} \right)} \right\} - 1}}{{1 - \exp \left\{ { - {\lambda _{\rm{U}}}\left( {\left( {\frac{{2{{\left( {{R_{{\rm{co}}}}} \right)}^2}}}{\pi } - \frac{{{b^2}}}{\pi }} \right)\arcsin \left( {\frac{b}{{2{R_{{\rm{co}}}}}}} \right) - \frac{{{{\left( {{R_{{\rm{co}}}}} \right)}^2}}}{2}\sin \left( {4\arcsin \left( {\frac{b}{{2{R_{{\rm{co}}}}}}} \right)} \right) + \frac{{{b^2}}}{2}\left( {1 - \sin \left( {\pi  - 2\arcsin \left( {\frac{b}{{2{R_{{\rm{co}}}}}}} \right)} \right)} \right)} \right)} \right\}}}} \right)$}.
\end{equation*}
To bound the expectation of this term,
 consider $W$ as a poisson random variable with parameter $\lambda$, whose distribution is $P_{\lambda}(w)=\frac{{{\lambda ^w}{e^{ - \lambda }}}}{{w!}}$, then, we have:
\begin{equation}
E\left\{ {\frac{1}{{W + 1}}} \right\} = \sum\limits_{k = 0}^\infty  {\frac{{{\lambda ^k}{e^{ - \lambda }}}}{{\left( {k + 1} \right)!}}}  = \sum\limits_{k = 1}^\infty  {\frac{{{\lambda ^u}{e^{ - \lambda }}}}{{\lambda u!}}}  = \frac{{{e^{ - \lambda }}\left( {{e^\lambda } - 1} \right)}}{\lambda } = \frac{{1 - {e^{ - \lambda }}}}{\lambda }.
\label{t55}
\end{equation}
Therefore, we obtain:
\begin{equation*}
E\left\{ {\frac{{{\pmb{\mathbb{I}}}\left( {{N_{{\rm{BS}}}}\left( {{\cal C}\left( {{{\mathbf{x}}_{\rm{U}}},b} \right)} \right)} \right)\left( {1 + \left( {\hat h - 1} \right){p_{{\rm{b}}}}} \right){{\left( {{\lambda _{{\rm{wave}}}}} \right)}^2}{\delta ^2} \sigma _{{\rm{d}}}^2}}{{\left( {1 + {N_{{\rm{BS}}}}\left( {{\cal C}\left( {{{\mathbf{x}}_{\rm{U}}},2{R_{{\rm{co}}}}} \right) - {\cal C}\left( {{{\mathbf{x}}_{\rm{U}}},b} \right)} \right)} \right){{\left( {4\pi } \right)}^2}\left( {{{({h_{{\rm{BS}}}})}^2} + {{\left( {{R_{{\rm{co}}}}} \right)}^2}} \right)}} \times } \right.
\end{equation*}
\begin{equation*}
   \scalebox{.85}[1]{$\left( {\frac{{{\lambda _{\rm{U}}}\left( {\left( {\frac{{2{{\left( {{R_{{\rm{co}}}}} \right)}^2}}}{\pi } - \frac{{{b^2}}}{\pi }} \right)\arcsin \left( {\frac{b}{{2{R_{{\rm{co}}}}}}} \right) - \frac{{{{\left( {{R_{{\rm{co}}}}} \right)}^2}}}{2}\sin \left( {4\arcsin \left( {\frac{b}{{2{R_{{\rm{co}}}}}}} \right)} \right) + \frac{{{b^2}}}{2}\left( {1 - \sin \left( {\pi  - 2\arcsin \left( {\frac{b}{{2{R_{{\rm{co}}}}}}} \right)} \right)} \right)} \right)}}{{1 - \exp \left\{ { - {\lambda _{\rm{U}}}\left( {\left( {\frac{{2{{\left( {{R_{{\rm{co}}}}} \right)}^2}}}{\pi } - \frac{{{b^2}}}{\pi }} \right)\arcsin \left( {\frac{b}{{2{R_{{\rm{co}}}}}}} \right) - \frac{{{{\left( {{R_{{\rm{co}}}}} \right)}^2}}}{2}\sin \left( {4\arcsin \left( {\frac{b}{{2{R_{{\rm{co}}}}}}} \right)} \right) + \frac{{{b^2}}}{2}\left( {1 - \sin \left( {\pi  - 2\arcsin \left( {\frac{b}{{2{R_{{\rm{co}}}}}}} \right)} \right)} \right)} \right)} \right\}}} + } \right.$}
\end{equation*}
\begin{equation*}
   \scalebox{.85}[1]{$\left. {\left. {\frac{{\exp \left\{ { - {\lambda _{\rm{U}}}\left( {\left( {\frac{{2{{\left( {{R_{{\rm{co}}}}} \right)}^2}}}{\pi } - \frac{{{b^2}}}{\pi }} \right)\arcsin \left( {\frac{b}{{2{R_{{\rm{co}}}}}}} \right) - \frac{{{{\left( {{R_{{\rm{co}}}}} \right)}^2}}}{2}\sin \left( {4\arcsin \left( {\frac{b}{{2{R_{{\rm{co}}}}}}} \right)} \right) + \frac{{{b^2}}}{2}\left( {1 - \sin \left( {\pi  - 2\arcsin \left( {\frac{b}{{2{R_{{\rm{co}}}}}}} \right)} \right)} \right)} \right)} \right\} - 1}}{{1 - \exp \left\{ { - {\lambda _{\rm{U}}}\left( {\left( {\frac{{2{{\left( {{R_{{\rm{co}}}}} \right)}^2}}}{\pi } - \frac{{{b^2}}}{\pi }} \right)\arcsin \left( {\frac{b}{{2{R_{{\rm{co}}}}}}} \right) - \frac{{{{\left( {{R_{{\rm{co}}}}} \right)}^2}}}{2}\sin \left( {4\arcsin \left( {\frac{b}{{2{R_{{\rm{co}}}}}}} \right)} \right) + \frac{{{b^2}}}{2}\left( {1 - \sin \left( {\pi  - 2\arcsin \left( {\frac{b}{{2{R_{{\rm{co}}}}}}} \right)} \right)} \right)} \right)} \right\}}}} \right)} \right\} = $}
\end{equation*}
\begin{equation*}
   \scalebox{.85}[1]{$\frac{{\left( {1 - \exp \left( { - {\lambda _{{\rm{BS}}}}\pi \left( {4{{\left( {{R_{{\rm{co}}}}} \right)}^2} - {b^2}} \right)} \right)} \right)\left( {1 - \exp \left( { - {\lambda _{{\rm{BS}}}}\pi {b^2}} \right)} \right)\left( {1 + \left( {\hat h - 1} \right){p_{{\rm{b}}}}} \right){{\left( {{\lambda _{{\rm{wave}}}}} \right)}^2}{\delta ^2} \sigma _{{\rm{d}}}^2}}{{{\lambda _{{\rm{BS}}}}\pi \left( {4{{\left( {{R_{{\rm{co}}}}} \right)}^2} - {b^2}} \right){{\left( {4\pi } \right)}^2}\left( {{{({h_{{\rm{BS}}}})}^2} + {{\left( {{R_{{\rm{co}}}}} \right)}^2}} \right)}} \times $}
\end{equation*}
\begin{equation*}
   \scalebox{.85}[1]{$\left( {\frac{{{\lambda _{\rm{U}}}\left( {\left( {\frac{{2{{\left( {{R_{{\rm{co}}}}} \right)}^2}}}{\pi } - \frac{{{b^2}}}{\pi }} \right)\arcsin \left( {\frac{b}{{2{R_{{\rm{co}}}}}}} \right) - \frac{{{{\left( {{R_{{\rm{co}}}}} \right)}^2}}}{2}\sin \left( {4\arcsin \left( {\frac{b}{{2{R_{{\rm{co}}}}}}} \right)} \right) + \frac{{{b^2}}}{2}\left( {1 - \sin \left( {\pi  - 2\arcsin \left( {\frac{b}{{2{R_{{\rm{co}}}}}}} \right)} \right)} \right)} \right)}}{{1 - \exp \left\{ { - {\lambda _{\rm{U}}}\left( {\left( {\frac{{2{{\left( {{R_{{\rm{co}}}}} \right)}^2}}}{\pi } - \frac{{{b^2}}}{\pi }} \right)\arcsin \left( {\frac{b}{{2{R_{{\rm{co}}}}}}} \right) - \frac{{{{\left( {{R_{{\rm{co}}}}} \right)}^2}}}{2}\sin \left( {4\arcsin \left( {\frac{b}{{2{R_{{\rm{co}}}}}}} \right)} \right) + \frac{{{b^2}}}{2}\left( {1 - \sin \left( {\pi  - 2\arcsin \left( {\frac{b}{{2{R_{{\rm{co}}}}}}} \right)} \right)} \right)} \right)} \right\}}} + } \right.$}
\end{equation*}
\begin{equation}
   \scalebox{.85}[1]{$\left. {\frac{{\exp \left\{ { - {\lambda _{\rm{U}}}\left( {\left( {\frac{{2{{\left( {{R_{{\rm{co}}}}} \right)}^2}}}{\pi } - \frac{{{b^2}}}{\pi }} \right)\arcsin \left( {\frac{b}{{2{R_{{\rm{co}}}}}}} \right) - \frac{{{{\left( {{R_{{\rm{co}}}}} \right)}^2}}}{2}\sin \left( {4\arcsin \left( {\frac{b}{{2{R_{{\rm{co}}}}}}} \right)} \right) + \frac{{{b^2}}}{2}\left( {1 - \sin \left( {\pi  - 2\arcsin \left( {\frac{b}{{2{R_{{\rm{co}}}}}}} \right)} \right)} \right)} \right)} \right\} - 1}}{{1 - \exp \left\{ { - {\lambda _{\rm{U}}}\left( {\left( {\frac{{2{{\left( {{R_{{\rm{co}}}}} \right)}^2}}}{\pi } - \frac{{{b^2}}}{\pi }} \right)\arcsin \left( {\frac{b}{{2{R_{{\rm{co}}}}}}} \right) - \frac{{{{\left( {{R_{{\rm{co}}}}} \right)}^2}}}{2}\sin \left( {4\arcsin \left( {\frac{b}{{2{R_{{\rm{co}}}}}}} \right)} \right) + \frac{{{b^2}}}{2}\left( {1 - \sin \left( {\pi  - 2\arcsin \left( {\frac{b}{{2{R_{{\rm{co}}}}}}} \right)} \right)} \right)} \right)} \right\}}}} \right).$}
\label{t67}
\end{equation}

To bound term (\ref{t42}), we have:
\begin{equation*}
\frac{\kappa }{{\kappa  + 1}} \times\sum\limits_{{{\mathbf{x}}_{{\rm{IRS}}}} \in {\Psi _{{\rm{IRS}}}}({{\mathbf{x}}_{\rm{U}}})} {\begin{array}{*{20}{c}}
  {}&{} 
\end{array}\sum\limits_{{{{\mathbf{x'}}}_{{\rm{IRS}}}} \in {\Psi _{{\rm{IRS}}}}({{\mathbf{x}}_{\rm{U}}})} {\begin{array}{*{20}{c}}
  {}&{} 
\end{array}\sum\limits_{{{{\mathbf{x'}}}_{\rm{U}}} \in {\Psi _{\rm{U}}}({{{\mathbf{\tilde x}}}_{{\rm{BS}}}}({{\mathbf{x}}_{\rm{U}}})):{{{\mathbf{x'}}}_{\rm{U}}} \ne {{\mathbf{x}}_{\rm{U}}},{{{\mathbf{\tilde x}}}_{{\rm{BS}}}}({{{\mathbf{x'}}}_{\rm{U}}}) = {{{\mathbf{\tilde x}}}_{{\rm{BS}}}}({{\mathbf{x}}_{\rm{U}}})} {} } } 
\end{equation*}
\begin{equation*}
   \scalebox{.9}[1]{$\frac{{\left( {1 - {\pmb{\mathbb{I}}}\left( {\left\| {{{{\mathbf{\tilde x}}}_{\rm{U}}}({{\mathbf{x}}_{{\rm{IRS}}}}) - {{\mathbf{x}}_{\rm{U}}}} \right\|} \right)} \right)\left( {1 - {\pmb{\mathbb{I}}}\left( {\left\| {{{{\mathbf{\tilde x}}}_{\rm{U}}}({{{\mathbf{x'}}}_{{\rm{IRS}}}}) - {{\mathbf{x}}_{\rm{U}}}} \right\|} \right)} \right)Q(Q - 1)\left( {1 - {\delta _{{{\mathbf{x}}_{{\rm{IRS}}}} - {{\mathbf{x}}^\prime }_{{\rm{IRS}}}}}} \right){{\left( {{\lambda _{{\rm{wave}}}}} \right)}^4}{{\left( {\frac{\pi }{4}} \right)}^2}{\delta ^2} \sigma _{{\rm{d}}}^2}}{{{{\left( {4\pi } \right)}^4}\sqrt {{{({h_{{\rm{IRS}}}})}^2} + \left\| {{{\mathbf{x}}_{{\rm{IRS}}}} - {{\mathbf{x}}_{\rm{U}}}} \right\|_2^2}  \times \sqrt {{{({h_{{\rm{IRS}}}})}^2} + \left\| {{{{\mathbf{x'}}}_{{\rm{IRS}}}} - {{\mathbf{x}}_{\rm{U}}}} \right\|_2^2}  \times \sqrt {{{({h_{{\rm{BS}}}} - {h_{{\rm{IRS}}}})}^2} + \left\| {{{{\mathbf{\tilde x}}}_{{\rm{BS}}}}({{\mathbf{x}}_{\rm{U}}}) - {{\mathbf{x}}_{{\rm{IRS}}}}} \right\|_2^2}  \times \sqrt {{{({h_{{\rm{BS}}}} - {h_{{\rm{IRS}}}})}^2} + \left\| {{{{\mathbf{\tilde x}}}_{{\rm{BS}}}}({{\mathbf{x}}_{\rm{U}}}) - {{{\mathbf{x'}}}_{{\rm{IRS}}}}} \right\|_2^2} }} = $}
\end{equation*}
\begin{equation*}
\frac{\kappa }{{\kappa  + 1}} \times\sum\limits_{{{\mathbf{x}}_{{\rm{IRS}}}} \in {\Psi _{{\rm{IRS}}}}({{\mathbf{x}}_{\rm{U}}})} {\begin{array}{*{20}{c}}
  {}&{} 
\end{array}\sum\limits_{{{{\mathbf{x'}}}_{{\rm{IRS}}}} \in {\Psi _{{\rm{IRS}}}}({{\mathbf{x}}_{\rm{U}}})} {\begin{array}{*{20}{c}}
  {}&{} 
\end{array}\sum\limits_{{{{\mathbf{x'}}}_{\rm{U}}} \in {\Psi _{\rm{U}}}({{{\mathbf{\tilde x}}}_{{\rm{BS}}}}({{\mathbf{x}}_{\rm{U}}})):{{{\mathbf{x'}}}_{\rm{U}}} \ne {{\mathbf{x}}_{\rm{U}}}} {} } } 
\end{equation*}
\begin{equation*}
   \scalebox{.9}[1]{$\frac{{\left( {1 - {\pmb{\mathbb{I}}}\left( {\left\| {{{{\mathbf{\tilde x}}}_{\rm{U}}}({{\mathbf{x}}_{{\rm{IRS}}}}) - {{\mathbf{x}}_{\rm{U}}}} \right\|} \right)} \right)\left( {1 - {\pmb{\mathbb{I}}}\left( {\left\| {{{{\mathbf{\tilde x}}}_{\rm{U}}}({{{\mathbf{x'}}}_{{\rm{IRS}}}}) - {{\mathbf{x}}_{\rm{U}}}} \right\|} \right)} \right)\left( {1 - {\pmb{\mathbb{I}}}\left( {\left\| {{{{\mathbf{\tilde x}}}_{{\rm{BS}}}}({{{\mathbf{x'}}}_{\rm{U}}}) - {{{\mathbf{\tilde x}}}_{{\rm{BS}}}}({{\mathbf{x}}_{\rm{U}}})} \right\|} \right)} \right)Q(Q - 1)\left( {1 - {\delta _{{{\mathbf{x}}_{{\rm{IRS}}}} - {{\mathbf{x}}^\prime }_{{\rm{IRS}}}}}} \right){{\left( {{\lambda _{{\rm{wave}}}}} \right)}^4}{{\left( {\frac{\pi }{4}} \right)}^2}{\delta ^2} \sigma _{{\rm{d}}}^2}}{{{{\left( {4\pi } \right)}^4}\sqrt {{{({h_{{\rm{IRS}}}})}^2} + \left\| {{{\mathbf{x}}_{{\rm{IRS}}}} - {{\mathbf{x}}_{\rm{U}}}} \right\|_2^2}  \times \sqrt {{{({h_{{\rm{IRS}}}})}^2} + \left\| {{{{\mathbf{x'}}}_{{\rm{IRS}}}} - {{\mathbf{x}}_{\rm{U}}}} \right\|_2^2}  \times \sqrt {{{({h_{{\rm{BS}}}} - {h_{{\rm{IRS}}}})}^2} + \left\| {{{{\mathbf{\tilde x}}}_{{\rm{BS}}}}({{\mathbf{x}}_{\rm{U}}}) - {{\mathbf{x}}_{{\rm{IRS}}}}} \right\|_2^2}  \times \sqrt {{{({h_{{\rm{BS}}}} - {h_{{\rm{IRS}}}})}^2} + \left\| {{{{\mathbf{\tilde x}}}_{{\rm{BS}}}}({{\mathbf{x}}_{\rm{U}}}) - {{{\mathbf{x'}}}_{{\rm{IRS}}}}} \right\|_2^2} }} \ge$}
\end{equation*}
\begin{equation*}
\frac{\kappa }{{\kappa  + 1}} \times\sum\limits_{{{\mathbf{x}}_{{\rm{IRS}}}} \in {\Psi _{{\rm{IRS}}}}({{\mathbf{x}}_{\rm{U}}})} {\begin{array}{*{20}{c}}
  {}&{} 
\end{array}\sum\limits_{{{{\mathbf{x'}}}_{{\rm{IRS}}}} \in {\Psi _{{\rm{IRS}}}}({{\mathbf{x}}_{\rm{U}}})} {\begin{array}{*{20}{c}}
  {}&{} 
\end{array}\sum\limits_{{{{\mathbf{x'}}}_{\rm{U}}} \in {\Psi _{\rm{U}}}({{{\mathbf{\tilde x}}}_{{\rm{BS}}}}({{\mathbf{x}}_{\rm{U}}})):{{{\mathbf{x'}}}_{\rm{U}}} \ne {{\mathbf{x}}_{\rm{U}}}} {} } } 
\end{equation*}
\begin{equation*}
   \scalebox{.9}[1]{$\frac{{\left( {1 - {\pmb{\mathbb{I}}}\left( {\left\| {{{{\mathbf{\tilde x}}}_{\rm{U}}}({{\mathbf{x}}_{{\rm{IRS}}}}) - {{\mathbf{x}}_{\rm{U}}}} \right\|} \right)} \right)\left( {1 - {\pmb{\mathbb{I}}}\left( {\left\| {{{{\mathbf{\tilde x}}}_{\rm{U}}}({{{\mathbf{x'}}}_{{\rm{IRS}}}}) - {{\mathbf{x}}_{\rm{U}}}} \right\|} \right)} \right)\left( {1 - {\pmb{\mathbb{I}}}\left( {\left\| {{{{\mathbf{\tilde x}}}_{{\rm{BS}}}}({{{\mathbf{x'}}}_{\rm{U}}}) - {{{\mathbf{\tilde x}}}_{{\rm{BS}}}}({{\mathbf{x}}_{\rm{U}}})} \right\|} \right)} \right)Q(Q - 1)\left( {1 - {\delta _{{{\mathbf{x}}_{{\rm{IRS}}}} - {{\mathbf{x}}^\prime }_{{\rm{IRS}}}}}} \right){{\left( {{\lambda _{{\rm{wave}}}}} \right)}^4}{{\left( {\frac{\pi }{4}} \right)}^2}{\delta ^2} \sigma _{{\rm{d}}}^2}}{{{{\left( {4\pi } \right)}^4}\left( {\max \left\{ {{{({h_{{\rm{IRS}}}})}^2},{{({h_{{\rm{BS}}}} - {h_{{\rm{IRS}}}})}^2}} \right\} + 3{{\left( {{R_{{\rm{co}}}}} \right)}^2} + \left\| {{{\mathbf{x}}_{{\rm{IRS}}}} - {{\mathbf{x}}_{\rm{U}}}} \right\|_2^2} \right) \times \left( {\max \left\{ {{{({h_{{\rm{IRS}}}})}^2},{{({h_{{\rm{BS}}}} - {h_{{\rm{IRS}}}})}^2}} \right\} + 3{{\left( {{R_{{\rm{co}}}}} \right)}^2} + \left\| {{{{\mathbf{x'}}}_{{\rm{IRS}}}} - {{\mathbf{x}}_{\rm{U}}}} \right\|_2^2} \right)}},$}
\end{equation*}
thus, we obtain:
\begin{equation*}
E\left\{\frac{\kappa }{{\kappa  + 1}} \times {\sum\limits_{{{\mathbf{x}}_{{\rm{IRS}}}} \in {\Psi _{{\rm{IRS}}}}({{\mathbf{x}}_{\rm{U}}})} {\begin{array}{*{20}{c}}
  {}&{} 
\end{array}\sum\limits_{{{{\mathbf{x'}}}_{{\rm{IRS}}}} \in {\Psi _{{\rm{IRS}}}}({{\mathbf{x}}_{\rm{U}}})} {\begin{array}{*{20}{c}}
  {}&{} 
\end{array}\sum\limits_{{{{\mathbf{x'}}}_{\rm{U}}} \in {\Psi _{\rm{U}}}({{{\mathbf{\tilde x}}}_{{\rm{BS}}}}({{\mathbf{x}}_{\rm{U}}})):{{{\mathbf{x'}}}_{\rm{U}}} \ne {{\mathbf{x}}_{\rm{U}}}} {} } } } \right.
\end{equation*}
\begin{equation*}
   \scalebox{.84}[1]{$\left. {\left. {\frac{{\left( {1 -{\pmb{\mathbb{I}}}\left( {\left\| {{{{\mathbf{\tilde x}}}_{\rm{U}}}({{\mathbf{x}}_{{\rm{IRS}}}}) - {{\mathbf{x}}_{\rm{U}}}} \right\|} \right)} \right)\left( {1 - {\pmb{\mathbb{I}}}\left( {\left\| {{{{\mathbf{\tilde x}}}_{\rm{U}}}({{{\mathbf{x'}}}_{{\rm{IRS}}}}) - {{\mathbf{x}}_{\rm{U}}}} \right\|} \right)} \right)\left( {1 - {\pmb{\mathbb{I}}}\left( {\left\| {{{{\mathbf{\tilde x}}}_{{\rm{BS}}}}({{{\mathbf{x'}}}_{\rm{U}}}) - {{{\mathbf{\tilde x}}}_{{\rm{BS}}}}({{\mathbf{x}}_{\rm{U}}})} \right\|} \right)} \right)Q(Q - 1)\left( {1 - {\delta _{{{\mathbf{x}}_{{\rm{IRS}}}} - {{\mathbf{x}}^\prime }_{{\rm{IRS}}}}}} \right){{\left( {{\lambda _{{\rm{wave}}}}} \right)}^4}{\delta ^2} \sigma _{{\rm{d}}}^2}}{{{4^6}{\pi ^2}\left( {\max \left\{ {{{({h_{{\rm{IRS}}}})}^2},{{({h_{{\rm{BS}}}} - {h_{{\rm{IRS}}}})}^2}} \right\} + 3{{\left( {{R_{{\rm{co}}}}} \right)}^2} + \left\| {{{\mathbf{x}}_{{\rm{IRS}}}} - {{\mathbf{x}}_{\rm{U}}}} \right\|_2^2} \right)\left( {\max \left\{ {{{({h_{{\rm{IRS}}}})}^2},{{({h_{{\rm{BS}}}} - {h_{{\rm{IRS}}}})}^2}} \right\} + 3{{\left( {{R_{{\rm{co}}}}} \right)}^2} + \left\| {{{{\mathbf{x'}}}_{{\rm{IRS}}}} - {{\mathbf{x}}_{\rm{U}}}} \right\|_2^2} \right)}}} \right|{\Phi _{\rm{U}}},{\Phi _{{\rm{BS}}}},{\Phi _{{\rm{IRS}}}}} \right\}$}
\end{equation*}
\begin{equation*}
 =\frac{\kappa }{{\kappa  + 1}} \times \sum\limits_{{{\mathbf{x}}_{{\rm{IRS}}}} \in {\Psi _{{\rm{IRS}}}}({{\mathbf{x}}_{\rm{U}}})} {\begin{array}{*{20}{c}}
  {}&{} 
\end{array}\sum\limits_{{{{\mathbf{x'}}}_{{\rm{IRS}}}} \in {\Psi _{{\rm{IRS}}}}({{\mathbf{x}}_{\rm{U}}})} {\begin{array}{*{20}{c}}
  {}&{} 
\end{array}\sum\limits_{{{{\mathbf{x'}}}_{\rm{U}}} \in {\Psi _{\rm{U}}}({{{\mathbf{\tilde x}}}_{{\rm{BS}}}}({{\mathbf{x}}_{\rm{U}}})):{{{\mathbf{x'}}}_{\rm{U}}} \ne {{\mathbf{x}}_{\rm{U}}}} {} } } 
\end{equation*}
\begin{equation*}
   \scalebox{.85}[1]{$\frac{{{\pmb{\mathbb{I}}}\left( {{n_{{\rm{BS}}}}({{\mathbf{x}}_{\rm{U}}})} \right)Q(Q - 1)\left( {1 - {\delta _{{{\mathbf{x}}_{{\rm{IRS}}}} - {{\mathbf{x}}^\prime }_{{\rm{IRS}}}}}} \right){{\left( {{\lambda _{{\rm{wave}}}}} \right)}^4}{\delta ^2} \sigma _{{\rm{d}}}^2}}{{{4^6}{\pi ^2}{n_{\rm{U}}}({{\mathbf{x}}_{{\rm{IRS}}}}){n_{\rm{U}}}({{{\mathbf{x'}}}_{{\rm{IRS}}}}){n_{{\rm{BS}}}}({{{\mathbf{x'}}}_{\rm{U}}})\left( {\max \left\{ {{{({h_{{\rm{IRS}}}})}^2},{{({h_{{\rm{BS}}}} - {h_{{\rm{IRS}}}})}^2}} \right\} + 3{{\left( {{R_{{\rm{co}}}}} \right)}^2} + \left\| {{{\mathbf{x}}_{{\rm{IRS}}}} - {{\mathbf{x}}_{\rm{U}}}} \right\|_2^2} \right)\left( {\max \left\{ {{{({h_{{\rm{IRS}}}})}^2},{{({h_{{\rm{BS}}}} - {h_{{\rm{IRS}}}})}^2}} \right\} + 3{{\left( {{R_{{\rm{co}}}}} \right)}^2} + \left\| {{{{\mathbf{x'}}}_{{\rm{IRS}}}} - {{\mathbf{x}}_{\rm{U}}}} \right\|_2^2} \right)}}$}
\end{equation*}
\begin{equation*}
 \ge\frac{\kappa }{{\kappa  + 1}} \times \sum\limits_{{{\mathbf{x}}_{{\rm{IRS}}}} \in {\Psi _{{\rm{IRS}}}}({{\mathbf{x}}_{\rm{U}}})} {\begin{array}{*{20}{c}}
  {}&{} 
\end{array}\sum\limits_{{{{\mathbf{x'}}}_{{\rm{IRS}}}} \in {\Psi _{{\rm{IRS}}}}({{\mathbf{x}}_{\rm{U}}})} {\begin{array}{*{20}{c}}
  {}&{} 
\end{array}} } 
\end{equation*}
\begin{equation*}
   \scalebox{.9}[1]{$\frac{{{\pmb{\mathbb{I}}}\left( {{n_{{\rm{BS}}}}({{\mathbf{x}}_{\rm{U}}})} \right)\left( {{n_{\rm{U}}}({{{\mathbf{\tilde x}}}_{{\rm{BS}}}}({{\mathbf{x}}_{\rm{U}}})) - 1} \right)Q(Q - 1)\left( {1 - {\delta _{{{\mathbf{x}}_{{\rm{IRS}}}} - {{\mathbf{x}}^\prime }_{{\rm{IRS}}}}}} \right){{\left( {{\lambda _{{\rm{wave}}}}} \right)}^4}{\delta ^2} \sigma _{{\rm{d}}}^2}}{{{4^6}{\pi ^2}{{\left( {{{\tilde n}_{\rm{U}}}({{\mathbf{x}}_{\rm{U}}})} \right)}^2}{{\tilde n}_{{\rm{BS}}}}({{\mathbf{x}}_{\rm{U}}})\left( {\max \left\{ {{{({h_{{\rm{IRS}}}})}^2},{{({h_{{\rm{BS}}}} - {h_{{\rm{IRS}}}})}^2}} \right\} + 3{{\left( {{R_{{\rm{co}}}}} \right)}^2} + \left\| {{{\mathbf{x}}_{{\rm{IRS}}}} - {{\mathbf{x}}_{\rm{U}}}} \right\|_2^2} \right)\left( {\max \left\{ {{{({h_{{\rm{IRS}}}})}^2},{{({h_{{\rm{BS}}}} - {h_{{\rm{IRS}}}})}^2}} \right\} + 3{{\left( {{R_{{\rm{co}}}}} \right)}^2} + \left\| {{{{\mathbf{x'}}}_{{\rm{IRS}}}} - {{\mathbf{x}}_{\rm{U}}}} \right\|_2^2} \right)}}.$}
\end{equation*}
\begin{equation*}
\ge \frac{\kappa }{{\kappa  + 1}} \times\sum\limits_{{{\mathbf{x}}_{{\rm{IRS}}}} \in {\Psi _{{\rm{IRS}}}}({{\mathbf{x}}_{\rm{U}}})} {\begin{array}{*{20}{c}}
  {}&{} 
\end{array}\sum\limits_{{{{\mathbf{x'}}}_{{\rm{IRS}}}} \in {\Psi _{{\rm{IRS}}}}({{\mathbf{x}}_{\rm{U}}})} {\begin{array}{*{20}{c}}
  {}&{} 
\end{array}} } 
\end{equation*}
\begin{equation*}
   \scalebox{1}[1]{$\frac{{{\pmb{\mathbb{I}}}\left( {{n_{{\rm{BS}}}}({{\mathbf{x}}_{\rm{U}}})} \right)\left( {{N_{\rm{U}}}\left( {{\cal C}\left( {{{\mathbf{x}}_{\rm{U}}},\min \left\{ {{R_{{\rm{co}}}} - \left\| {{{{\mathbf{\tilde x}}}_{{\rm{BS}}}}({{\mathbf{x}}_{\rm{U}}}) - {{\mathbf{x}}_{\rm{U}}}} \right\|,\left\| {{{{\mathbf{\tilde x}}}_{{\rm{BS}}}}({{\mathbf{x}}_{\rm{U}}}) - {{\mathbf{x}}_{\rm{U}}}} \right\|} \right\}} \right)} \right) - 1} \right)Q(Q - 1)\left( {1 - {\delta _{{{\mathbf{x}}_{{\rm{IRS}}}} - {{\mathbf{x}}^\prime }_{{\rm{IRS}}}}}} \right)}}{{{4^6}{\pi ^2}{{\left( {{N_{\rm{U}}}\left( {{\cal C}\left( {{{\mathbf{x}}_{\rm{U}}},2{R_{{\rm{co}}}}} \right) - {\cal C}\left( {{{\mathbf{x}}_{\rm{U}}},\min \left\{ {{R_{{\rm{co}}}} - \left\| {{{{\mathbf{\tilde x}}}_{{\rm{BS}}}}({{\mathbf{x}}_{\rm{U}}}) - {{\mathbf{x}}_{\rm{U}}}} \right\|,\left\| {{{{\mathbf{\tilde x}}}_{{\rm{BS}}}}({{\mathbf{x}}_{\rm{U}}}) - {{\mathbf{x}}_{\rm{U}}}} \right\|} \right\}} \right)} \right) + 1} \right)}^2}\left( {{N_{{\rm{BS}}}}\left( {{\cal C}\left( {{{\mathbf{x}}_{\rm{U}}},2{R_{{\rm{co}}}}} \right) - {\cal C}\left( {{{\mathbf{x}}_{\rm{U}}},{R_{{\rm{co}}}}} \right)} \right) + 1} \right)}} \times $}
\end{equation*}
\begin{equation*}
   \scalebox{1}[1]{$\frac{{{{\left( {{\lambda _{{\rm{wave}}}}} \right)}^4}{\delta ^2} \sigma _{{\rm{d}}}^2}}{{\left( {\max \left\{ {{{({h_{{\rm{IRS}}}})}^2},{{({h_{{\rm{BS}}}} - {h_{{\rm{IRS}}}})}^2}} \right\} + 3{{\left( {{R_{{\rm{co}}}}} \right)}^2} + \left\| {{{\mathbf{x}}_{{\rm{IRS}}}} - {{\mathbf{x}}_{\rm{U}}}} \right\|_2^2} \right)\left( {\max \left\{ {{{({h_{{\rm{IRS}}}})}^2},{{({h_{{\rm{BS}}}} - {h_{{\rm{IRS}}}})}^2}} \right\} + 3{{\left( {{R_{{\rm{co}}}}} \right)}^2} + \left\| {{{{\mathbf{x'}}}_{{\rm{IRS}}}} - {{\mathbf{x}}_{\rm{U}}}} \right\|_2^2} \right)}}.$}
\end{equation*}
Then, we have:
\begin{equation*}
E\left\{\frac{\kappa }{{\kappa  + 1}} \times {\sum\limits_{{{\mathbf{x}}_{{\rm{IRS}}}} \in {\Psi _{{\rm{IRS}}}}({{\mathbf{x}}_{\rm{U}}})} {\begin{array}{*{20}{c}}
  {}&{} 
\end{array}\sum\limits_{{{{\mathbf{x'}}}_{{\rm{IRS}}}} \in {\Psi _{{\rm{IRS}}}}({{\mathbf{x}}_{\rm{U}}})} {\begin{array}{*{20}{c}}
  {}&{} 
\end{array}} } } \right.
\end{equation*}
\begin{equation*}
   \scalebox{.9}[1]{$\frac{{{\pmb{\mathbb{I}}}\left( {{n_{{\rm{BS}}}}({{\mathbf{x}}_{\rm{U}}})} \right)\left( {{N_{\rm{U}}}\left( {{\cal C}\left( {{{\mathbf{x}}_{\rm{U}}},\min \left\{ {{R_{{\rm{co}}}} - \left\| {{{{\mathbf{\tilde x}}}_{{\rm{BS}}}}({{\mathbf{x}}_{\rm{U}}}) - {{\mathbf{x}}_{\rm{U}}}} \right\|,\left\| {{{{\mathbf{\tilde x}}}_{{\rm{BS}}}}({{\mathbf{x}}_{\rm{U}}}) - {{\mathbf{x}}_{\rm{U}}}} \right\|} \right\}} \right)} \right) - 1} \right)Q(Q - 1)\left( {1 - {\delta _{{{\mathbf{x}}_{{\rm{IRS}}}} - {{\mathbf{x}}^\prime }_{{\rm{IRS}}}}}} \right)}}{{{4^6}{\pi ^2}{{\left( {{N_{\rm{U}}}\left( {{\cal C}\left( {{{\mathbf{x}}_{\rm{U}}},2{R_{{\rm{co}}}}} \right) - {\cal C}\left( {{{\mathbf{x}}_{\rm{U}}},\min \left\{ {{R_{{\rm{co}}}} - \left\| {{{{\mathbf{\tilde x}}}_{{\rm{BS}}}}({{\mathbf{x}}_{\rm{U}}}) - {{\mathbf{x}}_{\rm{U}}}} \right\|,\left\| {{{{\mathbf{\tilde x}}}_{{\rm{BS}}}}({{\mathbf{x}}_{\rm{U}}}) - {{\mathbf{x}}_{\rm{U}}}} \right\|} \right\}} \right)} \right) + 1} \right)}^2}\left( {{N_{{\rm{BS}}}}\left( {{\cal C}\left( {{{\mathbf{x}}_{\rm{U}}},2{R_{{\rm{co}}}}} \right) - {\cal C}\left( {{{\mathbf{x}}_{\rm{U}}},{R_{{\rm{co}}}}} \right)} \right) + 1} \right)}} \times $}
\end{equation*}
\begin{equation*}
   \scalebox{.9}[1]{$\left. {\left. {\frac{{{{\left( {{\lambda _{{\rm{wave}}}}} \right)}^4}{\delta ^2} \sigma _{{\rm{d}}}^2}}{{\left( {\max \left\{ {{{({h_{{\rm{IRS}}}})}^2},{{({h_{{\rm{BS}}}} - {h_{{\rm{IRS}}}})}^2}} \right\} + 3{{\left( {{R_{{\rm{co}}}}} \right)}^2} + \left\| {{{\mathbf{x}}_{{\rm{IRS}}}} - {{\mathbf{x}}_{\rm{U}}}} \right\|_2^2} \right)\left( {\max \left\{ {{{({h_{{\rm{IRS}}}})}^2},{{({h_{{\rm{BS}}}} - {h_{{\rm{IRS}}}})}^2}} \right\} + 3{{\left( {{R_{{\rm{co}}}}} \right)}^2} + \left\| {{{{\mathbf{x'}}}_{{\rm{IRS}}}} - {{\mathbf{x}}_{\rm{U}}}} \right\|_2^2} \right)}}} \right|{\Phi _{{\rm{IRS}}}},{\Phi _{{\rm{BS}}}}} \right\} = $}
\end{equation*}
\begin{equation*}
\frac{\kappa }{{\kappa  + 1}} \times\sum\limits_{{{\mathbf{x}}_{{\rm{IRS}}}} \in {\Psi _{{\rm{IRS}}}}({{\mathbf{x}}_{\rm{U}}})} {\begin{array}{*{20}{c}}
  {}&{} 
\end{array}\sum\limits_{{{{\mathbf{x'}}}_{{\rm{IRS}}}} \in {\Psi _{{\rm{IRS}}}}({{\mathbf{x}}_{\rm{U}}})} {\begin{array}{*{20}{c}}
  {}&{} 
\end{array}} } 
\end{equation*}
\begin{equation*}
   \scalebox{1}[1]{$\frac{{{\pmb{\mathbb{I}}}\left( {{n_{{\rm{BS}}}}({{\mathbf{x}}_{\rm{U}}})} \right)Q(Q - 1)\left( {1 - {\delta _{{{\mathbf{x}}_{{\rm{IRS}}}} - {{\mathbf{x}}^\prime }_{{\rm{IRS}}}}}} \right){{\left( {{\lambda _{{\rm{wave}}}}} \right)}^4}{\delta ^2} \sigma _{{\rm{d}}}^2}}{{{4^6}{\pi ^2}\left( {\max \left\{ {{{({h_{{\rm{IRS}}}})}^2},{{({h_{{\rm{BS}}}} - {h_{{\rm{IRS}}}})}^2}} \right\} + 3{{\left( {{R_{{\rm{co}}}}} \right)}^2} + \left\| {{{\mathbf{x}}_{{\rm{IRS}}}} - {{\mathbf{x}}_{\rm{U}}}} \right\|_2^2} \right)\left( {\max \left\{ {{{({h_{{\rm{IRS}}}})}^2},{{({h_{{\rm{BS}}}} - {h_{{\rm{IRS}}}})}^2}} \right\} + 3{{\left( {{R_{{\rm{co}}}}} \right)}^2} + \left\| {{{{\mathbf{x'}}}_{{\rm{IRS}}}} - {{\mathbf{x}}_{\rm{U}}}} \right\|_2^2} \right)}} \times $}
\end{equation*}
\begin{equation*}
   \scalebox{.94}[1]{$\frac{{\left( {{\lambda _{\rm{U}}}\pi {{\left( {\min \left\{ {{R_{{\rm{co}}}} - \left\| {{{{\mathbf{\tilde x}}}_{{\rm{BS}}}}({{\mathbf{x}}_{\rm{U}}}) - {{\mathbf{x}}_{\rm{U}}}} \right\|,\left\| {{{{\mathbf{\tilde x}}}_{{\rm{BS}}}}({{\mathbf{x}}_{\rm{U}}}) - {{\mathbf{x}}_{\rm{U}}}} \right\|} \right\}} \right)}^2} + \exp \left\{ { - {\lambda _{\rm{U}}}\pi {{\left( {\min \left\{ {{R_{{\rm{co}}}} - \left\| {{{{\mathbf{\tilde x}}}_{{\rm{BS}}}}({{\mathbf{x}}_{\rm{U}}}) - {{\mathbf{x}}_{\rm{U}}}} \right\|,\left\| {{{{\mathbf{\tilde x}}}_{{\rm{BS}}}}({{\mathbf{x}}_{\rm{U}}}) - {{\mathbf{x}}_{\rm{U}}}} \right\|} \right\}} \right)}^2}} \right\} - 1} \right)}}{{\left( {1 - \exp \left\{ { - {\lambda _{\rm{U}}}\pi {{\left( {\min \left\{ {{R_{{\rm{co}}}} - \left\| {{{{\mathbf{\tilde x}}}_{{\rm{BS}}}}({{\mathbf{x}}_{\rm{U}}}) - {{\mathbf{x}}_{\rm{U}}}} \right\|,\left\| {{{{\mathbf{\tilde x}}}_{{\rm{BS}}}}({{\mathbf{x}}_{\rm{U}}}) - {{\mathbf{x}}_{\rm{U}}}} \right\|} \right\}} \right)}^2}} \right\}} \right)\left( {{N_{{\rm{BS}}}}\left( {{\cal C}\left( {{{\mathbf{x}}_{\rm{U}}},2{R_{{\rm{co}}}}} \right) - {\cal C}\left( {{{\mathbf{x}}_{\rm{U}}},{R_{{\rm{co}}}}} \right)} \right) + 1} \right)}} \times $}
\end{equation*}
\begin{equation*}
   \scalebox{1}[1]{$\left( {\frac{{1 - \exp \left\{ { - {\lambda _{\rm{U}}}\pi \left( {4{{\left( {{R_{{\rm{co}}}}} \right)}^2} - {{\left( {\min \left\{ {{R_{{\rm{co}}}} - \left\| {{{{\mathbf{\tilde x}}}_{{\rm{BS}}}}({{\mathbf{x}}_{\rm{U}}}) - {{\mathbf{x}}_{\rm{U}}}} \right\|,\left\| {{{{\mathbf{\tilde x}}}_{{\rm{BS}}}}({{\mathbf{x}}_{\rm{U}}}) - {{\mathbf{x}}_{\rm{U}}}} \right\|} \right\}} \right)}^2}} \right)} \right\}}}{{{{\left( {{\lambda _{\rm{U}}}\pi \left( {4{{\left( {{R_{{\rm{co}}}}} \right)}^2} - {{\left( {\min \left\{ {{R_{{\rm{co}}}} - \left\| {{{{\mathbf{\tilde x}}}_{{\rm{BS}}}}({{\mathbf{x}}_{\rm{U}}}) - {{\mathbf{x}}_{\rm{U}}}} \right\|,\left\| {{{{\mathbf{\tilde x}}}_{{\rm{BS}}}}({{\mathbf{x}}_{\rm{U}}}) - {{\mathbf{x}}_{\rm{U}}}} \right\|} \right\}} \right)}^2}} \right)} \right)}^2}}} - } \right.$}
\end{equation*}
\begin{equation*}
   \scalebox{.8}[1]{$\left. {\frac{{{\lambda _{\rm{U}}}\pi \left( {4{{\left( {{R_{{\rm{co}}}}} \right)}^2} - {{\left( {\min \left\{ {{R_{{\rm{co}}}} - \left\| {{{{\mathbf{\tilde x}}}_{{\rm{BS}}}}({{\mathbf{x}}_{\rm{U}}}) - {{\mathbf{x}}_{\rm{U}}}} \right\|,\left\| {{{{\mathbf{\tilde x}}}_{{\rm{BS}}}}({{\mathbf{x}}_{\rm{U}}}) - {{\mathbf{x}}_{\rm{U}}}} \right\|} \right\}} \right)}^2}} \right)\exp \left\{ { - {\lambda _{\rm{U}}}\pi \left( {4{{\left( {{R_{{\rm{co}}}}} \right)}^2} - {{\left( {\min \left\{ {{R_{{\rm{co}}}} - \left\| {{{{\mathbf{\tilde x}}}_{{\rm{BS}}}}({{\mathbf{x}}_{\rm{U}}}) - {{\mathbf{x}}_{\rm{U}}}} \right\|,\left\| {{{{\mathbf{\tilde x}}}_{{\rm{BS}}}}({{\mathbf{x}}_{\rm{U}}}) - {{\mathbf{x}}_{\rm{U}}}} \right\|} \right\}} \right)}^2}} \right)} \right\}}}{{{{\left( {{\lambda _{\rm{U}}}\pi \left( {4{{\left( {{R_{{\rm{co}}}}} \right)}^2} - {{\left( {\min \left\{ {{R_{{\rm{co}}}} - \left\| {{{{\mathbf{\tilde x}}}_{{\rm{BS}}}}({{\mathbf{x}}_{\rm{U}}}) - {{\mathbf{x}}_{\rm{U}}}} \right\|,\left\| {{{{\mathbf{\tilde x}}}_{{\rm{BS}}}}({{\mathbf{x}}_{\rm{U}}}) - {{\mathbf{x}}_{\rm{U}}}} \right\|} \right\}} \right)}^2}} \right)} \right)}^2}}}} \right) \ge$}
\end{equation*}
\begin{equation*}
\frac{\kappa }{{\kappa  + 1}} \times\sum\limits_{{{\mathbf{x}}_{{\rm{IRS}}}} \in {\Psi _{{\rm{IRS}}}}({{\mathbf{x}}_{\rm{U}}})} {\begin{array}{*{20}{c}}
  {}&{} 
\end{array}\sum\limits_{{{{\mathbf{x'}}}_{{\rm{IRS}}}} \in {\Psi _{{\rm{IRS}}}}({{\mathbf{x}}_{\rm{U}}})} {\begin{array}{*{20}{c}}
  {}&{} 
\end{array}} } 
\end{equation*}
\begin{equation*}
   \scalebox{1}[1]{$\frac{{{\pmb{\mathbb{I}}}\left( {{n_{{\rm{BS}}}}({{\mathbf{x}}_{\rm{U}}})} \right)Q(Q - 1)\left( {1 - {\delta _{{{\mathbf{x}}_{{\rm{IRS}}}} - {{\mathbf{x}}^\prime }_{{\rm{IRS}}}}}} \right){{\left( {{\lambda _{{\rm{wave}}}}} \right)}^4}{\delta ^2} \sigma _{{\rm{d}}}^2}}{{{4^6}{\pi ^2}\left( {\max \left\{ {{{({h_{{\rm{IRS}}}})}^2},{{({h_{{\rm{BS}}}} - {h_{{\rm{IRS}}}})}^2}} \right\} + 3{{\left( {{R_{{\rm{co}}}}} \right)}^2} + \left\| {{{\mathbf{x}}_{{\rm{IRS}}}} - {{\mathbf{x}}_{\rm{U}}}} \right\|_2^2} \right)\left( {\max \left\{ {{{({h_{{\rm{IRS}}}})}^2},{{({h_{{\rm{BS}}}} - {h_{{\rm{IRS}}}})}^2}} \right\} + 3{{\left( {{R_{{\rm{co}}}}} \right)}^2} + \left\| {{{{\mathbf{x'}}}_{{\rm{IRS}}}} - {{\mathbf{x}}_{\rm{U}}}} \right\|_2^2} \right)}} \times $}
\end{equation*}
\begin{equation*}
   \scalebox{.95}[1]{$\frac{{\left( {{\lambda _{\rm{U}}}\pi {{\left( {\min \left\{ {{R_{{\rm{co}}}} - \left\| {{{{\mathbf{\tilde x}}}_{{\rm{BS}}}}({{\mathbf{x}}_{\rm{U}}}) - {{\mathbf{x}}_{\rm{U}}}} \right\|,\left\| {{{{\mathbf{\tilde x}}}_{{\rm{BS}}}}({{\mathbf{x}}_{\rm{U}}}) - {{\mathbf{x}}_{\rm{U}}}} \right\|} \right\}} \right)}^2} + \exp \left\{ { - {\lambda _{\rm{U}}}\pi {{\left( {\min \left\{ {{R_{{\rm{co}}}} - \left\| {{{{\mathbf{\tilde x}}}_{{\rm{BS}}}}({{\mathbf{x}}_{\rm{U}}}) - {{\mathbf{x}}_{\rm{U}}}} \right\|,\left\| {{{{\mathbf{\tilde x}}}_{{\rm{BS}}}}({{\mathbf{x}}_{\rm{U}}}) - {{\mathbf{x}}_{\rm{U}}}} \right\|} \right\}} \right)}^2}} \right\} - 1} \right)}}{{\left( {1 - \exp \left\{ { - {\lambda _{\rm{U}}}\pi {{\left( {\min \left\{ {{R_{{\rm{co}}}} - \left\| {{{{\mathbf{\tilde x}}}_{{\rm{BS}}}}({{\mathbf{x}}_{\rm{U}}}) - {{\mathbf{x}}_{\rm{U}}}} \right\|,\left\| {{{{\mathbf{\tilde x}}}_{{\rm{BS}}}}({{\mathbf{x}}_{\rm{U}}}) - {{\mathbf{x}}_{\rm{U}}}} \right\|} \right\}} \right)}^2}} \right\}} \right)\left( {{N_{{\rm{BS}}}}\left( {{\cal C}\left( {{{\mathbf{x}}_{\rm{U}}},2{R_{{\rm{co}}}}} \right) - {\cal C}\left( {{{\mathbf{x}}_{\rm{U}}},{R_{{\rm{co}}}}} \right)} \right) + 1} \right)}} \times $}
\end{equation*}
\begin{equation*}
\max \left\{ {0,\frac{{1 - \exp \left\{ { - {\lambda _{\rm{U}}}\pi \left( {3{{\left( {{R_{{\rm{co}}}}} \right)}^2}} \right)} \right\} - {\lambda _{\rm{U}}}\pi \left( {4{{\left( {{R_{{\rm{co}}}}} \right)}^2}} \right)\exp \left\{ { - {\lambda _{\rm{U}}}\pi \left( {3{{\left( {{R_{{\rm{co}}}}} \right)}^2}} \right)} \right\}}}{{{{\left( {{\lambda _{\rm{U}}}\pi \left( {4{{\left( {{R_{{\rm{co}}}}} \right)}^2}} \right)} \right)}^2}}}} \right\},
\end{equation*}
which follows from the following fact for poisson variable $W$:
\begin{equation*}
E\left\{ {\frac{1}{{{{\left( {W + 1} \right)}^2}}}} \right\} \ge \sum\limits_{k = 0}^\infty  {\frac{{{\lambda ^k}{e^{ - \lambda }}}}{{\left( {k + 2} \right)!}}}  = \sum\limits_{k = 2}^\infty  {\frac{{{\lambda ^u}{e^{ - \lambda }}}}{{{\lambda ^2}u!}}}  = \frac{{{e^{ - \lambda }}\left( {{e^\lambda } - 1 - \lambda } \right)}}{{{\lambda ^2}}} = \frac{{1 - {e^{ - \lambda }} - \lambda {e^{ - \lambda }}}}{{{\lambda ^2}}}.
\end{equation*}
Then, we obtain:
\begin{equation*}
E\left\{\frac{\kappa }{{\kappa  + 1}} \times {\sum\limits_{{{\mathbf{x}}_{{\rm{IRS}}}} \in {\Psi _{{\rm{IRS}}}}({{\mathbf{x}}_{\rm{U}}})} {\begin{array}{*{20}{c}}
  {}&{} 
\end{array}\sum\limits_{{{{\mathbf{x'}}}_{{\rm{IRS}}}} \in {\Psi _{{\rm{IRS}}}}({{\mathbf{x}}_{\rm{U}}})} {\begin{array}{*{20}{c}}
  {}&{} 
\end{array}} } } \right.
\end{equation*}
\begin{equation*}
   \scalebox{1}[1]{$\frac{{{\pmb{\mathbb{I}}}\left( {{n_{{\rm{BS}}}}({{\mathbf{x}}_{\rm{U}}})} \right)Q(Q - 1)\left( {1 - {\delta _{{{\mathbf{x}}_{{\rm{IRS}}}} - {{\mathbf{x}}^\prime }_{{\rm{IRS}}}}}} \right){{\left( {{\lambda _{{\rm{wave}}}}} \right)}^4}{\delta ^2} \sigma _{{\rm{d}}}^2}}{{{4^6}{\pi ^2}\left( {\max \left\{ {{{({h_{{\rm{IRS}}}})}^2},{{({h_{{\rm{BS}}}} - {h_{{\rm{IRS}}}})}^2}} \right\} + 3{{\left( {{R_{{\rm{co}}}}} \right)}^2} + \left\| {{{\mathbf{x}}_{{\rm{IRS}}}} - {{\mathbf{x}}_{\rm{U}}}} \right\|_2^2} \right)\left( {\max \left\{ {{{({h_{{\rm{IRS}}}})}^2},{{({h_{{\rm{BS}}}} - {h_{{\rm{IRS}}}})}^2}} \right\} + 3{{\left( {{R_{{\rm{co}}}}} \right)}^2} + \left\| {{{{\mathbf{x'}}}_{{\rm{IRS}}}} - {{\mathbf{x}}_{\rm{U}}}} \right\|_2^2} \right)}} \times $}
\end{equation*}
\begin{equation*}
   \scalebox{.95}[1]{$\frac{{\left( {{\lambda _{\rm{U}}}\pi {{\left( {\min \left\{ {{R_{{\rm{co}}}} - \left\| {{{{\mathbf{\tilde x}}}_{{\rm{BS}}}}({{\mathbf{x}}_{\rm{U}}}) - {{\mathbf{x}}_{\rm{U}}}} \right\|,\left\| {{{{\mathbf{\tilde x}}}_{{\rm{BS}}}}({{\mathbf{x}}_{\rm{U}}}) - {{\mathbf{x}}_{\rm{U}}}} \right\|} \right\}} \right)}^2} + \exp \left\{ { - {\lambda _{\rm{U}}}\pi {{\left( {\min \left\{ {{R_{{\rm{co}}}} - \left\| {{{{\mathbf{\tilde x}}}_{{\rm{BS}}}}({{\mathbf{x}}_{\rm{U}}}) - {{\mathbf{x}}_{\rm{U}}}} \right\|,\left\| {{{{\mathbf{\tilde x}}}_{{\rm{BS}}}}({{\mathbf{x}}_{\rm{U}}}) - {{\mathbf{x}}_{\rm{U}}}} \right\|} \right\}} \right)}^2}} \right\} - 1} \right)}}{{\left( {1 - \exp \left\{ { - {\lambda _{\rm{U}}}\pi {{\left( {\min \left\{ {{R_{{\rm{co}}}} - \left\| {{{{\mathbf{\tilde x}}}_{{\rm{BS}}}}({{\mathbf{x}}_{\rm{U}}}) - {{\mathbf{x}}_{\rm{U}}}} \right\|,\left\| {{{{\mathbf{\tilde x}}}_{{\rm{BS}}}}({{\mathbf{x}}_{\rm{U}}}) - {{\mathbf{x}}_{\rm{U}}}} \right\|} \right\}} \right)}^2}} \right\}} \right)\left( {{N_{{\rm{BS}}}}\left( {{\cal C}\left( {{{\mathbf{x}}_{\rm{U}}},2{R_{{\rm{co}}}}} \right) - {\cal C}\left( {{{\mathbf{x}}_{\rm{U}}},{R_{{\rm{co}}}}} \right)} \right) + 1} \right)}} \times $}
\end{equation*}
\begin{equation*}
   \scalebox{.95}[1]{$\left. {\max \left\{ {0,\frac{{1 - \exp \left\{ { - {\lambda _{\rm{U}}}\pi \left( {3{{\left( {{R_{{\rm{co}}}}} \right)}^2}} \right)} \right\} - {\lambda _{\rm{U}}}\pi \left( {4{{\left( {{R_{{\rm{co}}}}} \right)}^2}} \right)\exp \left\{ { - {\lambda _{\rm{U}}}\pi \left( {3{{\left( {{R_{{\rm{co}}}}} \right)}^2}} \right)} \right\}}}{{{{\left( {{\lambda _{\rm{U}}}\pi \left( {4{{\left( {{R_{{\rm{co}}}}} \right)}^2}} \right)} \right)}^2}}}} \right\}} \right|$}
\end{equation*}
\begin{equation*}
   \scalebox{.9}[1]{$\left. {\begin{array}{*{20}{c}}
  {} \\ 
  {} 
\end{array}{\Phi _{{\rm{IRS}}}},{N_{{\rm{BS}}}}\left( {{\cal C}\left( {{{\mathbf{x}}_{\rm{U}}},2{R_{{\rm{co}}}}} \right) - {\cal C}\left( {{{\mathbf{x}}_{\rm{U}}},{R_{{\rm{co}}}}} \right)} \right),{n_{{\rm{BS}}}}({{\mathbf{x}}_{\rm{U}}}),\left\{ {\frac{{{R_{{\rm{co}}}}}}{2} - d < \left\| {{{{\mathbf{\tilde x}}}_{{\rm{BS}}}}({{\mathbf{x}}_{\rm{U}}}) - {{\mathbf{x}}_{\rm{U}}}} \right\| < \frac{{{R_{{\rm{co}}}}}}{2} + d} \right\}} \right\}$}
\end{equation*}
\begin{equation*}
\ge\frac{\kappa }{{\kappa  + 1}} \times \sum\limits_{{{\mathbf{x}}_{{\rm{IRS}}}} \in {\Psi _{{\rm{IRS}}}}({{\mathbf{x}}_{\rm{U}}})} {\begin{array}{*{20}{c}}
  {}&{} 
\end{array}\sum\limits_{{{{\mathbf{x'}}}_{{\rm{IRS}}}} \in {\Psi _{{\rm{IRS}}}}({{\mathbf{x}}_{\rm{U}}})} {\begin{array}{*{20}{c}}
  {}&{} 
\end{array}} } 
\end{equation*}
\begin{equation*}
   \scalebox{1}[1]{$\frac{{{\pmb{\mathbb{I}}}\left( {{n_{{\rm{BS}}}}({{\mathbf{x}}_{\rm{U}}})} \right)Q(Q - 1)\left( {1 - {\delta _{{{\mathbf{x}}_{{\rm{IRS}}}} - {{\mathbf{x}}^\prime }_{{\rm{IRS}}}}}} \right){{\left( {{\lambda _{{\rm{wave}}}}} \right)}^4}{\delta ^2} \sigma _{{\rm{d}}}^2}}{{{4^6}{\pi ^2}\left( {\max \left\{ {{{({h_{{\rm{IRS}}}})}^2},{{({h_{{\rm{BS}}}} - {h_{{\rm{IRS}}}})}^2}} \right\} + 3{{\left( {{R_{{\rm{co}}}}} \right)}^2} + \left\| {{{\mathbf{x}}_{{\rm{IRS}}}} - {{\mathbf{x}}_{\rm{U}}}} \right\|_2^2} \right)\left( {\max \left\{ {{{({h_{{\rm{IRS}}}})}^2},{{({h_{{\rm{BS}}}} - {h_{{\rm{IRS}}}})}^2}} \right\} + 3{{\left( {{R_{{\rm{co}}}}} \right)}^2} + \left\| {{{{\mathbf{x'}}}_{{\rm{IRS}}}} - {{\mathbf{x}}_{\rm{U}}}} \right\|_2^2} \right)}} \times $}
\end{equation*}
\begin{equation*}
\frac{{\max \left\{ {\left( {{\lambda _{\rm{U}}}\pi {{\left( {\frac{{{R_{{\rm{co}}}}}}{2} - d} \right)}^2} + \exp \left\{ { - {\lambda _{\rm{U}}}\pi {{\left( {\frac{{{R_{{\rm{co}}}}}}{2} + d} \right)}^2}} \right\} - 1} \right),0} \right\}}}{{\left( {1 - \exp \left\{ { - {\lambda _{\rm{U}}}\pi {{\left( {\frac{{{R_{{\rm{co}}}}}}{2} + d} \right)}^2}} \right\}} \right)\left( {{N_{{\rm{BS}}}}\left( {{\cal C}\left( {{{\mathbf{x}}_{\rm{U}}},2{R_{{\rm{co}}}}} \right) - {\cal C}\left( {{{\mathbf{x}}_{\rm{U}}},{R_{{\rm{co}}}}} \right)} \right) + 1} \right)}} \times 
\end{equation*}
\begin{equation*}
\max \left\{ {0,\frac{{1 - \exp \left\{ { - {\lambda _{\rm{U}}}\pi \left( {3{{\left( {{R_{{\rm{co}}}}} \right)}^2}} \right)} \right\} - {\lambda _{\rm{U}}}\pi \left( {4{{\left( {{R_{{\rm{co}}}}} \right)}^2}} \right)\exp \left\{ { - {\lambda _{\rm{U}}}\pi \left( {3{{\left( {{R_{{\rm{co}}}}} \right)}^2}} \right)} \right\}}}{{{{\left( {{\lambda _{\rm{U}}}\pi \left( {4{{\left( {{R_{{\rm{co}}}}} \right)}^2}} \right)} \right)}^2}}}} \right\},
\end{equation*}
hence, we have:
\begin{equation*}
E\left\{\frac{\kappa }{{\kappa  + 1}} \times {\sum\limits_{{{\mathbf{x}}_{{\rm{IRS}}}} \in {\Psi _{{\rm{IRS}}}}({{\mathbf{x}}_{\rm{U}}})} {\begin{array}{*{20}{c}}
  {}&{} 
\end{array}\sum\limits_{{{{\mathbf{x'}}}_{{\rm{IRS}}}} \in {\Psi _{{\rm{IRS}}}}({{\mathbf{x}}_{\rm{U}}})} {\begin{array}{*{20}{c}}
  {}&{} 
\end{array}} } } \right.
\end{equation*}
\begin{equation*}
   \scalebox{1}[1]{$\frac{{{\pmb{\mathbb{I}}}\left( {{n_{{\rm{BS}}}}({{\mathbf{x}}_{\rm{U}}})} \right)Q(Q - 1)\left( {1 - {\delta _{{{\mathbf{x}}_{{\rm{IRS}}}} - {{\mathbf{x}}^\prime }_{{\rm{IRS}}}}}} \right){{\left( {{\lambda _{{\rm{wave}}}}} \right)}^4}{\delta ^2} \sigma _{{\rm{d}}}^2}}{{{4^6}{\pi ^2}\left( {\max \left\{ {{{({h_{{\rm{IRS}}}})}^2},{{({h_{{\rm{BS}}}} - {h_{{\rm{IRS}}}})}^2}} \right\} + 3{{\left( {{R_{{\rm{co}}}}} \right)}^2} + \left\| {{{\mathbf{x}}_{{\rm{IRS}}}} - {{\mathbf{x}}_{\rm{U}}}} \right\|_2^2} \right)\left( {\max \left\{ {{{({h_{{\rm{IRS}}}})}^2},{{({h_{{\rm{BS}}}} - {h_{{\rm{IRS}}}})}^2}} \right\} + 3{{\left( {{R_{{\rm{co}}}}} \right)}^2} + \left\| {{{{\mathbf{x'}}}_{{\rm{IRS}}}} - {{\mathbf{x}}_{\rm{U}}}} \right\|_2^2} \right)}} \times $}
\end{equation*}
\begin{equation*}
\frac{{\max \left\{ {\left( {{\lambda _{\rm{U}}}\pi {{\left( {\frac{{{R_{{\rm{co}}}}}}{2} - d} \right)}^2} + \exp \left\{ { - {\lambda _{\rm{U}}}\pi {{\left( {\frac{{{R_{{\rm{co}}}}}}{2} + d} \right)}^2}} \right\} - 1} \right),0} \right\}}}{{\left( {1 - \exp \left\{ { - {\lambda _{\rm{U}}}\pi {{\left( {\frac{{{R_{{\rm{co}}}}}}{2} + d} \right)}^2}} \right\}} \right)\left( {{N_{{\rm{BS}}}}\left( {{\cal C}\left( {{{\mathbf{x}}_{\rm{U}}},2{R_{{\rm{co}}}}} \right) - {\cal C}\left( {{{\mathbf{x}}_{\rm{U}}},{R_{{\rm{co}}}}} \right)} \right) + 1} \right)}} \times 
\end{equation*}
\begin{equation*}
\left. {\max \left\{ {0,\frac{{1 - \exp \left\{ { - {\lambda _{\rm{U}}}\pi \left( {3{{\left( {{R_{{\rm{co}}}}} \right)}^2}} \right)} \right\} - {\lambda _{\rm{U}}}\pi \left( {4{{\left( {{R_{{\rm{co}}}}} \right)}^2}} \right)\exp \left\{ { - {\lambda _{\rm{U}}}\pi \left( {3{{\left( {{R_{{\rm{co}}}}} \right)}^2}} \right)} \right\}}}{{{{\left( {{\lambda _{\rm{U}}}\pi \left( {4{{\left( {{R_{{\rm{co}}}}} \right)}^2}} \right)} \right)}^2}}}} \right\}} \right|
\end{equation*}
\begin{equation*}
\left. {\begin{array}{*{20}{c}}
  {} \\ 
  {} 
\end{array}{\Phi _{{\rm{IRS}}}},{N_{{\rm{BS}}}}\left( {{\cal C}\left( {{{\mathbf{x}}_{\rm{U}}},2{R_{{\rm{co}}}}} \right) - {\cal C}\left( {{{\mathbf{x}}_{\rm{U}}},{R_{{\rm{co}}}}} \right)} \right),{n_{{\rm{BS}}}}({{\mathbf{x}}_{\rm{U}}})} \right\} \ge
\end{equation*}
\begin{equation*}
\frac{\kappa }{{\kappa  + 1}} \times\sum\limits_{{{\mathbf{x}}_{{\rm{IRS}}}} \in {\Psi _{{\rm{IRS}}}}({{\mathbf{x}}_{\rm{U}}})} {\begin{array}{*{20}{c}}
  {}&{} 
\end{array}\sum\limits_{{{{\mathbf{x'}}}_{{\rm{IRS}}}} \in {\Psi _{{\rm{IRS}}}}({{\mathbf{x}}_{\rm{U}}})} {\begin{array}{*{20}{c}}
  {}&{} 
\end{array}} } 
\end{equation*}
\begin{equation*}
   \scalebox{1}[1]{$\frac{{{\pmb{\mathbb{I}}}\left( {{n_{{\rm{BS}}}}({{\mathbf{x}}_{\rm{U}}})} \right)Q(Q - 1)\left( {1 - {\delta _{{{\mathbf{x}}_{{\rm{IRS}}}} - {{\mathbf{x}}^\prime }_{{\rm{IRS}}}}}} \right){{\left( {{\lambda _{{\rm{wave}}}}} \right)}^4}{\delta ^2} \sigma _{{\rm{d}}}^2}}{{{4^6}{\pi ^2}\left( {\max \left\{ {{{({h_{{\rm{IRS}}}})}^2},{{({h_{{\rm{BS}}}} - {h_{{\rm{IRS}}}})}^2}} \right\} + 3{{\left( {{R_{{\rm{co}}}}} \right)}^2} + \left\| {{{\mathbf{x}}_{{\rm{IRS}}}} - {{\mathbf{x}}_{\rm{U}}}} \right\|_2^2} \right)\left( {\max \left\{ {{{({h_{{\rm{IRS}}}})}^2},{{({h_{{\rm{BS}}}} - {h_{{\rm{IRS}}}})}^2}} \right\} + 3{{\left( {{R_{{\rm{co}}}}} \right)}^2} + \left\| {{{{\mathbf{x'}}}_{{\rm{IRS}}}} - {{\mathbf{x}}_{\rm{U}}}} \right\|_2^2} \right)}} \times $}
\end{equation*}
\begin{equation*}
\frac{{\max \left\{ {\left( {{\lambda _{\rm{U}}}\pi {{\left( {\frac{{{R_{{\rm{co}}}}}}{2} - d} \right)}^2} + \exp \left\{ { - {\lambda _{\rm{U}}}\pi {{\left( {\frac{{{R_{{\rm{co}}}}}}{2} + d} \right)}^2}} \right\} - 1} \right),0} \right\}}}{{\left( {1 - \exp \left\{ { - {\lambda _{\rm{U}}}\pi {{\left( {\frac{{{R_{{\rm{co}}}}}}{2} + d} \right)}^2}} \right\}} \right)\left( {{N_{{\rm{BS}}}}\left( {{\cal C}\left( {{{\mathbf{x}}_{\rm{U}}},2{R_{{\rm{co}}}}} \right) - {\cal C}\left( {{{\mathbf{x}}_{\rm{U}}},{R_{{\rm{co}}}}} \right)} \right) + 1} \right)}} \times 
\end{equation*}
\begin{equation*}
\max \left\{ {0,\frac{{1 - \exp \left\{ { - {\lambda _{\rm{U}}}\pi \left( {3{{\left( {{R_{{\rm{co}}}}} \right)}^2}} \right)} \right\} - {\lambda _{\rm{U}}}\pi \left( {4{{\left( {{R_{{\rm{co}}}}} \right)}^2}} \right)\exp \left\{ { - {\lambda _{\rm{U}}}\pi \left( {3{{\left( {{R_{{\rm{co}}}}} \right)}^2}} \right)} \right\}}}{{{{\left( {{\lambda _{\rm{U}}}\pi \left( {4{{\left( {{R_{{\rm{co}}}}} \right)}^2}} \right)} \right)}^2}}}} \right\} \times 
\end{equation*}
\begin{equation*}
\Pr \left\{ {\left. {\left\{ {\frac{{{R_{{\rm{co}}}}}}{2} - d < \left\| {{{{\mathbf{\tilde x}}}_{{\rm{BS}}}}({{\mathbf{x}}_{\rm{U}}}) - {{\mathbf{x}}_{\rm{U}}}} \right\| < \frac{{{R_{{\rm{co}}}}}}{2} + d} \right\}} \right|{n_{{\rm{BS}}}}({{\mathbf{x}}_{\rm{U}}})} \right\} = 
\end{equation*}
\begin{equation*}
\frac{\kappa }{{\kappa  + 1}} \times\sum\limits_{{{\mathbf{x}}_{{\rm{IRS}}}} \in {\Psi _{{\rm{IRS}}}}({{\mathbf{x}}_{\rm{U}}})} {\begin{array}{*{20}{c}}
  {}&{} 
\end{array}\sum\limits_{{{{\mathbf{x'}}}_{{\rm{IRS}}}} \in {\Psi _{{\rm{IRS}}}}({{\mathbf{x}}_{\rm{U}}})} {\begin{array}{*{20}{c}}
  {}&{} 
\end{array}} } 
\end{equation*}
\begin{equation*}
   \scalebox{1}[1]{$\frac{{{\pmb{\mathbb{I}}}\left( {{n_{{\rm{BS}}}}({{\mathbf{x}}_{\rm{U}}})} \right)Q(Q - 1)\left( {1 - {\delta _{{{\mathbf{x}}_{{\rm{IRS}}}} - {{\mathbf{x}}^\prime }_{{\rm{IRS}}}}}} \right){{\left( {{\lambda _{{\rm{wave}}}}} \right)}^4}{\delta ^2} \sigma _{{\rm{d}}}^2}}{{{4^6}{\pi ^2}\left( {\max \left\{ {{{({h_{{\rm{IRS}}}})}^2},{{({h_{{\rm{BS}}}} - {h_{{\rm{IRS}}}})}^2}} \right\} + 3{{\left( {{R_{{\rm{co}}}}} \right)}^2} + \left\| {{{\mathbf{x}}_{{\rm{IRS}}}} - {{\mathbf{x}}_{\rm{U}}}} \right\|_2^2} \right)\left( {\max \left\{ {{{({h_{{\rm{IRS}}}})}^2},{{({h_{{\rm{BS}}}} - {h_{{\rm{IRS}}}})}^2}} \right\} + 3{{\left( {{R_{{\rm{co}}}}} \right)}^2} + \left\| {{{{\mathbf{x'}}}_{{\rm{IRS}}}} - {{\mathbf{x}}_{\rm{U}}}} \right\|_2^2} \right)}} \times $}
\end{equation*}
\begin{equation*}
\frac{{2d\max \left\{ {\left( {{\lambda _{\rm{U}}}\pi {{\left( {\frac{{{R_{{\rm{co}}}}}}{2} - d} \right)}^2} + \exp \left\{ { - {\lambda _{\rm{U}}}\pi {{\left( {\frac{{{R_{{\rm{co}}}}}}{2} + d} \right)}^2}} \right\} - 1} \right),0} \right\}}}{{{R_{{\rm{co}}}}\left( {1 - \exp \left\{ { - {\lambda _{\rm{U}}}\pi {{\left( {\frac{{{R_{{\rm{co}}}}}}{2} + d} \right)}^2}} \right\}} \right)\left( {{N_{{\rm{BS}}}}\left( {{\cal C}\left( {{{\mathbf{x}}_{\rm{U}}},2{R_{{\rm{co}}}}} \right) - {\cal C}\left( {{{\mathbf{x}}_{\rm{U}}},{R_{{\rm{co}}}}} \right)} \right) + 1} \right)}} \times 
\end{equation*}
\begin{equation*}
\max \left\{ {0,\frac{{1 - \exp \left\{ { - {\lambda _{\rm{U}}}\pi \left( {3{{\left( {{R_{{\rm{co}}}}} \right)}^2}} \right)} \right\} - {\lambda _{\rm{U}}}\pi \left( {4{{\left( {{R_{{\rm{co}}}}} \right)}^2}} \right)\exp \left\{ { - {\lambda _{\rm{U}}}\pi \left( {3{{\left( {{R_{{\rm{co}}}}} \right)}^2}} \right)} \right\}}}{{{{\left( {{\lambda _{\rm{U}}}\pi \left( {4{{\left( {{R_{{\rm{co}}}}} \right)}^2}} \right)} \right)}^2}}}} \right\}.
\end{equation*}
Then, by (\ref{t55}), we obtain:
\begin{equation*}
E\left\{ \frac{\kappa }{{\kappa  + 1}} \times{\sum\limits_{{{\mathbf{x}}_{{\rm{IRS}}}} \in {\Psi _{{\rm{IRS}}}}({{\mathbf{x}}_{\rm{U}}})} {\begin{array}{*{20}{c}}
  {}&{} 
\end{array}\sum\limits_{{{{\mathbf{x'}}}_{{\rm{IRS}}}} \in {\Psi _{{\rm{IRS}}}}({{\mathbf{x}}_{\rm{U}}})} {\begin{array}{*{20}{c}}
  {}&{} 
\end{array}} } } \right.
\end{equation*}
\begin{equation*}
   \scalebox{1}[1]{$\frac{{{\pmb{\mathbb{I}}}\left( {{n_{{\rm{BS}}}}({{\mathbf{x}}_{\rm{U}}})} \right)Q(Q - 1)\left( {1 - {\delta _{{{\mathbf{x}}_{{\rm{IRS}}}} - {{\mathbf{x}}^\prime }_{{\rm{IRS}}}}}} \right){{\left( {{\lambda _{{\rm{wave}}}}} \right)}^4}{\delta ^2} \sigma _{{\rm{d}}}^2}}{{{4^6}{\pi ^2}\left( {\max \left\{ {{{({h_{{\rm{IRS}}}})}^2},{{({h_{{\rm{BS}}}} - {h_{{\rm{IRS}}}})}^2}} \right\} + 3{{\left( {{R_{{\rm{co}}}}} \right)}^2} + \left\| {{{\mathbf{x}}_{{\rm{IRS}}}} - {{\mathbf{x}}_{\rm{U}}}} \right\|_2^2} \right)\left( {\max \left\{ {{{({h_{{\rm{IRS}}}})}^2},{{({h_{{\rm{BS}}}} - {h_{{\rm{IRS}}}})}^2}} \right\} + 3{{\left( {{R_{{\rm{co}}}}} \right)}^2} + \left\| {{{{\mathbf{x'}}}_{{\rm{IRS}}}} - {{\mathbf{x}}_{\rm{U}}}} \right\|_2^2} \right)}} \times $}
\end{equation*}
\begin{equation*}
\frac{{2d\max \left\{ {\left( {{\lambda _{\rm{U}}}\pi {{\left( {\frac{{{R_{{\rm{co}}}}}}{2} - d} \right)}^2} + \exp \left\{ { - {\lambda _{\rm{U}}}\pi {{\left( {\frac{{{R_{{\rm{co}}}}}}{2} + d} \right)}^2}} \right\} - 1} \right),0} \right\}}}{{{R_{{\rm{co}}}}\left( {1 - \exp \left\{ { - {\lambda _{\rm{U}}}\pi {{\left( {\frac{{{R_{{\rm{co}}}}}}{2} + d} \right)}^2}} \right\}} \right)\left( {{N_{{\rm{BS}}}}\left( {{\cal C}\left( {{{\mathbf{x}}_{\rm{U}}},2{R_{{\rm{co}}}}} \right) - {\cal C}\left( {{{\mathbf{x}}_{\rm{U}}},{R_{{\rm{co}}}}} \right)} \right) + 1} \right)}} \times 
\end{equation*}
\begin{equation*}
\scalebox{.95}[1]{$\left. {\left. {\max \left\{ {0,\frac{{1 - \exp \left\{ { - {\lambda _{\rm{U}}}\pi \left( {3{{\left( {{R_{{\rm{co}}}}} \right)}^2}} \right)} \right\} - {\lambda _{\rm{U}}}\pi \left( {4{{\left( {{R_{{\rm{co}}}}} \right)}^2}} \right)\exp \left\{ { - {\lambda _{\rm{U}}}\pi \left( {3{{\left( {{R_{{\rm{co}}}}} \right)}^2}} \right)} \right\}}}{{{{\left( {{\lambda _{\rm{U}}}\pi \left( {4{{\left( {{R_{{\rm{co}}}}} \right)}^2}} \right)} \right)}^2}}}} \right\}} \right|{\Phi _{{\rm{IRS}}}}} \right\} = $}
\end{equation*}
\begin{equation*}
\frac{\kappa }{{\kappa  + 1}} \times\sum\limits_{{{\mathbf{x}}_{{\rm{IRS}}}} \in {\Psi _{{\rm{IRS}}}}({{\mathbf{x}}_{\rm{U}}})} {\begin{array}{*{20}{c}}
  {}&{} 
\end{array}\sum\limits_{{{{\mathbf{x'}}}_{{\rm{IRS}}}} \in {\Psi _{{\rm{IRS}}}}({{\mathbf{x}}_{\rm{U}}})} {\begin{array}{*{20}{c}}
  {}&{} 
\end{array}} } 
\end{equation*}
\begin{equation*}
   \scalebox{1}[1]{$\frac{{\left( {1 - \exp \left\{ { - {\lambda _{{\rm{BS}}}}\pi {{\left( {{R_{{\rm{co}}}}} \right)}^2}} \right\}} \right)Q(Q - 1)\left( {1 - {\delta _{{{\mathbf{x}}_{{\rm{IRS}}}} - {{\mathbf{x}}^\prime }_{{\rm{IRS}}}}}} \right){{\left( {{\lambda _{{\rm{wave}}}}} \right)}^4}{\delta ^2} \sigma _{{\rm{d}}}^2}}{{{4^6}{\pi ^2}\left( {\max \left\{ {{{({h_{{\rm{IRS}}}})}^2},{{({h_{{\rm{BS}}}} - {h_{{\rm{IRS}}}})}^2}} \right\} + 3{{\left( {{R_{{\rm{co}}}}} \right)}^2} + \left\| {{{\mathbf{x}}_{{\rm{IRS}}}} - {{\mathbf{x}}_{\rm{U}}}} \right\|_2^2} \right)\left( {\max \left\{ {{{({h_{{\rm{IRS}}}})}^2},{{({h_{{\rm{BS}}}} - {h_{{\rm{IRS}}}})}^2}} \right\} + 3{{\left( {{R_{{\rm{co}}}}} \right)}^2} + \left\| {{{{\mathbf{x'}}}_{{\rm{IRS}}}} - {{\mathbf{x}}_{\rm{U}}}} \right\|_2^2} \right)}} \times $}
\end{equation*}
\begin{equation*}
\scalebox{.95}[1]{$\frac{{2d\max \left\{ {\left( {{\lambda _{\rm{U}}}\pi {{\left( {\frac{{{R_{{\rm{co}}}}}}{2} - d} \right)}^2} + \exp \left\{ { - {\lambda _{\rm{U}}}\pi {{\left( {\frac{{{R_{{\rm{co}}}}}}{2} + d} \right)}^2}} \right\} - 1} \right),0} \right\}\left( {1 - \exp \left\{ { - {\lambda _{{\rm{BS}}}}3\pi {{\left( {{R_{{\rm{co}}}}} \right)}^2}} \right\}} \right)}}{{\left( {1 - \exp \left\{ { - {\lambda _{\rm{U}}}\pi {{\left( {\frac{{{R_{{\rm{co}}}}}}{2} + d} \right)}^2}} \right\}} \right){\lambda _{{\rm{BS}}}}3\pi {{\left( {{R_{{\rm{co}}}}} \right)}^3}}} \times $}
\end{equation*}
\begin{equation*}
\max \left\{ {0,\frac{{1 - \exp \left\{ { - {\lambda _{\rm{U}}}\pi \left( {3{{\left( {{R_{{\rm{co}}}}} \right)}^2}} \right)} \right\} - {\lambda _{\rm{U}}}\pi \left( {4{{\left( {{R_{{\rm{co}}}}} \right)}^2}} \right)\exp \left\{ { - {\lambda _{\rm{U}}}\pi \left( {3{{\left( {{R_{{\rm{co}}}}} \right)}^2}} \right)} \right\}}}{{{{\left( {{\lambda _{\rm{U}}}\pi \left( {4{{\left( {{R_{{\rm{co}}}}} \right)}^2}} \right)} \right)}^2}}}} \right\},
\end{equation*}
which yields:
\begin{equation*}
E\left\{\frac{\kappa }{{\kappa  + 1}} \times {\sum\limits_{{{\mathbf{x}}_{{\rm{IRS}}}} \in {\Psi _{{\rm{IRS}}}}({{\mathbf{x}}_{\rm{U}}})} {\begin{array}{*{20}{c}}
  {}&{} 
\end{array}\sum\limits_{{{{\mathbf{x'}}}_{{\rm{IRS}}}} \in {\Psi _{{\rm{IRS}}}}({{\mathbf{x}}_{\rm{U}}})} {\begin{array}{*{20}{c}}
  {}&{} 
\end{array}} } } \right.
\end{equation*}
\begin{equation*}
   \scalebox{1}[1]{$\frac{{\left( {1 - \exp \left\{ { - {\lambda _{{\rm{BS}}}}\pi {{\left( {{R_{{\rm{co}}}}} \right)}^2}} \right\}} \right)Q(Q - 1)\left( {1 - {\delta _{{{\mathbf{x}}_{{\rm{IRS}}}} - {{\mathbf{x}}^\prime }_{{\rm{IRS}}}}}} \right){{\left( {{\lambda _{{\rm{wave}}}}} \right)}^4}{\delta ^2} \sigma _{{\rm{d}}}^2}}{{{4^6}{\pi ^2}\left( {\max \left\{ {{{({h_{{\rm{IRS}}}})}^2},{{({h_{{\rm{BS}}}} - {h_{{\rm{IRS}}}})}^2}} \right\} + 3{{\left( {{R_{{\rm{co}}}}} \right)}^2} + \left\| {{{\mathbf{x}}_{{\rm{IRS}}}} - {{\mathbf{x}}_{\rm{U}}}} \right\|_2^2} \right)\left( {\max \left\{ {{{({h_{{\rm{IRS}}}})}^2},{{({h_{{\rm{BS}}}} - {h_{{\rm{IRS}}}})}^2}} \right\} + 3{{\left( {{R_{{\rm{co}}}}} \right)}^2} + \left\| {{{{\mathbf{x'}}}_{{\rm{IRS}}}} - {{\mathbf{x}}_{\rm{U}}}} \right\|_2^2} \right)}} \times $}
\end{equation*}
\begin{equation*}
\scalebox{.95}[1]{$\frac{{2d\max \left\{ {\left( {{\lambda _{\rm{U}}}\pi {{\left( {\frac{{{R_{{\rm{co}}}}}}{2} - d} \right)}^2} + \exp \left\{ { - {\lambda _{\rm{U}}}\pi {{\left( {\frac{{{R_{{\rm{co}}}}}}{2} + d} \right)}^2}} \right\} - 1} \right),0} \right\}\left( {1 - \exp \left\{ { - {\lambda _{{\rm{BS}}}}3\pi {{\left( {{R_{{\rm{co}}}}} \right)}^2}} \right\}} \right)}}{{\left( {1 - \exp \left\{ { - {\lambda _{\rm{U}}}\pi {{\left( {\frac{{{R_{{\rm{co}}}}}}{2} + d} \right)}^2}} \right\}} \right){\lambda _{{\rm{BS}}}}3\pi {{\left( {{R_{{\rm{co}}}}} \right)}^3}}} \times $}
\end{equation*}
\begin{equation*}
\scalebox{.95}[1]{$\left. {\left. {\max \left\{ {0,\frac{{1 - \exp \left\{ { - {\lambda _{\rm{U}}}\pi \left( {3{{\left( {{R_{{\rm{co}}}}} \right)}^2}} \right)} \right\} - {\lambda _{\rm{U}}}\pi \left( {4{{\left( {{R_{{\rm{co}}}}} \right)}^2}} \right)\exp \left\{ { - {\lambda _{\rm{U}}}\pi \left( {3{{\left( {{R_{{\rm{co}}}}} \right)}^2}} \right)} \right\}}}{{{{\left( {{\lambda _{\rm{U}}}\pi \left( {4{{\left( {{R_{{\rm{co}}}}} \right)}^2}} \right)} \right)}^2}}}} \right\}} \right|{n_{{\rm{IRS}}}}({{\mathbf{x}}_{\rm{U}}})} \right\} = $}
\end{equation*}
\begin{equation*}
\frac{\kappa }{{\kappa  + 1}} \times\frac{{{n_{{\rm{IRS}}}}({{\mathbf{x}}_{\rm{U}}})\left( {{n_{{\rm{IRS}}}}({{\mathbf{x}}_{\rm{U}}}) - 1} \right)\left( {1 - \exp \left\{ { - {\lambda _{{\rm{BS}}}}\pi {{\left( {{R_{{\rm{co}}}}} \right)}^2}} \right\}} \right)Q(Q - 1){{\left( {{\lambda _{{\rm{wave}}}}} \right)}^4}{\delta ^2} \sigma _{{\rm{d}}}^2}}{{{4^6}{\pi ^4}{{\left( {{R_{{\rm{co}}}}} \right)}^4}}} \times 
\end{equation*}
\begin{equation*}
\scalebox{1}[1]{$\frac{{2d\max \left\{ {\left( {{\lambda _{\rm{U}}}\pi {{\left( {\frac{{{R_{{\rm{co}}}}}}{2} - d} \right)}^2} + \exp \left\{ { - {\lambda _{\rm{U}}}\pi {{\left( {\frac{{{R_{{\rm{co}}}}}}{2} + d} \right)}^2}} \right\} - 1} \right),0} \right\}\left( {1 - \exp \left\{ { - {\lambda _{{\rm{BS}}}}3\pi {{\left( {{R_{{\rm{co}}}}} \right)}^2}} \right\}} \right)}}{{\left( {1 - \exp \left\{ { - {\lambda _{\rm{U}}}\pi {{\left( {\frac{{{R_{{\rm{co}}}}}}{2} + d} \right)}^2}} \right\}} \right){\lambda _{{\rm{BS}}}}3\pi {{\left( {{R_{{\rm{co}}}}} \right)}^3}}} \times $}
\end{equation*}
\begin{equation*}
\max \left\{ {0,\frac{{1 - \exp \left\{ { - {\lambda _{\rm{U}}}\pi \left( {3{{\left( {{R_{{\rm{co}}}}} \right)}^2}} \right)} \right\} - {\lambda _{\rm{U}}}\pi \left( {4{{\left( {{R_{{\rm{co}}}}} \right)}^2}} \right)\exp \left\{ { - {\lambda _{\rm{U}}}\pi \left( {3{{\left( {{R_{{\rm{co}}}}} \right)}^2}} \right)} \right\}}}{{{{\left( {{\lambda _{\rm{U}}}\pi \left( {4{{\left( {{R_{{\rm{co}}}}} \right)}^2}} \right)} \right)}^2}}}} \right\}
\end{equation*}
\begin{equation*}
{\left( {\int\limits_{{\cal C}\left( {{{\mathbf{x}}_{\rm{U}}},{R_{{\rm{co}}}}} \right)} {\frac{{d{\mathbf{x}}}}{{\left( {\max \left\{ {{{({h_{{\rm{IRS}}}})}^2},{{({h_{{\rm{BS}}}} - {h_{{\rm{IRS}}}})}^2}} \right\} + 3{{\left( {{R_{{\rm{co}}}}} \right)}^2} + \left\| {{\mathbf{x}} - {{\mathbf{x}}_{\rm{U}}}} \right\|_2^2} \right)}}} } \right)^2} = 
\end{equation*}
\begin{equation*}
\frac{\kappa }{{\kappa  + 1}} \times\frac{{{n_{{\rm{IRS}}}}({{\mathbf{x}}_{\rm{U}}})\left( {{n_{{\rm{IRS}}}}({{\mathbf{x}}_{\rm{U}}}) - 1} \right)\left( {1 - \exp \left\{ { - {\lambda _{{\rm{BS}}}}\pi {{\left( {{R_{{\rm{co}}}}} \right)}^2}} \right\}} \right)Q(Q - 1){{\left( {{\lambda _{{\rm{wave}}}}} \right)}^4}{\delta ^2} \sigma _{{\rm{d}}}^2}}{{{4^6}{\pi ^2}{{\left( {{R_{{\rm{co}}}}} \right)}^4}}} \times 
\end{equation*}
\begin{equation*}
\scalebox{1}[1]{$\frac{{2d\max \left\{ {\left( {{\lambda _{\rm{U}}}\pi {{\left( {\frac{{{R_{{\rm{co}}}}}}{2} - d} \right)}^2} + \exp \left\{ { - {\lambda _{\rm{U}}}\pi {{\left( {\frac{{{R_{{\rm{co}}}}}}{2} + d} \right)}^2}} \right\} - 1} \right),0} \right\}\left( {1 - \exp \left\{ { - {\lambda _{{\rm{BS}}}}3\pi {{\left( {{R_{{\rm{co}}}}} \right)}^2}} \right\}} \right)}}{{\left( {1 - \exp \left\{ { - {\lambda _{\rm{U}}}\pi {{\left( {\frac{{{R_{{\rm{co}}}}}}{2} + d} \right)}^2}} \right\}} \right){\lambda _{{\rm{BS}}}}3\pi {{\left( {{R_{{\rm{co}}}}} \right)}^3}}} \times $}
\end{equation*}
\begin{equation*}
\max \left\{ {0,\frac{{1 - \exp \left\{ { - {\lambda _{\rm{U}}}\pi \left( {3{{\left( {{R_{{\rm{co}}}}} \right)}^2}} \right)} \right\} - {\lambda _{\rm{U}}}\pi \left( {4{{\left( {{R_{{\rm{co}}}}} \right)}^2}} \right)\exp \left\{ { - {\lambda _{\rm{U}}}\pi \left( {3{{\left( {{R_{{\rm{co}}}}} \right)}^2}} \right)} \right\}}}{{{{\left( {{\lambda _{\rm{U}}}\pi \left( {4{{\left( {{R_{{\rm{co}}}}} \right)}^2}} \right)} \right)}^2}}}} \right\} \times 
\end{equation*}
\begin{equation*}
{\left( {\ln \left( {\frac{{\max \left\{ {{{({h_{{\rm{IRS}}}})}^2},{{({h_{{\rm{BS}}}} - {h_{{\rm{IRS}}}})}^2}} \right\} + 4{{\left( {{R_{{\rm{co}}}}} \right)}^2}}}{{\max \left\{ {{{({h_{{\rm{IRS}}}})}^2},{{({h_{{\rm{BS}}}} - {h_{{\rm{IRS}}}})}^2}} \right\} + 3{{\left( {{R_{{\rm{co}}}}} \right)}^2}}}} \right)} \right)^2},
\end{equation*}
then, we conclude:
\begin{equation*}
E\left\{ \frac{\kappa }{{\kappa  + 1}} \times{\frac{{{n_{{\rm{IRS}}}}({{\mathbf{x}}_{\rm{U}}})\left( {{n_{{\rm{IRS}}}}({{\mathbf{x}}_{\rm{U}}}) - 1} \right)\left( {1 - \exp \left\{ { - {\lambda _{{\rm{BS}}}}\pi {{\left( {{R_{{\rm{co}}}}} \right)}^2}} \right\}} \right)Q(Q - 1){{\left( {{\lambda _{{\rm{wave}}}}} \right)}^4}{\delta ^2} \sigma _{{\rm{d}}}^2}}{{{4^6}{\pi ^2}{{\left( {{R_{{\rm{co}}}}} \right)}^4}}} \times } \right.
\end{equation*}
\begin{equation*}
\scalebox{1}[1]{$\frac{{2d\max \left\{ {\left( {{\lambda _{\rm{U}}}\pi {{\left( {\frac{{{R_{{\rm{co}}}}}}{2} - d} \right)}^2} + \exp \left\{ { - {\lambda _{\rm{U}}}\pi {{\left( {\frac{{{R_{{\rm{co}}}}}}{2} + d} \right)}^2}} \right\} - 1} \right),0} \right\}\left( {1 - \exp \left\{ { - {\lambda _{{\rm{BS}}}}3\pi {{\left( {{R_{{\rm{co}}}}} \right)}^2}} \right\}} \right)}}{{\left( {1 - \exp \left\{ { - {\lambda _{\rm{U}}}\pi {{\left( {\frac{{{R_{{\rm{co}}}}}}{2} + d} \right)}^2}} \right\}} \right){\lambda _{{\rm{BS}}}}3\pi {{\left( {{R_{{\rm{co}}}}} \right)}^3}}} \times $}
\end{equation*}
\begin{equation*}
\max \left\{ {0,\frac{{1 - \exp \left\{ { - {\lambda _{\rm{U}}}\pi \left( {3{{\left( {{R_{{\rm{co}}}}} \right)}^2}} \right)} \right\} - {\lambda _{\rm{U}}}\pi \left( {4{{\left( {{R_{{\rm{co}}}}} \right)}^2}} \right)\exp \left\{ { - {\lambda _{\rm{U}}}\pi \left( {3{{\left( {{R_{{\rm{co}}}}} \right)}^2}} \right)} \right\}}}{{{{\left( {{\lambda _{\rm{U}}}\pi \left( {4{{\left( {{R_{{\rm{co}}}}} \right)}^2}} \right)} \right)}^2}}}} \right\} \times 
\end{equation*}
\begin{equation*}
\left. {{{\left( {\ln \left( {\frac{{\max \left\{ {{{({h_{{\rm{IRS}}}})}^2},{{({h_{{\rm{BS}}}} - {h_{{\rm{IRS}}}})}^2}} \right\} + 4{{\left( {{R_{{\rm{co}}}}} \right)}^2}}}{{\max \left\{ {{{({h_{{\rm{IRS}}}})}^2},{{({h_{{\rm{BS}}}} - {h_{{\rm{IRS}}}})}^2}} \right\} + 3{{\left( {{R_{{\rm{co}}}}} \right)}^2}}}} \right)} \right)}^2}} \right\} = 
\end{equation*}
\begin{equation*}
\frac{\kappa }{{\kappa  + 1}} \times\frac{{{{\left( {{\lambda _{{\rm{IRS}}}}\pi {{\left( {{R_{{\rm{co}}}}} \right)}^2}} \right)}^2}\left( {1 - \exp \left\{ { - {\lambda _{{\rm{BS}}}}\pi {{\left( {{R_{{\rm{co}}}}} \right)}^2}} \right\}} \right)Q(Q - 1){{\left( {{\lambda _{{\rm{wave}}}}} \right)}^4}{\delta ^2} \sigma _{{\rm{d}}}^2}}{{{4^6}{\pi ^2}{{\left( {{R_{{\rm{co}}}}} \right)}^4}}} \times 
\end{equation*}
\begin{equation*}
\scalebox{1}[1]{$\frac{{2d\max \left\{ {\left( {{\lambda _{\rm{U}}}\pi {{\left( {\frac{{{R_{{\rm{co}}}}}}{2} - d} \right)}^2} + \exp \left\{ { - {\lambda _{\rm{U}}}\pi {{\left( {\frac{{{R_{{\rm{co}}}}}}{2} + d} \right)}^2}} \right\} - 1} \right),0} \right\}\left( {1 - \exp \left\{ { - {\lambda _{{\rm{BS}}}}3\pi {{\left( {{R_{{\rm{co}}}}} \right)}^2}} \right\}} \right)}}{{\left( {1 - \exp \left\{ { - {\lambda _{\rm{U}}}\pi {{\left( {\frac{{{R_{{\rm{co}}}}}}{2} + d} \right)}^2}} \right\}} \right){\lambda _{{\rm{BS}}}}3\pi {{\left( {{R_{{\rm{co}}}}} \right)}^3}}} \times $}
\end{equation*}
\begin{equation*}
\max \left\{ {0,\frac{{1 - \exp \left\{ { - {\lambda _{\rm{U}}}\pi \left( {3{{\left( {{R_{{\rm{co}}}}} \right)}^2}} \right)} \right\} - {\lambda _{\rm{U}}}\pi \left( {4{{\left( {{R_{{\rm{co}}}}} \right)}^2}} \right)\exp \left\{ { - {\lambda _{\rm{U}}}\pi \left( {3{{\left( {{R_{{\rm{co}}}}} \right)}^2}} \right)} \right\}}}{{{{\left( {{\lambda _{\rm{U}}}\pi \left( {4{{\left( {{R_{{\rm{co}}}}} \right)}^2}} \right)} \right)}^2}}}} \right\} \times 
\end{equation*}
\begin{equation}
{\left( {\ln \left( {\frac{{\max \left\{ {{{({h_{{\rm{IRS}}}})}^2},{{({h_{{\rm{BS}}}} - {h_{{\rm{IRS}}}})}^2}} \right\} + 4{{\left( {{R_{{\rm{co}}}}} \right)}^2}}}{{\max \left\{ {{{({h_{{\rm{IRS}}}})}^2},{{({h_{{\rm{BS}}}} - {h_{{\rm{IRS}}}})}^2}} \right\} + 3{{\left( {{R_{{\rm{co}}}}} \right)}^2}}}} \right)} \right)^2}.
\end{equation}
For (\ref{t44}), we have:
\begin{equation*}
\sum\limits_{{{\mathbf{x}}_{{\rm{IRS}}}} \in {\Psi _{{\rm{IRS}}}}({{\mathbf{x}}_{\rm{U}}})} {\begin{array}{*{20}{c}}
  {}&{} 
\end{array}\sum\limits_{{{{\mathbf{x'}}}_{{\rm{BS}}}} \in {\Psi _{{\rm{BS}}}}({{\mathbf{x}}_{{\rm{IRS}}}})} {\begin{array}{*{20}{c}}
  {}&{} 
\end{array}\sum\limits_{{{{\mathbf{x'}}}_{\rm{U}}} \in {\Psi _{\rm{U}}}({{{\mathbf{x'}}}_{{\rm{BS}}}}):{{{\mathbf{x'}}}_{\rm{U}}} \ne {{\mathbf{x}}_{\rm{U}}},{{{\mathbf{\tilde x}}}_{{\rm{BS}}}}({{{\mathbf{x'}}}_{\rm{U}}}) = {{{\mathbf{x'}}}_{{\rm{BS}}}}} {} } } 
\end{equation*}
\begin{equation*}
\frac{{Q{{\left( {{\lambda _{{\rm{wave}}}}} \right)}^4}{\delta ^2} \sigma _{{\rm{d}}}^2}}{{{{\left( {4\pi } \right)}^4}\left( {{{({h_{{\rm{IRS}}}})}^2} + \left\| {{{\mathbf{x}}_{{\rm{IRS}}}} - {{\mathbf{x}}_{\rm{U}}}} \right\|_2^2} \right)\left( {{{({h_{{\rm{BS}}}} - {h_{{\rm{IRS}}}})}^2} + \left\| {{{{\mathbf{x'}}}_{{\rm{BS}}}} - {{\mathbf{x}}_{{\rm{IRS}}}}} \right\|_2^2} \right)}} = 
\end{equation*}
\begin{equation*}
\sum\limits_{{{\mathbf{x}}_{{\rm{IRS}}}} \in {\Psi _{{\rm{IRS}}}}({{\mathbf{x}}_{\rm{U}}})} {\begin{array}{*{20}{c}}
  {}&{} 
\end{array}\sum\limits_{{{{\mathbf{x'}}}_{{\rm{BS}}}} \in {\Psi _{{\rm{BS}}}}({{\mathbf{x}}_{{\rm{IRS}}}})} {\begin{array}{*{20}{c}}
  {}&{} 
\end{array}\sum\limits_{{{{\mathbf{x'}}}_{\rm{U}}} \in {\Psi _{\rm{U}}}({{{\mathbf{x'}}}_{{\rm{BS}}}}):{{{\mathbf{x'}}}_{\rm{U}}} \ne {{\mathbf{x}}_{\rm{U}}}} {} } } 
\end{equation*}
\begin{equation*}
\frac{{\left( {1 - {\pmb{\mathbb{I}}}\left( {\left\| {{{{\mathbf{\tilde x}}}_{{\rm{BS}}}}({{{\mathbf{x'}}}_{\rm{U}}}) - {{{\mathbf{x'}}}_{{\rm{BS}}}}} \right\|} \right)} \right)Q{{\left( {{\lambda _{{\rm{wave}}}}} \right)}^4}{\delta ^2} \sigma _{{\rm{d}}}^2}}{{{{\left( {4\pi } \right)}^4}\left( {{{({h_{{\rm{IRS}}}})}^2} + \left\| {{{\mathbf{x}}_{{\rm{IRS}}}} - {{\mathbf{x}}_{\rm{U}}}} \right\|_2^2} \right)\left( {{{({h_{{\rm{BS}}}} - {h_{{\rm{IRS}}}})}^2} + \left\| {{{{\mathbf{x'}}}_{{\rm{BS}}}} - {{\mathbf{x}}_{{\rm{IRS}}}}} \right\|_2^2} \right)}}.
\end{equation*}
Then, we obtain:
\begin{equation*}
E\left\{ {\sum\limits_{{{\mathbf{x}}_{{\rm{IRS}}}} \in {\Psi _{{\rm{IRS}}}}({{\mathbf{x}}_{\rm{U}}})} {\begin{array}{*{20}{c}}
  {}&{} 
\end{array}\sum\limits_{{{{\mathbf{x'}}}_{{\rm{BS}}}} \in {\Psi _{{\rm{BS}}}}({{\mathbf{x}}_{{\rm{IRS}}}})} {\begin{array}{*{20}{c}}
  {}&{} 
\end{array}\sum\limits_{{{{\mathbf{x'}}}_{\rm{U}}} \in {\Psi _{\rm{U}}}({{{\mathbf{x'}}}_{{\rm{BS}}}}):{{{\mathbf{x'}}}_{\rm{U}}} \ne {{\mathbf{x}}_{\rm{U}}}} {} } } } \right.
\end{equation*}
\begin{equation*}
\left. {\left. {\frac{{\left( {1 -{\pmb{\mathbb{I}}}\left( {\left\| {{{{\mathbf{\tilde x}}}_{{\rm{BS}}}}({{{\mathbf{x'}}}_{\rm{U}}}) - {{{\mathbf{x'}}}_{{\rm{BS}}}}} \right\|} \right)} \right)Q{{\left( {{\lambda _{{\rm{wave}}}}} \right)}^4}{\delta ^2} \sigma _{{\rm{d}}}^2}}{{{{\left( {4\pi } \right)}^4}\left( {{{({h_{{\rm{IRS}}}})}^2} + \left\| {{{\mathbf{x}}_{{\rm{IRS}}}} - {{\mathbf{x}}_{\rm{U}}}} \right\|_2^2} \right)\left( {{{({h_{{\rm{BS}}}} - {h_{{\rm{IRS}}}})}^2} + \left\| {{{{\mathbf{x'}}}_{{\rm{BS}}}} - {{\mathbf{x}}_{{\rm{IRS}}}}} \right\|_2^2} \right)}}} \right|{\Phi _{\rm{U}}},{\Phi _{{\rm{BS}}}},{\Phi _{{\rm{IRS}}}}} \right\} = 
\end{equation*}
\begin{equation*}
\sum\limits_{{{\mathbf{x}}_{{\rm{IRS}}}} \in {\Psi _{{\rm{IRS}}}}({{\mathbf{x}}_{\rm{U}}})} {\begin{array}{*{20}{c}}
  {}&{} 
\end{array}\sum\limits_{{{{\mathbf{x'}}}_{{\rm{BS}}}} \in {\Psi _{{\rm{BS}}}}({{\mathbf{x}}_{{\rm{IRS}}}})} {\begin{array}{*{20}{c}}
  {}&{} 
\end{array}\sum\limits_{{{{\mathbf{x'}}}_{\rm{U}}} \in {\Psi _{\rm{U}}}({{{\mathbf{x'}}}_{{\rm{BS}}}}):{{{\mathbf{x'}}}_{\rm{U}}} \ne {{\mathbf{x}}_{\rm{U}}}} {} } } 
\end{equation*}
\begin{equation*}
\frac{{{\pmb{\mathbb{I}}}\left( {{n_{{\rm{BS}}}}({{{\mathbf{x'}}}_{\rm{U}}})} \right)Q{{\left( {{\lambda _{{\rm{wave}}}}} \right)}^4}{\delta ^2} \sigma _{{\rm{d}}}^2}}{{{n_{{\rm{BS}}}}({{{\mathbf{x'}}}_{\rm{U}}}){{\left( {4\pi } \right)}^4}\left( {{{({h_{{\rm{IRS}}}})}^2} + \left\| {{{\mathbf{x}}_{{\rm{IRS}}}} - {{\mathbf{x}}_{\rm{U}}}} \right\|_2^2} \right)\left( {{{({h_{{\rm{BS}}}} - {h_{{\rm{IRS}}}})}^2} + \left\| {{{{\mathbf{x'}}}_{{\rm{BS}}}} - {{\mathbf{x}}_{{\rm{IRS}}}}} \right\|_2^2} \right)}} \ge
\end{equation*}
\begin{equation*}
\sum\limits_{{{\mathbf{x}}_{{\rm{IRS}}}} \in {\Psi _{{\rm{IRS}}}}({{\mathbf{x}}_{\rm{U}}})} {\begin{array}{*{20}{c}}
  {}&{} 
\end{array}\sum\limits_{{{{\mathbf{x'}}}_{{\rm{BS}}}} \in {\Psi _{{\rm{BS}}}}({{\mathbf{x}}_{{\rm{IRS}}}})} {\begin{array}{*{20}{c}}
  {}&{} 
\end{array}\sum\limits_{{{{\mathbf{x'}}}_{\rm{U}}} \in {\Psi _{\rm{U}}}({{{\mathbf{x'}}}_{{\rm{BS}}}})\bigcap {{{\tilde {\tilde {\Psi }}}_{\rm{U}}}({{\mathbf{x}}_{\rm{U}}})} :{{{\mathbf{x'}}}_{\rm{U}}} \ne {{\mathbf{x}}_{\rm{U}}}} {} } } 
\end{equation*}
\begin{equation*}
\frac{{{\pmb{\mathbb{I}}}\left( {{n_{{\rm{BS}}}}({{{\mathbf{x'}}}_{\rm{U}}})} \right)Q{{\left( {{\lambda _{{\rm{wave}}}}} \right)}^4}{\delta ^2} \sigma _{{\rm{d}}}^2}}{{{n_{{\rm{BS}}}}({{{\mathbf{x'}}}_{\rm{U}}}){{\left( {4\pi } \right)}^4}\left( {{{({h_{{\rm{IRS}}}})}^2} + \left\| {{{\mathbf{x}}_{{\rm{IRS}}}} - {{\mathbf{x}}_{\rm{U}}}} \right\|_2^2} \right)\left( {{{({h_{{\rm{BS}}}} - {h_{{\rm{IRS}}}})}^2} + \left\| {{{{\mathbf{x'}}}_{{\rm{BS}}}} - {{\mathbf{x}}_{{\rm{IRS}}}}} \right\|_2^2} \right)}} \ge
\end{equation*}
\begin{equation*}
\sum\limits_{{{\mathbf{x}}_{{\rm{IRS}}}} \in {\Psi _{{\rm{IRS}}}}({{\mathbf{x}}_{\rm{U}}})} {\begin{array}{*{20}{c}}
  {}&{} 
\end{array}\sum\limits_{{{{\mathbf{x'}}}_{{\rm{BS}}}} \in {\Psi _{{\rm{BS}}}}({{\mathbf{x}}_{{\rm{IRS}}}})} {\begin{array}{*{20}{c}}
  {}&{} 
\end{array}\sum\limits_{{{{\mathbf{x'}}}_{\rm{U}}} \in {\Psi _{\rm{U}}}({{{\mathbf{x'}}}_{{\rm{BS}}}})\bigcap {{{{\tilde {\tilde {\Psi }}}}_{\rm{U}}}({{\mathbf{x}}_{\rm{U}}})} :{{{\mathbf{x'}}}_{\rm{U}}} \ne {{\mathbf{x}}_{\rm{U}}}} {} } } 
\end{equation*}
\begin{equation*}
\frac{{{\pmb{\mathbb{I}}}\left( {\prod\limits_{{{{\mathbf{x'}}}_{\rm{U}}} \in {\Psi _{\rm{U}}}({{{\mathbf{x'}}}_{{\rm{BS}}}})\bigcap {{{\tilde {\tilde {\Psi}} }_{\rm{U}}}({{\mathbf{x}}_{\rm{U}}})} :{{{\mathbf{x'}}}_{\rm{U}}} \ne {{\mathbf{x}}_{\rm{U}}}} {{n_{{\rm{BS}}}}({{{\mathbf{x'}}}_{\rm{U}}})} } \right)Q{{\left( {{\lambda _{{\rm{wave}}}}} \right)}^4}{\delta ^2} \sigma _{{\rm{d}}}^2}}{{{n_{{\rm{BS}}}}({{{\mathbf{x'}}}_{\rm{U}}}){{\left( {4\pi } \right)}^4}\left( {{{({h_{{\rm{IRS}}}})}^2} + \left\| {{{\mathbf{x}}_{{\rm{IRS}}}} - {{\mathbf{x}}_{\rm{U}}}} \right\|_2^2} \right)\left( {{{({h_{{\rm{BS}}}} - {h_{{\rm{IRS}}}})}^2} + \left\| {{{{\mathbf{x'}}}_{{\rm{BS}}}} - {{\mathbf{x}}_{{\rm{IRS}}}}} \right\|_2^2} \right)}} \ge
\end{equation*}
\begin{equation*}
\sum\limits_{{{\mathbf{x}}_{{\rm{IRS}}}} \in {\Psi _{{\rm{IRS}}}}({{\mathbf{x}}_{\rm{U}}})} {\begin{array}{*{20}{c}}
  {}&{} 
\end{array}\sum\limits_{{{{\mathbf{x'}}}_{{\rm{BS}}}} \in {\Psi _{{\rm{BS}}}}({{\mathbf{x}}_{{\rm{IRS}}}})} {} } 
\end{equation*}
\begin{equation*}
\frac{{{\pmb{\mathbb{I}}}\left( {{N_{{\rm{BS}}}}\left( {{\cal C}\left( {{{\mathbf{x}}_{\rm{U}}},b} \right)} \right)} \right)\left( {{N_{\rm{U}}}\left( {{\cal C}\left( {{{\mathbf{x}}_{\rm{U}}},b} \right)\bigcap {{\cal C}\left( {{{{\mathbf{x'}}}_{{\rm{BS}}}},{R_{{\rm{co}}}}} \right)} } \right) - 1} \right)Q{{\left( {{\lambda _{{\rm{wave}}}}} \right)}^4}{\delta ^2} \sigma _{{\rm{d}}}^2}}{{{{\tilde n}_{{\rm{BS}}}}({{\mathbf{x}}_{\rm{U}}}){{\left( {4\pi } \right)}^4}\left( {{{({h_{{\rm{IRS}}}})}^2} + \left\| {{{\mathbf{x}}_{{\rm{IRS}}}} - {{\mathbf{x}}_{\rm{U}}}} \right\|_2^2} \right)\left( {{{({h_{{\rm{BS}}}} - {h_{{\rm{IRS}}}})}^2} + \left\| {{{{\mathbf{x'}}}_{{\rm{BS}}}} - {{\mathbf{x}}_{{\rm{IRS}}}}} \right\|_2^2} \right)}} \ge
\end{equation*}
\begin{equation*}
\sum\limits_{{{\mathbf{x}}_{{\rm{IRS}}}} \in {\Psi _{{\rm{IRS}}}}({{\mathbf{x}}_{\rm{U}}})} {\begin{array}{*{20}{c}}
  {}&{} 
\end{array}\sum\limits_{{{{\mathbf{x'}}}_{{\rm{BS}}}} \in {\Psi _{{\rm{BS}}}}({{\mathbf{x}}_{{\rm{IRS}}}})} {} } 
\end{equation*}
\begin{equation*}
   \scalebox{1}[1]{$\frac{{{\pmb{\mathbb{I}}}\left( {{N_{{\rm{BS}}}}\left( {{\cal C}\left( {{{\mathbf{x}}_{\rm{U}}},b} \right)} \right)} \right) \times \left( {{N_{\rm{U}}}\left( {{\cal C}\left( {{{\mathbf{x}}_{\rm{U}}},b} \right)\bigcap {{\cal C}\left( {{{\mathbf{x}}_{\rm{U}}} + \frac{{{R_{{\rm{co}}}}}}{{\left\| {{{{\mathbf{x'}}}_{{\rm{BS}}}} - {{\mathbf{x}}_{\rm{U}}}} \right\|}}\left( {{{{\mathbf{x'}}}_{{\rm{BS}}}} - {{\mathbf{x}}_{\rm{U}}}} \right),{R_{{\rm{co}}}}} \right)} } \right) - 1} \right)Q{{\left( {{\lambda _{{\rm{wave}}}}} \right)}^4}{\delta ^2} \sigma _{{\rm{d}}}^2}}{{\left( {{N_{{\rm{BS}}}}\left( {{\cal C}\left( {{{\mathbf{x}}_{\rm{U}}},2{R_{{\rm{co}}}}} \right) - {\cal C}\left( {{{\mathbf{x}}_{\rm{U}}},b} \right)} \right) + {N_{{\rm{BS}}}}\left( {{\cal C}\left( {{{\mathbf{x}}_{\rm{U}}},b} \right)} \right)} \right){{\left( {4\pi } \right)}^4}\left( {{{({h_{{\rm{IRS}}}})}^2} + \left\| {{{\mathbf{x}}_{{\rm{IRS}}}} - {{\mathbf{x}}_{\rm{U}}}} \right\|_2^2} \right)\left( {{{({h_{{\rm{BS}}}} - {h_{{\rm{IRS}}}})}^2} + \left\| {{{{\mathbf{x'}}}_{{\rm{BS}}}} - {{\mathbf{x}}_{{\rm{IRS}}}}} \right\|_2^2} \right)}} \ge$}
\end{equation*}
\begin{equation*}
\sum\limits_{{{\mathbf{x}}_{{\rm{IRS}}}} \in {{\tilde {\tilde {\Psi}} }_{{\rm{IRS}}}}({{\mathbf{x}}_{\rm{U}}})} {\begin{array}{*{20}{c}}
  {}&{} 
\end{array}\sum\limits_{{{{\mathbf{x'}}}_{{\rm{BS}}}} \in {\Psi _{{\rm{BS}}}}({{\mathbf{x}}_{{\rm{IRS}}}})} {} } 
\end{equation*}
\begin{equation*}
   \scalebox{1}[1]{$\frac{{{\pmb{\mathbb{I}}}\left( {{N_{{\rm{BS}}}}\left( {{\cal C}\left( {{{\mathbf{x}}_{\rm{U}}},b} \right)} \right)} \right) \times \left( {{N_{\rm{U}}}\left( {{\cal C}\left( {{{\mathbf{x}}_{\rm{U}}},b} \right)\bigcap {{\cal C}\left( {{{\mathbf{x}}_{\rm{U}}} + \frac{{{R_{{\rm{co}}}}}}{{\left\| {{{{\mathbf{x'}}}_{{\rm{BS}}}} - {{\mathbf{x}}_{\rm{U}}}} \right\|}}\left( {{{{\mathbf{x'}}}_{{\rm{BS}}}} - {{\mathbf{x}}_{\rm{U}}}} \right),{R_{{\rm{co}}}}} \right)} } \right) - 1} \right)Q{{\left( {{\lambda _{{\rm{wave}}}}} \right)}^4}{\delta ^2} \sigma _{{\rm{d}}}^2}}{{\left( {{N_{{\rm{BS}}}}\left( {{\cal C}\left( {{{\mathbf{x}}_{\rm{U}}},2{R_{{\rm{co}}}}} \right) - {\cal C}\left( {{{\mathbf{x}}_{\rm{U}}},b} \right)} \right) + {N_{{\rm{BS}}}}\left( {{\cal C}\left( {{{\mathbf{x}}_{\rm{U}}},b} \right)} \right)} \right){{\left( {4\pi } \right)}^4}\left( {{{({h_{{\rm{IRS}}}})}^2} + \left\| {{{\mathbf{x}}_{{\rm{IRS}}}} - {{\mathbf{x}}_{\rm{U}}}} \right\|_2^2} \right)\left( {{{({h_{{\rm{BS}}}} - {h_{{\rm{IRS}}}})}^2} + {{\left( {{R_{{\rm{co}}}}} \right)}^2}} \right)}} \ge$}
\end{equation*}
\begin{equation*}
\sum\limits_{{{\mathbf{x}}_{{\rm{IRS}}}} \in {{{\tilde {\tilde {\Psi}} }}_{{\rm{IRS}}}}({{\mathbf{x}}_{\rm{U}}})} {\begin{array}{*{20}{c}}
  {}&{} 
\end{array}\sum\limits_{{{{\mathbf{x'}}}_{{\rm{BS}}}} \in {\Phi _{{\rm{BS}}}}\bigcap {C\left( {{{\mathbf{x}}_{\rm{U}}},b} \right)} } {} } 
\end{equation*}
\begin{equation*}
   \scalebox{1}[1]{$\frac{{{\pmb{\mathbb{I}}}\left( {{N_{{\rm{BS}}}}\left( {{\cal C}\left( {{{\mathbf{x}}_{\rm{U}}},b} \right)} \right)} \right) \times \left( {{N_{\rm{U}}}\left( {{\cal C}\left( {{{\mathbf{x}}_{\rm{U}}},b} \right)\bigcap {{\cal C}\left( {{{\mathbf{x}}_{\rm{U}}} + \frac{{{R_{{\rm{co}}}}}}{{\left\| {{{{\mathbf{x'}}}_{{\rm{BS}}}} - {{\mathbf{x}}_{\rm{U}}}} \right\|}}\left( {{{{\mathbf{x'}}}_{{\rm{BS}}}} - {{\mathbf{x}}_{\rm{U}}}} \right),{R_{{\rm{co}}}}} \right)} } \right) - 1} \right)Q{{\left( {{\lambda _{{\rm{wave}}}}} \right)}^4}{\delta ^2} \sigma _{{\rm{d}}}^2}}{{\left( {{N_{{\rm{BS}}}}\left( {{\cal C}\left( {{{\mathbf{x}}_{\rm{U}}},2{R_{{\rm{co}}}}} \right) - {\cal C}\left( {{{\mathbf{x}}_{\rm{U}}},b} \right)} \right) + {N_{{\rm{BS}}}}\left( {{\cal C}\left( {{{\mathbf{x}}_{\rm{U}}},b} \right)} \right)} \right){{\left( {4\pi } \right)}^4}\left( {{{({h_{{\rm{IRS}}}})}^2} + \left\| {{{\mathbf{x}}_{{\rm{IRS}}}} - {{\mathbf{x}}_{\rm{U}}}} \right\|_2^2} \right)\left( {{{({h_{{\rm{BS}}}} - {h_{{\rm{IRS}}}})}^2} + {{\left( {{R_{{\rm{co}}}}} \right)}^2}} \right)}} = $}
\end{equation*}
\begin{equation*}
   \scalebox{.9}[1]{$\sum\limits_{{{\mathbf{x}}_{{\rm{IRS}}}} \in {{{\tilde {\tilde {\Psi}} } }_{{\rm{IRS}}}}({{\mathbf{x}}_{\rm{U}}})} {\frac{{{\pmb{\mathbb{I}}}\left( {{N_{{\rm{BS}}}}\left( {{\cal C}\left( {{{\mathbf{x}}_{\rm{U}}},b} \right)} \right)} \right){N_{{\rm{BS}}}}\left( {{\cal C}\left( {{{\mathbf{x}}_{\rm{U}}},b} \right)} \right)\left( {{N_{\rm{U}}}\left( {{\cal C}\left( {{{\mathbf{x}}_{\rm{U}}},b} \right)\bigcap {{\cal C}\left( {{{\mathbf{x}}_{\rm{U}}} + \frac{{{R_{{\rm{co}}}}}}{{\left\| {{{{\mathbf{x'}}}_{{\rm{BS}}}} - {{\mathbf{x}}_{\rm{U}}}} \right\|}}\left( {{{{\mathbf{x'}}}_{{\rm{BS}}}} - {{\mathbf{x}}_{\rm{U}}}} \right),{R_{{\rm{co}}}}} \right)} } \right) - 1} \right)Q{{\left( {{\lambda _{{\rm{wave}}}}} \right)}^4}{\delta ^2} \sigma _{{\rm{d}}}^2}}{{\left( {{N_{{\rm{BS}}}}\left( {{\cal C}\left( {{{\mathbf{x}}_{\rm{U}}},2{R_{{\rm{co}}}}} \right) - {\cal C}\left( {{{\mathbf{x}}_{\rm{U}}},b} \right)} \right) + {N_{{\rm{BS}}}}\left( {{\cal C}\left( {{{\mathbf{x}}_{\rm{U}}},b} \right)} \right)} \right){{\left( {4\pi } \right)}^4}\left( {{{({h_{{\rm{IRS}}}})}^2} + \left\| {{{\mathbf{x}}_{{\rm{IRS}}}} - {{\mathbf{x}}_{\rm{U}}}} \right\|_2^2} \right)\left( {{{({h_{{\rm{BS}}}} - {h_{{\rm{IRS}}}})}^2} + {{\left( {{R_{{\rm{co}}}}} \right)}^2}} \right)}}}  = $}
\end{equation*}
\begin{equation*}
   \scalebox{1}[1]{$\sum\limits_{{{\mathbf{x}}_{{\rm{IRS}}}} \in {{{\tilde {\tilde {\Psi}} } }_{{\rm{IRS}}}}({{\mathbf{x}}_{\rm{U}}})} {\frac{{{\pmb{\mathbb{I}}}\left( {{N_{{\rm{BS}}}}\left( {{\cal C}\left( {{{\mathbf{x}}_{\rm{U}}},b} \right)} \right)} \right)\left( {{N_{\rm{U}}}\left( {{\cal C}\left( {{{\mathbf{x}}_{\rm{U}}},b} \right)\bigcap {{\cal C}\left( {{{\mathbf{x}}_{\rm{U}}} + \frac{{{R_{{\rm{co}}}}}}{{\left\| {{{{\mathbf{x'}}}_{{\rm{BS}}}} - {{\mathbf{x}}_{\rm{U}}}} \right\|}}\left( {{{{\mathbf{x'}}}_{{\rm{BS}}}} - {{\mathbf{x}}_{\rm{U}}}} \right),{R_{{\rm{co}}}}} \right)} } \right) - 1} \right)Q{{\left( {{\lambda _{{\rm{wave}}}}} \right)}^4}{\delta ^2} \sigma _{{\rm{d}}}^2}}{{\left( {{N_{{\rm{BS}}}}\left( {{\cal C}\left( {{{\mathbf{x}}_{\rm{U}}},2{R_{{\rm{co}}}}} \right) - {\cal C}\left( {{{\mathbf{x}}_{\rm{U}}},b} \right)} \right) + 1} \right){{\left( {4\pi } \right)}^4}\left( {{{({h_{{\rm{IRS}}}})}^2} + \left\| {{{\mathbf{x}}_{{\rm{IRS}}}} - {{\mathbf{x}}_{\rm{U}}}} \right\|_2^2} \right)\left( {{{({h_{{\rm{BS}}}} - {h_{{\rm{IRS}}}})}^2} + {{\left( {{R_{{\rm{co}}}}} \right)}^2}} \right)}}} ,$}
\end{equation*}
where:
\begin{equation}
{{\tilde {\tilde{ \Psi}} }_{\rm{IRS}}}({{\mathbf{x}}_{\rm{U}}}) = {\Phi _{\rm{IRS}}}\bigcap {{\cal C}\left( {{{\mathbf{x}}_{\rm{U}}},{R_{{\rm{co}}}} - b} \right)} .
\end{equation}
Then, we obtain:
\begin{equation*}
   \scalebox{.85}[1]{$E\left\{ {\left. {\sum\limits_{{{\mathbf{x}}_{{\rm{IRS}}}} \in {{\tilde {\tilde {\Psi}} }_{{\rm{IRS}}}}({{\mathbf{x}}_{\rm{U}}})} {\frac{{{\pmb{\mathbb{I}}}\left( {{N_{{\rm{BS}}}}\left( {{\cal C}\left( {{{\mathbf{x}}_{\rm{U}}},b} \right)} \right)} \right)\left( {{N_{\rm{U}}}\left( {{\cal C}\left( {{{\mathbf{x}}_{\rm{U}}},b} \right)\bigcap {{\cal C}\left( {{{\mathbf{x}}_{\rm{U}}} + \frac{{{R_{{\rm{co}}}}}}{{\left\| {{{{\mathbf{x'}}}_{{\rm{BS}}}} - {{\mathbf{x}}_{\rm{U}}}} \right\|}}\left( {{{{\mathbf{x'}}}_{{\rm{BS}}}} - {{\mathbf{x}}_{\rm{U}}}} \right),{R_{{\rm{co}}}}} \right)} } \right) - 1} \right)Q{{\left( {{\lambda _{{\rm{wave}}}}} \right)}^4}{\delta ^2} \sigma _{{\rm{d}}}^2}}{{\left( {{N_{{\rm{BS}}}}\left( {{\cal C}\left( {{{\mathbf{x}}_{\rm{U}}},2{R_{{\rm{co}}}}} \right) - {\cal C}\left( {{{\mathbf{x}}_{\rm{U}}},b} \right)} \right) + 1} \right){{\left( {4\pi } \right)}^4}\left( {{{({h_{{\rm{IRS}}}})}^2} + \left\| {{{\mathbf{x}}_{{\rm{IRS}}}} - {{\mathbf{x}}_{\rm{U}}}} \right\|_2^2} \right)\left( {{{({h_{{\rm{BS}}}} - {h_{{\rm{IRS}}}})}^2} + {{\left( {{R_{{\rm{co}}}}} \right)}^2}} \right)}}} } \right|{\Phi _{{\rm{IRS}}}}} \right\} = $}
\end{equation*}
\begin{equation*}
   \scalebox{.89}[1]{$\left( {\frac{{{\lambda _{\rm{U}}}\left( {\left( {\frac{{2{{\left( {{R_{{\rm{co}}}}} \right)}^2}}}{\pi } - \frac{{{b^2}}}{\pi }} \right)\arcsin \left( {\frac{b}{{2{R_{{\rm{co}}}}}}} \right) - \frac{{{{\left( {{R_{{\rm{co}}}}} \right)}^2}}}{2}\sin \left( {4\arcsin \left( {\frac{b}{{2{R_{{\rm{co}}}}}}} \right)} \right) + \frac{{{b^2}}}{2}\left( {1 - \sin \left( {\pi  - 2\arcsin \left( {\frac{b}{{2{R_{{\rm{co}}}}}}} \right)} \right)} \right)} \right)}}{{1 - \exp \left\{ { - {\lambda _{\rm{U}}}\left( {\left( {\frac{{2{{\left( {{R_{{\rm{co}}}}} \right)}^2}}}{\pi } - \frac{{{b^2}}}{\pi }} \right)\arcsin \left( {\frac{b}{{2{R_{{\rm{co}}}}}}} \right) - \frac{{{{\left( {{R_{{\rm{co}}}}} \right)}^2}}}{2}\sin \left( {4\arcsin \left( {\frac{b}{{2{R_{{\rm{co}}}}}}} \right)} \right) + \frac{{{b^2}}}{2}\left( {1 - \sin \left( {\pi  - 2\arcsin \left( {\frac{b}{{2{R_{{\rm{co}}}}}}} \right)} \right)} \right)} \right)} \right\}}} + } \right.$}
\end{equation*}
\begin{equation*}
   \scalebox{.89}[1]{$\left. {\frac{{\exp \left\{ { - {\lambda _{\rm{U}}}\left( {\left( {\frac{{2{{\left( {{R_{{\rm{co}}}}} \right)}^2}}}{\pi } - \frac{{{b^2}}}{\pi }} \right)\arcsin \left( {\frac{b}{{2{R_{{\rm{co}}}}}}} \right) - \frac{{{{\left( {{R_{{\rm{co}}}}} \right)}^2}}}{2}\sin \left( {4\arcsin \left( {\frac{b}{{2{R_{{\rm{co}}}}}}} \right)} \right) + \frac{{{b^2}}}{2}\left( {1 - \sin \left( {\pi  - 2\arcsin \left( {\frac{b}{{2{R_{{\rm{co}}}}}}} \right)} \right)} \right)} \right)} \right\} - 1}}{{1 - \exp \left\{ { - {\lambda _{\rm{U}}}\left( {\left( {\frac{{2{{\left( {{R_{{\rm{co}}}}} \right)}^2}}}{\pi } - \frac{{{b^2}}}{\pi }} \right)\arcsin \left( {\frac{b}{{2{R_{{\rm{co}}}}}}} \right) - \frac{{{{\left( {{R_{{\rm{co}}}}} \right)}^2}}}{2}\sin \left( {4\arcsin \left( {\frac{b}{{2{R_{{\rm{co}}}}}}} \right)} \right) + \frac{{{b^2}}}{2}\left( {1 - \sin \left( {\pi  - 2\arcsin \left( {\frac{b}{{2{R_{{\rm{co}}}}}}} \right)} \right)} \right)} \right)} \right\}}}} \right) \times $}
\end{equation*}
\begin{equation*}
\scalebox{1}[1]{$\sum\limits_{{{\mathbf{x}}_{{\rm{IRS}}}} \in {{\tilde {\tilde {\Psi }}}_{{\rm{IRS}}}}({{\mathbf{x}}_{\rm{U}}})} {\frac{{\left( {1 - \exp \left( { - {\lambda _{{\rm{BS}}}}\pi \left( {4{{\left( {{R_{{\rm{co}}}}} \right)}^2} - {b^2}} \right)} \right)} \right)\left( {1 - \exp \left( { - {\lambda _{{\rm{BS}}}}\pi {b^2}} \right)} \right)Q{{\left( {{\lambda _{{\rm{wave}}}}} \right)}^4}{\delta ^2} \sigma _{{\rm{d}}}^2}}{{{\lambda _{{\rm{BS}}}}\pi \left( {4{{\left( {{R_{{\rm{co}}}}} \right)}^2} - {b^2}} \right){{\left( {4\pi } \right)}^4}\left( {{{({h_{{\rm{IRS}}}})}^2} + \left\| {{{\mathbf{x}}_{{\rm{IRS}}}} - {{\mathbf{x}}_{\rm{U}}}} \right\|_2^2} \right)\left( {{{({h_{{\rm{BS}}}} - {h_{{\rm{IRS}}}})}^2} + {{\left( {{R_{{\rm{co}}}}} \right)}^2}} \right)}}} ,$}
\end{equation*}
which is obtained similar to (\ref{t67}).
Finally, Campbell’s theorem \cite[Theorem 4.1]{Haenggi}, we obtain:
\begin{equation*}
   \scalebox{.86}[1]{$E\left\{ {\left( {\frac{{{\lambda _{\rm{U}}}\left( {\left( {\frac{{2{{\left( {{R_{{\rm{co}}}}} \right)}^2}}}{\pi } - \frac{{{b^2}}}{\pi }} \right)\arcsin \left( {\frac{b}{{2{R_{{\rm{co}}}}}}} \right) - \frac{{{{\left( {{R_{{\rm{co}}}}} \right)}^2}}}{2}\sin \left( {4\arcsin \left( {\frac{b}{{2{R_{{\rm{co}}}}}}} \right)} \right) + \frac{{{b^2}}}{2}\left( {1 - \sin \left( {\pi  - 2\arcsin \left( {\frac{b}{{2{R_{{\rm{co}}}}}}} \right)} \right)} \right)} \right)}}{{1 - \exp \left\{ { - {\lambda _{\rm{U}}}\left( {\left( {\frac{{2{{\left( {{R_{{\rm{co}}}}} \right)}^2}}}{\pi } - \frac{{{b^2}}}{\pi }} \right)\arcsin \left( {\frac{b}{{2{R_{{\rm{co}}}}}}} \right) - \frac{{{{\left( {{R_{{\rm{co}}}}} \right)}^2}}}{2}\sin \left( {4\arcsin \left( {\frac{b}{{2{R_{{\rm{co}}}}}}} \right)} \right) + \frac{{{b^2}}}{2}\left( {1 - \sin \left( {\pi  - 2\arcsin \left( {\frac{b}{{2{R_{{\rm{co}}}}}}} \right)} \right)} \right)} \right)} \right\}}} + } \right.} \right.$}
\end{equation*}
\begin{equation*}
   \scalebox{.86}[1]{$\left. {\frac{{\exp \left\{ { - {\lambda _{\rm{U}}}\left( {\left( {\frac{{2{{\left( {{R_{{\rm{co}}}}} \right)}^2}}}{\pi } - \frac{{{b^2}}}{\pi }} \right)\arcsin \left( {\frac{b}{{2{R_{{\rm{co}}}}}}} \right) - \frac{{{{\left( {{R_{{\rm{co}}}}} \right)}^2}}}{2}\sin \left( {4\arcsin \left( {\frac{b}{{2{R_{{\rm{co}}}}}}} \right)} \right) + \frac{{{b^2}}}{2}\left( {1 - \sin \left( {\pi  - 2\arcsin \left( {\frac{b}{{2{R_{{\rm{co}}}}}}} \right)} \right)} \right)} \right)} \right\} - 1}}{{1 - \exp \left\{ { - {\lambda _{\rm{U}}}\left( {\left( {\frac{{2{{\left( {{R_{{\rm{co}}}}} \right)}^2}}}{\pi } - \frac{{{b^2}}}{\pi }} \right)\arcsin \left( {\frac{b}{{2{R_{{\rm{co}}}}}}} \right) - \frac{{{{\left( {{R_{{\rm{co}}}}} \right)}^2}}}{2}\sin \left( {4\arcsin \left( {\frac{b}{{2{R_{{\rm{co}}}}}}} \right)} \right) + \frac{{{b^2}}}{2}\left( {1 - \sin \left( {\pi  - 2\arcsin \left( {\frac{b}{{2{R_{{\rm{co}}}}}}} \right)} \right)} \right)} \right)} \right\}}}} \right) \times $}
\end{equation*}
\begin{equation*}
   \scalebox{.86}[1]{$\sum\limits_{{{\mathbf{x}}_{{\rm{IRS}}}} \in {{\tilde {\tilde {\Psi }}}_{{\rm{IRS}}}}({{\mathbf{x}}_{\rm{U}}})} {\left. {\frac{{\left( {1 - \exp \left( { - {\lambda _{{\rm{BS}}}}\pi \left( {4{{\left( {{R_{{\rm{co}}}}} \right)}^2} - {b^2}} \right)} \right)} \right)\left( {1 - \exp \left( { - {\lambda _{{\rm{BS}}}}\pi {b^2}} \right)} \right)Q{{\left( {{\lambda _{{\rm{wave}}}}} \right)}^4}{\delta ^2} \sigma _{{\rm{d}}}^2}}{{{\lambda _{{\rm{BS}}}}\pi \left( {4{{\left( {{R_{{\rm{co}}}}} \right)}^2} - {b^2}} \right){{\left( {4\pi } \right)}^4}\left( {{{({h_{{\rm{IRS}}}})}^2} + \left\| {{{\mathbf{x}}_{{\rm{IRS}}}} - {{\mathbf{x}}_{\rm{U}}}} \right\|_2^2} \right)\left( {{{({h_{{\rm{BS}}}} - {h_{{\rm{IRS}}}})}^2} + {{\left( {{R_{{\rm{co}}}}} \right)}^2}} \right)}}} \right\}}  = $}
\end{equation*}
\begin{equation*}
   \scalebox{.86}[1]{$\left( {\frac{{{\lambda _{\rm{U}}}\left( {\left( {\frac{{2{{\left( {{R_{{\rm{co}}}}} \right)}^2}}}{\pi } - \frac{{{b^2}}}{\pi }} \right)\arcsin \left( {\frac{b}{{2{R_{{\rm{co}}}}}}} \right) - \frac{{{{\left( {{R_{{\rm{co}}}}} \right)}^2}}}{2}\sin \left( {4\arcsin \left( {\frac{b}{{2{R_{{\rm{co}}}}}}} \right)} \right) + \frac{{{b^2}}}{2}\left( {1 - \sin \left( {\pi  - 2\arcsin \left( {\frac{b}{{2{R_{{\rm{co}}}}}}} \right)} \right)} \right)} \right)}}{{1 - \exp \left\{ { - {\lambda _{\rm{U}}}\left( {\left( {\frac{{2{{\left( {{R_{{\rm{co}}}}} \right)}^2}}}{\pi } - \frac{{{b^2}}}{\pi }} \right)\arcsin \left( {\frac{b}{{2{R_{{\rm{co}}}}}}} \right) - \frac{{{{\left( {{R_{{\rm{co}}}}} \right)}^2}}}{2}\sin \left( {4\arcsin \left( {\frac{b}{{2{R_{{\rm{co}}}}}}} \right)} \right) + \frac{{{b^2}}}{2}\left( {1 - \sin \left( {\pi  - 2\arcsin \left( {\frac{b}{{2{R_{{\rm{co}}}}}}} \right)} \right)} \right)} \right)} \right\}}} + } \right.$}
\end{equation*}
\begin{equation*}
   \scalebox{.86}[1]{$\left. {\frac{{\exp \left\{ { - {\lambda _{\rm{U}}}\left( {\left( {\frac{{2{{\left( {{R_{{\rm{co}}}}} \right)}^2}}}{\pi } - \frac{{{b^2}}}{\pi }} \right)\arcsin \left( {\frac{b}{{2{R_{{\rm{co}}}}}}} \right) - \frac{{{{\left( {{R_{{\rm{co}}}}} \right)}^2}}}{2}\sin \left( {4\arcsin \left( {\frac{b}{{2{R_{{\rm{co}}}}}}} \right)} \right) + \frac{{{b^2}}}{2}\left( {1 - \sin \left( {\pi  - 2\arcsin \left( {\frac{b}{{2{R_{{\rm{co}}}}}}} \right)} \right)} \right)} \right)} \right\} - 1}}{{1 - \exp \left\{ { - {\lambda _{\rm{U}}}\left( {\left( {\frac{{2{{\left( {{R_{{\rm{co}}}}} \right)}^2}}}{\pi } - \frac{{{b^2}}}{\pi }} \right)\arcsin \left( {\frac{b}{{2{R_{{\rm{co}}}}}}} \right) - \frac{{{{\left( {{R_{{\rm{co}}}}} \right)}^2}}}{2}\sin \left( {4\arcsin \left( {\frac{b}{{2{R_{{\rm{co}}}}}}} \right)} \right) + \frac{{{b^2}}}{2}\left( {1 - \sin \left( {\pi  - 2\arcsin \left( {\frac{b}{{2{R_{{\rm{co}}}}}}} \right)} \right)} \right)} \right)} \right\}}}} \right) \times $}
\end{equation*}
\begin{equation*}
   \scalebox{.86}[1]{$\frac{{{\lambda _{{\rm{IRS}}}}\left( {1 - \exp \left( { - {\lambda _{{\rm{BS}}}}\pi \left( {4{{\left( {{R_{{\rm{co}}}}} \right)}^2} - {b^2}} \right)} \right)} \right)\left( {1 - \exp \left( { - {\lambda _{{\rm{BS}}}}\pi {b^2}} \right)} \right)Q{{\left( {{\lambda _{{\rm{wave}}}}} \right)}^4}{\delta ^2} \sigma _{{\rm{d}}}^2}}{{{\lambda _{{\rm{BS}}}}\pi \left( {4{{\left( {{R_{{\rm{co}}}}} \right)}^2} - {b^2}} \right){{\left( {4\pi } \right)}^4}\left( {{{({h_{{\rm{BS}}}} - {h_{{\rm{IRS}}}})}^2} + {{\left( {{R_{{\rm{co}}}}} \right)}^2}} \right)}} \times \int\limits_{{\cal C}\left( {{{\mathbf{x}}_{\rm{U}}},b} \right)} {\frac{{d{\mathbf{x}}}}{{{{({h_{{\rm{IRS}}}})}^2} + \left\| {{\mathbf{x}} - {{\mathbf{x}}_{\rm{U}}}} \right\|_2^2}}}  = $}
\end{equation*}
\begin{equation*}
   \scalebox{.86}[1]{$\left( {\frac{{{\lambda _{\rm{U}}}\left( {\left( {\frac{{2{{\left( {{R_{{\rm{co}}}}} \right)}^2}}}{\pi } - \frac{{{b^2}}}{\pi }} \right)\arcsin \left( {\frac{b}{{2{R_{{\rm{co}}}}}}} \right) - \frac{{{{\left( {{R_{{\rm{co}}}}} \right)}^2}}}{2}\sin \left( {4\arcsin \left( {\frac{b}{{2{R_{{\rm{co}}}}}}} \right)} \right) + \frac{{{b^2}}}{2}\left( {1 - \sin \left( {\pi  - 2\arcsin \left( {\frac{b}{{2{R_{{\rm{co}}}}}}} \right)} \right)} \right)} \right)}}{{1 - \exp \left\{ { - {\lambda _{\rm{U}}}\left( {\left( {\frac{{2{{\left( {{R_{{\rm{co}}}}} \right)}^2}}}{\pi } - \frac{{{b^2}}}{\pi }} \right)\arcsin \left( {\frac{b}{{2{R_{{\rm{co}}}}}}} \right) - \frac{{{{\left( {{R_{{\rm{co}}}}} \right)}^2}}}{2}\sin \left( {4\arcsin \left( {\frac{b}{{2{R_{{\rm{co}}}}}}} \right)} \right) + \frac{{{b^2}}}{2}\left( {1 - \sin \left( {\pi  - 2\arcsin \left( {\frac{b}{{2{R_{{\rm{co}}}}}}} \right)} \right)} \right)} \right)} \right\}}} + } \right.$}
\end{equation*}
\begin{equation*}
   \scalebox{.86}[1]{$\left. {\frac{{\exp \left\{ { - {\lambda _{\rm{U}}}\left( {\left( {\frac{{2{{\left( {{R_{{\rm{co}}}}} \right)}^2}}}{\pi } - \frac{{{b^2}}}{\pi }} \right)\arcsin \left( {\frac{b}{{2{R_{{\rm{co}}}}}}} \right) - \frac{{{{\left( {{R_{{\rm{co}}}}} \right)}^2}}}{2}\sin \left( {4\arcsin \left( {\frac{b}{{2{R_{{\rm{co}}}}}}} \right)} \right) + \frac{{{b^2}}}{2}\left( {1 - \sin \left( {\pi  - 2\arcsin \left( {\frac{b}{{2{R_{{\rm{co}}}}}}} \right)} \right)} \right)} \right)} \right\} - 1}}{{1 - \exp \left\{ { - {\lambda _{\rm{U}}}\left( {\left( {\frac{{2{{\left( {{R_{{\rm{co}}}}} \right)}^2}}}{\pi } - \frac{{{b^2}}}{\pi }} \right)\arcsin \left( {\frac{b}{{2{R_{{\rm{co}}}}}}} \right) - \frac{{{{\left( {{R_{{\rm{co}}}}} \right)}^2}}}{2}\sin \left( {4\arcsin \left( {\frac{b}{{2{R_{{\rm{co}}}}}}} \right)} \right) + \frac{{{b^2}}}{2}\left( {1 - \sin \left( {\pi  - 2\arcsin \left( {\frac{b}{{2{R_{{\rm{co}}}}}}} \right)} \right)} \right)} \right)} \right\}}}} \right) \times $}
\end{equation*}
\begin{equation}
   \scalebox{.86}[1]{$\frac{{{\lambda _{{\rm{IRS}}}}\left( {1 - \exp \left( { - {\lambda _{{\rm{BS}}}}\pi \left( {4{{\left( {{R_{{\rm{co}}}}} \right)}^2} - {b^2}} \right)} \right)} \right)\left( {1 - \exp \left( { - {\lambda _{{\rm{BS}}}}\pi {b^2}} \right)} \right)Q{{\left( {{\lambda _{{\rm{wave}}}}} \right)}^4}{\delta ^2} \sigma _{{\rm{d}}}^2}}{{{\lambda _{{\rm{BS}}}}\pi \left( {4{{\left( {{R_{{\rm{co}}}}} \right)}^2} - {b^2}} \right){4^4}{\pi ^3}\left( {{{({h_{{\rm{BS}}}} - {h_{{\rm{IRS}}}})}^2} + {{\left( {{R_{{\rm{co}}}}} \right)}^2}} \right)}} \times \ln \left( {1 + {{\left( {\frac{b}{{{h_{{\rm{IRS}}}}}}} \right)}^2}} \right).$}
\end{equation}

\section{Proof of Theorem 5}

\begin{equation*}
\mathbb{P}\left( {\mathbb{I}\left( {{{\mathbf{x}}_{\rm{U}}}} \right)} \right) = 
\end{equation*}
\begin{equation*}
   \scalebox{1}[1]{$E\left\{ {\left. {{{\left| {\mathbb{I}\left( {{{\mathbf{x}}_{\rm{U}}}} \right)} \right|}^2}} \right|{\Phi _{\rm{U}}},{\Phi _{{\rm{BS}}}},{\Phi _{{\rm{IRS}}}},{{{\mathbf{\tilde x}}}_{{\rm{BS}}}}({{\mathbf{x}}_{\rm{U}}}),{{{\mathbf{\tilde x}}}_{\rm{U}}}({{\mathbf{x}}_{{\rm{IRS}}}}),{{\mathbb{H}}^{[q]}}({{\mathbf{x}}_{{\rm{BS}}}},{{\mathbf{x}}_{{\rm{IRS}}}}),{{\mathbb{H}}^{[q]}}({{\mathbf{x}}_{{\rm{IRS}}}},{{\mathbf{x}}_{\rm{U}}}),{\mathbb{H}}({{\mathbf{x}}_{{\rm{BS}}}},{{\mathbf{x}}_{\rm{U}}})} \right\} \le$}
\end{equation*}
\begin{equation*}
 \scalebox{.89}[1]{$\sum\limits_{{{\mathbf{x}}_{{\rm{IRS}}}} \in {\Psi _{{\rm{IRS}}}}({{\mathbf{x}}_{\rm{U}}})} {\sum\limits_{q = 1}^Q {\sum\limits_{{{{\mathbf{x'}}}_{{\rm{BS}}}} \in {{\tilde \Psi }_{{\rm{BS}}}}({{\mathbf{x}}_{\rm{U}}})} {\sum\limits_{{{{\mathbf{x'}}}_{\rm{U}}} \in {{\tilde \Psi }_{\rm{U}}}({{\mathbf{x}}_{\rm{U}}}):{{{\mathbf{x'}}}_{\rm{U}}} \ne {{\mathbf{x}}_{\rm{U}}}} {\sum\limits_{{{{\mathbf{x''}}}_{{\rm{IRS}}}} \in {\Psi _{{\rm{IRS}}}}({{\mathbf{x}}_{\rm{U}}})} {\sum\limits_{q'' = 1}^Q {\sum\limits_{{{{\mathbf{x''}}}_{{\rm{BS}}}} \in {{\tilde \Psi }_{{\rm{BS}}}}({{\mathbf{x}}_{\rm{U}}})} {\sum\limits_{{{{\mathbf{x''}}}_{\rm{U}}} \in {{\tilde \Psi }_{\rm{U}}}({{\mathbf{x}}_{\rm{U}}}):{{{\mathbf{x''}}}_{\rm{U}}} \ne {{\mathbf{x}}_{\rm{U}}}} {4{{\left( {1 + {{\tilde n}_{{\rm{IRS}}}}({{\mathbf{x}}_{\rm{U}}})} \right)}^2} \times } } } } } } } } $}
\end{equation*}
\begin{equation*}
 \scalebox{1}[1]{$\left( {\left| {{{\mathbb{H}}^{[q]}}({{\mathbf{x}}_{{\rm{IRS}}}},{{\mathbf{x}}_{\rm{U}}}){{\mathbb{H}}^{[q]}}({{{\mathbf{x'}}}_{{\rm{BS}}}},{{\mathbf{x}}_{{\rm{IRS}}}})} \right|  +\left| {{\mathbb{H}}({{{\mathbf{x'}}}_{{\rm{BS}}}},{{\mathbf{x}}_{\rm{U}}})} \right|} \right) \times $}
\end{equation*}
\begin{equation*}
 \scalebox{1}[1]{$\left( {\left| {{{\mathbb{H}}^{[q'']}}({{{\mathbf{x''}}}_{{\rm{IRS}}}},{{\mathbf{x}}_{\rm{U}}}){{\mathbb{H}}^{[q'']}}({{{\mathbf{x''}}}_{{\rm{BS}}}},{{{\mathbf{x''}}}_{{\rm{IRS}}}})} \right| + \left| {{\mathbb{H}}({{{\mathbf{x''}}}_{{\rm{BS}}}},{{\mathbf{x}}_{\rm{U}}})} \right|} \right) \times  \sigma _{{\rm{d}}}^2 .$}
\end{equation*}
From \cite[Eq. 5-76]{Papoulis}, we can see that for a Rayleigh random variable $W$ with parameter $\sigma$, we have:
\begin{equation}
E\left\{ {{{\left| W \right|}^p}} \right\} \le {2^{\frac{p}{2} + 1}}\left( {\left\lceil {\frac{{p + 1}}{2}} \right\rceil } \right)!{\sigma ^p}\sqrt {\frac{\pi }{2}} .
\end{equation}
Thus, we have:
\begin{equation*}
E\left\{ {{{\left| {\mathbb{H}({{\mathbf{x}}_{{\text{BS}}}},{{\mathbf{x}}_{\text{U}}})} \right|}^p}} \right\} \le {2^{\frac{{3p}}{2} + 1}}\left( {\left\lceil {\frac{{p + 1}}{2}} \right\rceil } \right)!{\left( {\max \left\{ {1,\frac{{{\lambda _{{\text{wave}}}}}}{{4\pi {h_{{\text{BS}}}}}}} \right\}} \right)^p}\sqrt {\frac{\pi }{2}}, 
\end{equation*}
\begin{equation*}
E\left\{ {{{\left| {{\mathbb{H}^{[q]}}({{\mathbf{x}}_{{\text{IRS}}}},{{\mathbf{x}}_{\text{U}}})} \right|}^p}} \right\} \le {2^{\frac{{3p}}{2} + 1}}\left( {\left\lceil {\frac{{p + 1}}{2}} \right\rceil } \right)!{\left( {\max \left\{ {1,\frac{{{\lambda _{{\text{wave}}}}}}{{4\pi {h_{{\text{IRS}}}}}}} \right\}} \right)^p}\sqrt {\frac{\pi }{2}} ,
\end{equation*}
\begin{equation*}
E\left\{ {{{\left| {{\mathbb{H}^{[q]}}({{\mathbf{x}}_{{\text{BS}}}},{{\mathbf{x}}_{{\text{IRS}}}})} \right|}^p}} \right\} \le {2^{\frac{{3p}}{2} + 1}}\left( {\left\lceil {\frac{{p + 1}}{2}} \right\rceil } \right)!{\left( {\max \left\{ {1,\frac{{{\lambda _{{\text{wave}}}}}}{{4\pi \left| {{h_{{\text{BS}}}} - {h_{{\text{IRS}}}}} \right|}}} \right\}} \right)^p}\sqrt {\frac{\pi }{2}} .
\end{equation*}

Moreover, by Lyapunov inequality \cite[Inequality 5-93]{Papoulis}, for two random variables $X$ and $Y$ with identical distribution and finite $n$-th moment, we have: 
\begin{equation}
E\left\{ {{{\left| X \right|}^k}} \right\}E\left\{ {{{\left| Y \right|}^{n - k}}} \right\} \le E\left\{ {{{\left| X \right|}^n}} \right\} = E\left\{ {{{\left| Y \right|}^n}} \right\},k \in \{ 0,1,...,n\} .
\end{equation}
Hence, we obtain:
\begin{equation*}
E\left\{ {\left. {{{\left( {\mathbb{P}\left( {\mathbb{I}\left( {{{\mathbf{x}}_{\rm{U}}}} \right)} \right)} \right)}^p}} \right|{\Phi _{\rm{U}}},{\Phi _{{\rm{BS}}}},{\Phi _{{\rm{IRS}}}}} \right\} \le {\left( {4} \right)^p}{\left( {1 + {{\tilde n}_{{\rm{IRS}}}}({{\mathbf{x}}_{\rm{U}}})} \right)^{4p}} \times {\left( {{{\tilde n}_{{\rm{BS}}}}({{\mathbf{x}}_{\rm{U}}})} \right)^{2p}} \times {\left( {{{\tilde n}_{\rm{U}}}({{\mathbf{x}}_{\rm{U}}})} \right)^{2p}}{Q^{2p}}\sigma _{{\rm{d}}}^{2p}
\end{equation*}
\begin{equation*}
{2^{2p}} \times {\left( {{2^{2p + 1}}\left( {\left\lceil {\frac{{2p + 1}}{2}} \right\rceil } \right)!\sqrt {\frac{\pi }{2}} } \right)^3} \times
\end{equation*}
\begin{equation*}
 \scalebox{1}[1]{${\left( {\max \left\{ {1,\frac{{{\lambda _{{\text{wave}}}}}}{{4\pi {h_{{\text{BS}}}}}}} \right\} \times \max \left\{ {1,\frac{{{\lambda _{{\text{wave}}}}}}{{4\pi \left| {{h_{{\text{BS}}}} - {h_{{\text{IRS}}}}} \right|}}} \right\} \times \max \left\{ {1,\frac{{{\lambda _{{\text{wave}}}}}}{{4\pi {h_{{\text{IRS}}}}}}} \right\}} \right)^{2p}} \le$}
\end{equation*}
\begin{equation*}
{2^{6p}}{\left( {1 + {{\tilde n}_{{\rm{IRS}}}}({{\mathbf{x}}_{\rm{U}}})} \right)^{4p}} \times {\left( {{{\tilde n}_{{\rm{BS}}}}({{\mathbf{x}}_{\rm{U}}})} \right)^{2p}} \times {\left( {{{\tilde n}_{\rm{U}}}({{\mathbf{x}}_{\rm{U}}})} \right)^{2p}}{Q^{2p}}\sigma _{{\rm{d}}}^{2p}\times
\end{equation*}
\begin{equation*}
 \scalebox{1}[1]{${\left( {\max \left\{ {1,\frac{{{\lambda _{{\text{wave}}}}}}{{4\pi {h_{{\text{BS}}}}}}} \right\} \times \max \left\{ {1,\frac{{{\lambda _{{\text{wave}}}}}}{{4\pi \left| {{h_{{\text{BS}}}} - {h_{{\text{IRS}}}}} \right|}}} \right\} \times \max \left\{ {1,\frac{{{\lambda _{{\text{wave}}}}}}{{4\pi {h_{{\text{IRS}}}}}}} \right\}} \right)^{2p}} \times$}
\end{equation*}
\begin{equation*}
 \scalebox{1}[1]{${\left( {2\pi {{\left( {\frac{{\left\lceil {\frac{{2p + 1}}{2}} \right\rceil }}{e}} \right)}^{\left\lceil {\frac{{2p + 1}}{2}} \right\rceil  + \frac{1}{2}}}\exp \left\{ {\frac{1}{2} + \frac{1}{{12\left\lceil {\frac{{2p + 1}}{2}} \right\rceil }}} \right\}} \right)^3}  \le $}
\end{equation*}
\begin{equation*}
{2^{6p}}{\left( {1 + {{\tilde n}_{{\rm{IRS}}}}({{\mathbf{x}}_{\rm{U}}})} \right)^{4p}} \times {\left( {{{\tilde n}_{{\rm{BS}}}}({{\mathbf{x}}_{\rm{U}}})} \right)^{2p}} \times {\left( {{{\tilde n}_{\rm{U}}}({{\mathbf{x}}_{\rm{U}}})} \right)^{2p}}{Q^{2p}}\times \sigma _{{\rm{d}}}^{2p}
\end{equation*}
\begin{equation*}
 \scalebox{1}[1]{${\left( {\max \left\{ {1,\frac{{{\lambda _{{\text{wave}}}}}}{{4\pi {h_{{\text{BS}}}}}}} \right\} \times \max \left\{ {1,\frac{{{\lambda _{{\text{wave}}}}}}{{4\pi \left| {{h_{{\text{BS}}}} - {h_{{\text{IRS}}}}} \right|}}} \right\} \times \max \left\{ {1,\frac{{{\lambda _{{\text{wave}}}}}}{{4\pi {h_{{\text{IRS}}}}}}} \right\}} \right)^{2p}} \times$}
\end{equation*}
\begin{equation*}
{\left( {2\pi {{\left( {\frac{{p + 1}}{e}} \right)}^{p + \frac{3}{2}}}\exp \left\{ {\frac{7}{{12}}} \right\}} \right)^3} \times  \le
\end{equation*}
\begin{equation*}
{2^{6p}}{\left( {1 + {{\tilde n}_{{\rm{IRS}}}}({{\mathbf{x}}_{\rm{U}}})} \right)^{4p}} \times {\left( {{{\tilde n}_{{\rm{BS}}}}({{\mathbf{x}}_{\rm{U}}})} \right)^{2p}} \times {\left( {{{\tilde n}_{\rm{U}}}({{\mathbf{x}}_{\rm{U}}})} \right)^{2p}}{Q^{2p}} \times \sigma _{{\rm{d}}}^{2p}\times
\end{equation*}
\begin{equation*}
 \scalebox{1}[1]{${\left( {\max \left\{ {1,\frac{{{\lambda _{{\text{wave}}}}}}{{4\pi {h_{{\text{BS}}}}}}} \right\} \times \max \left\{ {1,\frac{{{\lambda _{{\text{wave}}}}}}{{4\pi \left| {{h_{{\text{BS}}}} - {h_{{\text{IRS}}}}} \right|}}} \right\} \times \max \left\{ {1,\frac{{{\lambda _{{\text{wave}}}}}}{{4\pi {h_{{\text{IRS}}}}}}} \right\}} \right)^{2p}} \times$}
\end{equation*}
\begin{equation*}
{\left( {2\pi {{\left( {\frac{{2p}}{e}} \right)}^{p + \frac{3}{2}}}\exp \left\{ {\frac{7}{{12}}} \right\}} \right)^3} \le
\end{equation*}
\begin{equation*}
{2^{6p}}{\left( {1 + {{\tilde n}_{{\rm{IRS}}}}({{\mathbf{x}}_{\rm{U}}})} \right)^{4p}} \times {\left( {{{\tilde n}_{{\rm{BS}}}}({{\mathbf{x}}_{\rm{U}}})} \right)^{2p}} \times {\left( {{{\tilde n}_{\rm{U}}}({{\mathbf{x}}_{\rm{U}}})} \right)^{2p}}{Q^{2p}} \times  \sigma _{{\rm{d}}}^{2p}\times
\end{equation*}
\begin{equation*}
 \scalebox{1}[1]{${\left( {\max \left\{ {1,\frac{{{\lambda _{{\text{wave}}}}}}{{4\pi {h_{{\text{BS}}}}}}} \right\} \times \max \left\{ {1,\frac{{{\lambda _{{\text{wave}}}}}}{{4\pi \left| {{h_{{\text{BS}}}} - {h_{{\text{IRS}}}}} \right|}}} \right\} \times \max \left\{ {1,\frac{{{\lambda _{{\text{wave}}}}}}{{4\pi {h_{{\text{IRS}}}}}}} \right\}} \right)^{2p}} \times$}
\end{equation*}
\begin{equation*}
{\left( {{2^{11}}{\pi ^5}{{\left( 3 \right)}^{\frac{3}{2}}}\exp \left\{ {\frac{{35}}{{12}}} \right\}} \right)^p}{\left( {\frac{{18}}{{\exp \{ 3\} }}} \right)^p}{p^{3p}}.
\end{equation*}

On the other hand, by (\ref{lambda-k}), we have:
\begin{equation*}
E\left\{ {{{\left( {1 + {{\tilde n}_{{\rm{IRS}}}}({{\mathbf{x}}_{\rm{U}}})} \right)}^p}} \right\} = E\left\{ {\sum\limits_{i = 0}^p {\left( {\begin{array}{*{20}{c}}
  p \\ 
  i 
\end{array}} \right){{\left( {{{\tilde n}_{{\rm{IRS}}}}({{\mathbf{x}}_{\rm{U}}})} \right)}^i}} } \right\} \le
\end{equation*}
\begin{equation*}
1 + \sum\limits_{i = 1}^p {\left( {\begin{array}{*{20}{c}}
  p \\ 
  i 
\end{array}} \right){{\left( {\frac{i}{{\log \left( {1 + \frac{1}{{{\lambda _{{\rm{IRS}}}}4\pi {{\left( {{R_{{\rm{co}}}}} \right)}^2}}}} \right)}}} \right)}^i} \le } 1 + \sum\limits_{i = 1}^p {\left( {\begin{array}{*{20}{c}}
  p \\ 
  i 
\end{array}} \right){{\left( {\frac{p}{{\log \left( {1 + \frac{1}{{{\lambda _{{\rm{IRS}}}}4\pi {{\left( {{R_{{\rm{co}}}}} \right)}^2}}}} \right)}}} \right)}^i} = } 
\end{equation*}
\begin{equation}
{\left( {1 + \frac{p}{{\log \left( {1 + \frac{1}{{{\lambda _{{\rm{IRS}}}}4\pi {{\left( {{R_{{\rm{co}}}}} \right)}^2}}}} \right)}}} \right)^p} \le {\left( {2\max \left\{ {1,\frac{1}{{\log \left( {1 + \frac{1}{{{\lambda _{{\rm{IRS}}}}4\pi {{\left( {{R_{{\rm{co}}}}} \right)}^2}}}} \right)}}} \right\}} \right)^p}{p^p},
\label{eqx1}
\end{equation}
\begin{equation}
E\left\{ {{{\left( {{{\tilde n}_{{\rm{BS}}}}({{\mathbf{x}}_{\rm{U}}})} \right)}^p}} \right\} \le {\left( {\frac{p}{{\log \left( {1 + \frac{1}{{{\lambda _{{\rm{BS}}}}\pi {{\left( {{R_{{\rm{co}}}}} \right)}^2}}}} \right)}}} \right)^p},
\label{eqx2}
\end{equation}
\begin{equation}
E\left\{ {{{\left( {{{\tilde n}_{\rm{U}}}({{\mathbf{x}}_{\rm{U}}})} \right)}^p}} \right\} \le {\left( {\frac{p}{{\log \left( {1 + \frac{1}{{{\lambda _{\rm{U}}}4\pi {{\left( {{R_{{\rm{co}}}}} \right)}^2}}}} \right)}}} \right)^p} \times \frac{1}{{1 - \exp \left\{ { - {\lambda _{\rm{U}}}4\pi {{\left( {{R_{{\rm{co}}}}} \right)}^2}} \right\}}}.
\label{eqx3}
\end{equation}
Then, we obtain:
\begin{equation*}
E\left\{ {{{\left( {\mathbb{P}\left( {\mathbb{I}\left( {{{\mathbf{x}}_{\rm{U}}}} \right)} \right)} \right)}^p}} \right\} \le 
\end{equation*}
\begin{equation*}
{2^{6p}}{\left( {1 + {{\tilde n}_{{\rm{IRS}}}}({{\mathbf{x}}_{\rm{U}}})} \right)^{4p}} \times {\left( {{{\tilde n}_{{\rm{BS}}}}({{\mathbf{x}}_{\rm{U}}})} \right)^{2p}} \times {\left( {{{\tilde n}_{\rm{U}}}({{\mathbf{x}}_{\rm{U}}})} \right)^{2p}}{Q^{2p}} \times  \sigma _{{\rm{d}}}^{2p}\times
\end{equation*}
\begin{equation*}
 \scalebox{1}[1]{${\left( {\max \left\{ {1,\frac{{{\lambda _{{\text{wave}}}}}}{{4\pi {h_{{\text{BS}}}}}}} \right\} \times \max \left\{ {1,\frac{{{\lambda _{{\text{wave}}}}}}{{4\pi \left| {{h_{{\text{BS}}}} - {h_{{\text{IRS}}}}} \right|}}} \right\} \times \max \left\{ {1,\frac{{{\lambda _{{\text{wave}}}}}}{{4\pi {h_{{\text{IRS}}}}}}} \right\}} \right)^{2p}} \times$}
\end{equation*}
\begin{equation*}
{\left( {{2^{11}}{\pi ^5}{{\left( 3 \right)}^{\frac{3}{2}}}\exp \left\{ {\frac{{35}}{{12}}} \right\}} \right)^p}{\left( {\frac{{18}}{{\exp \{ 3\} }}} \right)^p}{p^{3p}}\times
\end{equation*}
\begin{equation*}
{\left( {8\max \left\{ {1,\frac{1}{{\log \left( {1 + \frac{1}{{{\lambda _{{\rm{IRS}}}}4\pi {{\left( {{R_{{\rm{co}}}}} \right)}^2}}}} \right)}}} \right\}} \right)^{4p}}{p^{4p}} \times {\left( {\frac{{2p}}{{\log \left( {1 + \frac{1}{{{\lambda _{{\rm{BS}}}}4\pi {{\left( {{R_{{\rm{co}}}}} \right)}^2}}}} \right)}}} \right)^{2p}} \times 
\end{equation*}
\begin{equation*}
{\left( {\frac{{2p}}{{\log \left( {1 + \frac{1}{{{\lambda _{\rm{U}}}4\pi {{\left( {{R_{{\rm{co}}}}} \right)}^2}}}} \right)}}} \right)^{2p}} \times \frac{1}{{1 - \exp \left\{ { - {\lambda _{\rm{U}}}4\pi {{\left( {{R_{{\rm{co}}}}} \right)}^2}} \right\}}} \le
\end{equation*}
\begin{equation*}
 {K^p} \times {p^{11p}},
\end{equation*}
where:
\begin{equation*}
K = Q^2{2^{6}}\left( {\frac{{ \sigma _{{\rm{d}}}^2 }}{{1 - \exp \left\{ { - {\lambda _{\rm{U}}}4\pi {{\left( {{R_{{\rm{co}}}}} \right)}^2}} \right\}}}} \right) \times 
\end{equation*}
\begin{equation*}
 \scalebox{1}[1]{${\left( {\max \left\{ {1,\frac{{{\lambda _{{\text{wave}}}}}}{{4\pi {h_{{\text{BS}}}}}}} \right\} \times \max \left\{ {1,\frac{{{\lambda _{{\text{wave}}}}}}{{4\pi \left| {{h_{{\text{BS}}}} - {h_{{\text{IRS}}}}} \right|}}} \right\} \times \max \left\{ {1,\frac{{{\lambda _{{\text{wave}}}}}}{{4\pi {h_{{\text{IRS}}}}}}} \right\}} \right)^{2p}} \times$}
\end{equation*}
\begin{equation*}
\left( {{2^{11}}{\pi ^5}{{\left( 3 \right)}^{\frac{3}{2}}}\exp \left\{ {\frac{{35}}{{12}}} \right\}} \right) \times \left( {\frac{{18}}{{\exp \{ 3\} }}} \right) \times {\left( {8\max \left\{ {1,\frac{1}{{\log \left( {1 + \frac{1}{{{\lambda _{{\rm{IRS}}}}4\pi {{\left( {{R_{{\rm{co}}}}} \right)}^2}}}} \right)}}} \right\}} \right)^4} \times 
\end{equation*}
\begin{equation}
{\left( {\frac{2}{{\log \left( {1 + \frac{1}{{{\lambda _{{\rm{BS}}}}4\pi {{\left( {{R_{{\rm{co}}}}} \right)}^2}}}} \right)}}} \right)^2} \times {\left( {\frac{2}{{\log \left( {1 + \frac{1}{{{\lambda _{\rm{U}}}4\pi {{\left( {{R_{{\rm{co}}}}} \right)}^2}}}} \right)}}} \right)^2}.
\end{equation}

\section{Proof of Theorem 6}

First, we obtain:
\begin{equation*}
E\left\{ {\mathbb{G}\left( {\tau \mathbb{P}\left( {\mathbb{I}\left( {{{\mathbf{x}}_{\rm{U}}}} \right)} \right)} \right)} \right\}= 
\end{equation*}
\begin{equation*}
E\left\{ {\sum\limits_{p = 0}^\infty  {\frac{{{{\left| {\tau \mathbb{P}\left( {\mathbb{I}\left( {{{\mathbf{x}}_{\rm{U}}}} \right)} \right)} \right|}^p}}}{{\left( {11p} \right)!}}} } \right\} \le \sum\limits_{p = 0}^\infty  {\frac{{\left( {E\left\{ {{{\left| {\mathbb{P}\left( {\mathbb{I}\left( {{{\mathbf{x}}_{\rm{U}}}} \right)} \right)} \right|}^p}} \right\}} \right){\tau ^p}}}{{\left( {11p} \right)!}}}  \le \sum\limits_{p = 0}^\infty  {\frac{{\left( {E\left\{ {{{\left| {\mathbb{P}\left( {\mathbb{I}\left( {{{\mathbf{x}}_{\rm{U}}}} \right)} \right)} \right|}^p}} \right\}} \right){\tau ^p}}}{{\left( {11p} \right)!}}}  \le
\end{equation*}
\begin{equation}
1 + \sum\limits_{p = 1}^\infty  {\frac{{{K^p}{p^{11p}}{\tau ^p}}}{{\left( {11p} \right)!}}}  \le 1 + \sum\limits_{p = 1}^\infty  {\frac{{{K^p}{p^{11p}}{\tau ^p}}}{{{{\left( {\frac{{11p}}{e}} \right)}^{11p}}}}}  = 1 + \sum\limits_{p = 1}^\infty  {{{\left( {\frac{{K{e^{11}}\tau }}{{{{11}^{11}}}}} \right)}^p}}  = \frac{1}{{1 - \frac{{K{e^{11}}\tau }}{{{{11}^{11}}}}}}.
\end{equation}
which is obtained by Fatou's Lemma \cite{Stein} and Stirling's Formula \cite{Maria}. Then, we have:
\begin{equation*}
\Pr \left\{ {\left| {\mathbb{P}\left( {\mathbb{I}\left( {{{\mathbf{x}}_{\rm{U}}}} \right)} \right)} \right| > t} \right\} \le \Pr \left\{ {\left| {\mathbb{P}\left( {\mathbb{I}\left( {{{\mathbf{x}}_{\rm{U}}}} \right)} \right)} \right| > t} \right\} = \Pr \left\{ {\mathbb{G}\left( {\tau \left| {\mathbb{P}\left( {\mathbb{I}\left( {{{\mathbf{x}}_{\rm{U}}}} \right)} \right)} \right|} \right) > \mathbb{G}\left( {\tau t} \right)} \right\} \le
\end{equation*}
\begin{equation}
\frac{{E\left\{ {\mathbb{G}\left( {\tau \left| {\mathbb{P}\left( {\mathbb{I}\left( {{{\mathbf{x}}_{\rm{U}}}} \right)} \right)} \right|} \right)} \right\}}}{{\mathbb{G}\left( {\tau t} \right)}} \le \frac{1}{{\left( {1 - \frac{{K{e^{11}}\tau }}{{{{11}^{11}}}}} \right)\mathbb{G}\left( {\tau t} \right)}},
\end{equation}
which is obtained by Markov inequality.

\section{Proof of Theorem 7}

We have:
\begin{equation*}
\mathbb{P}\left( {\mathbb{S}\left( {{{\mathbf{x}}_{\rm{U}}}} \right)} \right) = 
\end{equation*}
\begin{equation*}
 \scalebox{1}[1]{$E\left\{ {\left. {{{\left| {\mathbb{S}({{\mathbf{x}}_{\rm{U}}})} \right|}^2}} \right|{\Phi _{\rm{U}}},{\Phi _{{\rm{BS}}}},{\Phi _{{\rm{IRS}}}},{{{\mathbf{\tilde x}}}_{{\rm{BS}}}}({{\mathbf{x}}_{\rm{U}}}),{{{\mathbf{\tilde x}}}_{\rm{U}}}({{\mathbf{x}}_{{\rm{IRS}}}}),{{\mathbb{H}}^{[q]}}({{\mathbf{x}}_{{\rm{BS}}}},{{\mathbf{x}}_{{\rm{IRS}}}}),{{\mathbb{H}}^{[q]}}({{\mathbf{x}}_{{\rm{IRS}}}},{{\mathbf{x}}_{\rm{U}}}),{\mathbb{H}}({{\mathbf{x}}_{{\rm{BS}}}},{{\mathbf{x}}_{\rm{U}}})} \right\}$}
\end{equation*}
\begin{equation*}
\le \sum\limits_{{{\mathbf{x}}_{{\rm{BS}}}} \in {\Psi _{{\rm{BS}}}}({{\mathbf{x}}_{\rm{U}}})} {\sum\limits_{{{\mathbf{x}}_{{\rm{IRS}}}} \in {\Psi _{{\rm{IRS}}}}({{\mathbf{x}}_{\rm{U}}})} {\sum\limits_{q = 1}^Q {\sum\limits_{{{{\mathbf{x'}}}_{{\rm{BS}}}} \in {\Psi _{{\rm{BS}}}}({{\mathbf{x}}_{\rm{U}}})} {\sum\limits_{{{{\mathbf{x'}}}_{{\rm{IRS}}}} \in {\Psi _{{\rm{IRS}}}}({{\mathbf{x}}_{\rm{U}}})} {\sum\limits_{q' = 1}^Q {{{\left( {1 + {{\tilde n}_{{\rm{IRS}}}}({{\mathbf{x}}_{\rm{U}}})} \right)}^2} \sigma _{{\rm{d}}}^2} } } } } } 
\end{equation*}
\begin{equation*}
\left( {\left| {{{\mathbb{H}}^{[q]}}({{\mathbf{x}}_{{\rm{IRS}}}},{{\mathbf{x}}_{\rm{U}}}){{\mathbb{H}}^{[q]}}({{\mathbf{x}}_{{\rm{BS}}}},{{\mathbf{x}}_{{\rm{IRS}}}})} \right| + \left| {{\mathbb{H}}({{\mathbf{x}}_{{\rm{BS}}}},{{\mathbf{x}}_{\rm{U}}})} \right|} \right) \times 
\end{equation*}
\begin{equation*}
\left( {\left| {{{\mathbb{H}}^{[q']}}({{{\mathbf{x'}}}_{{\rm{IRS}}}},{{\mathbf{x}}_{\rm{U}}}){{\mathbb{H}}^{[q']}}({{{\mathbf{x'}}}_{{\rm{BS}}}},{{{\mathbf{x'}}}_{{\rm{IRS}}}})} \right| + \left| {{\mathbb{H}}({{{\mathbf{x'}}}_{{\rm{BS}}}},{{\mathbf{x}}_{\rm{U}}})} \right|} \right).
\end{equation*}

Similar to the proof of Theorem \ref{II-LP}, we have:
\begin{equation*}
E\left\{ {{{\left( {\mathbb{P}\left( {\mathbb{S}\left( {{{\mathbf{x}}_{\rm{U}}}} \right)} \right)} \right)}^p}} \right\} \le 
\end{equation*}
\begin{equation*}
{2^{6p}}{\left( {1 + {{\tilde n}_{{\rm{IRS}}}}({{\mathbf{x}}_{\rm{U}}})} \right)^{4p}} \times {\left( {{{\tilde n}_{{\rm{BS}}}}({{\mathbf{x}}_{\rm{U}}})} \right)^{2p}} \times {Q^{2p}} \times  \sigma _{{\rm{d}}}^{2p}\times
\end{equation*}
\begin{equation*}
 \scalebox{1}[1]{${\left( {\max \left\{ {1,\frac{{{\lambda _{{\text{wave}}}}}}{{4\pi {h_{{\text{BS}}}}}}} \right\} \times \max \left\{ {1,\frac{{{\lambda _{{\text{wave}}}}}}{{4\pi \left| {{h_{{\text{BS}}}} - {h_{{\text{IRS}}}}} \right|}}} \right\} \times \max \left\{ {1,\frac{{{\lambda _{{\text{wave}}}}}}{{4\pi {h_{{\text{IRS}}}}}}} \right\}} \right)^{2p}} \times$}
\end{equation*}
\begin{equation*}
{\left( {{2^{11}}{\pi ^5}{{\left( 3 \right)}^{\frac{3}{2}}}\exp \left\{ {\frac{{35}}{{12}}} \right\}} \right)^p}{\left( {\frac{{18}}{{\exp \{ 3\} }}} \right)^p}{p^{3p}}\times
\end{equation*}
\begin{equation*}
{\left( {8\max \left\{ {1,\frac{1}{{\log \left( {1 + \frac{1}{{{\lambda _{{\rm{IRS}}}}4\pi {{\left( {{R_{{\rm{co}}}}} \right)}^2}}}} \right)}}} \right\}} \right)^{4p}}{p^{4p}} \times {\left( {\frac{{2p}}{{\log \left( {1 + \frac{1}{{{\lambda _{{\rm{BS}}}}4\pi {{\left( {{R_{{\rm{co}}}}} \right)}^2}}}} \right)}}} \right)^{2p}} \le  {L^p} \times {p^{9p}}
\end{equation*}
where:
\begin{equation*}
L = Q^2{2^{6}}\left( {{{ \sigma _{{\rm{d}}}^2 }}} \right) \times 
\end{equation*}
\begin{equation*}
 \scalebox{1}[1]{${\left( {\max \left\{ {1,\frac{{{\lambda _{{\text{wave}}}}}}{{4\pi {h_{{\text{BS}}}}}}} \right\} \times \max \left\{ {1,\frac{{{\lambda _{{\text{wave}}}}}}{{4\pi \left| {{h_{{\text{BS}}}} - {h_{{\text{IRS}}}}} \right|}}} \right\} \times \max \left\{ {1,\frac{{{\lambda _{{\text{wave}}}}}}{{4\pi {h_{{\text{IRS}}}}}}} \right\}} \right)^{2p}} \times$}
\end{equation*}
\begin{equation*}
\left( {{2^{11}}{\pi ^5}{{\left( 3 \right)}^{\frac{3}{2}}}\exp \left\{ {\frac{{35}}{{12}}} \right\}} \right) \times \left( {\frac{{18}}{{\exp \{ 3\} }}} \right) \times {\left( {8\max \left\{ {1,\frac{1}{{\log \left( {1 + \frac{1}{{{\lambda _{{\rm{IRS}}}}4\pi {{\left( {{R_{{\rm{co}}}}} \right)}^2}}}} \right)}}} \right\}} \right)^4} \times 
\end{equation*}
\begin{equation}
{\left( {\frac{2}{{\log \left( {1 + \frac{1}{{{\lambda _{{\rm{BS}}}}4\pi {{\left( {{R_{{\rm{co}}}}} \right)}^2}}}} \right)}}} \right)^2} .
\end{equation}

\section{Proof of Theorem 8}

First, we obtain:
\begin{equation*}
E\left\{ {\mathbb{H}\left( {\tau \mathbb{P}\left( {\mathbb{S}\left( {{{\mathbf{x}}_{\rm{U}}}} \right)} \right)} \right)} \right\}= 
\end{equation*}
\begin{equation*}
E\left\{ {\sum\limits_{p = 0}^\infty  {\frac{{{{\left| {\tau \mathbb{P}\left( {\mathbb{S}\left( {{{\mathbf{x}}_{\rm{U}}}} \right)} \right)} \right|}^p}}}{{\left( {9p} \right)!}}} } \right\} \le \sum\limits_{p = 0}^\infty  {\frac{{\left( {E\left\{ {{{\left| {\mathbb{P}\left( {\mathbb{S}\left( {{{\mathbf{x}}_{\rm{U}}}} \right)} \right)} \right|}^p}} \right\}} \right){\tau ^p}}}{{\left( {9p} \right)!}}}  \le \sum\limits_{p = 0}^\infty  {\frac{{\left( {E\left\{ {{{\left| {\mathbb{P}\left( {\mathbb{S}\left( {{{\mathbf{x}}_{\rm{U}}}} \right)} \right)} \right|}^p}} \right\}} \right){\tau ^p}}}{{\left( {9p} \right)!}}}  \le
\end{equation*}
\begin{equation}
1 + \sum\limits_{p = 1}^\infty  {\frac{{{L^p}{p^{9p}}{\tau ^p}}}{{\left( {9p} \right)!}}}  \le 1 + \sum\limits_{p = 1}^\infty  {\frac{{{L^p}{p^{9p}}{\tau ^p}}}{{{{\left( {\frac{{9p}}{e}} \right)}^{9p}}}}}  = 1 + \sum\limits_{p = 1}^\infty  {{{\left( {\frac{{L{e^9}\tau }}{{{9^9}}}} \right)}^p}}  = \frac{1}{{1 - \frac{{L{e^9}\tau }}{{{9^9}}}}}.
\end{equation}
which is obtained by Fatou's Lemma \cite{Stein} and Stirling's Formula \cite{Maria}. Then, we have:
\begin{equation*}
\Pr \left\{ {\left| {\mathbb{P}\left( {\mathbb{S}\left( {{{\mathbf{x}}_{\rm{U}}}} \right)} \right)} \right| > t} \right\} = \Pr \left\{ {\mathbb{H}\left( {\tau \left| {\mathbb{P}\left( {\mathbb{S}\left( {{{\mathbf{x}}_{\rm{U}}}} \right)} \right)} \right|} \right) > \mathbb{H}\left( {\tau t} \right)} \right\} \le
\end{equation*}
\begin{equation}
\frac{{E\left\{ {\mathbb{H}\left( {\tau \left| {\mathbb{P}\left( {\mathbb{I}\left( {{{\mathbf{x}}_{\rm{U}}}} \right)} \right)} \right|} \right)} \right\}}}{{\mathbb{H}\left( {\tau t} \right)}} \le \frac{1}{{\left( {1 - \frac{{L{e^{9}}\tau }}{{{{9}^{9}}}}} \right)\mathbb{H}\left( {\tau t} \right)}},
\end{equation}
which is obtained by Markov inequality.

\section{Proof of Theorem 9}

By Theorem 8, we obtain:
\begin{equation*}
\Pr \left\{ {\mathbb{C}\left( {{{\mathbf{x}}_{\rm{U}}}} \right) > \alpha } \right\} = \Pr \left\{ {\ln \left( {1 + \frac{{\mathbb{P}\left( {\mathbb{S}\left( {{{\mathbf{x}}_{\rm{U}}}} \right)} \right)}}{{\mathbb{P}\left( {\mathbb{I}\left( {{{\mathbf{x}}_{\rm{U}}}} \right)} \right) + {N_0}}}} \right) > \alpha } \right\} = 
\end{equation*}
\begin{equation*}
\Pr \left\{ {1 + \frac{{\mathbb{P}\left( {\mathbb{S}\left( {{{\mathbf{x}}_{\rm{U}}}} \right)} \right)}}{{\mathbb{P}\left( {\mathbb{I}\left( {{{\mathbf{x}}_{\rm{U}}}} \right)} \right) + {N_0}}} > {e^\alpha }} \right\} = \Pr \left\{ {\mathbb{P}\left( {\mathbb{S}\left( {{{\mathbf{x}}_{\rm{U}}}} \right)} \right) + \left( {1 - {e^\alpha }} \right)\mathbb{P}\left( {\mathbb{I}\left( {{{\mathbf{x}}_{\rm{U}}}} \right)} \right) > {N_0}\left( {{e^\alpha } - 1} \right)} \right\} \le
\end{equation*}
\begin{equation*}
\Pr \left\{ {\mathbb{P}\left( {\mathbb{S}\left( {{{\mathbf{x}}_{\rm{U}}}} \right)} \right) > {N_0}\left( {{e^\alpha } - 1} \right)} \right\} \le \frac{1}{{\left( {1 - \frac{{L{e^9}\tau }}{{{9^9}}}} \right)\mathbb{H}\left( {\tau {N_0}\left( {{e^\alpha } - 1} \right)} \right)}}.
\end{equation*}
where
\begin{equation*}
0 < \tau  < \frac{{{{9}^{9}}}}{{\left( {{ L} } \right){e^{9}}}}.
\end{equation*}

On the other hand, we have:
\begin{equation*}
\Pr \left\{ {\mathbb{C}\left( {{{\mathbf{x}}_{\rm{U}}}} \right) > \alpha } \right\} = \Pr \left\{ {\ln \left( {1 + \frac{{\mathbb{P}\left( {\mathbb{S}\left( {{{\mathbf{x}}_{\rm{U}}}} \right)} \right)}}{{\mathbb{P}\left( {\mathbb{I}\left( {{{\mathbf{x}}_{\rm{U}}}} \right)} \right) + {N_0}}}} \right) > \alpha } \right\} = 
\end{equation*}
\begin{equation*}
\Pr \left\{ {1 + \frac{{\mathbb{P}\left( {\mathbb{S}\left( {{{\mathbf{x}}_{\rm{U}}}} \right)} \right)}}{{\mathbb{P}\left( {\mathbb{I}\left( {{{\mathbf{x}}_{\rm{U}}}} \right)} \right) + {N_0}}} > {e^\alpha }} \right\} = \Pr \left\{ {\mathbb{P}\left( {\mathbb{S}\left( {{{\mathbf{x}}_{\rm{U}}}} \right)} \right) + \left( {1 - {e^\alpha }} \right)\mathbb{P}\left( {\mathbb{I}\left( {{{\mathbf{x}}_{\rm{U}}}} \right)} \right) > {N_0}\left( {{e^\alpha } - 1} \right)} \right\} \le
\end{equation*}
\begin{equation*}
\Pr \left\{ {\mathbb{P}\left( {\mathbb{S}\left( {{{\mathbf{x}}_{\rm{U}}}} \right)} \right) > {N_0}\left( {{e^\alpha } - 1} \right)} \right\} \le \frac{{E\left\{ {\mathbb{P}\left( {\mathbb{S}\left( {{{\mathbf{x}}_{\rm{U}}}} \right)} \right)} \right\}}}{{{N_0}\left( {{e^\alpha } - 1} \right)}} \le \frac{{P_{\max }^\mathbb{S}}}{{{N_0}\left( {{e^\alpha } - 1} \right)}}.
\end{equation*}

\end{appendices}

\end{document}